\documentclass[modern]{aastex62}
\usepackage{newtxtext,newtxmath} % this is causing the etoolbox warning
\usepackage[USenglish]{babel}
\usepackage[utf8]{inputenc}
\usepackage[T1]{fontenc}
\usepackage{comment}
\usepackage{enumitem}
\usepackage[numberedsection]{glossaries}
\usepackage{todonotes}
\usepackage{collect}
\usepackage{wrapfig}
\usepackage{bclogo}
\usepackage{environ}
\usepackage{lipsum}
\usepackage[titles]{tocloft}
\usepackage{makecell}

\setlist{noitemsep}

\graphicspath{{./}{./figures/}{./images}{./tabs}}

\newcommand{\Contact}[1]{}
\newcommand{\Contributors}[2]{\noindent {\bfseries Contributors:} #1 } % ignoring date for now in second param
\newcommand{\mail}[1]{\href{mailto:#1}{#1}}

\newcommand{\Comment}[3]{}
\newcommand{\WOM}[1]{\Comment{red}{WOM}{#1}} % Wil comments
\newcommand{\cleanedup}[1]{\Comment{magenta}{This section has been cleaned up by}{#1}}

\providecommand{\secref}[1]{\hyperref[#1]{Section~\ref{#1}}}
\providecommand{\appref}[1]{\hyperref[#1]{Appendix~\ref{#1}}}
\providecommand{\tabref}[1]{\hyperref[#1]{Table~\ref{#1}}}
\providecommand{\figref}[1]{\hyperref[#1]{Figure~\ref{#1}}}
\providecommand{\eqnref}[1]{\hyperref[#1]{Eq.~\ref{#1}}}
\providecommand{\recref}[1]{\hyperref[#1]{REC-\ref{#1}}}

\makeatletter
\renewcommand{\section}{\newpage{} \@startsection
{section}
{1}
{0mm}
{-3.5ex \@plus -1ex \@minus -.2ex}
{0.2ex \@plus.2ex}
{\normalfont\bfseries}}
\makeatother
\newacronym{1D} {1D} {One-dimensional}
\newacronym{2D} {2D} {Two-dimensional}
\newacronym{2MASS} {2MASS} {Two-Micron All Sky Survey}
\newacronym{3D} {3D} {Three-dimensional}
\newacronym{AAS} {AAS} {American Astronomical Society}
\newacronym{AAVSO} {AAVSO} {American Association of Variable Star Observers}
\newacronym{AC} {AC} {Alternating Current}
\newacronym{ACT} {ACT} {Atacama Cosmology Telescope, for \gls{CMB} observations}
\newacronym{ADAM} {ADAM} {\gls{Asteroid Discovery, Analysis, and Mapping}}
\newacronym{ADQL} {ADQL} {Astronomical Data Query Language}
\newacronym{AGN} {AGN} {Active Galactic Nuclei}
\newacronym{AI} {AI} {Artificial Intelligence}
\newacronym{AIC} {AIC} {\gls{Akaike Information Criterion}}
\newacronym{ALMA} {ALMA} {Atacama Large Millimeter Array (\gls{ESO})}
\newacronym{ALeRCE} {ALeRCE} {\gls{Automatic Learning for the Rapid Classification of Events}}
\newacronym{AMPEL} {AMPEL} {Alert Management, Photometry, and Evaluation of Light curves}
\newacronym{ANTARES} {ANTARES} {\gls{Arizona-NOIRLab Temporal Analysis and Response to Events System}}
\newacronym{API} {API} {Application Programming Interface}
\newacronym{ARAS} {ARAS} {Astronomical Ring for Access to Spectroscopy}
\newacronym{ASAP} {ASAP} {As Soon As Possible}
\newacronym{ASAS-SN} {ASAS-SN} {All-Sky Automated Survey for Supernovae}
\newacronym{AST} {AST} {NSF Division of Astronomical Sciences}
\newacronym{ATLAS} {ATLAS} {The Asteroid Terrestrial-impact Last }
\newacronym{AU} {AU} {deprecated acronym for astronomical unit; use \gls{au} instead}
\newacronym{AURA} {AURA} {\gls{Association of Universities for Research in Astronomy}}
\newacronym{AXS} {AXS} {Astronomy eXtensions for Spark}
\newglossaryentry{Akaike Information Criterion} {name={Akaike Information Criterion}, description={an estimator of prediction error and thereby relative quality of statistical models for a given set of data}}
\newglossaryentry{Alert} {name={Alert}, description={A packet of information for each source detected with signal-to-noise ratio > 5 in a difference image by Alert Production, containing measurement and characterization parameters based on the past 12 months of LSST observations plus small cutouts of the single-visit, template, and difference images, distributed via the internet}}
\newglossaryentry{Alert Production} {name={Alert Production}, description={Executing on the Prompt Processing system, the Alert Production payload processes and calibrates incoming images, performs Difference Image Analysis to identify DIASources and DIAObjects, and then packages the resulting alerts for distribution.}}
\newglossaryentry{Apache Parquet} {name={Apache Parquet}, description={A columnar storage data persistence format maintained by the Apache project}}
\newglossaryentry{Archive} {name={Archive}, description={The repository for documents required by the NSF to be kept. These include documents related to design and development, construction, integration, test, and operations of the LSST observatory system. The archive is maintained using the enterprise content management system DocuShare, which is accessible through a link on the project website www.project.lsst.org}}
\newglossaryentry{Archive Center} {name={Archive Center}, description={Part of the LSST Data Management System, the LSST archive center is a data center at NCSA that hosts the LSST Archive, which includes released science data and metadata, observatory and engineering data, and supporting software such as the LSST Software Stack}}
\newglossaryentry{Arizona-NOIRLab Temporal Analysis and Response to Events System} {name={Arizona-NOIRLab Temporal Analysis and Response to Events System}, description={ANTARES is a real-time astronomy system under development at NOIRLab. \url{https://antares.noirlab.edu}}}
\newglossaryentry{Association Pipeline} {name={Association Pipeline}, description={An application that matches detected Sources or DIASources or generated Objects to an existing catalog of Objects, producing a (possibly many-to-many) set of associations and a list of unassociated inputs. Association Pipelines are used in Alert Production after DIASource generation and in the final stages of Data Release processing to ensure continuity of Object identifiers}}
\newglossaryentry{Association of Universities for Research in Astronomy} {name={Association of Universities for Research in Astronomy}, description={ consortium of US institutions and international affiliates that operates world-class astronomical observatories, AURA is the legal entity responsible for managing what it calls independent operating Centers, including LSST, under respective cooperative agreements with the National Science Foundation. AURA assumes fiducial responsibility for the funds provided through those cooperative agreements. AURA also is the legal owner of the AURA Observatory properties in Chile}}
\newglossaryentry{Asteroid Discovery, Analysis, and Mapping} {name={Asteroid Discovery, Analysis, and Mapping}, description={a cloud-based astrodynamics platform in development by the Asteroid Institute, a program of the B612 Foundation }}
\newglossaryentry{Automatic Learning for the Rapid Classification of Events} {name={Automatic Learning for the Rapid Classification of Events}, description={The ALeRCE broker is a Chilean-led broker which is processing the alert stream from the ZTF and a Community Broker for the Vera C. Rubin Observatory and its LSST, as well as other large etendue survey telescopes. \url{http://alerce.science/}}}
\newacronym{Avro} {Avro} {is a row-oriented remote procedure call and data serialization framework developed within Apache's Hadoop project}
\newacronym{B} {B} {Byte (8 bit)}
\newacronym{BCE} {BCE} {Before Common Era}
\newglossaryentry{BEAMS} {name={BEAMS}, description={Bayesian Estimation Applied to Multiple Species (software for classification of light curves based on photometry)
}}
\newacronym{BH} {BH} {Black Hole}
\newacronym{BHB} {BHB} {Blue Horizontal Branch}
\newacronym{BHNS} {BHNS} {Black hole-neutron star}
\newacronym{BNS} {BNS} {Binary Neutron Star}
\newacronym{BOSS} {BOSS} {Baryon Oscillation Spectroscopic Survey}
\newglossaryentry{BlackGEM} {name={BlackGEM}, description={is a wide-field array of optical telescopes to be located at ESO’s La Silla Observatory in Chile’s Atacama desert. }}
\newglossaryentry{Blazhko} {name={Blazhko}, description={the phenomenon of amplitude or phase modulation. Associated with some RRL}}
\newglossaryentry{Broker} {name={Broker}, description={Software which receives and redistributes Alerts, and may also perform processing such as filtering for certain characteristics, cross-matching with non-LSST catalogs, and/or light-curve classification, in order to identify and prioritize targets for follow-up and/or make scientific analyses. }}
\newglossaryentry{Butler} {name={Butler}, description={A middleware component for persisting and retrieving image datasets (raw or processed), calibration reference data, and catalogs}}
\newacronym{CA} {CA} {Control (or Cost) Account}
\newacronym{CADC} {CADC} {Canadian Astronomy Data Centre}
\newglossaryentry{CARMA} {name={CARMA}, description={Continuous time autoregressive moving average process, standard way to describe optical AGN variability}}
\newacronym{CARMENES} {CARMENES} {Calar Alto high-Resolution search for M dwarfs with Exoearths with Near-infrared and optical Echelle Spectrographs}
\newacronym{CAS} {CAS} {\gls{Central Administrative Services}}
\newacronym{CASA} {CASA} {Common Astronomy Software Applications (for \gls{ALMA})}
\newacronym{CASLEO} {CASLEO} {Complejo Astron\'omico El Leoncito}
\newacronym{CCD} {CCD} {\gls{Charge-Coupled Device}}
\newacronym{CCL} {CCL} {Core Cosmology Library, \url{https://github.com/LSSTDESC/CCL}}
\newacronym{CDF} {CDF} {Cumulative Distribution Function}
\newacronym{CDMX} {CDMX} {Ciudad de Mexico}
\newacronym{CDS} {CDS} {Centre de Donnes astronomiques de Strasbourg}
\newacronym{CHIME} {CHIME} {Canadian Hydrogen Intensity Mapping Experiment}
\newacronym{CI} {CI} {\gls{Cyber Infrastructure}}
\newacronym{CIGALE} {CIGALE} {a python Code Investigating GALaxy Emission}
\newglossaryentry{CMASS} {name={CMASS}, description={constant mass, a spectroscopic galaxy sample as part of the BOSS survey}}
\newacronym{CMB} {CMB} {Cosmic Microwave Background}
\newacronym{CMB-S4} {CMB-S4} {Cosmic Microwave Background Stage 4}
\newacronym{CMNN} {CMNN} {Color-Matched Nearest Neighbors}
\newacronym{CNN} {CNN} {Convolutional Neural Network}
\newacronym{CNP} {CNP} {Conditional Neural Processes}
\newacronym{CNRS} {CNRS} {Centre national de la recherche scientifique}
\newacronym{CO} {CO} {Carbon Monoxide}
\newacronym{CPU} {CPU} {Central Processing Unit}
\newacronym{CRTS} {CRTS} {Catalina Real-Time Transient Survey}
\newacronym{CSA} {CSA} {Cooperative Support Agreement}
\newacronym{CSM} {CSM} {Circum-Stellar Material}
\newacronym{CSQ} {CSQ} {Changing state quasar or \gls{AGN}}
\newacronym{CTTS} {CTTS} {Classical T Tauri stars}
\newglossaryentry{Camera} {name={Camera}, description={The LSST subsystem responsible for the 3.2-gigapixel LSST camera, which will take more than 800 panoramic images of the sky every night. SLAC leads a consortium of Department of Energy laboratories to design and build the camera sensors, optics, electronics, cryostat, filters and filter exchange mechanism, and camera control system}}
\newglossaryentry{Center} {name={Center}, description={An entity managed by AURA that is responsible for execution of a federally funded project}}
\newglossaryentry{Central Administrative Services} {name={Central Administrative Services}, description={AURA corporate division responsible for providing accounting, procurement, and business IT support services to AURA centers}}
\newglossaryentry{Charge-Coupled Device} {name={Charge-Coupled Device}, description={a particular kind of solid-state sensor for detecting optical-band photons. It is composed of a 2-D array of pixels, and one or more read-out amplifiers}}
\newglossaryentry{Citizen Science} {name={Citizen Science}, description={the collection and analysis of data relating to the natural world by members of the general public, typically as part of a collaborative project with professional scientists.}}
\newacronym{CoRoT} {CoRoT} {Convection, Rotation et Transits plan\'{e}taires}
\newacronym{ComCam} {ComCam} {The commissioning \gls{camera} is a single-raft, 9-CCD \gls{camera} that will be installed in LSST during commissioning, before the final \gls{camera} is ready.}
\newglossaryentry{Commissioning} {name={Commissioning}, description={A two-year phase at the end of the Construction project during which a technical team a) integrates the various technical components of the three subsystems; b) shows their compliance with ICDs and system-level requirements as detailed in the LSST Observatory System Specifications document (OSS, LSE-30); and c) performs science verification to show compliance with the survey performance specifications as detailed in the LSST Science Requirements Document (SRD, LPM-17)}}
\newglossaryentry{Construction} {name={Construction}, description={The period during which LSST observatory facilities, components, hardware, and software are built, tested, integrated, and commissioned. Construction follows design and development and precedes operations. The LSST construction phase is funded through the NSF MREFC account}}
\newglossaryentry{Cyber Infrastructure} {name={Cyber Infrastructure}, description={Sometimes denoted CI, A term first used by the US NSF, and it typically is used to refer to information technology systems that provide particularly powerful and advanced capabilities.}}
\newacronym{DAC} {DAC} {\gls{Data Access Center}}
\newacronym{DB} {DB} {DataBase}
\newacronym{DC2} {DC2} {Data Challenge 2 (\gls{DESC})}
\newacronym{DCR} {DCR} {\gls{Differential Chromatic Refraction}}
\newacronym{DDF} {DDF} {Deep Drilling Field}
\newacronym{DE} {DE} {dark energy}
\newacronym{DEC} {DEC} {Declination}
\newacronym{DECAT} {DECAT} {DECam Alliance for Transients}
\newacronym{DECaLS} {DECaLS} {The Dark Energy \gls{Camera} Legacy Survey}
\newacronym{DECam} {DECam} {Dark Energy \gls{Camera}}
\newacronym{DEEP} {DEEP} {Deep Extragalactic Evolutionary Probe}
\newacronym{DELVE} {DELVE} {DECam Local Volume Exploration Survey}
\newacronym{DES} {DES} {Dark Energy Survey}
\newacronym{DESC} {DESC} {Dark Energy \gls{Science Collaboration}}
\newacronym{DESI} {DESI} {Dark Energy Spectroscopic Instrument}
\newacronym{DHO} {DHO} {damped harmonic oscillator}
\newacronym{DIA} {DIA} {\gls{Difference Image Analysis}}
\newglossaryentry{DIAObject} {name={DIAObject}, description={A DIAObject is the association of DIASources, by coordinate, that have been detected with signal-to-noise ratio greater than 5 in at least one difference image. It is distinguished from a regular Object in that its brightness varies in time, and from a SSObject in that it is stationary (non-moving)}}
\newglossaryentry{DIASource} {name={DIASource}, description={A DIASource is a detection with signal-to-noise ratio greater than 5 in a difference image}}
\newacronym{DM} {DM} {\gls{Data Management}}
\newacronym{DMS} {DMS} {\gls{Data Management Subsystem}}
\newacronym{DMTN} {DMTN} {DM Technical Note}
\newacronym{DOE} {DOE} {\gls{Department of Energy}}
\newacronym{DP0} {DP0} {Data Preview 0}
\newacronym{DP2} {DP2} {Data Preview 2}
\newacronym{DR} {DR} {\gls{Data Release}}
\newacronym{DR1} {DR1} {Data \gls{Release} 1}
\newacronym{DR2} {DR2} {Data \gls{Release} 2}
\newacronym{DR3} {DR3} {Data \gls{Release} 3}
\newacronym{DRP} {DRP} {\gls{Data Release Production}}
\newacronym{DRW} {DRW} {damped random walk}
\newacronym{DS9} {DS9} {Deep Space 9 (specific astronomical data visualisation application; SAOImage)}
\newglossaryentry{Data Access Center} {name={Data Access Center}, description={Part of the LSST Data Management System, the US and Chilean DACs will provide authorized access to the released LSST data products, software such as the Science Platform, and computational resources for data analysis. The US DAC also includes a service for distributing bulk data on daily and annual (Data Release) timescales to partner institutions, collaborations, and LSST Education and Public Outreach (EPO). }}
\newglossaryentry{Data Management} {name={Data Management}, description={The LSST Subsystem responsible for the Data Management System (DMS), which will capture, store, catalog, and serve the LSST dataset to the scientific community and public. The DM team is responsible for the DMS architecture, applications, middleware, infrastructure, algorithms, and Observatory Network Design. DM is a distributed team working at LSST and partner institutions, with the DM Subsystem Manager located at LSST headquarters in Tucson}}
\newglossaryentry{Data Management Subsystem} {name={Data Management Subsystem}, description={The Data Management Subsystem is one of the four subsystems which constitute the LSST Construction Project. The Data Management Subsystem is responsible for developing and delivering the LSST Data Management System to the LSST Operations Project}}
\newglossaryentry{Data Management System} {name={Data Management System}, description={The computing infrastructure, middleware, and applications that process, store, and enable information extraction from the LSST dataset; the DMS will process peta-scale data volume, convert raw images into a faithful representation of the universe, and archive the results in a useful form. The infrastructure layer consists of the computing, storage, networking hardware, and system software. The middleware layer handles distributed processing, data access, user interface, and system operations services. The applications layer includes the data pipelines and the science data archives' products and services}}
\newglossaryentry{Data Release} {name={Data Release}, description={The approximately annual reprocessing of all LSST data, and the installation of the resulting data products in the LSST Data Access Centers, which marks the start of the two-year proprietary period}}
\newglossaryentry{Data Release Production} {name={Data Release Production}, description={An episode of (re)processing all of the accumulated LSST images, during which all output DR data products are generated. These episodes are planned to occur annually during the LSST survey, and the processing will be executed at the Archive Center. This includes Difference Imaging Analysis, generating deep Coadd Images, Source detection and association, creating Object and Solar System Object catalogs, and related metadata}}
\newglossaryentry{Department of Energy} {name={Department of Energy}, description={cabinet department of the United States federal government; the DOE has assumed technical and financial responsibility for providing the LSST camera. The DOE's responsibilities are executed by a collaboration led by SLAC National Accelerator Laboratory}}
\newglossaryentry{Difference Image} {name={Difference Image}, description={Refers to the result formed from the pixel-by-pixel difference of two images of the sky, after warping to the same pixel grid, scaling to the same photometric response, matching to the same PSF shape, and applying a correction for Differential Chromatic Refraction. The pixels in a difference thus formed should be zero (apart from noise) except for sources that are new, or have changed in brightness or position. In the LSST context, the difference is generally taken between a visit image and template. }}
\newglossaryentry{Difference Image Analysis} {name={Difference Image Analysis}, description={The detection and characterization of sources in the Difference Image that are above a configurable threshold, done as part of Alert Generation Pipeline}}
\newglossaryentry{Differential Chromatic Refraction} {name={Differential Chromatic Refraction}, description={The refraction of incident light by Earth's atmosphere causes the apparent position of objects to be shifted, and the size of this shift depends on both the wavelength of the source and its airmass at the time of observation. DCR corrections are done as a part of DIA}}
\newglossaryentry{DocuShare} {name={DocuShare}, description={The trade name for the enterprise management software used by LSST to archive and manage documents}}
\newglossaryentry{Document} {name={Document}, description={Any object (in any application supported by DocuShare or design archives such as PDMWorks or GIT) that supports project management or records milestones and deliverables of the LSST Project}}
\newacronym{EC2} {EC2} {Amazon Elastic Compute Cloud}
\newacronym{EDR3} {EDR3} {Early Data \gls{Release} 3}
\newacronym{ELAsTiCC} {ELAsTiCC} {Extended \gls{LSST} Astronomical Time Series Classification Challenge}
\newacronym{ELG} {ELG} {Emission-Line Galaxies}
\newacronym{ELM} {ELM} {Extremely Low Mass(Survey)}
\newacronym{ENE} {ENE} {East North East}
\newacronym{EPO} {EPO} {\gls{Education and Public Outreach}}
\newacronym{ESA} {ESA} {European Space Agency}
\newacronym{ESO} {ESO} {European Southern Observatory}
\newglossaryentry{Education and Public Outreach} {name={Education and Public Outreach}, description={The LSST subsystem responsible for the cyberinfrastructure, user interfaces, and outreach programs necessary to connect educators, planetaria, citizen scientists, amateur astronomers, and the general public to the transformative LSST dataset}}
\newacronym{FBOT} {FBOT} {Fast blue optical \gls{transient}}
\newacronym{FBOTs} {FBOTs} {Fast blue optical transients}
\newacronym{FELTs} {FELTs} {Fast-Evolving Luminous Transients}
\newacronym{FITS} {FITS} {\gls{Flexible Image Transport System}}
\newacronym{FOV} {FOV} {field of view}
\newacronym{FWHM} {FWHM} {Full Width at Half-Maximum}
\newglossaryentry{Filter} {name={Filter}, description={A filter in astronomy is an optical element used to restrict the passband of light reaching the focal plane, it transmits a selected range of wavelengths. Filters elements are often named after standard photometric passbands, such as those used in the SDSS survey: u, g, r, i, z}}
\newglossaryentry{Fink} {name={Fink}, description={Fink is a community driven project, open to anyone, that processes time-domains alert streams and connects them with follow-up facilities and science teams. \url{https://fink-broker.org}}}
\newglossaryentry{Firefly} {name={Firefly}, description={A framework of software components written by IPAC for building web-based user interfaces to astronomical archives, through which data may be searched and retrieved, and viewed as FITS images, catalogs, and/or plots. Firefly tools will be integrated into the Science Platform}}
\newglossaryentry{Flexible Image Transport System} {name={Flexible Image Transport System}, description={an international standard in astronomy for storing images, tables, and metadata in disk files. See the IAU FITS Standard for details}}
\newacronym{FoV} {FoV} {Field of View (also denoted \gls{FOV})}
\newglossaryentry{ForcedSource} {name={ForcedSource}, description={DRP table resulting from forced photometry}}
\newacronym{GALAH} {GALAH} {GALactic Archaeology with \gls{HERMES}}
\newacronym{GAMA} {GAMA} {Galaxy And Mass Assembly (survey)}
\newacronym{GB} {GB} {Gigabyte}
\newacronym{GLADE} {GLADE} {Galaxy List for the Advanced Detector Era}
\newacronym{GPU} {GPU} {Graphics Processing Unit}
\newacronym{GR} {GR} {General Relativity}
\newacronym{GRB} {GRB} {Gamma-Ray Burst}
\newacronym{GSE} {GSE} {Gaia Sausage-Enceladus}
\newacronym{GUI} {GUI} {Graphical User Interface}
\newacronym{GW} {GW} {Gravitational Wave}
\newacronym{GZ} {GZ} {Galaxy Zoo}
\newglossaryentry{Gaia} {name={Gaia}, description={a space observatory of the European Space Agency, launched in 2013 and expected to operate until 2025. The spacecraft is designed for astrometry: measuring the positions, distances and motions of stars with unprecedented precision}}
\newacronym{Gb} {Gb} {Gigabit}
\newacronym{HB} {HB} {Horizontal Branch}
\newacronym{HEALPix} {HEALPix} {Hierarchical Equal-Area iso-Latitude Pixelisation}
\newacronym{HELP} {HELP} { Herschel Extragalactic Legacy Project}
\newacronym{HERMES} {HERMES} {a high-resolution fibre-fed spectrograph for the 1.2m Mercator telescope}
\newacronym{HI} {HI} {Hydrogen iodide}
\newacronym{HITS} {HITS} {High Cadence Transient Survey}
\newacronym{HPC} {HPC} {High Performance Computing}
\newacronym{HR} {HR} {Human Resources}
\newacronym{HSC} {HSC} {Hyper Suprime-Cam}
\newacronym{HST} {HST} {Hubble Space Telescope}
\newacronym{HTTP} {HTTP} {HyperText Transfer Protocol}
\newglossaryentry{Handle} {name={Handle}, description={The unique identifier assigned to a document uploaded to DocuShare}}
\newacronym{IA} {IA} {intrinsic alignments of galaxy shapes}
\newacronym{IAU} {IAU} {International Astronomical Union}
\newacronym{IDAC} {IDAC} {\gls{Independent Data Access Center}}
\newacronym{IMBH} {IMBH} {Intermediate Mass Black Hole}
\newacronym{IMF} {IMF} {Initial Mass Function}
\newacronym{IN2P3} {IN2P3} {Institut National de Physique Nucléaire et de Physique des Particules}
\newacronym{INAF} {INAF} {Istituto Nazionale di Astrofisica}
\newacronym{IPAC} {IPAC} {No longer an acronym; science and data center at Caltech}
\newacronym{IPHAS} {IPHAS} {INT Photometric H$\alpha$ survey}
\newacronym{IR} {IR} {infrared}
\newacronym{IRAF} {IRAF} {\gls{Image Reduction and Analysis Facility}}
\newacronym{ISIS} {ISIS} {Interactive Spectral Interpretation System, \url{https://space.mit.edu/cxc/isis/}}
\newacronym{ISO} {ISO} {Interstellar \gls{Object}}
\newacronym{IT} {IT} {Information Technology}
\newacronym{IVOA} {IVOA} {International Virtual-Observatory Alliance}
\newglossaryentry{Image Reduction and Analysis Facility} {name={Image Reduction and Analysis Facility}, description={  a collection of software written at the National Optical Astronomy Observatory (now NOIRLab) geared towards the reduction of astronomical images in pixel array form.}}
\newglossaryentry{Independent Data Access Center} {name={Independent Data Access Center}, description={Externally supported and administered versions of the DAC to serve the full, or a limited subset of, the LSST data products and/or software to authorized users. }}
\newglossaryentry{J2000} {name={J2000}, description={Julian Date referring to the instant of 12 noon (midday) on January 1, 2000. IAU standard equinox.}}
\newacronym{JD} {JD} {\gls{Julian Date}}
\newacronym{JPL} {JPL} {Jet Propulsion Laboratory (\gls{DE} ephemerides)}
\newacronym{JWST} {JWST} {James Webb Space Telescope (formerly known as NGST)}
\newglossaryentry{Julian Date} {name={Julian Date}, description={The Julian Date (JD) of any instant is the Julian day number for the preceding noon (UTC), plus the fraction of the day elapsed since that instant. The Julian day number is a running sequence of integral days, starting at noon, since the beginning of the Julian Period; JD 0.0 corresponds to noon on 1 January 4713 BCE. Various Julian Date converters are available on the Web. For example, 18h 00m 00.0s UT on 2014-July-01 (near the start of LSST construction) corresponds to JD 2456840.25}}
\newglossaryentry{K2} {name={K2}, description={ NASA mission that provides precise photometric data from numerous target fields in the ecliptic.}}
\newacronym{KBMOD} {KBMOD} {Kernel-Based Moving \gls{Object} Detection, \url{https://github.com/dirac-institute/kbmod}}
\newacronym{KBO} {KBO} {Kuiper-Belt \gls{Object}}
\newacronym{KiDS} {KiDS} {Kilo-Degree Survey}
\newacronym{LAMOST} {LAMOST} {Large Sky Area Multi-Object Fibre Spectroscopic Telescope, also known as the Guo Shoujing Telescope}
\newacronym{LC} {LC} {Light Curve}
\newacronym{LCDM} {LCDM} { $\Lambda$ Cold Dark Matter; cosmological model}
\newacronym{LCO} {LCO} {Las Cumbres Observatories}
\newacronym{LDM} {LDM} {LSST Data Management (Document \gls{Handle})}
\newacronym{LEs} {LEs} {Light Echoes}
\newacronym{LF} {LF} {luminosity function}
\newacronym{LG} {LG} {Local Group}
\newacronym{LIGO} {LIGO} {Laser Interferometer Gravitational-Wave Observatory}
\newacronym{LINCC} {LINCC} {LSST Interdisciplinary Network for Collaboration and Computing}
\newacronym{LISA} {LISA} {Laser Interferometer Space Antenna}
\newacronym{LMC} {LMC} {Large Magellanic Cloud}
\newacronym{LPM} {LPM} {LSST Project Management (Document \gls{Handle})}
\newacronym{LRG} {LRG} {Luminous Red Galaxies}
\newacronym{LSB} {LSB} {Low Surface Brightness}
\newacronym{LSE} {LSE} {LSST \gls{Systems Engineering} (Document Handle)}
\newacronym{LSS} {LSS} {Large Scale Structure}
\newacronym{LSST} {LSST} {Legacy Survey of Space and Time (formerly Large Synoptic Survey Telescope)}
\newglossaryentry{LSST Corporation} {name={LSST Corporation}, description={An Arizona 501(c)3 not-for-profit corporation formed in 2003 for the purpose of designing, constructing, and operating the LSST System. During design and development, the Corporation stewarded private funding used for such essential contributions as early site preparation, mirror construction, and early data management system development. During construction, LSSTC will secure private operations funding from international affiliates and play a key role in preparing the scientific community to use the LSST dataset}}
\newglossaryentry{LSST Project Office} {name={LSST Project Office}, description={Official name of the stand-alone AURA operating center responsible for execution of the LSST construction project under the NSF MREFC account}}
\newacronym{LSSTC} {LSSTC} {\gls{LSST Corporation}}
\newacronym{LSSTPO} {LSSTPO} {\gls{LSST Project Office}}
\newacronym{LV} {LV} {Local Volume}
\newglossaryentry{Lasair} {name={Lasair}, description={a broker for astronomers studying transient and variable astrophysical sources. It is being developed as a collaboration between the University of Edinburgh and Queen's University, Belfast to build a broker service for alerts generated by the LSST at the Vera Rubin Observatory. \url{https://lasair.roe.ac.uk/}}}
\newacronym{M3} {M3} {tertiary mirror}
\newglossaryentry{M31} {name={M31}, description={also known as the Andromeda galaxy, can be seen with the naked eye in the constellation of Andromeda.}}
\newacronym{MANOS} {MANOS} {Mission Accessible Near-Earth Objects Survey}
\newacronym{MCMC} {MCMC} {Monte Carlo Markov Chain}
\newacronym{ML} {ML} {Machine Learning}
\newacronym{MLP} {MLP} {Multi-Layer Perceptron}
\newacronym{MMA} {MMA} {Multi Messenger Astronomy}
\newacronym{MNRAS} {MNRAS} {Monthly Notices of the Royal Astronomical Society}
\newacronym{MOA} {MOA} {Memo Of Agreement}
\newacronym{MOC} {MOC} {Multi Ordered Catalogue}
\newacronym{MPC} {MPC} {Minor Planet \gls{Center}}
\newacronym{MREFC} {MREFC} {\gls{Major Research Equipment and Facility Construction}}
\newacronym{MW} {MW} {Milky Way}
\newglossaryentry{Major Research Equipment and Facility Construction} {name={Major Research Equipment and Facility Construction}, description={the NSF account through which large facilities construction projects such as LSST are funded}}
\newglossaryentry{Mapper} {name={Mapper}, description={A piece of software that abstracts persisting and unpersisting data; specifically, it knows how to navigate a data repository to locate data that match selection criteria that are relevant for data obtained with a particular camera. Used by the Butler}}
\newacronym{NASA} {NASA} {National Aeronautics and Space Administration}
\newacronym{NCSA} {NCSA} {National \gls{Center} for Supercomputing Applications}
\newacronym{NEO} {NEO} {Near-Earth \gls{Object}}
\newacronym{NGC} {NGC} {\gls{New General Catalogue}}
\newacronym{NOIRLab} {NOIRLab} {NSF's National Optical-Infrared Astronomy Research Laboratory; \url{https://nationalastro.org}}
\newacronym{NRAO} {NRAO} {National Radio Astronomy Observatory}
\newacronym{NSF} {NSF} {\gls{National Science Foundation}}
\newglossaryentry{National Science Foundation} {name={National Science Foundation}, description={primary federal agency supporting research in all fields of fundamental science and engineering; NSF selects and funds projects through competitive, merit-based review}}
\newglossaryentry{New General Catalogue} {name={New General Catalogue}, description={an astronomical catalogue of deep-sky objects compiled by John Louis Emil Dreyer in 1888}}
\newacronym{OSS} {OSS} {Observatory System Specifications; \gls{LSE}-30}
\newglossaryentry{Object} {name={Object}, description={In LSST nomenclature this refers to an astronomical object, such as a star, galaxy, or other physical entity. E.g., comets, asteroids are also Objects but typically called a Moving Object or a Solar System Object (SSObject). One of the DRP data products is a table of Objects detected by LSST which can be static, or change brightness or position with time}}
\newglossaryentry{Operations} {name={Operations}, description={The 10-year period following construction and commissioning during which the LSST Observatory conducts its survey}}
\newglossaryentry{Opportunity} {name={Opportunity}, description={The degree of exposure to an event that might happen to the benefit of a program, project, or other activity. It is described by a combination of the probability that the opportunity event will occur and the consequence of the extent of gain from the occurrence, or impact. There are two levels of opportunities. At the macro level, a project itself is the manifestation of the pursuit of an opportunity. At the element level, tactical opportunities exist, whereby certain events, if realized, provide a cost or schedule savings to the project or increase technical performance}}
\newacronym{PASP} {PASP} {Publications of the Astronomical Society of the Pacific}
\newacronym{PB} {PB} {PetaByte}
\newacronym{PCA} {PCA} {Principal Component Analysis}
\newacronym{PDF} {PDF} {Probability Density Function}
\newacronym{PDS} {PDS} {Planetary Data System}
\newacronym{PNG} {PNG} {Portable Network Graphics}
\newacronym{PRIN} {PRIN} {Progetti di Ricerca di Rilevante Interesse Nazionale}
\newacronym{PS1-MDS} {PS1-MDS} {PS1 Medium Deep Survey}
\newacronym{PSD} {PSD} {power spectral density}
\newacronym{PSF} {PSF} {Point Spread Function}
\newacronym{PTF} {PTF} {Palomar Transient Factory}
\newacronym{Pan-STARRS} {Pan-STARRS} {Panoramic Survey Telescope and Rapid Response System}
\newglossaryentry{Pan-STARRS1} {name={Pan-STARRS1}, description={the first telescope of the Panoramic Survey Telescope and Rapid Response System}}
\newacronym{Parsl} {Parsl} {Parallel Scripting Library \url{http://parsl-project.org/}}
\newglossaryentry{Project Manager} {name={Project Manager}, description={The person responsible for exercising leadership and oversight over the entire LSST project; he or she controls schedule, budget, and all contingency funds}}
\newglossaryentry{Prompt Processing} {name={Prompt Processing}, description={The data processing which occurs at the Archive Center based on the stream of images coming from the telescope. This includes both Alert Production, which scans the image stream to identify and send alerts on transient and variable sources, and Solar System Processing, which identifies and characterizes objects in our solar system. It also includes specialized rapid calibration and Commissioning processing. Prompt Processing generates the Prompt Data Products.}}
\newacronym{QA} {QA} {\gls{Quality Assurance}}
\newacronym{QC} {QC} {\gls{Quality Control}}
\newacronym{QSO} {QSO} {Quasi-Stellar \gls{Object} (Quasar) }
\newglossaryentry{Qserv} {name={Qserv}, description={LSST's distributed parallel database. This database system is used for collecting, storing, and serving LSST Data Release Catalogs and Project metadata, and is part of the Software Stack}}
\newglossaryentry{Quality Assurance} {name={Quality Assurance}, description={All activities, deliverables, services, documents, procedures or artifacts which are designed to ensure the quality of DM deliverables. This may include QC systems, in so far as they are covered in the charge described in LDM-622. Note that contrasts with the LDM-522 definition of “QA” as “Quality Analysis”, a manual process which occurs only during commissioning and operations. See also: Quality Control}}
\newglossaryentry{Quality Control} {name={Quality Control}, description={Services and processes which are aimed at measuring and monitoring a system to verify and characterize its performance (as in LDM-522). Quality Control systems run autonomously, only notifying people when an anomaly has been detected. See also Quality Assurance}}
\newacronym{R1} {R1} {Doctoral Universities – Very high research activity}
\newacronym{RA} {RA} {Right Ascension}
\newacronym{RAIL} {RAIL} {Redshift Assessment Infrastructure Layers, \url{https://github.com/LSSTDESC/RAIL}}
\newacronym{RAM} {RAM} {Random Access Memory}
\newacronym{RDBMS} {RDBMS} {Relational Database Management System }
\newacronym{RESSPECT} {RESSPECT} {Recommendation System for Spectroscopic Follow-up}
\newacronym{RGB} {RGB} {Red Giant Branch}
\newacronym{RMS} {RMS} {Root-Mean-Square}
\newacronym{RNADE} {RNADE} {Real-valued Neural Autoregressive Distribution Estimation}
\newacronym{RNN} {RNN} {Recurrent Neural Network}
\newacronym{ROSAT} {ROSAT} {R\"{o}ntgensatellit X-ray telescope}
\newacronym{RRL} {RRL} {RR Lyrae stars}
\newglossaryentry{RRab} {name={RRab}, description={RRL subgroup of fundamental-mode pulsators, most common and display the steep rises in brightness typical of RRL}}
\newglossaryentry{RRc} {name={RRc}, description={RRL subgroup with shorter periods and more sinusoidal variation. These are the less common population of RRL}}
\newglossaryentry{RRd} {name={RRd}, description={RRL subgroup of double mode pulsars and are the most rare RRL }}
\newglossaryentry{RSP} {name={RSP}, description={Rubin Science Platform}}
\newacronym{RTA} {RTA} {Real Time Analysis}
\newglossaryentry{Release} {name={Release}, description={Publication of a new version of a document, software, or data product. Depending on context, releases may require approval from Project- or DM-level change control boards, and then form part of the formal project baseline}}
\newglossaryentry{Review} {name={Review}, description={Programmatic and/or technical audits of a given component of the project, where a preferably independent committee advises further project decisions, based on the current status and their evaluation of it. The reviews assess technical performance and maturity, as well as the compliance of the design and end product with the stated requirements and interfaces}}
\newacronym{SACC} {SACC} {Save All Correlations and Covariances}
\newacronym{SAGA} {SAGA} {Satellites Around Galactic Analogs (Survery)}
\newacronym{SB} {SB} {Surface Brightness}
\newacronym{SC} {SC} {\gls{Science Collaboration}}
\newacronym{SCIPPR} {SCIPPR} {Supernova Cosmology Inference with Probabilistic Photometric Redshifts}
\newacronym{SDSS} {SDSS} {\gls{Sloan Digital Sky Survey}}
\newacronym{SED} {SED} {\gls{Spectral Energy Distribution}}
\newacronym{SF} {SF} {\gls{Structure Function}}
\newacronym{SFR} {SFR} {Star Formation Rate}
\newacronym{SIA} {SIA} {Simple Image Access}
\newacronym{SKA} {SKA} {Square Kilometer Array}
\newacronym{SLAC} {SLAC} {\gls{SLAC National Accelerator Laboratory}}
\newglossaryentry{SLAC National Accelerator Laboratory} {name={SLAC National Accelerator Laboratory}, description={ A national laboratory funded by the US Department of Energy (DOE); SLAC leads a consortium of DOE laboratories that has assumed responsibility for providing the LSST camera. Although the Camera project manages its own schedule and budget, including contingency, the Camera team’s schedule and requirements are integrated with the larger Project.  The camera effort is accountable to the LSSTPO.}}
\newacronym{SLSN} {SLSN} {super luminous supernova(e)}
\newacronym{SMBH} {SMBH} {Supermassive Black Hole}
\newacronym{SMF} {SMF} {Stellar Mass Function}
\newacronym{SN} {SN} {SuperNovae}
\newacronym{SNANA} {SNANA} {SuperNova ANAlysis (\url{https://snana.uchicago.edu/})}
\newacronym{SNR} {SNR} {Signal to Noise Ratio}
\newacronym{SO} {SO} {Simons Observatory}
\newacronym{SPT} {SPT} {South Pole Telescope}
\newacronym{SQL} {SQL} {Structured Query Language}
\newacronym{SRD} {SRD} {LSST Science Requirements; \gls{LPM}-17}
\newacronym{SS} {SS} {\gls{Subsystem Scientist}}
\newacronym{SSI} {SSI} {\gls{Synthetic Source Injection}}
\newacronym{SSO} {SSO} {\gls{Solar System Object}}
\newacronym{SSOIS} {SSOIS} {Solar System \gls{Object} Image Search (\url{https://www.cadc-ccda.hia-iha.nrc-cnrc.gc.ca/en/ssois/})}
\newacronym{SSP} {SSP} {\gls{Solar System Processing}}
\newacronym{STEM} {STEM} {Science, Technology, Engineering and Math}
\newacronym{SU} {SU} {Stanford University}
\newacronym{SVOM} {SVOM} {Space Variable Objects Monitor}
\newglossaryentry{Science Collaboration} {name={Science Collaboration}, description={An autonomous body of scientists interested in a particular area of science enabled by the LSST dataset, which through precursor studies, simulations, and algorithm development lays the groundwork for the large-scale science projects the LSST will enable.  In addition to preparing their members to take full advantage of LSST early in its operations phase, the science collaborations have helped to define the system's science requirements, refine and promote the science case, and quality check design and development work}}
\newglossaryentry{Science Pipelines} {name={Science Pipelines}, description={The library of software components and the algorithms and processing pipelines assembled from them that are being developed by DM to generate science-ready data products from LSST images. The Pipelines may be executed at scale as part of LSST Prompt or Data Release processing, or pieces of them may be used in a standalone mode or executed through the LSST Science Platform. The Science Pipelines are one component of the LSST Software Stack}}
\newglossaryentry{Science Platform} {name={Science Platform}, description={A set of integrated web applications and services deployed at the LSST Data Access Centers (DACs) through which the scientific community will access, visualize, and perform next-to-the-data analysis of the LSST data products}}
\newglossaryentry{Sloan Digital Sky Survey} {name={Sloan Digital Sky Survey}, description={is a digital survey of roughly 10,000 square degrees of sky around the north Galactic pole, plus a ~300 square degree stripe along the celestial equator}}
\newglossaryentry{Software Stack} {name={Software Stack}, description={Often referred to as the LSST Stack, or just The Stack, it is the collection of software written by the LSST Data Management Team to process, generate, and serve LSST images, transient alerts, and catalogs. The Stack includes the LSST Science Pipelines, as well as packages upon which the DM software depends. It is open source and publicly available}}
\newglossaryentry{Solar System Object} {name={Solar System Object}, description={A solar system object is an astrophysical object that is identified as part of the Solar System: planets and their satellites, asteroids, comets, etc. This class of object had historically been referred to within the LSST Project as Moving Objects}}
\newglossaryentry{Solar System Processing} {name={Solar System Processing}, description={A component of the Prompt Processing system, Solar System Processing identifies new SSObjects using unassociated DIASources.}}
\newglossaryentry{Source} {name={Source}, description={A single detection of an astrophysical object in an image, the characteristics for which are stored in the Source Catalog of the DRP database. The association of Sources that are non-moving lead to Objects; the association of moving Sources leads to Solar System Objects. (Note that in non-LSST usage "source" is often used for what LSST calls an Object.)}}
\newglossaryentry{Specification} {name={Specification}, description={One or more performance parameter(s) being established by a requirement that the delivered system or subsystem must meet}}
\newglossaryentry{Spectral Energy Distribution} {name={Spectral Energy Distribution}, description={the radiated energy of an astrophysical object as a function of energy (or wavelength) across the entire spectrum of light}}
\newglossaryentry{Stripe 82} {name={Stripe 82}, description={A 2.5° wide equatorial band of sky covering roughly 300 square degrees that was observed repeatedly in 5 passbands during the course of the SDSS, In part for calibration purposes}}
\newglossaryentry{Structure Function} {name={Structure Function}, description={measure of variance of observations separated in time}}
\newglossaryentry{Subsystem} {name={Subsystem}, description={A set of elements comprising a system within the larger LSST system that is responsible for a key technical deliverable of the project}}
\newglossaryentry{Subsystem Manager} {name={Subsystem Manager}, description={responsible manager for an LSST subsystem; he or she exercises authority, within prescribed limits and under scrutiny of the Project Manager, over the relevant subsystem's cost, schedule, and work plans}}
\newglossaryentry{Subsystem Scientist} {name={Subsystem Scientist}, description={The principal science advisor  to a Subsystem Manager; he or she ensures that the subsystem specifications are appropriated for achieving the project's goals}}
\newglossaryentry{Synthetic Source Injection} {name={Synthetic Source Injection}, description={injecting fake objects onto images to test the detection and measurement process}}
\newglossaryentry{Systems Engineering} {name={Systems Engineering}, description={an interdisciplinary field of engineering that focuses on how to design and manage complex engineering systems over their life cycles. Issues such as requirements engineering, reliability, logistics, coordination of different teams, testing and evaluation, maintainability and many other disciplines necessary for successful system development, design, implementation, and ultimate decommission become more difficult when dealing with large or complex projects. Systems engineering deals with work-processes, optimization methods, and risk management tools in such projects. It overlaps technical and human-centered disciplines such as industrial engineering, control engineering, software engineering, organizational studies, and project management. Systems engineering ensures that all likely aspects of a project or system are considered, and integrated into a whole}}
\newacronym{TAP} {TAP} {Table Access Protocol}
\newacronym{TB} {TB} {TeraByte}
\newacronym{TDE} {TDE} {Tidal Disruption Event}
\newacronym{TDEs} {TDEs} {Tidal Disruption Events}
\newacronym{TESS} {TESS} {Transiting Exoplanet Survey Satellite}
\newacronym{TGAS} {TGAS} {Tycho-Gaia Astrometric Solution}
\newacronym{THOR} {THOR} {Tracklet-less Heliocentric Orbit Recovery, an \gls{algorithm} described in \citet{2021AJ....162..143M}}
\newacronym{TNO} {TNO} {trans-Neptunian object}
\newacronym{TNS} {TNS} {Transient Name Server}
\newacronym{TOM} {TOM} {Target and Observation Manager}
\newacronym{TOPCAT} {TOPCAT} {Tool for OPerations on Catalogues And Tables}
\newacronym{TS} {TS} {Test \gls{Specification}}
\newacronym{TVS} {TVS} {Transients and Variable Stars \gls{Science Collaboration}}
\newacronym{TVSSC} {TVSSC} {Transients and Variable Stars \gls{Science Collaboration}}
\newglossaryentry{Task} {name={Task}, description={Tasks are the basic unit of code re-use in the LSST Stack. They perform a well defined, logically contained piece of functionality. Tasks come standard with configuration, logging, processing metadata, and debugging features. For further details, see How to Write a Task in the source code documentation.  Tasks can be nested, providing a natural way to structure - and configure - high level algorithms that delegate work to lower-level algorithms}}
\newglossaryentry{Template} {name={Template}, description={A co-added, single-band image of the sky that is deep, and created in a manner to remove transient or fast moving objects from the final image. Constituent images for templates may be selected from a limited range of quality parameters, such as PSF size or airmass. Such images are used to perform Difference Image Analysis in order to detect variable, transient, and Solar System astrophysical objects}}
\newacronym{ToO} {ToO} {Target of \gls{Opportunity}}
\newacronym{UCL} {UCL} {University College London (\gls{UK})}
\newacronym{UDF} {UDF} {User Defined Function}
\newacronym{UK} {UK} {United Kingdom}
\newacronym{UKIDSS} {UKIDSS} {UKIRT Infrared Deep Sky Survey}
\newacronym{UKIRT} {UKIRT} {United Kingdom Infrared Telescope}
\newacronym{UMAP} {UMAP} {Uniform Manifold Approximation and Projection for dimension reduction}
\newacronym{UML} {UML} {unified modeling language}
\newacronym{UNIONS} {UNIONS} {Ultraviolet Near- Infrared Optical Northern Survey}
\newacronym{US} {US} {United States}
\newacronym{UT} {UT} {Universal Time}
\newacronym{UTC} {UTC} {Coordinated Universal Time}
\newacronym{UV} {UV} {Ultraviolet}
\newacronym{UW} {UW} {University of Washington}
\newacronym{VISTA} {VISTA} {Visible and Infrared Survey Telescope for Astronomy}
\newacronym{VLA} {VLA} {Very Large Array (\gls{NRAO})}
\newacronym{VLASS} {VLASS} {The Very Large Array Sky Survey carried out by \gls{VLA}}
\newacronym{VLT} {VLT} {Very Large Telescope (\gls{ESO})}
\newacronym{VPHAS} {VPHAS} {VST/OmegaCAM Photometric H-Alpha Survey}
\newacronym{VST} {VST} {VLT Survey Telescope}
\newglossaryentry{Validation} {name={Validation}, description={A process of confirming that the delivered system will provide its desired functionality; overall, a validation process includes the evaluation, integration, and test activities carried out at the system level to ensure that the final developed system satisfies the intent and performance of that system in operations}}
\newacronym{W3C} {W3C} {World Wide Web Consortium}
\newacronym{WBS} {WBS} {\gls{Work Breakdown Structure}}
\newacronym{WCS} {WCS} {\gls{World Coordinate System}}
\newacronym{WFD} {WFD} {Wide Fast Deep}
\newacronym{WFIRST} {WFIRST} {Wide Field Infrared Survey Telescope}
\newacronym{WISE} {WISE} {Wide-field Survey Explorer}
\newacronym{WL} {WL} {Weak gravitational Lens cosmic shear}
\newacronym{WTTS} {WTTS} {Weak-lined T Tauri stars}
\newglossaryentry{Work Breakdown Structure} {name={Work Breakdown Structure}, description={a tool that defines and organizes the LSST project's total work scope through the enumeration and grouping of the project's discrete work elements}}
\newglossaryentry{World Coordinate System} {name={World Coordinate System}, description={a mapping from image pixel coordinates to physical coordinates; in the case of images the mapping is to sky coordinates, generally in an equatorial (RA, Dec) system. The WCS is expressed in FITS file extensions as a collection of header keyword=value pairs (basically, the values of parameters for a selected functional representation of the mapping) that are specified in the FITS Standard}}
\newacronym{XMM} {XMM} {ESA X-ray Multi-mirror Mission}
\newacronym{XMM-Newton} {XMM-Newton} { \gls{ESA} X-ray Multi-mirror Mission}
\newacronym{XRISM} {XRISM} {X-ray Imaging and Spectroscopy Mission}
\newacronym{YSO} {YSO} {Young Stellar \gls{Object}}
\newacronym{ZTF} {ZTF} {Zwicky Transient Facility}
\newglossaryentry{active asteroid} {name={active asteroid}, description={small Solar System bodies that have asteroid-like orbits but show comet-like visual characteristics}}
\newglossaryentry{aggregate metric} {name={aggregate metric}, description={An aggregation of multiple point metrics. For example, the overall photometric repeatability for a particular tract given given the repeatability of multiple individual stars in the tract. See also: “metric”}}
\newglossaryentry{aggregation} {name={aggregation}, description={The process of reducing multiple input values to a single output, e.g., a metric value, computed from a collection of input values. For example, a sum or average of a metric computed over patches to produce an aggregate metric at tract level. See also: “metric”, “aggregate metric”}}
\newglossaryentry{airmass} {name={airmass}, description={The pathlength of light from an astrophysical source through the Earth's atmosphere. It is given approximately by sec z, where z is the angular distance from the zenith (the point directly overhead, where airmass = 1.0) to the source}}
\newglossaryentry{algorithm} {name={algorithm}, description={A computational implementation of a calculation or some method of processing}}
\newacronym{arcmin} {arcmin} {arcminute minute of arc (unit of angle)}
\newacronym{arcsec} {arcsec} {arcsecond second of arc (unit of angle)}
\newglossaryentry{astrometry} {name={astrometry}, description={In astronomy, the sub-discipline of astrometry concerns precision measurement of positions (at a reference epoch), and real and apparent motions of astrophysical objects. Real motion means 3-D motions of the object with respect to an inertial reference frame; apparent motions are an artifact of the motion of the Earth. Astrometry per se is sometimes confused with the act of determining a World Coordinate System (WCS), which is a functional characterization of the mapping from pixels in an image or spectrum to world coordinate such as (RA, Dec) or wavelength}}
\newglossaryentry{astronomical object} {name={astronomical object}, description={A star, galaxy, asteroid, or other physical object of astronomical interest. Beware: in non-LSST usage, these are often known as sources}}
\newacronym{au} {au} {astronomical unit}
\newglossaryentry{background} {name={background}, description={In an image, the background consists of contributions from the sky (e.g., clouds or scattered moonlight), and from the telescope and camera optics, which must be distinguished from the astrophysical background. The sky and instrumental backgrounds are characterized and removed by the LSST processing software using a low-order spatial function whose coefficients are recorded in the image metadata}}
\newglossaryentry{cadence} {name={cadence}, description={The sequence of pointings, visit exposures, and exposure durations performed over the course of a survey}}
\newglossaryentry{calibration} {name={calibration}, description={The process of translating signals produced by a measuring instrument such as a telescope and camera into physical units such as flux, which are used for scientific analysis. Calibration removes most of the contributions to the signal from environmental and instrumental factors, such that only the astronomical component remains}}
\newglossaryentry{camera} {name={camera}, description={An imaging device mounted at a telescope focal plane, composed of optics, a shutter, a set of filters, and one or more sensors arranged in a focal plane array}}
\newglossaryentry{cloud} {name={cloud}, description={A visible mass of condensed water vapor floating in the atmosphere, typically high above the ground or in interstellar space acting as the birthplace for stars.  Also a way of computing (on other peoples computers leveraging their services and availability).}}
\newglossaryentry{configuration} {name={configuration}, description={A task-specific set of configuration parameters, also called a 'config'. The config is read-only; once a task is constructed, the same configuration will be used to process all data. This makes the data processing more predictable: it does not depend on the order in which items of data are processed. This is distinct from arguments or options, which are allowed to vary from one task invocation to the next}}
\newglossaryentry{data collection} {name={data collection}, description={A data collection in the second-generation (Gen2) Butler (referred to as a data repository in earlier generations) consists of hierarchically organized data files, an inventory or registry of the contents (i.e., metadata from the data files) stored in an sqlite3 file, and a Mapper file that specifies to the LSST Stack software the camera model to apply when accessing the data in the data repository}}
\newglossaryentry{data repository} {name={data repository}, description={A data repository consists of hierarchically organized data files, an inventory or registry of the contents (i.e., metadata from the data files) stored in an sqlite3 file, and a Mapper file that specifies to the LSST Stack software the camera model to apply when accessing the data in the repository. With the second-generation (Gen2) Butler, the term repository will be replaced by data collection}}
\newglossaryentry{deblend} {name={deblend}, description={Deblending is the act of inferring the intensity profiles of two or more overlapping sources from a single footprint within an image. Source footprints may overlap in crowded fields, or where the astrophysical phenomena intrinsically overlap (e.g., a supernova embedded in an external galaxy), or by spatial co-incidence (e.g., an asteroid passing in front of a star). Deblending may make use of a priori information from images (e.g., deep CoAdds or visit images obtained in good seeing), from catalogs, or from models. A 'deblend' is commonly referred to in terms of 'parent' (total) and 'child' (component) objects}}
\newacronym{deg} {deg} {degree; unit of angle}
\newglossaryentry{drill down} {name={drill down}, description={Move from a higher level aggregation of data to its inputs. For example, given data describing a tract, to drill down to constituent patches and then to objects. Also refers to the act of identifying an issue in a high-level summary of the data (e.g. an aberrant metric value) and interactively investigating its inputs to find the source of the problem}}
\newglossaryentry{element} {name={element}, description={A node in the hierarchical project WBS}}
\newglossaryentry{ephemeris} {name={ephemeris}, description={An ephemeris (pl: ephemerides) gives the predicted positions of astronomical objects or artificial satellites in the sky with time. The ephemerides are computed from mathematical models of motion of the object and the Earth. In LSST Solar System Processing, it refers to a predicted position (RA/Dec/time/etc) of a Solar System Object (SSObject)}}
\newglossaryentry{epoch} {name={epoch}, description={Sky coordinate reference frame, e.g., J2000. Alternatively refers to a single observation (usually photometric, can be multi-band) of a variable source}}
\newglossaryentry{flux} {name={flux}, description={Shorthand for radiative flux, it is a measure of the transport of radiant energy per unit area per unit time. In astronomy this is usually expressed in cgs units: erg/cm2/s}}
\newglossaryentry{footprint} {name={footprint}, description={See 'source footprint', 'instrumental footprint', or 'survey footprint', `Footprint` is a Python class representing a source footprint}}
\newglossaryentry{forced photometry} {name={forced photometry}, description={A measurement of the photometric properties of a source, or expected source, with one or more parameters held fixed. Most often this means fixing the location of the center of the brightness profile (which may be known or predicted in advance), and measuring other properties such as total brightness, shape, and orientation. Forced photometry will be done for all Objects in the Data Release Production}}
\newglossaryentry{instrumental footprint} {name={instrumental footprint}, description={The size and shape of a region on the sky that is covered by the field of view of an instrument, or part of an instrument, e.g., the LSST Camera, or ComCam, or a single LSST CCD.  Often represented by a geometric region defined in field-angle space}}
\newglossaryentry{interoperability} {name={interoperability}, description={the ability of systems or software to exchange and make use of information between them.}}
\newacronym{kSZ} {kSZ} {kinetic Sunyaev-Zeldovich effect}
\newglossaryentry{metadata} {name={metadata}, description={General term for data about data, e.g., attributes of astronomical objects (e.g. images, sources, astroObjects, etc.) that are characteristics of the objects themselves, and facilitate the organization, preservation, and query of data sets. (E.g., a FITS header contains metadata)}}
\newglossaryentry{metric} {name={metric}, description={A measurable quantity which may be tracked. A metric has a name, description, unit, references, and tags (which are used for grouping). A metric is a scalar by definition. See also: aggregate metric, model metric, point metric}}
\newglossaryentry{metric value} {name={metric value}, description={The result of computing a particular metric on some given data. Note that metric values are typically computed rather than measured. See also: metric}}
\newglossaryentry{middleware} {name={middleware}, description={Software that acts as a bridge between other systems or software usually a database or network. Specifically in the Data Management System this refers to Butler for data access and Workflow management for distributed processing.}}
\newglossaryentry{model metric} {name={model metric}, description={A metric describing a model related to the data. For example, the coefficients of a 2D polynomial fit to the background of a single CCD exposure}}
\newglossaryentry{monitoring} {name={monitoring}, description={In DM QA, this refers to the process of collecting, storing, aggregating and visualizing metrics}}
\newglossaryentry{parquet} {name={parquet}, description={see Apache Parquet}}
\newglossaryentry{passband} {name={passband}, description={The window of wavelength or the energy range admitted by an optical system; specifically the transmission as a function of wavelength or energy. Typically the passband is limited by a filter. The width of the passband may be characterized in a variety of ways, including the width of the half-power points of the transmission curve, or by the equivalent width of a filter with 100\% transmission within the passband, and zero elsewhere}}
\newglossaryentry{patch} {name={patch}, description={An quadrilateral sub-region of a sky tract, with a size in pixels chosen to fit easily into memory on desktop computers}}
\newacronym{photo-z} {photo-z} {\gls{photometric redshift}}
\newglossaryentry{photometric redshift} {name={photometric redshift}, description={Often abbreviated to photo-z, this is an estimate of the true redshift (of a galaxy) determined from multi-band photometry. Generally determined from a fit of source colors to grid of model SEDs with redshift}}
\newglossaryentry{pipeline} {name={pipeline}, description={A configured sequence of software tasks (Stages) to process data and generate data products. Example: Association Pipeline}}
\newglossaryentry{point metric} {name={point metric}, description={A metric that is associated with a single entry in a catalog. Examples include the shape of a source, the standard deviation of the flux of an object detected on a Coadd, the flux of an source detected on a difference image}}
\newglossaryentry{postage stamp} {name={postage stamp}, description={Image cutouts that are ~30x30 arcseconds, centered on an Object, and included in every Alert}}
\newglossaryentry{provenance} {name={provenance}, description={Information about how LSST images, Sources, and Objects were created (e.g., versions of pipelines, algorithmic components, or templates) and how to recreate them}}
\newglossaryentry{seeing} {name={seeing}, description={An astronomical term for characterizing the stability of the atmosphere, as measured by the width of the point-spread function on images. The PSF width is also affected by a number of other factors, including the airmass, passband, and the telescope and camera optics}}
\newglossaryentry{shape} {name={shape}, description={In reference to a Source or Object, the shape is a functional characterization of its spatial intensity distribution, and the integral of the shape is the flux. Shape characterizations are a data product in the DIASource, DIAObject, Source, and Object catalogs}}
\newglossaryentry{sky map} {name={sky map}, description={A sky tessellation for LSST. The Stack includes software to define a geometric mapping from the representation of World Coordinates in input images to the LSST sky map. This tessellation is comprised of individual tracts which are, in turn, comprised of patches}}
\newglossaryentry{software} {name={software}, description={The programs and other operating information used by a computer.}}
\newglossaryentry{source footprint} {name={source footprint}, description={A set of pixels that are determined to be part of a Source (or DIASource). It is implemented as a list of spans. A span contains coordinates of a stripe of pixels: row (y) given span belongs to, and a section of a column (xStart, xEnd). In DM code, the term 'footprint' refers to a 'source footprint'}}
\newglossaryentry{sqlite3} {name={sqlite3}, description={A software package external to DM, sqlite3 provides a SQL interface compliant with the DB-API 2.0 specification for SQLite, a self-contained public-domain SQL database engine}}
\newglossaryentry{stack} {name={stack}, description={a grouping, usually in layers (hence stack), of software packages and services to achieve a common goal. Often providing a higher level set of end user oriented services and tools}}
\newglossaryentry{survey footprint} {name={survey footprint}, description={The portion of the sky covered by data from an astronomical survey, e.g., the main wide-fast-deep LSST 10-year survey, the LSST deep drilling fields, or the Science Validation data taken during commissioning.  Sometimes represented by Boolean maps or other summary statistics in an all-sky representation, e.g., the IVOA MOC standard}}
\newacronym{tSZ} {tSZ} {thermal Sunyaev-Zeldovich effect}
\newglossaryentry{tracklet} {name={tracklet}, description={Links between unassociated DIASources within one night to identify moving objects}}
\newglossaryentry{tract} {name={tract}, description={A portion of sky, a spherical convex polygon, within the LSST all-sky tessellation (sky map). Each tract is subdivided into sky patches}}
\newglossaryentry{transient} {name={transient}, description={A transient source is one that has been detected on a difference image, but has not been associated with either an astronomical object or a solar system body}}
\newglossaryentry{warp} {name={warp}, description={(noun) The pixels from a single CCD Exposure that overlap a given coadd patch, trimmed and resampled into the patch's coordinate system; in other words, an image that has been astrometrically registered to the common coordinate system of a tract}}
\newglossaryentry{zBEAMS} {name={zBEAMS}, description={Extension of BEAMS light curve classification method to include redshift ($z$) information}}
\makeglossaries
\begin{document}

\pagenumbering{roman}

\title{\Large From Data to Software to Science with the Rubin Observatory LSST}
\shorttitle{Data to Software to Science}

%% DO NOT EDIT THIS FILE. IT IS GENERATED FROM db2authors.py"
%% Regenerate using:
%%    python $LSST_TEXMF_DIR/bin/db2authors.py > authors.tex

\author[0000-0001-5228-6598]{Katelyn~Breivik}
\affiliation{Center for Computational Astrophysics, Flatiron Institute, 162 Fifth Ave, New York, NY, 10010, USA}

\author[0000-0001-5576-8189]{Andrew~J.~Connolly}
\affiliation{Department of Astronomy and the DIRAC Institute, University of Washington, 3910 15th Avenue NE, Seattle, WA 98195, USA}

\author[0000-0002-5956-851X]{K.~E.~Saavik~Ford}
\affiliation{Department of Science, CUNY Borough of Manhattan Community College, 199 Chambers Street, New York, NY 10007, USA}
\affiliation{Department of Astrophysics, American Museum of Natural History, New York, NY 10028, USA}
\affiliation{The Graduate Center of the City University of New York, 365 Fifth Avenue, New York, NY 10016, USA}
\affiliation{Center for Computational Astrophysics, Flatiron Institute, 162 Fifth Ave, New York, NY, 10010, USA}

\author[0000-0003-1996-9252]{Mario~Juri\'{c}}
\affiliation{Department of Astronomy and the DIRAC Institute, University of Washington, 3910 15th Avenue NE, Seattle, WA 98195, USA}

\author[0000-0003-2271-1527]{Rachel~Mandelbaum}
\affiliation{McWilliams Center for Cosmology, Department of Physics, Carnegie Mellon University, Pittsburgh, PA 15213, USA}

\author[0000-0001-9515-478X]{Adam~A.~Miller}
\affiliation{Center for Interdisciplinary Exploration and Research in Astrophysics (CIERA) and Department of Physics and Astronomy, Northwestern University, 1800 Sherman Road, Evanston, IL 60201, USA}

\author[0000-0001-8452-9574]{Dara~Norman}
\affiliation{National Optical-Infrared Astronomy Research Laboratory, 950 N.\ Cherry Ave., Tucson, AZ 85719, USA}

\author[0000-0002-7134-8296]{Knut~Olsen}
\affiliation{National Optical-Infrared Astronomy Research Laboratory, 950 N.\ Cherry Ave., Tucson, AZ 85719, USA}

\author[0000-0003-4141-6195]{William~O'Mullane}
\affiliation{Rubin Observatory Project Office, 950 N.\ Cherry Ave., Tucson, AZ  85719, USA}

\author[0000-0003-0872-7098]{Adrian~Price-Whelan}
\affiliation{Center for Computational Astrophysics, Flatiron Institute, 162 Fifth Ave, New York, NY, 10010, USA}

% ...

\begin{abstract}

The Vera C. Rubin Observatory Legacy Survey of Space and Time (LSST) dataset will dramatically alter our understanding of the Universe, from the origins of the Solar System to the nature of dark matter and dark energy. Much of this research will depend on the existence of robust, tested, and scalable algorithms, software, and services. Identifying and developing such tools ahead of time has the potential to significantly accelerate the delivery of early science from LSST. Developing these collaboratively, and making them broadly available, can enable more inclusive and equitable collaboration on LSST science.

To facilitate such opportunities, a community workshop entitled ``From Data to Software to Science with the Rubin Observatory LSST'' was organized by the LSST Interdisciplinary Network for Collaboration and Computing (LINCC) and partners, and held at the Flatiron Institute in New York, March 28-30th 2022. The workshop included over 50 in-person attendees invited from over 300 applications. It identified seven key software areas of need: (i) scalable cross-matching and distributed joining of catalogs, (ii) robust photometric redshift determination, (iii) software for determination of selection functions, (iv) frameworks for scalable time-series analyses, (v) services for image access and reprocessing at scale, (vi) object image access (cutouts) and analysis at scale, and (vii) scalable job execution systems.

This white paper summarizes the discussions of this workshop. It considers the motivating science use cases, identified cross-cutting algorithms, software, and services, their high-level technical specifications, and the principles of inclusive collaborations needed to develop them. We provide it as a useful roadmap of needs, as well as to spur action and collaboration between groups and individuals looking to develop reusable software for early LSST science.

\vspace{1cm}
\end{abstract}

\maketitle
\vspace{4mm}
\noindent\textbf{Author list:} The organizing committee are listed in alphabetical order first. Other contributors and attendees are listed in alphabetical order after these. Endorsers, who did not contribute text but who support the overall goals and message of the white paper,  are listed separately.\\
\textbf{Editors:} William O'Mullane and Mario Juri\'{c}

\vspace{1cm}
\subsection*{Endorsers}

\noindent Camille Avestruz (University of Michigan),
Massimo~Brescia (University of Napoli Federico II),
James J.~Buchanan (Lawrence Livermore National Laboratory),
Jeffrey L.~Carlin (Rubin Observatory),
Aleksandra \'Ciprijanovi\'c (Fermi National Accelerator Laboratory),
Kristen C.\ Dage (McGill University),
Tansu Daylan (Princeton University),
Mariano~Dominguez (IATE-OAC-UNC and CONICET),
Melissa L.~Graham (University of Washington),
Akhtar Mahmood (Bellarmine University),
Martin~Makler (ICAS \& ICIFI, UNSAM - CONICET, Argentina and CBPF, Brazil),
Eniko Regos (Konkoly Observatory), Bruno O.~S\'anchez (Duke University),
R\'obert~Szab\'o (Konkoly Observatory),
Christopher Theissen (UC San Diego), 
Jos\'e Alberto V\'azquez Gonz\'alez ( Instituto de Ciencias F\'isicas UNAM),
Yuanyuan Zhang (Texas A\&M University)

\addcontentsline{toc}{section}{Executive Summary}

\section*{Executive Summary} \label{sec:execsum}

The Vera C.\ Rubin Observatory and the \gls{LSST}
will dramatically alter our understanding of the Universe, from the
origins of the Solar System to the nature of dark matter and dark
energy.  Many of the diverse science cases LSST will enable rely on the existence of robust, tested, and scalable algorithms and
software. Concentrating on the development of a small set of crucial
cross-cutting software components and services has the potential to enable many
early science cases for the LSST science community.

This document presents the conclusions from a three day workshop
entitled ``From Data to Software to Science with the Rubin Observatory
LSST'', organized by \gls{LINCC} Frameworks working with Rubin
Observatory and other partners such as \glspl{IDAC} and \gls{LSST}
\glspl{SC} with a goal of identifying key cross-cutting software
components that can accelerate LSST science. The meeting was held at
the Flatiron institute in New York from March 28$^\text{th}$ to
30$^\text{th}$ 2022. Due to logistical restrictions, attendance was
restricted to around fifty in-person attendees, but plenaries were
open through Zoom to the $>$300 applicants for the workshop. Care
was taken to have diverse groups and all LSST~\glspl{SC} represented at
the workshop, including participants from traditionally underrepresented
groups and institutions. A focus of the meeting was the development of
inclusive collaborations both within the \gls{LINCC} initiative and
more broadly in the \glspl{SC}.

The meeting was structured around discussion of science use cases
(\appref{sec:usecases}) that could be undertaken with the first 1-2
years of Rubin data. All use cases had a common structure, focusing
on  computational and software challenges that could limit the science
community's ability to undertake their science. In total 41 use cases
were developed across seven broad research areas: Solar System,
Local Universe Static Science, Local Universe Variable and Transient
Science, Extragalactic Static Science, Extragalactic Variable Science,
Extragalactic Transient Science, and Cosmology.

From an evaluation of these science cases, six technical areas
(\secref{sec:techneeds}) were identified where substantial development
of cross-cutting software, algorithms, and computational infrastructure would need
to be developed. These included: the ability to rapidly cross-match
external data sets (e.g., at different wavelengths) with the Rubin
alert stream and data release catalogs; the development of robust
photometric redshift estimates (and their uncertainties) optimized for
a range of science cases; software for characterizing the
selection functions associated with Rubin data; support for time
series analyses including the development of new light curve
classification algorithms and the ability to scale these algorithms to
the volume of data from Rubin; the ability to reprocess the Rubin
images with algorithms optimized for specific science (e.g., detecting
low surface brightness features in images); and custom generation
of postage stamp images for sources detected by Rubin and other
telescopes and the application of novel analysis algorithms to these
postage stamp images.

The strengths and shortcomings of existing systems need to be
understood in order to develop concrete plans to rectify gaps in these
areas.  Awareness of developments within Rubin is important to enhance
and not duplicate tools in the making. Enabling interoperation of multiple
data sets either by co-location or by building effective streaming tools
is an area of interest where \glspl{IDAC} could take the lead.

Development and implementation of these technical and science applications will require continued engagement around common software infrastructure by the science community. An inclusive approach to this work will need to be developed that addresses the needs of the science community as a whole (\secref{sec:collab}) and that employs best practices for fostering and maintaining engagement with those members of the research community who are traditionally underrepresented, whether as marginalized communities or as smaller or minority serving institutions.  In \secref{sec:nextsteps}, we describe how \gls{LINCC} Frameworks will facilitate additional community discussions aimed at fostering broad and inclusive collaboration on cross-cutting software components that enable a large number of early science cases, building upon the technical areas identified in \secref{sec:techneeds} and using the strategies and best practices from \secref{sec:collab}.

\tableofcontents
\newpage
\clearpage

\pagenumbering{arabic}
\setcounter{page}{1}

\section{Introduction} \label{sec:intro}

The \gls{LSST}, which will be carried out by the Vera C. Rubin Observatory (henceforth “Rubin Observatory” or “Rubin”), is the flagship ground-based astronomical survey of the 2020s.  Every night, \gls{LSST} will process 20 TB of images and deliver a stream of 10 million alerts for transient, variable, and moving objects in the sky. Each year, Rubin Observatory will reprocess all LSST images with state-of-the-art image processing software to build improved and deeper composite images of the southern sky, detecting tens of billions of objects and characterizing their properties. The scientific reach of the LSST will be extraordinary, addressing questions about the makeup of the Universe as fundamental as: how did the Solar System form; what processes govern the birth and death of stars; how does the dark matter in the Universe sculpt the shape of our own Galaxy; what is the nature of the dark energy that drives the accelerated expansion of our Universe?

In order to take advantage of this once-in-a-generation opportunity to transform our knowledge of the Universe,  the astronomical community will need access to state-of-the-art analysis techniques that work at the scale and complexity of the \gls{LSST} data.  \gls{LSST}’s data products will include catalogs of sources, calibrated images, and a science platform to access and analyze these data. These resources must be supplemented with key algorithms and code that can enable petabyte-scale analyses to search for one-in-a-million or one-in-a-billion events in continuous streams of data, to identify trends and features within billions of sources, and to undertake sophisticated population analyses of the astronomical data.
%outline what the observatory delivers and what is on shoulders of research community
%add acronyms

These computational challenges are not generally unique to single areas of astrophysics. For example, the challenge of analyzing large samples of time series data are common, whether measuring periods of RR Lyrae stars to study the 3D distribution of stars in our Galaxy or classifying Type Ia supernovae to estimate their distance. Each requires access to multi-band time series data, the ability to run bespoke analyses on these data, and the ability to store and share the results of these analyses. If we can address these challenges as a community,  we can  improve and advance the science from Rubin data as a whole.

Equitable collaborations are as important for the progress of good science. This requires facilitated conversations on how to foster good communication, trust, and inclusion among research teams. By centering  topics for equitable collaborations early in the definition of scientific collaborations, we can lay a foundation for building mutually beneficial partnerships with  groups traditionally marginalized from the research process and create team environments that leverage diversity for scientific productivity.

The \gls{LINCC} Frameworks initiative is a program supported by the Schmidt Futures Foundation %{\bf note we will need approval from Schmidt Futures on any mentions of them} 
to develop science-focused astronomical software that can enable a broad range of research projects. The goal of LINCC Frameworks is to accelerate scientific research through a combination of developing productionized, reusable analysis frameworks and to further inclusive collaboration through outreach and education. The LINCC Frameworks team will be responsible for identifying and enabling common computational needs within the science use cases, providing tools for adoption by the LSST \glspl{SC} and other groups to use in the course of their analyses.

% discuss the science, identify common technical challenges in accomplishing this science, discuss how these science themes might map to software development.

This whitepaper represents the results of the \gls{LINCC} Frameworks ``Data to Software to Science'' meeting held at the Center for Computational Astrophysics at the Flatiron Institute between March 28th and March 30th 2022. In the context of LINCC Frameworks, the motivation for this whitepaper is to bring together a diverse group of researchers to discuss scientific use cases that could be undertaken with data from the first two years of the \gls{LSST} survey, and to identify the common computational challenges in undertaking this research. Workshop participants were selected across a broad range of science interests, career stages, and demographics, and to represent researchers at large and small institutions together with those institutions that are historically underrepresented in astronomy. In this paper we discuss the science cases (\autoref{sec:motivation}) and the technical challenges (\autoref{sec:techneeds}) that emerged as we discussed how to undertake early Rubin science. We describe best practises when establishing research collaborations that comprise a broad and diverse community in \autoref{sec:collab}.  The detailed science use cases presented at the meeting are described in \autoref{sec:usecases} and the technical use cases in \autoref{sec:techdetails}. The technical themes that were systematized from the science use cases and plans to develop these themes into scientific software development are outlined in \autoref{sec:nextsteps}.

This is the first step in a process to identify and develop scientifically cross-cutting analysis software infrastructure to meet the science needs of Rubin, augmenting existing efforts within Rubin Observatory and the LSST \glspl{SC}. The needs identified in this white paper will be further developed into software requirements and design documents in collaboration with the broader Rubin community.

%Dara and Tim will add a section about DEI  here seemed better to add this section up above LINCC Frameworks so was moved there.
%discuss science cases - > technical

%The objective was not to design software needed for science with Rubin but to identify the primary areas where software should be developed and why. Will translate to software design docs

\section{Motivating Science} \label{sec:motivation}

Rubin Observatory and \gls{LSST} were designed around four main science pillars \citep{2019ApJ...873..111I} covering fairly broad areas. The  impact of \gls{LSST} will, however, extend well beyond the topics for which it was designed, enabling the scientific activities of a broad community.  Currently, the primary organizational units within that community are the \gls{LSST} \glspl{SC}\footnote{\url{https://www.lsstcorporation.org/science-collaborations}}, which operate independently from each other but are connected within a federation via a charter.  The \gls{LSST} \glspl{SC} vary in scope and size, have membership drawn from the international \gls{LSST} science community, and when taken together have organized efforts to prepare for many of the scientific applications of \gls{LSST}.  Some \glspl{SC} have written science roadmaps outlining their planned  activities, and the steps to achieve them, using the data releases and other products (e.g., alerts) from Rubin Observatory.

The ``Data to Software to Science'' workshop identified seven primary scientific areas, defined based on expected commonalities. These were not explicitly tied to the LSST \glspl{SC} -- e.g., some \glspl{SC} have an interest in multiple of the seven scientific areas, and some areas relate to the interests of several \glspl{SC}.  Organizers ensured representation at the workshop from all LSST \glspl{SC}. Participants worked within the seven science areas to flesh out the high priority and high urgency analyses they would do within their SC with early LSST data. Some science use cases could fit into more than one of these seven areas, as they are not fully distinct; in that case, participants made an arbitrary choice of where to discuss the use case.  A template was provided to guide participants in defining the steps needed for their analysis, while also identifying the associated computational needs and required software tools (\secref{sec:SciTemplate}).  The goal was to enable the identification of common software needs, which might foster collaborations between scientific communities on common infrastructure and could drive \gls{LINCC} development priorities (i.e., to provide tools that those \glspl{SC} would use in their analyses).

The use cases were not intended to be comprehensive, but rather to be examples of research that was of interest to the workshop participants. Many interesting science analyses may, therefore, not be represented in this whitepaper. Given the diversity of the science cases described in \autoref{sec:usecases} we believe, however, that the technical cases that arose from our evaluation of the science cases is reflective of the computational and \gls{software} challenges that the community will face in delivering on the potential of LSST. The use cases are compiled in \autoref{sec:usecases}; below we outline some of the main themes that emerged.

%\rachel{Note: a lot of this is based on breakout session report-outs in \url{https://docs.google.com/presentation/d/15hJQoklRJtZ7WuRiLXAhKKY8BiI4vNNMqA7PGl9-LZA/edit?usp=sharing}.  All summaries could use a check from an expert.}

\textbf{Cosmology}: Use cases in cosmology (\secref{sec:cosmology}) include both static  (e.g., cosmological weak lensing, clustering, or galaxy cluster analysis) and \gls{transient}  (e.g., Type Ia supernovae) science.  Some key use cases also incorporate other survey data, such as from the \gls{CMB}, or targeted follow-up observations  (e.g., for supernova cosmology). For static use cases, key software infrastructure needs include photometric redshifts and associated tools, catalog-level cross-matching capabilities, and the ability to quantify the survey selection function.  For \gls{transient} science cases, time series analysis (light curve representation and classification algorithms) is crucial.  In a subset of cases, image reprocessing may be needed.

\textbf{Extragalactic static}: Use cases in this area  (\secref{sec:ess}) span a range of applications, from extragalactic stellar streams to dwarf galaxy populations, galaxy morphology and physical parameter studies, and galaxy cluster science.  Key enabling technology includes photometric redshifts, custom \gls{postage stamp} processing (e.g., to apply custom sky subtraction or deblending algorithms), cross-matching capabilities against high-resolution or multi-wavelength  datasets, and software to track the selection function and observational effects across the survey.  In a subset of cases, the ability to apply custom analysis or classification software is also needed. For example, solely morphological object classification is likely to be insufficient to discriminate between very distant galaxies and nearby stars of similar colors (brown dwarfs).

\textbf{Extragalactic transients}: Extragalactic \gls{transient} science spans a range of use cases  (\secref{sec:ets}), from those that will use LSST alone or LSST in conjunction with follow-up data (based on events to which classifiers have been applied), to those that are more explicitly cross-observatory, such as the follow-up of detected gravitational wave events.  Detection of transients relies crucially on the template creation process used to produce difference images, and on the ability to effectively represent light curves to enable time series analysis such as classification algorithms.  Many use cases also rely on cross-matching capabilities and photometric redshifts.  Depending on how templates are produced, additional image reprocessing capabilities may be needed.

\textbf{Extragalactic variables}: These use cases (\secref{sec:evs}) typically relate to science with \gls{AGN}, including strong gravitational lensing of quasars. Among the challenges are that  variability is across a wide range of time scales, resulting in a need to analyze light curves over long time baselines. The optimal methodology to quantify the variability (e.g., structure functions) using the light curves is still to be determined. The ability to recognize variability is important to the triggering of follow-up observations (e.g., spectroscopy, and high-cadence photometry). Some \gls{AGN} science also relies on observations other than optical ones, requiring cross-matching capabilities (along with cross-calibration) to include data from other observatories. Photometric redshifts for \gls{AGN} pose challenges since some template-based \gls{photometric redshift} methods may not include \gls{AGN} among their templates.

\textbf{Local Universe \gls{transient} and variable:}  Science use cases in this category (\secref{sec:luts}) relate to variable stars, and physical phenomena such as microlensing and binary systems.  Because these are all non-static phenomena, efficient light curve analysis technology is essential to enabling these use cases.  Many of the use cases require catalog-level cross-matching against multiwavelength datasets of a variety of sizes, from small (e.g., Gaia) to large (e.g., Roman Space Telescope), and some require a good understanding of selection functions.

\textbf{Local Universe static:} Science use cases in this area (\secref{sec:luss}) range from understanding static phenomena in the Milky Way (e.g., stellar science), probing Milky Way structure, and understanding populations of local dwarf galaxies and other aspects of the Local Group. The key enabling technologies identified across these use cases included catalog-level cross-matching, and software to understand the survey selection function.  Image reprocessing and image-based analysis or classification algorithms, primarily used for dwarf galaxy and stellar stream identification, were identified as less common needs. Some use cases involving understanding \gls{MW} structure might rely as well on variable stars and hence have the same dependencies as the Local Universe transient \& variable science cases.

\textbf{Solar system}: The solar system use cases (\secref{sec:solar}) primarily focus on the detection and characterization of populations of objects within our solar system, though a subset also touch on solar system science outside of our own solar system (e.g., interstellar populations ejected from other solar systems).  These use cases often involve use of both catalogs and images for the analysis, resulting in a need for large-scale image access (i.e., bulk reprocessing of calibrated images and the analysis of cutouts). The largest computational challenges arise in use cases that require additional image analysis and  catalog/time-series post-processing. Data and \gls{software} for computing selection functions is also needed.

The identified cross-cutting \gls{software} needs from these areas are summarized in \secref{sec:techneeds}.

\section{Cross-cutting Software and Infrastructure to Enable Early Science} \label{sec:techneeds}

Though highly scientifically diverse, the science areas described in the previous section revealed a significant degree of commonality in terms of technical needs. This presents a significant opportunity: a development of a small set of components or services has the potential to enable a large number of early science cases. We identified seven distinct technical need areas:
\begin{enumerate}
\item Scalable Cross-matching
\item Photometric redshifts
\item Selection function determination
\item Time series analysis support infrastructure
\item Sky image access and reprocessing at scale
\item Object image access and analysis at scale, and the need for
\item Scalable job execution systems.
\end{enumerate}

All but the last of these were discussed within scientifically-diverse breakout groups to better understand developments in each technical area that would support the various science use cases\footnote{Discussion of the scalable job execution system was postponed for a later date because not enough technical stakeholders in this area were present.}, with the identified connections between scientific and technical areas shown in \autoref{tab:scitech}. 

\begin{table}[h]
\begin{tabular}{|c||c|c|c|c|c|c|}
\hline
 & \hspace{-1cm}\makecell{Cross-\\matching} & Photo-$z$ & \hspace{-1cm}\makecell{Selection\\ functions} & \hspace{-1cm}\makecell{Time\\series} & \hspace{-1cm}\makecell{Image\\ reprocessing}  & \hspace{-1cm}\makecell{Image\\ analysis}  \\
\hline\hline
Cosmology & \checkmark\checkmark & \checkmark\checkmark & \checkmark\checkmark & \checkmark\checkmark & \checkmark & \checkmark \\ \hline
Extragalactic static & \checkmark\checkmark & \checkmark\checkmark & \checkmark\checkmark & & \checkmark\checkmark & \checkmark \\ \hline
Extragalactic transient & \checkmark\checkmark & \checkmark\checkmark & \checkmark & \checkmark\checkmark & \checkmark & \checkmark \\ \hline
Extragalactic variable & \checkmark\checkmark & \checkmark & \checkmark & \checkmark\checkmark & \checkmark & \checkmark \\ \hline
\hspace{-1cm}\makecell{Local Universe \\ transient \& variable} & \checkmark\checkmark & & \checkmark  & \checkmark\checkmark & & \\ \hline
Local Universe static &\checkmark\checkmark & & \checkmark\checkmark& & \checkmark & \checkmark\\ \hline
Solar system & \checkmark & & \checkmark\checkmark  & \checkmark\checkmark & \checkmark & \checkmark\checkmark \\ \hline
   \end{tabular}
\caption{Table highlighting the connection between scientific and technical areas discussed at the workshop.  Rows are science areas while columns are for infrastructure capabilities.  A double checkmark (\checkmark\checkmark) signifies that some infrastructure capability is essential to enable a particular scientific area, while a single checkmark (\checkmark) signifies that the infrastructure capability would enhance or expand scientific discovery within that area but is not necessary to enable all of it. \label{tab:scitech}
}
\end{table}

The summary of each technical area is as follows:
\\

\noindent \textbf{Scalable Cross-matching:}  Nearly all science use cases presented at the Workshop require the ability to (generally positionally) cross-correlate the detections in the LSST catalog with one or more other catalogs – an operation commonly known as “cross-matching”. This capability would enable enrichment of LSST data with information taken in other wavelengths, at other times, different resolutions, or in general different characteristics. The capability is needed in two regimes:
\begin{enumerate}
    \item Real-time – low-latency matching of $O(10{\rm k})$ sources to $O(10)$ catalogs each with O(1Bn) objects (to support adding information to alert streams from other catalogs), and for
    \item Offline data analytics – the ability to match O(10Bn) x O(1Bn) object catalogs, followed by joining data from both catalogs (e.g., full time series of observations, multi-wavelength studies) for analysis at scale.
\end{enumerate}
In both regimes, the cross-matching capability {\bf must} be easy to use for the end-user. For example, it may be provided at the community broker level for real-time cases, or accessible as simple Python (e.g. Pandas-like) calls or SQL-like statements callable from Python notebooks for the offline-level.

Importantly, cross-matching is {\em generally just the first step in a longer analysis process}, nearly always followed by fetching additional data from catalogs being cross-matched. Therefore, this operation is {\em better thought of as a (distributed) join of (large) tables on a spatial (user-defined) index, followed by (potentially heavy) computation on the result of the join}. This argues for the cross-matching capability to either be a part or seamlessly interface to a larger scalable analytics system. 

This may also have implication for computation and storage service providers (facilities). For example, more emphasis may be needed on co-locating large datasets, enabling caching of frequently used remote datasets, as well as possible hardware considerations to enable efficient I/O.
% ADDRESSED: \rachel{I think it is worth mentioning that the high demand for cross-matching may also have implications for hardware and the general design and support plan for user compute facilities (how are they thinking about the need to potentially co-host datasets etc.?).}
\\

\noindent \textbf{Photometric redshift determination:} A photometric redshift, or {\em photo-z}, is an estimate of the redshift (or distance) of an object made from available (generally multi-band) photometric information. Photo-z estimation is an algorithmically and computationally complex problem, and one that -- depending on the science case -- may not have a uniquely optimal solution. In some cases, photo-z estimation may incorporate some elements of classification (e.g., to distinguish between different extragalactic sources based on their SED).
% ADDRESSED: \rachel{Next sentence is written as if this is a unique or new proposal.  Since the workshop material is necessarily a synthesis of a variety of inputs, I suggest avoiding such statements in general, and focusing on something like `Consensus among participants across science use cases was that we need to move beyond\dots' or whatever. (I see other places like this below.)}
There was consensus among participants across all science areas about the need to move beyond simple errors in quantifying photo-z vs. spectroscopic redshifts, towards metrics targeting science cases (Section~\ref{sec:PhotozMetrics}).
%ADDRESSED: \rachel{I propose adding the following sentence to connect this to an infrastructure need:}
Currently, different scientific communities typically develop such metrics independently and apply them to specific photo-z codes in common use; in the era of LSST, shared infrastructure permitting application of different science-driven metrics to the outputs of multiple photo-z codes may be essential input to Rubin Observatory's choice of a single photo-z estimator to apply to a given data release, and could also help drive the scientific community's work on developing improved photo-z algorithms.

From a computational capability perspective, photo-z estimation codes need to be able to i) run on $O(5{\rm Bn})$ galaxies at least as often as every data release, ii) allow for nightly calculation using multi-wavelength data for the Broker-filtered extragalactic alerts in any given night, and iii) be able to update their estimates if additional information becomes available (e.g., spectra). Photo-z codes are expected to be highly data-parallel. The outputs are expected to be on order of $O(100)$ numbers per object (roughly $O(10{\rm TB})$ for an LSST-scale dataset). Depending on the algorithm, they may be data intensive in terms of input (e.g., if they require cutouts of individual objects, and not just photometry). Significant prior art already exists, including codes such as RAIL and qp. More detail can be found in Section~\ref{sec:Uncertainty}.

While numerous science cases require photo-z estimates, we recognize they do so at varying degrees of complexity -- from full posterior $p(z)$ PDFs, to simple point estimates. For many science cases, we would like to know things like i) Is there a single “peak” in the p(z), or is it multimodal?, ii) What is the “best-fit” redshift or the peak redshifts in case of multimodality? iii) What is the spread (2nd and higher moments)? or iv) Is there any other statistical property of the p(z) that correlates with object classification? Existing photo-z codes do not have a standardized way of reporting these things; we encourage some of this standardization to occur. We discuss these approaches in Section~\ref{sec:PhotozRep}.
\\

\noindent \textbf{Selection Functions:} Selection functions are core components of any modeling procedure that aims to quantify the population statistics or density distribution of sources or objects. A selection function for a given modeling method may contain things like the detection efficiency of sources with a given brightness or shape, the classification accuracy of sources, the cadence of observations, the Milky Way and intergalactic dust distribution, or the crowdedness (in source counts) of a field.

The Rubin \gls{DM} system will provide the core selection function data product: the detection efficiency of an ideal point source given as a function of position within the focal plane for each visit (direct and differenced), and the same quantity computed for each delivered coadd.

However, each specific science case, classification, or detection algorithm will need a specialized selection function which depends on the science question or model being studied. An example may be the detectability of objects within the Solar System as a function of their orbital and physical parameters, or the detectability of galaxies as a function of their morphology or surface brightness.

Building these will require additional software, and may require external catalog data or additional (potentially even pixel-level) processing. The detailed requirements are not clear at this point: we therefore recommend to engage the community in constructing worked examples of how to build and use selection functions of varied complexity for different use cases. We discuss some known options in more detail in Section~\ref{sec:tSelect}. %\rachel{Reading that, I'm a bit concerned that it doesn't seem to refer to healsparse, which I thought was meant to solve some of these issues for cosmology (and could discuss whether it might meet broader needs).}
%\mjuric{I'm interpreting that section as focusing on the difficulty of building the selection functions for particular types of objects -- once built, healsparse may solve the question of how to efficiently store them?}
\\

\noindent \textbf{Time Series analysis support infrastructure:} The LSST will deliver a uniquely large volume of high-quality multi-epoch measurements for each of the $O(40~{\rm Bn})$ objects it is expected to detect. Converting this rich dataset into a wealth of science results will depend of software infrastructure to i) extract features and classify the captured time-series, ii) enable parametric fitting, and iii) enable anomaly detection.

Variable sky astrophysics is a vast field requiring many period-finding and feature-identifying tools. Thus, to efficiently identify all sources of variability in (nearly) real-time for classification and catalog-creation, a complex multistage \gls{algorithm} is required. Currently, users must make use of many other tools to handle these different types of variability, resulting in running many similar analyses on the same data set to tease out different features. And while many well-built and useful tools are available for feature extraction and classification, running them efficiently on LSST-scale datasets is challenging. We therefore explore the creation of a classification algorithm designed to handle the entirety of the variable-sky database generated by \gls{LSST}, in combination with other southern-sky surveys. This tool would quickly and automatically classify transients and variables based on features in a multi-band light curve (shape, period, filter-specific amplitude and decay, etc) and the available color, magnitude, parallax, angular diameter information from \gls{LSST}. The details are discussed in Section~\ref{sec:FeatureExtract}.

Beyond classification, many analyses involve fitting a pre-specified model to data where the model parameters have semantic content (Section~\ref{sec:TSparametricFit}). Given the numerous and diverse objects that Rubin LSST will observe the models which are fit to the observations must be flexible while also including physical information about each object. Commonly used tools in parametric time series analysis should be automatically computed (or trained) on all objects on a regular basis, along with specific subsets being analyzed with targeted tools as needed; all provided through a unified interface with a common data structure. Additionally, model selection criteria such as AIC or Bayes factors shall also be pre-computed to enable comparison between models, along with uncertainties and ranges of validity for model parameters. Providing the data alongside informative statistics in a networked and unified interface should help maximize the potential of LSST.

Finally, there will be some objects that are difficult to classify or otherwise unusual -- {\em anomalies}. We discuss tools to detect them in Section~\ref{sec:AnomalyDetect}, primarily concentrating on real-time detection in LSST alert stream. To optimize the use of Rubin Observatory as a discovery machine for rare and high-priority events, infrastructure must be developed to efficiently identify anomalies among massive datasets in a timely manner. This will require synergy between machine learning tools for anomaly detection and visualization techniques for interactive and low-latency high-level analysis. These proposed tools should be sufficiently scalable and fast enough to enable prioritization and follow-up of rapidly-evolving events before they dim (early \gls{SN} interaction, cometary outbursts or breakup, rapidly changing \gls{AGN}, microlensing events/\gls{TDEs}/KNe/other unknown phenomena and extreme cases of known types).
\\

\noindent \textbf{Sky Image Access and Reprocessing at Scale:} While much of the science that will be done with LSST will rely entirely on the catalogs delivered by the science pipelines, a significant number of science use cases will also require analysis of the image data. This includes a need for reprocessing of subsets ranging from cutouts of individual objects (covered next in this section) to larger image cutouts, full-focal plane data, or survey-scale pixel-level reprocessing as discussed in Section~\ref{sec:ImageReproc} and summarized here.

The analysis of presented use cases point to the need for scalable i) data access services ii) processing infrastructure, and iii) processing software. For all these, the defining factor is the scale of the problem: $O(100{\rm TB+})$ to $O(10{\rm PB+})$ of image data. Supporting these use cases will require close collaboration between the Rubin Operations team (making sure the data is available in practice), infrastructure providers (whether HPC facilities, IDACs, or the public Cloud), and the users wishing to execute large-scale campaigns and write custom processing software.

We especially stress the importance of the first one of these, {\em performant data access}: services to quickly -- O(seconds) -- deliver a large number of custom-sized or -shaped cutouts up to the full dataset are critical to enabling this technological element. See Section~\ref{sec:ImageReproc} for detailed discussion.
\\

\noindent \textbf{Object Image Access and Analysis at Scale:} Finally, we look at the kinds of \gls{software} tools that are likely needed to enable image-based analyses done at the level of {\em individual objects}. While there is overlap with the previous use case (`Sky Image Access and Reprocessing at Scale'), we discussed this use case separately as it is both scientifically different -- we assume the positions of objects are known -- but also as it may be technically different: here the access patterns are fundamentally {\em object-aligned}, and may benefit from a different storage strategy to be made performant.

A core element for this type of analysis is  an (object-level) scalable image cutout service. This service must be able to i) make image cutouts of any input image over scales ranging from $\sim10$arcsec to $\sim10$arcmin, ii) optionally deblend objects in the scene, iii) provide the ability to link the items on the resulting image to archival data (both catalogs and images) from external sources, and iv) allow for filtering of selected cutouts (e.g., different bands). Ideally, to support alert-triggered follow-up, such a service would also be able to return results in real-time (order of seconds), at least for most recent imaging data. The details are discussed in Section~\ref{sec:ImageAnalysis}.

\section{Inclusive collaboration} \label{sec:collab}

%\subsection{Introduction} \label{sec:colintro}

This section builds on  discussions held as part of a breakout session at the LSSTC From Data to Software to Science Workshop. The session focused on having participants think about how to make more equitable collaborations. Participants were separated into small breakout groups and given use cases that touched on themes of the inclusivity of collaborations, specifically focused on issues of  cross-institutional partnerships, the allocation of credit, resource access, and working with students. Each use case was paired with a set of questions that prompted participants to discuss challenges to the collaborative research process, as well as potential solutions to those challenges (see \appref{sec:DeiScenarios} for a full set of use cases and questions provided to breakout groups).

Below, we present key themes that arose during these breakout discussions, but rather than present all points that participants mentioned, we have instead chosen to highlight the themes that emerged that align with the social science literature on inclusive collaboration. First, we detail various challenges that face research collaboration, and then we outline solutions that could be implemented to ensure research collaborations are more inclusive.
Many of the challenges and solutions below center on specific policies or practices, such as how to incorporate students more effectively into teams, how to build cross-institutional partnerships, or how to structure team environments at an organizational level in ways that will create more inclusive outcomes for all team members. These recommendations must be considered in the broader context of team diversity, specifically demographic diversity. Astronomy and astrophysics has become increasingly diverse along lines of race and gender in recent decades  \citep{Merner2017,Porter2019}.
Research on team diversity has demonstrated that more diverse teams are typically more creative, innovative, productive, and impactful than homogeneous teams \citep{doi:10.1177/000312240907400203,Kalev2009,Freeman2014}. Fully integrating underrepresented researchers in the research team allows for the level exchange of resources like funding, expertise, or information, and thus aids in equity in the field while also creating the potential for diversity to have positive outcomes for research teams.
However, beyond the fact that incorporating more underrepresented researchers into astronomical research collaborations may make the outcomes of research more impactful, there is a moral responsibility to have research teams be more representative of the population regardless of their impact. This means more than just increasing the representation of underrepresented groups. Sociological research has demonstrated that too much focus on increasing representation can lead to dynamics of tokenism, which can mean potentially more opportunity but often leads to more work and harsher evaluations for underrepresented groups than their colleagues \citep[e.g.,][]{10.2307/2777808,https://doi.org/10.1111/soc4.12206,doi:10.1177/0891243219835737}. Instead, we have a responsibility to structure collaborative teams in ways that will fully integrate traditionally underrepresented scholars \citep{doi.org/10.17351/ests2017.142}.

In order to integrate underrepresented researchers into collaborative teams means we must actively work to organize our teams in ways that will allow these folks to share their experiences and expertise in meaningful ways. Others have noted that intentionally structuring research collaborations is a key strategy for diversifying astronomy in the coming decade \citep{2019BAAS...51g.216B}. It takes intentional efforts to design inclusive teams or enact authorship processes that will fully integrate underrepresented groups in meaningful ways. The recommendations put forth below are organizational strategies for how to better structure our research collaborations, with the intent of better incorporating the diversity of astronomy and astrophysics into the research process.

\subsection{Challenges in research collaborations} \label{sec:colchal}
Research collaborations have become increasingly common across \gls{STEM} disciplines \citep{2013EJSE...17S...1B,doi:10.1146/annurev-soc-081715-074219}. While collaborations are often noted for producing more innovative and high-impact research \citep{Frickel2016}, collaborative research often faces several challenges that can undermine the efficacy of the project. For instance, collaborators from different types of institutions\footnote{There is a wide variety of types of institutions, especially when considering the differing landscapes in different countries.  Even just within the US, there are many types of institutes - research and teaching universities, government labs, minority-serving institutions, and more.  However, for simplicity we use just two examples, research and teaching institutions, that already illustrate some of the challenges that can arise due to differing priorities and levels of research infrastructure available.} (research, teaching) often face differing evaluative pressures around productivity or efficiency. University administrators at research institutions often primarily evaluate scientific faculty by the grants they secure, the prestige of the journals in which they publish, and the impact of their research programs. In contrast, administrators at teaching institutions likely expect faculty to juggle higher teaching loads.  Astronomers at teaching institutions are evaluated by course evaluations and student mentorship, but are also expected to publish, although it is often not weighed as heavily in tenure or promotion. Thus, these divergent expectations of astronomers from research and teaching institutions may lead to tensions in the collaborative process.

Time pressures are intertwined with evaluative pressures for promotion or tenure, especially for collaborators that are housed at different types of institutions. Past scholarship has shown time to be an important ingredient for successful collaborations as well as a consistent source of tension among collaborators \citep{doi:10.1177/0003122411433763,https://doi.org/10.1111/socf.12591}. Teaching institution astronomers with higher teaching loads than astronomers at research institutions have their research times constrained by the rhythms of semester schedules. More teaching translates to less time for research and vice versa. The teaching loads at teaching institutions may also prohibit or severely restrict attendance at professional conferences; for instance, the winter \gls{AAS} meetings held in January of each year are less accessible to astronomers who are tied to the semester schedule. Thus, expectations that collaborators will prioritize research or be able to disseminate findings at key conferences in the field often have astronomers at research institutions in mind. This can lead to challenges between astronomers from different types of higher education institutions.

Authorship credit is consistently a source of tension in research collaborations (\citealt{Tsai2014}; \citealt{Bozeman2017}; \citealt{Youtie2014}; \citealt{https://doi.org/10.1111/socf.12591}; \citealt{Misra2020}). The uneven allocation of credit or unclear processes for allocating credit may contribute to a team environment that does not promote inclusion. Credit on publications is the coin of the realm in science; scholars at both teaching and research institutions are expected to publish to meet tenure and promotion milestones. In addition, students looking to be competitive on the job market after graduating are expected to publish as well. Research collaboration has become more common in recent years \citep{doi:10.1146/annurev-soc-081715-074219}, which on one hand opens up more opportunities for scholars to publish, but also creates scenarios in which collaborators are competing for authorship status on the fruits of their collaboration. Tensions around authorship are particularly challenging for junior researchers, who rely on publications in order to move their careers forward \citep{Tsai2016}.  It is common for senior or more well-known researchers to gain undue credit or be listed higher in authorship order than is often warranted \citep{merton1968social,zuckerman1977,rigney2010matthew,Tsai2016}.

There are several factors that can undermine the inclusivity of a team. For instance, the reliance on technology to support collaborations that span multiple institutions may undermine inclusion for some team members. Phone calls, Zoom, Slack, or other technologies that mediate non-face-to-face interaction between team members makes collaborations less personal. Research on the effects of virtual technology on work have found that these kinds of technologies can make individuals feel alienated from their team, and also cause some individuals to self-censor their opinions in ways they may not in face-to-face interaction \citep{Soga2021}. Virtual technology can make remote teamwork especially hard for women and other underrepresented minorities in the workplace \citep{Bolade-Ogunfodun2022}.

Other organizational dynamics of collaborative teams can undermine the inclusivity of a team environment as well. For instance, team inclusivity may be undermined as a result of when or where a meeting is held. Team meetings may be scheduled at times that are more convenient for some members than others. This may be especially true when there is a small group of leaders making decisions for the team, as a concentrated hierarchy is less likely to include the views of all team members. In addition, holding team meetings in physical locations like bars or holding meetings “after hours” may be exclusive to the lifestyles of some team members.

Another challenge is the incorporation of students onto collaborative projects. Scholars at research and teaching institutions also likely have differing access to student support for their research. Many research institution scholars have access to postdocs and graduate students (as well as interested undergraduates). The technical expertise and mentorship provided by postdocs and graduate students facilitates student support of research institution scholars’ research. In contrast, teaching institution astronomers may be working primarily with undergraduate students who have limited expertise and who require additional mentoring and instruction to carry out research.

\subsection{Recommendations} \label{sec:colrec}

\subsubsection{Building sustainable cross-institutional partnerships} \label{sec:colbuild}
In building sustainable cross-institutional partnerships, collaborative teams must work to actively manage expectations of what the team will accomplish and who will take on specific tasks. It is important for all collaborators to outline their expectations for the partnership before it is well underway as a way to avoid future conflicts. For instance, it would be fruitful for all team members to clearly define how they envision their role in the project as well as what they are hoping to get out of the collaboration. Each team member should outline their expectations for their workload, such as what their collaborators can expect of them, what a feasible workload looks like based on their other professional commitments, and task breakdowns. Team members should also lay out what they see  as a realistic timeline for research in relation to their teaching load and other pressures. This is especially important in collaborations that span different types of institutions, as research- and teaching-institution scholars face different evaluative pressures from their administrations. Cross-institutional partnerships that include both research- and teaching-institution scholars should consider the different ways in which semester schedules shape team members’ workloads.

In addition to laying out general expectations at the start of the project, team members should outline expectations for producing publications, the allocation of credit, and other types of dissemination. By allocation of credit, we refer to the arrangement of authorship on publications produced by the collaboration. Cross-institutional partnerships will benefit from detailing a clear division of labor upfront and articulating how this will translate to credit on publications or presentations. It is especially important to have these conversations upfront when a cross-institutional partnership spans both research- and teaching-institutions. Scholars should be assigned tasks that complement their strengths, but also what is valued by their institutions. For teaching-institution scholars, this may include  identifying positions that are more directly tied to teaching or outreach (if that is what the administrators at their institution value). This may also include dissemination of research findings in an array of journals or conferences that will be recognized as beneficial for the varying members of the team.

	All members of a cross-institutional partnership should be provided with the resources necessary for their success. This may be funding or computational resources that are necessary for their science to succeed. This could also be an exchange of personnel that would benefit the collaboration as well as the careers of those being exchanged. For instance, undergraduates from teaching institutions could be sent to work with research-institution scholars for closer mentorship from faculty members or graduate students as a way to prepare for the research process. Similarly, graduate students or postdocs housed at a research institution could go work with a partnering teaching institution as a way to gain experience with teaching, mentoring, or outreach.

	It is important for cross-institutional partners to maintain some degree of flexibility as well. Often, research does not go according to plan, the person who initially volunteered to draft a paper may no longer have time, or unforeseen institutional pressures may change a team members’ commitment to the project. Cross-institutional partnerships should periodically reassess priorities and goals of the project. This will allow collaborators to raise any issues or concerns they have, allow teaching institution astronomers to raise any unforeseen ways the collaboration conflicts with their other priorities, and allow the team to strategize ways to adapt to pressures or pitfalls as the project progress.

\subsubsection{Strategies for dealing with authorship} \label{sec:colstrat}

Authorship and credit allocation is one of the most contentious aspects of research collaboration (\citealt{Misra2020}; \citealt{https://doi.org/10.1111/socf.12591}). To mediate these tensions, collaborators must outline authorship expectations early on and make clear how credit will be allocated before they begin drafting papers. When determining authorship, it is often useful to clearly outline who contributed what to the paper, such as who developed an idea, who developed code, and who wrote what portions of the draft. It is also useful if all collaboration members agree in advance on a well-defined process for defining the authorship order. As a way to benefit junior researchers on a team, collaborators could potentially list junior people higher in authorship to benefit their careers. This is an option to be discussed by the team at the start of a collaboration. A team may also decide to write multiple papers so that everyone can benefit more equitably. Students and postdocs would benefit from designing mentoring plans with their advisors that train these junior scholars on co-authorship, and lay out expectations for authorship and publication.

\subsubsection{Strategies for fostering an inclusive team environment} \label{sec:colteams}

Several factors shape the inclusivity of a collaborative team environment. Some factors are difficult to control. For instance, in cross-institutional partnerships, it may be unrealistic to have the full team be able to meet face to face. Technology often must mediate these relationships, which can make things feel less tight knit with more people feeling detached. While virtual mediums like Slack, Zoom, or other technologies make connectivity easier, they are also less personal and can lead to certain team members feeling alienated.

There are several things that teams can do regardless of whether teams are meeting in person or virtually. At the most basic level, giving each member of the team a role or task to encourage buy-in will get people more invested in the outcome of a team. Team members (especially junior team members) should be given opportunities to  share their ideas. Regular meetings are important to the success of collaborative teams, but not all team members necessarily feel comfortable sharing in large group discussions. Thus, team leaders should encourage feedback and meet with individual team members on a regular basis to solicit input that members may not feel comfortable sharing in the large group. This will ensure that the diversity of voices on a team are incorporated into the decision making of team leadership.

Team members should also be given the opportunity to share work in progress to get feedback from collaborators. This can mean setting aside times during regularly scheduled meetings, or setting aside blocks of time specifically for “chalk talks” or work-in-progress talks. These opportunities to share work should specifically feature diverse members of the team. Providing opportunities for team members to share work in progress can benefit research by generating more collaborative opportunities, as well as make varying members of the team feel recognized as researchers (\citealt{Misra2020}).

Structurally, collaborative teams can improve the inclusivity of their environment by putting into place transparent processes for team decision making, like blind voting with written ballots rather than giving consent during team meetings. Teams should also work to improve by soliciting open and honest feedback from team members. To do this, teams should put into place structures for team members to critique aspects of the project they feel could improve without fear of retaliation. Teams should also enact feedback mechanisms to solicit input from all collaborators. Having all members read and agree to a collective code of conduct is another strategy for fostering an inclusive environment. Regularly communicating with one another over research successes is a third. For large collaborations, creating a shared team calendar can help team members keep track of meetings, and creates a shared norm around prioritizing team events.

Scholarship on collaboration shows that getting together to work face-to-face enhances the productivity of a team \citep{doi:10.1177/0003122411433763}. Social events both on campus as well as other places (grabbing coffee as a team, going out to dinner, likely at conferences for cross-institutional collaborators) help connect team members to the broader collaboration and fosters a sense of community among the team.

\subsubsection{Strategies for working with students}\label{sec:colstudents}
Collaborations face some unique challenges in incorporating students like undergraduates or early graduate students onto projects. As a result, collaborative teams should work to foster an inclusive environment in which students feel comfortable asking questions and proposing ideas without risk of being ridiculed by more senior members of the collaborative team. Virtual technology like Slack or one-on-one meetings with research team faculty can create opportunities for students to pose questions that they may feel uncomfortable asking in large team meetings. Another component of fostering an inclusive environment for students is being mindful of student familiarity with different aspects of the research process in team discussion. Students likely have less technical experience or language proficiency than more senior researchers, and a lack of mindfulness to this fact can create barriers to student participation or success in the process. Teams should also be mindful of student availability for research, and how the collaboration may come into conflict with other work or academic commitments they may have.

One potential strategy for incorporating undergraduates or early graduate students within collaborative teams is giving them tasks that will set them up for success. This could be defining small tasks that are appropriate for novices to undertake, that will not be too challenging but also help encourage confidence in their research abilities. Similarly, some attendees proposed creating a hierarchy of tasks, ranging from easy to more difficult, as a strategy for incorporating students into the research process progressively. As such, faculty members in the collaboration should be prepared to mentor more junior students based on the difficulty of the tasks at hand. Teams could even develop starter projects for students (building starter projects into the design of research proposals) with novice students in mind, rather than dropping them into the deep end of astronomical or astrophysics research.

Collaborative teams looking to work with undergraduate or early-career graduate students need to be mindful of how the research will benefit them. For instance, setting up structures that will help students develop skills, such as data analysis, coding and scientific writing, would be beneficial.  Working with students to have them assess their strengths, weaknesses and goals within the research environment can foster a sense of ownership.  Creating opportunities for students to expand their social networks is also important. Students should not be viewed as free or cheap labor. Part of this is ensuring that students will be credited for their work on a project, both in publications and in public presentations of research findings.

\subsection{Conclusion} \label{sec:colconc}
While astronomy and astrophysics have become increasingly diverse in recent decades, more efforts need to be made to integrate traditionally underrepresented researchers into the research process. We view the organizational structure of research collaborations to be a fruitful area for interventions that will ultimately make astronomy and astrophysics more inclusive and equitable in the coming decade. In this white paper, we have outlined what we see as some core challenges to structuring inclusive collaborations in astronomical research, as well as recommendations for addressing these challenges.

% how/where do we add the references?
% See README.txt

\section{Next Steps} \label{sec:nextsteps}

We begin with some top-level proposals for next steps after this workshop, then discuss follow-up activities in specific technical or domain areas.

\subsection{Top-level proposals for next steps}\label{subsec:general-followup}

As demonstrated in previous sections, the community would benefit from continued engagement on common software infrastructure needs that would enable them to effectively carry out their high-impact scientific analyses with LSST data.  Many software infrastructure areas would serve multiple scientific stakeholders, which is a point in favor of coordination across existing groups in the Rubin Observatory LSST scientific community, including Rubin Observatory, the LSST \glspl{SC}, \gls{LINCC} Frameworks, and teams developing \glspl{IDAC}.  This coordination will promote effective use of the resources available to the community, while respecting the differing goals and scopes of the above-mentioned groups.

As a rule, for each area of effort identified, we envision near-term and longer-term activities.  In the near term, a reasonable approach would be for the interested stakeholders (identified below in \secref{subsec:specific-followup}) to discuss the needs, elaborating on the initial understanding outlined in this white paper, and outline a pathway to determining how best to meet them.  Depending on the state of existing tools in this area, that could include steps to determine the performance of the existing tools and opportunities to improve them to meet the needs of the LSST science community, or it could involve sketching out the design for a new tool if improving an existing one does not seem like a viable pathway forward. In several of the technical areas, existing software may do some of what is needed, but scalability is an issue.  Working demonstrations of these analyses on existing datasets can be used to identify the computational tall poles and potential avenues for optimization. Identifying responsible parties with the resources to develop these tools, and any formal collaboration development, should be another part of those discussions, setting the stage for the longer-term phase (in which the work is carried out).  

A first opportunity at starting those discussions is the 2022 Rubin Project \& Community Workshop, where representatives from all of the stakeholders will be present. The \gls{LINCC} Frameworks team, in particular, has a session with the goal of presenting the use cases developed here and identifying areas for progress. There are also sessions aimed at matching use cases and communities to \gls{IDAC} resources, and specific sessions organized by the LSST \glspl{SC}, such that there will be several opportunities for productive discussions.  Depending on the outcomes of these discussions, there will be additional opportunities for virtual discussion outside of and after this meeting.   

In several of the areas for future work described herein, ongoing work is taking place within Rubin Observatory, LSST \glspl{SC}, or other groups, with a variety of collaboration norms and  publication policies.  These groups are encouraged to review the recommendations for inclusive collaboration in \secref{sec:colrec} and identify opportunities to adopt best practices described therein, which could be particularly important if collaborations form across groups.  For example, explicitly writing down expectations for resource allocation, publication of results and authorship of resulting publications, and mechanisms to permit full collaborative engagement at all career stages would be highly valuable to set the stage for those collaborations.

Finally, we note that some of the common infrastructure needs identified across science use cases can be satisfied purely through software development.  Others may motivate rethinking which datasets are hosted at major computing centers, and in what form, in addition to motivating software infrastructure development.  For example, catalog-level cross-matching needs fall into the latter category, and may deserve particular attention when considering the development of \glspl{IDAC} and other settings where users anticipate carrying out early LSST science analyses.

\subsection{Follow-up activities in specific technical or scientific areas}\label{subsec:specific-followup}

In this section, we highlight more specific follow-up activities, which could follow the pattern described in \secref{subsec:general-followup} of including a discussion amongst key stakeholders (and open to all interested members of the LSST science community) to outline a plan for assessing how existing or newly developed tools might meet the identified needs.  More concretely, a list of anticipated steps is as follows, though in practice some variation may be needed for each area.
\begin{enumerate}
\item Identify the key stakeholders and ensure they (and the broader community) have an opportunity to join the discussion.
\item Work with domain experts to refine the requirements, starting from the use cases and requirements outlined in this document, but potentially including others as well.
\item Identify existing software that could fill some of the needs, potentially with further development (again, in some cases this has already been done within this white paper).  If there is any, then next steps could involve carrying out analysis of precursor survey data to benchmark performance and understand the key challenges in extending the software to work at LSST scale and precision.
\item Depending on the outcome of the previous exercise, develop a plan for next steps, whether it involves modifying existing software or developing new software.
\item Circulate to the community via \url{https://community.lsst.org/} and communication channels used by the groups involved (e.g., Rubin DM, LSST SCs, LINCC Frameworks, etc.), to get feedback and further discussion. 
\item Secure resources and formalize collaborations needed to carry out the work.
\end{enumerate}
\autoref{tab:scitech}  highlights the connections between scientific areas and technical infrastructure, which can be used to identify relevant LSST \glspl{SC} that should be consulted for each technical area.

There are several cross cutting technologies that need future exploration with close coordination amongst members of Rubin Observatory \gls{DM}, \glspl{IDAC} teams, and stakeholders from the science community. Care needs to be taken to ensure any new development is complementary to the capabilities provided by and compatible with Rubin \gls{DM}. These cross cutting technologies include batch processing (as mentioned in \secref{sec:techneeds} this was not discussed in depth at the workshop, but is an important area), image reprocessing, and image-based analysis. For example with image-based analysis,  a natural step would be to assess whether the image cutout services provided with the \gls{RSP} meet the needs described for image-based analyses at the level of individual objects, based on this white paper.

%\jeremy{I rephrased the above paragraph a bit to emphasize that these areas will be about collaborating with the DM team. Feel free to revert the changes if they don't work.}

%As mentioned in \secref{sec:techneeds}, batch processing (or a scalable job execution system) was not discussed in depth at the workshop, but is an important area for future exploration, including especially amongst members of Rubin Observatory \gls{DM}, \gls{IDAC}s teams, and stakeholders from the science community who can provide example analyses that would exercise such a system in differing ways (embarrassingly parallel versus other). Image reprocessing is another area where any discussion and development should center around capabilities provided by and compatible with Rubin \gls{DM}. A third area that connects tightly to Rubin \gls{DM} is image-based analysis. There, a natural step would be to assess whether the image cutout services provided with the \gls{RSP} meet the needs described for image-based analyses at the level of individual objects, based on this white paper.

Communities working on extragalactic science and cosmology should identify opportunities for joint photometric redshift infrastructure, especially infrastructure for uncertainty quantification and science-driven metric development.  Given that Rubin Observatory will provide the results of running a single photometric redshift algorithm in each release\footnote{This will be chosen following rigorous selection between a number of candidates; for details cf. \url{https://community.lsst.org/t/pz-lor-a-summary-of-the-proposed-pz-estimators-dm-shortlist/6308}.}, the existence of a community testbed to enable further algorithm development and feed the algorithms back to Rubin Observatory for future releases (along with information about impact across the science community) will be important.  Work within the LSST \gls{DESC} on \gls{RAIL} may serve as the foundation for a community-wide photometric redshift testbed, and a first step may be to assess whether it can naturally be extended to the range of capabilities discussed in this white paper.

Selection functions is an area where the key functionality is expected to come from Rubin Observatory.  Therefore, in this area we propose discussions should focus on what Rubin \gls{DM} will provide, what additional functionality may be needed on top of that (based on worked examples), and what software capabilities may already exist that could fill that niche or be extended so as to fill it.

While the communities working on time domain science are currently distributed across nearly all LSST \glspl{SC}, they should explore the potential for joint software infrastructure for light curve storage and analysis that would meet their needs while reducing the software development burden on any one group.  Given the need for interoperability with the light curves provided by Rubin Observatory and by alert brokers, including those groups in the discussion is important.  Improving existing light curve routines in Astropy until they work at the needed scale would be one pathway for exploration.

Regarding catalog-level cross-matching, this discussion has not only software implications (to support a variety of scales) but also affects how and where ancillary datasets are stored and made available to the LSST science community.
As outlined in \secref{sec:techneeds}, cross-matching is important in different regimes for different types of analyses.  A reasonable next step may be to delve deeper into these use cases and develop a set of requirements for those tools, along with some example precursor survey analyses with a variety of datasets that would help drive development. 

The forms of collaborative work (who is involved, is it driven by many groups or few, etc.) will determine how the work is carried out.  However, the potentially new collaborations to develop software infrastructure following the above actions should follow the recommendations in \secref{sec:colrec} as appropriate for the form of that collaboration.  For example, this includes outlining roles and responsibilities, expectations for disseminating the work and  allocating credit, and fostering an inclusive team environment.

\clearpage

\subsection*{Acknowledgments}

This workshop and the resulting white paper were initiated as part of the \gls{LINCC} Frameworks program, which was supported through the generosity of Eric and Wendy Schmidt by recommendation of the Schmidt Futures program. We acknowledge workshop support from Heising-Simons Foundation award 2020-1916 to \gls{LSSTC}.  We thank the Center for Computational Astrophysics (CCA) at the Flatiron Institute (a part of the Simons Foundation) and especially the administrative staff (Kristen Camputaro and Fatima Fall) for hosting the ``Data to Software to Science'' workshop that provided the catalyst for this document.

% Below this point are individual co-authors' acknowledgements.  Should reorder to match auth list once all are in place?
C.O.C.: This material is based in part upon work supported by the National Science Foundation Graduate Research Fellowship Program under grant No.\ (2018258765). Any opinions, findings, and conclusions or recommendations expressed in this material are those of the author(s) and do not necessarily reflect the views of the National Science Foundation. 
M.J.: This material is based upon work supported by the National Science Foundation under Grant No. AST-2003196. MJ, JM, HRS, SS, JRAD, and AJC wish to acknowledge the support from the University of Washington College of Arts and Sciences and the DiRAC Institute. The DiRAC Institute is supported through generous gifts from the Charles and Lisa Simonyi Fund for Arts and Sciences and the Washington Research Foundation.
Y.-Y.M. was supported by NASA through the NASA Hubble Fellowship grant no.\ HST-HF2-51441.001 awarded by the Space Telescope Science Institute, which is operated by the Association of Universities for Research in Astronomy, Incorporated, under NASA contract NAS5-26555. AC acknowledges support from NSF awards AST-1715122, AST-2107800, and OAC-1739419 and DOE awards DE-SC0011665 
JAVM aknowledges financial suport from CONACyT 252531. ABK  acknowledges funding provided by University of Belgrade-Faculty of Mathematics  (the contract 451-03-68/2022-14/200104), through the grants by the Ministry of Education, Science, and Technological Development of the Republic of Serbia. YT acknowledges the support of DFG priority program SPP 1992 ``Exploring the Diversity of Extrasolar Planets'' (TS 356/3-1). L. {\v C}. P. is supported by the Ministry of Education, Science and Technical development of R. Serbia (Project No 451-03-68/2022-14/ 200002).  CONACyT M\'exico under Grants No. 286897 and the Instituto Avanzado de Cosmolog\'ia Collaboration. GF acknowledges the support of the European Research Council under the Marie Sk\l{}odowska Curie actions through the Individual Global Fellowship No.~892401 PiCOGAMBAS. Work by RS, RB, LV, and SU was supported by the Preparing for Astrophysics with LSST Program, funded by the Heising Simons Foundation through grant 2021-2975, and administered by Las Cumbres Observatory. RB   acknowledges support from the project PRIN-INAF 2019  ``Spectroscopically Tracing the Disk Dispersal Evolution''. \DJ{}S acknowledges the funding provided by  Astronomical Observatory (the Ministry of Education, science and technological development of Republic Serbia contract 451-03-68$/$2022-14$/$ 200002). CSA: This work was partially enabled by funding from the UCL Cosmoparticle Initiative. AIM acknowledges support from the Max Planck Society and the Alexander von Humboldt Foundation in the framework of the Max Planck-Humboldt Research Award endowed by the Federal Ministry of Education and Research.  The Flatiron Institute is supported by the Simons Foundation. DI acknowledges funding provided by University of Belgrade-Faculty of Mathematics  (the contract 451-03-68/2022-14/200104), through the grants by the Ministry of Education, Science, and Technological Development of the Republic of Serbia, and the support of the Alexander von Humboldt Foundation. XL acknowledges support of the National Science Foundation Grant No.~2108841: ``Detecting and studying light echoes in the era of Rubin and Artificial Intelligence''; University of Delaware General University Research award GUR20A00782. AG acknowledges the financial support from the Slovenian Research Agency (grants P1-0031, I0-0033, J1-8136, J1-2460).
S.D. is supported by NASA through Hubble Fellowship grant HST-HF2-51454.001-A awarded by the Space Telescope Science Institute, which is operated by the Association of Universities for Research in Astronomy, Incorporated, under NASA contract NAS5-26555. JLS acknowledges NSF award AST-1816100. T.A. acknowledges support from ANID-FONDECYT Regular 1190335, the Millennium Science Initiative ICN12\_009 and the ANID BASAL project FB210003. KESF is supported by NSF AST-1831415 and Simons Foundation Grant 533845 and by the Center for Computational Astrophysics of the Flatiron Institute. LNdC would like to acknowledge the support from the Heising-Simons Foundation and  the CNPq/FAPERJ  program INCT do e-Universo.

Rubin Observatory is a joint initiative of the National Science Foundation (NSF) and the Department of Energy (DOE). Its primary mission is to carry out the Legacy Survey of Space and Time, providing an unprecedented data set for scientific research supported by both agencies. Rubin is operated jointly by NSF’s NOIRLab and SLAC National Accelerator Laboratory (SLAC). NOIRLab is managed for NSF by the Association of Universities for Research in Astronomy (AURA) and SLAC is operated for DOE by Stanford University.

\clearpage
\addcontentsline{toc}{section}{References}
\bibliographystyle{yahapj}
\bibliography{main,lsst}

\clearpage
\appendix

\section{Use case templates}\label{sec:templates}

We provided science use case and technical use case template for people to start work from at the workshop. These are reproduced here for reference.

\subsection{Science Use Case Template} \label{sec:SciTemplate}

\noindent {\color{red} Please make a copy of this document before editing (and save as a separate file in Science Use Cases )}

\noindent Title: <Description of YOUR Science Use Case>\\
Author: <Name and email>\\
Date: <Date of whitepaper>\\

\noindent {\bfseries Abstract}\\
Description of your science case. Points to consider:
\begin{itemize}
\item Imagine you are writing an abstract for a paper that will be published based on this science use case. What will you discover or measure? Provide a brief \gls{background} to the science, describe the results that are expected, include any numerical constraints that will come from the analysis (e.g. the distance to which the sources can be detected/analyzed, constraints on cosmological parameters that will be derived etc.)
\item What is the size and depth of the dataset that you will use and what type of data will you analyze (e.g. images, catalogs, single band data, light curves etc)
\item Comment on why you can't answer your science question today with today’s datasets and surveys. What is unique about the \gls{LSST} data for this science?
\item Provide references for \gls{background} reading for the science and analysis (e.g a couple of references or a pointer to a review article)
\end{itemize}
{\bfseries Science Objectives}\\
 Provide a list of science objectives for the analysis. Points to consider:
\begin{itemize}
\item Are there a set of science milestones that are needed to get to the final result? For example, will you need to initially separate stars and galaxies, \gls{deblend} the photometry, measure photometric redshifts. What are the requirements for each of these individual steps? For example, how well must you measure photometric redshifts to undertake your science case.
\item What will limit the science that you can achieve (e.g. the size of the sample, the accuracy of the photometry or \gls{astrometry})?
\item Is there existing work in this area (e.g. the development of brokers that will analyze and classify the alert streams) that might complement or make use of this use case (e.g. a catalog of supernovae can be used for cosmological distance estimates).
\end{itemize}
{\bfseries Challenges (what makes it hard)}\\
 Describe what are the primary challenges in undertaking the analysis. Points to consider:
\begin{itemize}
\item What challenges need to be overcome to undertake this use case. This could be technical challenges (e.g. how to analyze 107 light curves), algorithmic (e.g. current period finding algorithms are slow and don't work well with poorly sampled data), scientific (e.g. a lack of good models to fit to the data), logistic (e.g. getting access to follow-up telescope time).
\item Will the quality of the \gls{LSST} data impact the analysis (e.g. the number of false positives in the alert stream) and what will be needed to overcome any of these limitations (e.g. writing a specialized real-bogus classifier).
\item How often the analysis will be run (e.g. will it be rerun for each Rubin data release or periodically with the alert stream).
\item Is there an important timing/urgency constraint on your analysis (e.g. finding candidate microlensing events and triggering space-based followup)? If so, quantitatively, what are those time constraints?
\end{itemize}
{\bfseries Running on LSST Datasets (for the first 2 years)}\\
 What data sets and \gls{LSST} data products will be analyzed. Points to consider:
\begin{itemize}
\item What data sets will you utilize? For example, the alert stream, calibrated images, data release catalogs, the deep drilling fields. How long must the survey be in operation before you will run your analysis (e.g. you need 20 points in a light curve)
\item Are the \gls{LSST} data products sufficient for your analysis or will you need to create value-added catalogs or other derived data products
\item What is the size of the data you will use (e.g. the number of light curves you will analyze). Does your science case require analyzing a subset of the population or will you use all galaxies/stars in the data release.
\end{itemize}
{\bfseries Precursor data sets}\\
\begin{itemize}
\item What data can be used today to develop and test these use cases (e.g. the \gls{ZTF} public data set). Is this data public?
\item Are there other data sets that need to be assembled/collected 'prelaunch' to achieve your science? Do you need a validation set? Will there need to be cross-calibration (e.g. of photometry or astrometry) with those precursor data sets (e.g. gaining longer time baselines from adding LSST to \gls{ZTF} to \gls{CRTS})?
\end{itemize}
{\bfseries Analysis Workflow}\\
 Provide a step-by-step description of how you will analyze the data including:
\begin{itemize}
\item Data cleaning (e.g. removing bad data)
\item Derived or intermediate data sets and how these will be stored and accessed
\item Matching to existing data sets
\item The types of analysis techniques or \gls{software} packages that will be applied to the data (with a reference)
\end{itemize}
{\bfseries Software Capabilities Needed}\\
 Describe the functionality needed in the \gls{software} to undertake the science use case. Points to consider:
\begin{itemize}
\item Will we need to be able to query the \gls{LSST} archive and what parameters will be used in the query (e.g. what columns in the database will be used)?
\item Are you planning to access your data through community alert brokers? Are there \gls{software} components (e.g. classification algorithms) needed to enhance these brokers.
\item Are there new algorithms that will need to be developed for the science use cases or are there existing \gls{software} packages that will need modifying for the science use case (e.g. to optimize for speed)?
\item Is there new \gls{software} infrastructure needed to run at the scale of the LSST data and estimates of how quickly these analyses should be run? This could include the ability to access other datasets or cross-match to other catalogs?
\item How big are the data sets that will be run on and how slow are current approaches?
\item What will need to be stored from these analyses (e.g. for the derived data products) and how much data will that be?
\item What functionality will be needed to visualize the data, or the results, or to debug the analysis if it fails?
\end{itemize}
{\bfseries References for Further Reading}\\
\begin{itemize}
\item Provide a short bibliography that provides reference material regarding the science or the analysis.

\end{itemize}

\pagebreak
\subsection{Technical Case Template} \label{sec:TechTemplate}

\noindent {\color{red} Please make a copy of this document before editing}

\noindent Title: <Description of YOUR Technical Use Case>\\
 Author: <Name and email>\\
 Date: <Date of this use case>\\

\noindent {\bfseries Abstract}\\
 Technical use cases differ from science use cases in that they focus on the development of a specific technique that can be used across multiple applications. Examples of this include photometric redshifts, machine learning tools for finding outliers, tools to predict the orbits of asteroids including uncertainty propagation. For these technical cases the description could include:
\begin{itemize}
\item What problems will this technical use case solve? What is the size and depth of the dataset that you will use and what type of data will you analyze (e.g. images, catalogs, single band data, light curves etc).
\item Comment on why you can't use existing tools and frameworks. What is unique about the technical development you are proposing?
\item Provide references for background reading for the analysis (e.g a couple of references or a pointer to a review article)
\end{itemize}

\noindent {\bfseries Science Cases Needing this Tool}\\
 Provide a list of science cases that would utilize this tool or framework.

\noindent {\bfseries Requirements for the software}\\
 Describe what are the primary requirements and challenges for the software. Points to consider and describe:

\begin{itemize}
\item Does the data exist from Rubin in a format that can be used by this software or will new data products need to be generated (see \citealt{LSE-163}, \citealt{LDM-153} and \citealt{DMTN-153}).
What are the inputs and outputs for the technical analysis (e.g. photometry in multiple bands). Are there specific data structures that would be needed (e.g. light curves)
\item What volume of data must be processed and how often
\item Is there new software infrastructure needed to run at the scale of the LSST data and what are the estimates of how quickly these analyses should be run? This could include the ability to access other datasets or cross-match to other catalogs?
\item Are there other computational challenges that must be addressed (e.g. will memory be an issue when processing the data, will the outputs need to be stored, do database architectures exist for the outputs, will the outputs need to be fed to other packages or to visualization tools).
\item What outputs will need to be stored from these analyses (e.g. for the derived data products) and how much data will that be?
\item How much temporary storage is needed for processing step outputs which can be deleted on completion?
\item Will new functionality be needed to visualize the data, or the results, or to debug the analysis if it fails?
\item Why can’t we build these tools today (i.e. what makes this use case hard)
\end{itemize}

\noindent {\bfseries Running on LSST and other Datasets}\\
 Which data sets and LSST data products will be analyzed. Points to consider:
\begin{itemize}
\item What data sets will you utilize? For example, the alert stream, calibrated images, data release catalogs, the deep drilling fields. How long must the survey be in operation before you will run your analysis (e.g. you need 20 points in a light curve)
\item Are there precursor data sets on which this analysis can be run and validated. Will analysis of these precursor datasets lead to new publications.
\end{itemize}

\noindent {\bfseries Existing Tools}\\
\begin{itemize}
\item What tools exist today to undertake these analyses
\item What functionality is missing from these tools and frameworks that would need to be developed, or is there some issue with their application to the dataset at LSST scale? (i.e. what does not work well with these tools, are there missing capabilities or are they too slow or memory-intensive for LSST, etc.).
\item If new tools were built what components from existing frameworks would be critical to keep? (e.g. what parts of existing tools work well)
\end{itemize}

\noindent {\bfseries Computational Workflow}\\
 Provide a step-by-step description of how you will analyze your data including:
\begin{itemize}
\item How will you access the input data? Will we need to be able to query the LSST archive and what parameters will be used in the query (e.g. what columns in the database will be used)?
\item Are you planning to access your data through community alert brokers? Are there software components (e.g. classification algorithms) needed to enhance these brokers?
\item If you need to use a computational workflow system or resource management  (e.g. Pegasus, and Condor) which one would you use and why?
\item Describe where existing software packages would be used in each stage of the processing. Are there new algorithms that will need to be developed for the use cases or are there existing software packages that will need modifying for the science use case (e.g. to optimize for speed)?
\item Describe whether an analysis will need to be distributed across multiple cores or machines (e.g. can the analysis be undertaken as an embarrassingly parallel application or does it require message passing, is the analysis iterative requiring many passes of the data).
\item Do you understand the memory to number of cores ratio needed for the analysis ?
\item Describe how the outputs will be stored or visualized
\item Describe places where in the workflow there are software tools are missing or won’t scale to what you need
\end{itemize}

\noindent {\bfseries References for Further Reading}\\
\begin{itemize}
\item Provide a short bibliography that provides a list of the tools or algorithms used in this use case

\end{itemize}

\section{Science use cases} \label{sec:usecases}
\subsection{Introduction}
During the workshop, participants contributed science use cases that  are included here for completeness and as a resource for future work, meant to represent an incomplete sampling of high-priority science for the first two years of LSST.
These were organised in major scientific areas; that structure is preserved here as subsections.
Within each subsection,  one or more use cases are outlined.
% Generated file 
In this section:\
\begin{longtable}{l l p{0.6\textwidth} r}\hline
\hyperref[sec:ess]{\ref{sec:ess} }  &  & \hyperref[sec:ess]{ Extragalactic static science}   & \hyperref[sec:ess]{ \pageref{sec:ess} }  \\
 & \hyperref[sec:LSBdwarfG]{\ref{sec:LSBdwarfG} }  & \hyperref[sec:LSBdwarfG]{ Low surface brightness dwarf galaxy (candidate) catalog out to 100 Mpc: Multiple Science Cases}   & \hyperref[sec:LSBdwarfG]{ \pageref{sec:LSBdwarfG} }  \\
 & \hyperref[sec:SSexternalG]{\ref{sec:SSexternalG} }  & \hyperref[sec:SSexternalG]{ Stellar streams around external galaxies}   & \hyperref[sec:SSexternalG]{ \pageref{sec:SSexternalG} }  \\
 & \hyperref[sec:GalMorphML]{\ref{sec:GalMorphML} }  & \hyperref[sec:GalMorphML]{ Galaxy morphologies for LSST using machine learning with application to photometric redshifts}   & \hyperref[sec:GalMorphML]{ \pageref{sec:GalMorphML} }  \\
 & \hyperref[sec:GalPPSED]{\ref{sec:GalPPSED} }  & \hyperref[sec:GalPPSED]{ Estimation of galaxy physical parameters with \gls{SED} fitting}   & \hyperref[sec:GalPPSED]{ \pageref{sec:GalPPSED} }  \\
\hyperref[sec:ets]{\ref{sec:ets} }  &  & \hyperref[sec:ets]{ Extragalactic \gls{transient} science}   & \hyperref[sec:ets]{ \pageref{sec:ets} }  \\
 & \hyperref[sec:ImmediateClassification]{\ref{sec:ImmediateClassification} }  & \hyperref[sec:ImmediateClassification]{ Immediate Classification of Astrophysical Transients}   & \hyperref[sec:ImmediateClassification]{ \pageref{sec:ImmediateClassification} }  \\
 & \hyperref[sec:IndepthFast]{\ref{sec:IndepthFast} }  & \hyperref[sec:IndepthFast]{ In-Depth Studies of Fast Phenomena}   & \hyperref[sec:IndepthFast]{ \pageref{sec:IndepthFast} }  \\
 & \hyperref[sec:TooBeyondLigo]{\ref{sec:TooBeyondLigo} }  & \hyperref[sec:TooBeyondLigo]{ ToO Science (Beyond LIGO \gls{GW} Triggers)}   & \hyperref[sec:TooBeyondLigo]{ \pageref{sec:TooBeyondLigo} }  \\
 & \hyperref[sec:TDEfiltering]{\ref{sec:TDEfiltering} }  & \hyperref[sec:TDEfiltering]{ \gls{TDE} filtering}   & \hyperref[sec:TDEfiltering]{ \pageref{sec:TDEfiltering} }  \\
 & \hyperref[sec:PhotoPerf]{\ref{sec:PhotoPerf} }  & \hyperref[sec:PhotoPerf]{ Understand real photometric classification performance}   & \hyperref[sec:PhotoPerf]{ \pageref{sec:PhotoPerf} }  \\
\hyperref[sec:evs]{\ref{sec:evs} }  &  & \hyperref[sec:evs]{ Extragalactic variable science}   & \hyperref[sec:evs]{ \pageref{sec:evs} }  \\
 & \hyperref[sec:AugmentAGN]{\ref{sec:AugmentAGN} }  & \hyperref[sec:AugmentAGN]{ Augmenting \gls{AGN} Variability}   & \hyperref[sec:AugmentAGN]{ \pageref{sec:AugmentAGN} }  \\
 & \hyperref[sec:CondNeuralProc]{\ref{sec:CondNeuralProc} }  & \hyperref[sec:CondNeuralProc]{ Conditional Neural Processes for learning \gls{AGN} light curves}   & \hyperref[sec:CondNeuralProc]{ \pageref{sec:CondNeuralProc} }  \\
 & \hyperref[sec:FindAllAGN]{\ref{sec:FindAllAGN} }  & \hyperref[sec:FindAllAGN]{ Find All the \gls{AGN} ASAP}   & \hyperref[sec:FindAllAGN]{ \pageref{sec:FindAllAGN} }  \\
 & \hyperref[sec:ConnectSLAGN]{\ref{sec:ConnectSLAGN} }  & \hyperref[sec:ConnectSLAGN]{ Connection between short term variability of \gls{AGN} and their long term behavior}   & \hyperref[sec:ConnectSLAGN]{ \pageref{sec:ConnectSLAGN} }  \\
 & \hyperref[sec:MLAGN]{\ref{sec:MLAGN} }  & \hyperref[sec:MLAGN]{ Developing machine learning methods for AGN selection and calculating \gls{photometric redshift}}   & \hyperref[sec:MLAGN]{ \pageref{sec:MLAGN} }  \\
 & \hyperref[sec:DwarfAGN]{\ref{sec:DwarfAGN} }  & \hyperref[sec:DwarfAGN]{ Dwarf \gls{AGN} variability for intermediate-mass black hole identification}   & \hyperref[sec:DwarfAGN]{ \pageref{sec:DwarfAGN} }  \\
 & \hyperref[sec:CaustingCrossing]{\ref{sec:CaustingCrossing} }  & \hyperref[sec:CaustingCrossing]{ Mapping SMBH Near Fields with Microlensing}   & \hyperref[sec:CaustingCrossing]{ \pageref{sec:CaustingCrossing} }  \\
\hyperref[sec:luss]{\ref{sec:luss} }  &  & \hyperref[sec:luss]{ Local universe static science}   & \hyperref[sec:luss]{ \pageref{sec:luss} }  \\
 & \hyperref[sec:MappingMWStellarPops]{\ref{sec:MappingMWStellarPops} }  & \hyperref[sec:MappingMWStellarPops]{ Mapping the Accreted and Intrinsic Stellar Populations in the Milky Way}   & \hyperref[sec:MappingMWStellarPops]{ \pageref{sec:MappingMWStellarPops} }  \\
 & \hyperref[sec:LGdwarf]{\ref{sec:LGdwarf} }  & \hyperref[sec:LGdwarf]{ Local Group Dwarf Galaxies Bound and Unbound}   & \hyperref[sec:LGdwarf]{ \pageref{sec:LGdwarf} }  \\
 & \hyperref[sec:FaintEnd]{\ref{sec:FaintEnd} }  & \hyperref[sec:FaintEnd]{ The properties of the faint end of the Main Sequence: the stellar/sub-stellar boundary}   & \hyperref[sec:FaintEnd]{ \pageref{sec:FaintEnd} }  \\
 & \hyperref[sec:LocalIMF]{\ref{sec:LocalIMF} }  & \hyperref[sec:LocalIMF]{  The local \gls{IMF} as inferred from nearby star forming regions and clusters}   & \hyperref[sec:LocalIMF]{ \pageref{sec:LocalIMF} }  \\
\hyperref[sec:luts]{\ref{sec:luts} }  &  & \hyperref[sec:luts]{ Local universe variable \& \gls{transient} science}   & \hyperref[sec:luts]{ \pageref{sec:luts} }  \\
 & \hyperref[sec:SymbioticBinaries]{\ref{sec:SymbioticBinaries} }  & \hyperref[sec:SymbioticBinaries]{ Identifying symbiotic binaries by their color and variability}   & \hyperref[sec:SymbioticBinaries]{ \pageref{sec:SymbioticBinaries} }  \\
 & \hyperref[sec:LightEchoes]{\ref{sec:LightEchoes} }  & \hyperref[sec:LightEchoes]{ Light Echoes: study the reflection of transients on interstellar medium in the LSST Era}   & \hyperref[sec:LightEchoes]{ \pageref{sec:LightEchoes} }  \\
 & \hyperref[sec:CompactWD]{\ref{sec:CompactWD} }  & \hyperref[sec:CompactWD]{ Compact White Dwarf Binaries in \gls{LSST}}   & \hyperref[sec:CompactWD]{ \pageref{sec:CompactWD} }  \\
 & \hyperref[sec:MicroLensingGP]{\ref{sec:MicroLensingGP} }  & \hyperref[sec:MicroLensingGP]{ Analysis of Microlensing events by stars and compact objects}   & \hyperref[sec:MicroLensingGP]{ \pageref{sec:MicroLensingGP} }  \\
 & \hyperref[sec:YSOvariability]{\ref{sec:YSOvariability} }  & \hyperref[sec:YSOvariability]{ Young stellar objects and their variability}   & \hyperref[sec:YSOvariability]{ \pageref{sec:YSOvariability} }  \\
 & \hyperref[sec:LongPeriosMD]{\ref{sec:LongPeriosMD} }  & \hyperref[sec:LongPeriosMD]{ Long Period M dwarf Variability}   & \hyperref[sec:LongPeriosMD]{ \pageref{sec:LongPeriosMD} }  \\
 & \hyperref[sec:SubstellarCompanions]{\ref{sec:SubstellarCompanions} }  & \hyperref[sec:SubstellarCompanions]{ Identifying Substellar Companions to White Dwarfs}   & \hyperref[sec:SubstellarCompanions]{ \pageref{sec:SubstellarCompanions} }  \\
 & \hyperref[sec:RRLyrae]{\ref{sec:RRLyrae} }  & \hyperref[sec:RRLyrae]{ RR Lyrae Catalogs}   & \hyperref[sec:RRLyrae]{ \pageref{sec:RRLyrae} }  \\
 & \hyperref[sec:ExceptionalVari]{\ref{sec:ExceptionalVari} }  & \hyperref[sec:ExceptionalVari]{ Exceptional Variability: New Astrophysics \& Technosignatures}   & \hyperref[sec:ExceptionalVari]{ \pageref{sec:ExceptionalVari} }  \\
\hyperref[sec:solar]{\ref{sec:solar} }  &  & \hyperref[sec:solar]{ Solar system science}   & \hyperref[sec:solar]{ \pageref{sec:solar} }  \\
 & \hyperref[sec:NonTracklet]{\ref{sec:NonTracklet} }  & \hyperref[sec:NonTracklet]{ Non-Tracklet Discovery for Small Body Populations}   & \hyperref[sec:NonTracklet]{ \pageref{sec:NonTracklet} }  \\
 & \hyperref[sec:SmallBodyActivity]{\ref{sec:SmallBodyActivity} }  & \hyperref[sec:SmallBodyActivity]{ Characterizing Populations of Active Small Bodies}   & \hyperref[sec:SmallBodyActivity]{ \pageref{sec:SmallBodyActivity} }  \\
 & \hyperref[sec:InterstellarEjecta]{\ref{sec:InterstellarEjecta} }  & \hyperref[sec:InterstellarEjecta]{ Constraining the Number Density and Mass of the Galactic Interstellar Small Body Reservoir}   & \hyperref[sec:InterstellarEjecta]{ \pageref{sec:InterstellarEjecta} }  \\
 & \hyperref[sec:Multifrequency]{\ref{sec:Multifrequency} }  & \hyperref[sec:Multifrequency]{ Multiwavelength studies of Solar System moons and asteroids}   & \hyperref[sec:Multifrequency]{ \pageref{sec:Multifrequency} }  \\
 & \hyperref[sec:SmallBodyPops]{\ref{sec:SmallBodyPops} }  & \hyperref[sec:SmallBodyPops]{ Small Bodies in Rubin/LSST Data for Population-Level Studies}   & \hyperref[sec:SmallBodyPops]{ \pageref{sec:SmallBodyPops} }  \\
 & \hyperref[sec:ShiftStack]{\ref{sec:ShiftStack} }  & \hyperref[sec:ShiftStack]{ Shift-and-Stack for faint object detection}   & \hyperref[sec:ShiftStack]{ \pageref{sec:ShiftStack} }  \\
\hyperref[sec:cosmology]{\ref{sec:cosmology} }  &  & \hyperref[sec:cosmology]{ Cosmology}   & \hyperref[sec:cosmology]{ \pageref{sec:cosmology} }  \\
 & \hyperref[sec:CosmicShear]{\ref{sec:CosmicShear} }  & \hyperref[sec:CosmicShear]{ Weak lensing cosmology analysis / cosmic shear}   & \hyperref[sec:CosmicShear]{ \pageref{sec:CosmicShear} }  \\
 & \hyperref[sec:Type1aCosmo]{\ref{sec:Type1aCosmo} }  & \hyperref[sec:Type1aCosmo]{ Probabilistic Type Ia supernova cosmology analysis}   & \hyperref[sec:Type1aCosmo]{ \pageref{sec:Type1aCosmo} }  \\
 & \hyperref[sec:OptimalSpecCosmo]{\ref{sec:OptimalSpecCosmo} }  & \hyperref[sec:OptimalSpecCosmo]{ Optimal spectroscopic follow-up algorithms for Type Ia supernova cosmology}   & \hyperref[sec:OptimalSpecCosmo]{ \pageref{sec:OptimalSpecCosmo} }  \\
 & \hyperref[sec:CMBprobes]{\ref{sec:CMBprobes} }  & \hyperref[sec:CMBprobes]{ Cross-correlation between LSST and \gls{CMB} probes of gas physics}   & \hyperref[sec:CMBprobes]{ \pageref{sec:CMBprobes} }  \\
 & \hyperref[sec:Constrain]{\ref{sec:Constrain} }  & \hyperref[sec:Constrain]{ Self-consistent cosmological parameter constraints from galaxy clustering and galaxy-galaxy lensing using the \gls{DESI} Y1 \gls{LRG} sample}   & \hyperref[sec:Constrain]{ \pageref{sec:Constrain} }  \\
 & \hyperref[sec:WLCosmo]{\ref{sec:WLCosmo} }  & \hyperref[sec:WLCosmo]{ Weak lensing cosmology analysis / 3x2pt}   & \hyperref[sec:WLCosmo]{ \pageref{sec:WLCosmo} }  \\
\hline \end{longtable}

\subsection{Extragalactic static science} \label{sec:ess}
\subsubsection{Low surface brightness dwarf galaxy (candidate) catalog out to 100 Mpc: Multiple Science Cases} \label{sec:LSBdwarfG}

\cleanedup{Yao-Yuan Mao (May 13)}
\Contributors{ Yao-Yuan Mao (\mail{yymao.astro@gmail.com}), Shany Danieli (\mail{sdanieli@astro.princeton.edu})}
{ Date: March 28, 2022}

\paragraph{Abstract}
We will compile a catalog of nearby \gls{LSB} dwarf galaxy candidates using Rubin \gls{LSST} Year 2 photometric data. These candidates will encompass the majority of dwarf galaxies out to 100 Mpc, down to an $r$-band apparent magnitude of 24, with $>$90\% purity and $>$90\% completeness. The catalog will include the remeasured photometric properties and  distance estimates from a combination of surface brightness fluctuations, spectroscopic follow-up, and photometric redshifts. This census of dwarf galaxies will enable a range of science cases, including mapping of their fundamental distribution functions (e.g., the size--mass plane, the stellar mass function, etc.), and understanding what role the environment plays in shaping these galaxies by comparing isolated (``field'') and satellite galaxies. It will also allow stacking of these candidates to enable weak lensing measurements that will map the aggregate dark matter profile of low mass galaxies, and will provide a list of potential hosts of newly observed supernova and gravitational wave emitters.

\paragraph{Science Objectives}
\begin{itemize}
\item {\bf Science Case 1} - To advance our understanding of dwarf galaxy formation and evolution through a nearly complete sample of dwarf galaxies out to 100 Mpc. In particular, the comparison between field and satellite low-mass dwarf galaxies will allow us to understand how dwarf galaxies evolve, and how quenching mechanisms (e.g., reionization, host galaxy interactions) affect their star formation.
\item {\bf Science Case 2} - Dwarf galaxy candidates as lenses: use weak lensing measurement to map the dark matter profile of dwarf galaxies \citep[Sec.~3.2.1 of][]{2019arXiv190201055D}.
\item {\bf Science Case 3} - Host candidates for transient objects.
\item {\bf Main Product}: A catalog of dwarf galaxy candidates within 100 Mpc. It should be highly complete (including most true nearby dwarf galaxies) down to a specific $r$-band apparent magnitude (e.g., $r \leq 24$). It should also have high purity (i.e., contain few galaxies outside the desired redshift limit). For each object in this catalog, we will aim to provide:
\begin{itemize}
\item Photometric properties optimized for \gls{LSB} dwarf galaxies
\item Distance measurement for a subset, using surface brightness fluctuations \citep[e.g.,][]{2019ApJ...879...13C} or spectroscopic follow-up \citep[e.g.,][]{2021ApJ...907...85M}
\item ``Photometric distance'' for all objects, using an algorithm that is optimized for this distance range (i.e., within 100 Mpc)

\end{itemize}
\end{itemize}

\paragraph{Challenges (what makes it hard) }

\begin{itemize}
\item Issues with the source identification and characterization: sky subtraction, deblending/shredding, galactic cirrus.
\begin{itemize}
\item A large fraction of these very nearby dwarf galaxies are very low surface brightness objects, on which the source extraction algorithm may not perform well. At the image level, the sky subtraction and nearby starlight can significantly affect the identification and characterization of these low surface brightness galaxies \citep[e.g.,][]{2018ApJ...857..104G}. The main issues include: (1) a galaxy that is not identified by the algorithm at all; (2) a galaxy that is identified as multiple sources (shreds), each of which contains inaccurate photometry; (3) a galaxy that is identified correctly as a single source, but the photometry significantly differs from the true value (e.g., underestimating the luminosity due to missing outskirt light, or overestimating the luminosity due to contamination, or inaccurate sky subtraction); (4) fake sources from galactic cirrus or wrong sky subtraction.
\item Galaxies that are very close ($< 5$~Mpc) may be partially resolved, that is, some of the stars are identified as point sources, and the rest are identified as diffuse sources \citep[e.g.,][]{Mutlu_Pakdil_2021}. This case is particularly tricky because the number of point sources may not be enough for algorithms that search for resolved galaxies (e.g., satellite dwarf galaxies in the Milky Way) to pick them up, and the diffused source may have incorrect photometry.
\end{itemize}
\item Needs an algorithm to estimate the distance of \gls{LSB} dwarf galaxy candidates using images or improved photometric properties.
\begin{itemize}
\item Existing photometric redshift algorithms do not perform very well in this very low-redshift ($z < 0.05$) regime, if we are aiming for a high completeness and high purity sample. But recent progress has been made on this front \citep[e.g.,][]{2022ApJ...927..121W,2021arXiv211203939D}.
\item This is difficult mostly due to the lack of training data (dwarf galaxies with known redshift) in the regime that we are aiming for ($z < 0.05$, $r > 20$).
\item To obtain training data, we will need spectroscopic follow-up, which would be expensive. We will need to be strategic about the set of objects that we follow up that can optimize the performance of the algorithm.
\item The mis-characterization in the object catalog (as discussed above) may also mean that the algorithm may need to be run on images (using \gls{CNN}, for example), rather than on catalog entries. Training and running the algorithm on images also requires significant computing resources.
\item We may need a simple algorithm that first selects a high completeness but low purity sample based on catalog-level information only, and another more sophisticated algorithm that runs on images to improve the purity of the sample.
\end{itemize}
\end{itemize}

\paragraph{Running on LSST Datasets (for the first 2 years)}
Data sets and \gls{LSST} data products that will be analyzed:
\begin{itemize}
\item Object catalog
\item Calibrated images
\item Flags from deblending process
\item Cross-matched map with a galactic dust map
\end{itemize}

\paragraph{Precursor data sets}
\begin{itemize}
\item \gls{DESI} Legacy Imaging Survey, \gls{DES}, \gls{DELVE} (these are not as deep, but cover large sky area)
\item \gls{HSC}
\item Spectroscopic redshift surveys: \gls{GAMA} \citep{2022MNRAS.513..439D}, \gls{DESI}, \gls{SAGA} \citep{2017ApJ...847....4G,2021ApJ...907...85M}, Merian Survey \citep{2020PDU....3000719L}

\end{itemize}
\paragraph{Analysis Workflow}
\begin{itemize}
\item Visual inspection to understand potential issues with the identification and characterization of the object catalog (hopefully most of this will be done during commissioning)
\item Select a high completeness but lower purity sample based on catalog-level information (magnitude, color, size, \gls{photo-z}) only. At this step, we might need to access some deblending flags, so that we can work around issues (shreds, missing sources) in the existing catalog.
\item Obtain calibrated cutouts for this sample, re-fit the photometry with models that are optimized for \gls{LSB} dwarf galaxies. Produce a new catalog with matched photometry and colors.
\item Measure distances for a subset of this sample, using surface brightness fluctuations (for sources where this approach is possible), and with spectroscopic cross-match/follow-up.
\item Produce “photometric distance” for all objects in the sample, using the improved photometry and images, with the training data from the last step. This “photometric distance” algorithm used here should be optimized for the very low-redshift regime. It may be a repurposed \gls{photo-z} algorithm, or a ML approach (e.g., \gls{CNN}) that runs on images directly.
\item Improve the purity of this catalog by removing or flagging objects whose distance is greater than 100\,Mpc .

\end{itemize}
\paragraph{Software Capabilities Needed}
\begin{itemize}
\item Image processing software should flag potential issues with sky subtraction, deblending/shredding, and galactic cirrus, so that an add-on program can revisit those regions to find potential \gls{LSB} galaxies or to re-measure their photometric properties.
\item An algorithm that can select a high completeness but low purity sample based on catalog-level information. This output will be used for follow-up distance measurement (e.g., surface brightness fluctuations, spectroscopic redshifts).
\item A more sophisticated algorithm that uses all available training data (distance information) to estimate the distance of \gls{LSB} dwarf galaxy candidates using images or improved photometric properties.

\end{itemize}
\paragraph{References for Further Reading}

\citet{2017ApJ...847....4G,2018ApJ...857..104G,2019ApJ...879...13C,2019arXiv190201055D,2020PDU....3000719L,2021arXiv211203939D,2021ApJ...907...85M,2022MNRAS.513..439D,2022ApJ...927..121W}

\pagebreak
\subsubsection{Stellar streams around external galaxies} \label{sec:SSexternalG}
%\WOM{ No workflow }\\
\cleanedup{Sarah Pearson}
\Contributors{Sarah Pearson (\mail{spearson@nyu.edu})}

\paragraph{Abstract}
 With Rubin, we can finally hope to detect stellar streams from accreted dwarf galaxies around external galaxies in a statistical sense. Several other surveys have detected stellar streams around external galaxies, but not to a complete depth nor in a uniform sense. With detection of streams in Rubin \gls{LSST}, we can measure stream frequency, length, shape, and color, as well as identify potential progenitors. This will allow us to match observations with the full picture of expectations from cosmological hierarchical galaxy formation simulations, thus helping to constrain galaxy formation theories. We will aim to decipher the full accretion histories of galaxies beyond the Milky Way as well as determine the low-mass luminosity function of accreted dwarfs. We will target both massive and dwarf external galaxy hosts, as dwarfs galaxies themselves may have accreted less massive dwarfs. With follow-up radial velocities, Rubin stellar streams may enable us to learn about the dark matter distribution and mass of these external galaxies. \cite{2018arXiv181204897L} described in detail a project aimed at detecting stellar streams with Rubin \gls{LSST}, combined with data from other telescopes to help with cirrus and star galaxy separation. This document builds on their science use case and explores what data, software, and analysis will be needed to optimize stellar stream science with Rubin.

\paragraph{Science Objectives}

\begin{itemize}
\item Detect \gls{LSB} features, such as streams, in Rubin images.
\begin{itemize}
\item Run stream/shell-finding/classification algorithms on \gls{LSST} data (e.g., \citealt{2019MNRAS.486.3604H}, \citealt{2022ApJ...926..166P})
\item Characterize the lengths, \gls{SB}, widths of detected streams
\item Search for potential progenitor remnants along streams
\item Measure the average color and color profiles of the streams
\item Determine whether there are multiple low surface brightness features (shells/streams) within the host galaxies
\end{itemize}
\item Once we have detections: compare findings to expectations from cosmological simulations of \gls{LSB} features (e.g., Auriga cosmological simulations, \citealt{2016diga.confE..34G}).
\item To learn about the potentials of the host galaxies of the streams we might need follow-up radial velocity measurements.
\end{itemize}

\paragraph{Challenges (what makes it hard) }

\begin{itemize}
\item How do we determine if something is a stream/shell? Detection depends on the contrast with the \gls{background}. Successful searches will need to comb through huge amounts of data (algorithms for finding such features exist).
\item At what points in the 10-year Rubin \gls{LSST} survey will we be able to detect these features (using estimates of the surface brightness of \gls{LSST} images + theoretical estimates from simulations)? Estimates of when we should see the brightest vs. when we should see a complete set of \gls{LSB} features would be interesting, e.g., comparing detection efficiency in the 1-, 2-, and 10-year stacked image depths.
\item We need to handle star galaxy separation in complex image morphology contexts, to avoid contamination of stream photometry by background objects.
\item How do we avoid contamination from Galactic cirrus? We may need to restrict targets to those well off the Galactic plane.
\item Subtraction of both background objects and the host galaxy is needed to improve the contrast of diffuse structures, and these will have complex image morphologies.
\item From \cite{2018arXiv181204897L}: “Reflections and scattered light in the optical path are a real concern for deep wide-field surface photometry…… static star subtraction models applied to stacked images will be insufficient to remove these features. Instead, active-subtraction techniques applied at the data reduction stage (e.g., \citealt{2009PASP..121.1267S}) must be used to deal with these reflections by modeling and removing them on a star-by-star basis from the individual raw image frames. This makes it imperative that \gls{LSST} data servers provide users with the raw images, not just the image stacks.”
\item For multiband images: variation in the number of visits for each band, which will affect the depth reached.
\end{itemize}

\paragraph{Running on \gls{LSST} Datasets (for the first 2 years)}
\begin{itemize}

\item While we will detect the brightest streams with just a single visit, more work is needed to characterize the completeness of potential stream detections as a function of stacked depth.
We can compare to the  expected \gls{SB} limits of substructure from simulations (e.g., \citealt{Bullock2005}).
\begin{itemize}
\item From \cite{2018arXiv181204897L}: “we have made an approximate estimate of the expected low surface brightness limit that the LSST can provide after the scheduled 825 visits to the same sky location. This corresponds to a total amount of time on source of 3.44h. The expected surface brightness limits will be ($3\sigma$; 10x10 arcsec$^2$ boxes): 29.9 (u), 31.1 (g), 30.6 (r), 30.1 (i), 28.7 (z) AB mag arcsec$^{-2}$ .
Each visit that consists of 30 seconds on-source will correspond to the following limits ($3\sigma$): 26.6 (u), 27.8 (g), 27.3 (r), 26.8 (i), 25.4 (z) AB mag arcsec$^{-2}$ .”
\end{itemize}
\item Need individual images: “individual images should be made available, not just coadds, as scattered light is much easier to remove from individual frames than from image stacks.”
\item Ideally, multiband images, maybe other datasets for comparison.
\item Multiple postage stamps stacked together, with sizes of \gls{postage stamp} appropriate for the target galaxy.
\item Parallel over stacks of images (process individual galaxy, but all of the images should be available, stacking \gls{algorithm} is a custom coaddition \gls{algorithm} (only highest quality data).
\end{itemize}

\paragraph{Precursor and contemporaneous data sets}
\cite{2018arXiv181204897L}: “Combined \gls{LSST}/Euclid/WFIRST data set will provide a broad wavelength baseline for the estimation of the ages, metallicities and masses of the stellar populations of disrupted companions”

\paragraph{Analysis Workflow}
\begin{itemize}
\item Remove scattered light from individual images
\item Background subtraction to see diffuse structures
\item Stacking of multiple postage stamps surrounding galaxies of interest to see connecting features
\item Search for streams in data with stream-finding algorithms or by eye (make sure the features are not cirrus, remove false positives)
\item Compare to multi-band images or other data sets (e.g, Euclid, SDSS, Roman)
\item Characterize lengths, morphology, colors of objects
\item Search for progenitor along the detected stream candidates
\end{itemize}

\paragraph{Software Capabilities Needed}
 Need to run stream-finding algorithms (\citealt{2022ApJ...926..166P}) on images as well as substructure classification algorithms (\citealt{2019MNRAS.486.3604H}) and determine which are false positives.

\paragraph{References for Further Reading}
 Spectroscopic follow-up to Rubin: \cite{newman19}\\ %\url{https://arxiv.org/pdf/1903.09325.pdf} \\
 Stellar streams LSST white paper by: \cite{2018arXiv181204897L}\\
 How to get radial velocities of \gls{LSB} features: \cite{toloba2016}\\
 Dwarf companions beyond MW with \gls{LSST}: \url{https://noirlab.edu/science/sites/default/files/media/archives/documents/scidoc1994.pdf} \\

\pagebreak
\subsubsection{Galaxy morphologies for LSST using machine learning with application to photometric redshifts} \label{sec:GalMorphML}

\cleanedup{Ilin Lazar}

\Contributors{Ilin Lazar (\mail{i.lazar@herts.ac.uk}), J.\ Antonio Vazquez (\mail{jvazquez@astro.unam.mx})}
{4/01/22}

\paragraph{Abstract}
Morphology is a fundamental parameter, not only essential for the full spectrum of extra-galactic LSST science but also as a valuable prior in photo-z pipelines that can significantly improve photo-z accuracy (e.g., \citealt{2019A&A...621A..26P}). LSST offers an unparalleled combination of depth, area and statistics, with  $\approx$20 billion galaxies expected from its 18,000 deg$^2$ footprint with a point-source depth of $\approx$27.5 mag in the full 10-year stack. This offers a game-changing opportunity to study galaxy morphologies with better precision and statistics than ever before. Since spectroscopy will be sparse, significant investment is being made in photo-z pipelines (e.g., via in-kind contributions by Hatfield/Hsieh et al.\ in Galaxies and DESC). Improved photo-z measurements (which are essential to LSST science), using morphologies as input, will bring fundamental benefits to the entire LSST community.

A rich literature exists on measuring morphologies in surveys, from visual inspection (e.g., \citealt{10.1093/mnras/stac635}), including those using citizen science systems like \gls{GZ} \citep{2013MNRAS.435.2835W}, to automated methods, either via simple measures (such as Sersic profile fits, measurement of Concentration/Asymmetry/Smoothness \citep{2003ApJS..147....1C}, etc.) or sophisticated supervised or unsupervised machine-learning (ML) techniques (e.g., \citealt{2018MNRAS.473.1108H}). However, the unprecedented size of LSST requires a radically different approach. Visual inspection, even using \gls{GZ}, will be prohibitively time-consuming. Furthermore, since the morphological detail in galaxies will increase as LSST becomes deeper, morphological catalogs will be needed at multiple depths (e.g., from every data release). This makes LSST also challenging for recent supervised ML, since it may be difficult to repeatedly produce large training sets on short timescales, so this has to be well structured in advance. Another solution is unsupervised ML (UML), which can autonomously group morphologically-similar objects into a small number of “morphological clusters”, without training sets. Each cluster (rather than millions of individual galaxies) can then be collectively labelled, e.g.\ into Hubble types, via supervised ML. A member of this project developed such a UML algorithm and validated it on Hyper Suprime-Cam (HSC) surveys \citep{10.1093/mnras/stz3006}, one of LSST’s precursor data sets.

The algorithm randomly samples a large number of patches (of sizes 3-4 pixels squared) in survey images and converts each patch into a ‘feature vector’ that holds information about its properties (e.g., color/texture). Patches are clustered via a growing neural gas network and hierarchical clustering and galaxies with similar patch properties grouped together. Arbitrarily large galaxy samples, for M > 10$^9$ M$_\odot$ and $z<1$ (beyond which classification is difficult from ground-based imaging) are compressed into $\sim$150 clusters (typical purity >90\%). These are easily visually associated with Hubble types and obey known trends in e.g.\ stellar mass vs star-formation rate. Approximately 150 clusters are needed because identical morphologies at very different redshifts occupy separate clusters (since galaxies change in size with redshift). Peculiar objects (e.g., mergers) naturally end up in separate clusters. We will provide a classification catalog based on HSC-SSP U/Deep Data Release 3  in 2022 (Lazar+ in prep).

\paragraph {Science/Technical Objectives}

\begin{itemize}
\item Use of ML algorithms to create morphological catalogs for LSST Data Preview 2 (DP2) and subsequent data releases (i.e., DR1, DR2) served through the \gls{RSP}, using supervised and unsupervised methods.
\item Create an RSP interface to enable users to morphologically classify any future LSST data (e.g., varying depth or sky coverage), extending the utility of this project to perpetuity.
\item Input the morphologies in the developed codes, demonstrate quality gains and create improved photo-z estimates that can act as a demonstrator for other Rubin photo-z efforts.
\item Investigate peculiar/rare objects and the filamentary nature of the Local Universe as a function of galaxy morphology in a statistical sense.

\end{itemize}
\paragraph{Challenges (what makes it hard) }
\begin{itemize}
\item The algorithm is computationally intensive, and currently takes 10 days to process 1000 square degrees for all objects with $z<0.5$ on a 4 core/8 GB RAM machine.  One could decrease the algorithm processing time and/or computational needs by exploring alternate data management or reduction plans such as using Principal Component Analysis or excluding any data which does not bring any benefit to the clustering procedure. This can be tested using precursor HSC data.
\item Propose for CPU time to assure quick classification for every data release. 200 processing cores per data release which amount to 10\% of the total LSST computing resources (that can be proposed for) would allow for the algorithm to finish processing in less than 10 days.
\item GPU implementation can be used for supervised ML, possibly in an \gls{IDAC}.
\item Need to build the RSP interface in the most robust and user friendly way possible.

\end{itemize}

\paragraph{Precursor data sets}
\begin{itemize}
\item HSC DR3 DEEP and WIDE (Tested in Lazar in prep.)
\item HSC DR2 DEEP (Tested in \citealt{10.1093/mnras/stz3006})
\item Hyper Supreme Cam (LSST precursor)
\item The HST datasets
\item The DESI Legacy Surveys \citep{2019AJ....157..168D}
\item The LSST Data Previews
\end{itemize}
These data sets will be beneficial to act as training sets. The algorithm can be run as soon as the first data release is online if unsupervised machine is used, and if training is done on precursor or other external data.

\paragraph{Analysis/Implementation Workflow}
\begin{itemize}
\item We will query galaxy cutouts and catalog properties from the LSST archive. Example catalog properties needed: ra, dec, photometric redshift, stellar mass, SFR, colors, object radius.
\item The classification/clustering will be done with algorithms provided by ScikitLearn and suitable processing parallelization will be used.
\item The initial training sets will consist of external catalogs and the DP2 images available through the RSP. DP2 will include $\sim$100 deg$^2$ 20-year-depth imaging and $\sim$1600 deg$^2$ in $g$ and $i$ bands to Year 1 depth.
\item We will test on, and create morphological catalogs for, DP2 and DR1. We will seek guidance about catalog contents from the Galaxies and Informatics and Statistics SCs (via telecons/meetings). The deliverables are DP2 and DR1 catalogs, served through the RSP and described in a tech note.
\item Once the first DR comes out, the algorithms (supervised and unsupervised) will be run on these images and morphological catalogs will be generated. About 1Tb of storage will be needed to save these catalogs.
\item If 100 cores are used the training timescales may last for a couple of days to a week depending on the depth of the data and sky coverage.
\item If GPUs at IDACs are used, the training timescale could be reduced by a factor of 2. However, It will be very important to estimate what is cheaper and more efficient, using more cores within the RSP or moving data to IDACs with GPU facilities.
\item RSP interface/documentation – The deliverables are (1) an RSP interface to run our algorithms on any LSST data e.g.\ by cloning the algorithm’s GitHub repository into the LSST structure (or the user’s RSP file system) as a library, which the user can import and use via a Jupyter Notebook or Python script, (2) detailed user documentation for using this interface.
\item Create catalog for DR2 and forthcoming releases  – We will deliver morphological catalogs for different data releases as the LSST survey progresses, served through the RSP.
\end{itemize}

\paragraph{Software Capabilities Needed}
\begin{itemize}
\item Existing tools:
\begin{itemize}
\item External tools to develop ML algorithms have been developed and optimized to do efficient calculation. Tensorflow, Keras, Pytorch are the most accessible and friendly libraries to work with.
\item The most popular technique used for galaxy classification is convolutional neural networks (CNNs) (e.g., \citealt{Huertas-Company15}, \citealt{2015MNRAS.450.1441D}, \citealt{2021MNRAS.503.4446C}, \citealt{2022MNRAS.509.3966W}) using training data mainly from the zooniverse. Other techniques are using random forest classifiers, Support Vector Machines (e.g., \citealt{Goulding2018}) or unsupervised algorithms (e.g., \citealt{2020MNRAS.494.3750C}). Most of these techniques require large amounts of training data, are time consuming and may not be able to operate efficiently at LSST scales. Therefore, looking for novel alternatives is necessary.
\end{itemize}
\item What is it needed?
\begin{itemize}
\item An \gls{RSP} cloud account with Jupyter Notebook and terminal.
\item The algorithm will run using parallel processing.
\item For general use: At least 4 cores, 8Gb of RAM and 10 Gb of storage for any user who wishes to use the ML algorithms on the \gls{RSP} (for faster processing times 30 cores or more if possible)
\item For large scale catalogue production: Possible accessibility to 200 cores for each LSST yearly release; maybe even have a portion of the LSST CPU capabilities reserved for ML based applications on a yearly basis.
\item Need an efficient architecture within the RSP to be able to import training data in large scales from other surveys.
\item Need for an efficient data transmission to local IDACs with GPU facilities for supervised ML.
\end{itemize}
\item Additional Requirements:
\begin{itemize}
\item Need for combined calibrated images in all bands to carry out the classification.
\item This can be done either by combining images in the \gls{RSP} during the training process or having previously combined cutouts in PNG format. Meanwhile the first one will require computer power to accelerate the process, the second one will require additional 2 Tb of storage for every 500 millions of 50x50 pix images.

\end{itemize}
\end{itemize}
The classification process is expected to be repeated at every data release to generate morphological catalogues.

\paragraph{References for Further Reading}
\cite{2021MNRAS.503.4446C}
Galaxy finder  \url{https://share.streamlit.io/georgestein/galaxy_search} ); \cite{Hayat_2021,10.1093/mnras/stz3006,2018MNRAS.473.1108H,10.1093/mnras/stz3006,10.1093/mnras/stac635,2013MNRAS.435.2835W}

\pagebreak
\subsubsection{Estimation of galaxy physical parameters with \gls{SED} fitting} \label{sec:GalPPSED}

\cleanedup{<Gabriele Riccio>}

\Contributors{ Gabriele Riccio (facilitator, \mail{gabriele.riccio@ncbj.gov.pl}), Charlotte Olsen, Raphael Shirley, Sam Schmidt, Julia Gschwend, Viviana Acquaviva}
{26/03/2022}

\paragraph{Abstract}
In the past 20 years, the study of the multi-wavelength emission of galaxies from X-rays to radio was found to be necessary to properly analyze the physical properties of galaxies. Because the \gls{SED} is the result of a complex interplay of several components, such as old and young stars, stellar remnants, the interstellar medium, dust, and supermassive black holes, only the  panchromatic view of galaxies can give the full information about their physical properties. To fully comprehend the interactions between these parts, the simultaneous use of different spectral ranges is needed.
As broad-band photometry is much less expensive than spectroscopy in terms of observation time, modeling the broad-band SED of galaxies has become one of the most commonly employed methods to evaluate and constrain the  physical properties. In this way, properties such as the \gls{SFR} and stellar mass, which are essential for a complete understanding of galaxy formation and evolution, can be evaluated.
However, modeling the \gls{SED} can be an intricate problem because galaxies with very different properties can look similar over some wavelength ranges: that is, a young dusty galaxy can appear to be an old dust-free galaxy because they both look red in the optical.
This is particularly the case when restricted wavelength ranges, instead of the full \gls{SED}, are considered. The full SED is rarely available. This makes estimating the physical properties with only a limited wavelength range a great challenge for SED modeling. Considering the large portion of sky that LSST will observe and the depth of forthcoming observations, it is expected that LSST will unveil a significant number of faint galaxies that have remained undetected in current wide-area surveys or that do not have any counterpart in the available multi-wavelengths catalogs. The question is: ``how can we use LSST optical observations to obtain estimates of the main physical properties of galaxies, and how realistic and reliable they would be?''

\paragraph{Science Objectives}

\begin{itemize}
\item Fitting of  galaxy \gls{SED}s
\item Estimation of galaxies main physical properties, such as \gls{SFR}, stellar mass, dust luminosity.
\item Test of the reliability of these estimates and flags for possible failures.

\end{itemize}

\paragraph{Challenges (what makes it hard) }
\begin{itemize}
\item A proper joint estimation of redshift and physical parameters is computationally intensive, and many methods ``cheat'' and separately estimate a fixed redshift before fitting for the physical parameters. At the very least, that external redshift must be probabilistic rather than deterministic to be meaningful for \gls{LSST} data, meaning the SED fitting procedures would need to be more advanced than they currently are.
\item Many photo-z codes fit with fixed matched aperture photometry, which gives consistent stellar populations within that aperture; however, this may be only a fraction of the entire galaxy, so fitting for the physical parameters of the total galaxy may be biased by such an aperture approach (this is mainly a problem for very large, extended galaxies with obvious multiple components)
 \item The limited wavelength coverage of the $ugrizy$ filter set limits the robustness of physical parameter estimates compared to those which include \gls{UV} and \gls{IR} coverage.  This will make comparisons with the more extensive data available in the deep drilling fields a key to calibrating SED fitting and physical parameter estimation methods.  Extensive documentation of the limitations will also be necessary, given that a large portion of the estimates may be poor for lower S/N objects and those detected in fewer bands.
\item Provenance tracking as these fits will be run multiple times (each data release and in between) and we need to store and link the outputs

\end{itemize}
\paragraph{Running on \gls{LSST} Datasets (for the first 2 years)}
What data sets and \gls{LSST} data products will be analyzed. Points to consider:

\begin{itemize}
\item Deep drilling fields for validation through multiwavelength counterparts, and general data release catalogs for estimation of the parameters.
\item We will be using the \glspl{DDF} and test on commissioning data. This must be supplemented with \gls{IR} photometry from legacy surveys in order to constrain galaxy properties (beyond $M_*$ and photo z)
\item We will be using a subset of the \glspl{DDF} such that the galaxies have sufficient \gls{SNR} and bands. The size of the dataset will be determined by the number of DDF galaxies for which we have reliable crossmatches in \gls{IR} datasets
\end{itemize}

\paragraph{Precursor data sets}
\begin{itemize}
\item As \gls{SED} fitting requires in general multi-wavelength photometry, up to now several legacy surveys (e.g., \gls{HELP}) provide mid-far \gls{IR} photometry of many fields around the sky.
\item \gls{HSC}
\item Spectroscopic surveys
 \end{itemize}

\paragraph{Analysis Workflow}

\begin{itemize}
\item Galaxies and stellar identification (to flag in the catalog)
\item Preparation of the data for \gls{SED} fitting (fluxes + photo-z estimates)
\item Iterate on quality of the data (visualization in color space to identify outliers)
\item Matching to existing data sets where complementary band info exists.
\item Undertake the \gls{SED} fitting with parametric and non-parametric codes (e.g., CIGALE, Prospector).
\item Determine quality flags to apply to the data

\end{itemize}

\paragraph{Software Capabilities Needed}
More sophisticated SED fitters are absolutely needed for \gls{LSST} to advance our understanding of galaxy populations beyond current capabilities. Physical parameters like star formation history and stellar mass cannot be separated from redshift inference, or at the very least, inference of the former must account for the nontrivial uncertainty inherent in the latter.

The functionality needed in the \gls{software}:

\begin{itemize}
\item We will query the \gls{LSST} archive to have measurements (flux, mag, color, size, PSF fit, date), model fit (e.g., point-source, bulge-disk), deblending parameters, aperture surface brightness measurement, photo-z (if available).
\item Best-fitting SEDs and tables containing results of the fit need to be stored, resulting in as much data as the \gls{LSST} data products.
\item Custom photometry could be required to adapt \gls{LSST} images to the same PSF of ancillary data images.
\end{itemize}

\paragraph{References for Further Reading}
CIGALE: a python Code Investigating GALaxy Emission - \cite{Boquien_2019} \\
Preparing for \gls{LSST} data. Estimating the physical properties of z < 2.5 main-sequence galaxies - \cite{Riccio_2021}

\pagebreak

\subsection{Extragalactic \gls{transient} science} \label{sec:ets}
\subsubsection{Immediate Classification of Astrophysical Transients} \label{sec:ImmediateClassification}
%\WOM{Citations to main.bib please, No challenges section, no early data}
\Contributors{Ann Zabludoff (\mail{aiz@arizona.edu})}
{March 27, 2022}
\cleanedup{Zabludoff}

\paragraph{Abstract}
The time-domain community expects millions of new alerts from the Rubin Observatory nightly. After initial filtering, many will be identified as extragalactic, terminal (i.e., explosive) transients. How do we identify which are  \gls{SN}e, \gls{GRB} afterglows, \gls{TDEs}, or more exotic phenomena, before they fade and can no longer be followed up? Tools are necessary to incorporate the properties of the host galaxy, including early LSST imaging, the evolving photometry of the \gls{transient}, the spatial offset of the \gls{transient} from its host, the environment of the host, and data from other surveys, to develop rapid probabilistic classifications, which can then be disseminated by LSST event brokers like \gls{ANTARES}.

\paragraph{Science Objectives}
Here we focus on classifying \gls{transient}s, perhaps even before they occur,
solely from prior measurements of their host galaxy properties. Several studies correlating the properties or types of transients with their host galaxies exist \citep{2021GHOST,2022ApJS..259...13Q}. In general, transients without hosts are cross-matched with galaxy catalogs to find host candidates, and host features are compiled from previous and on-going surveys. Exploration of \gls{transient}-host connections in these testing and training data will %likely
help to optimize large-survey \gls{transient} brokers by providing \gls{transient} classifications only from existing space- and ground-measured host galaxy properties. In prioritizing which transients to follow-up with additional observations, such pre-explosion classification is itself valuable, as well as a provider of priors for other classifiers, like those incorporating \gls{LSST} light-curves,  \gls{LSST} imaging, and data from elsewhere.

Utilizing the known properties of host galaxies is a path to classifying TDEs and other \gls{transient} types without waiting days, weeks, or months for \gls{LSST} to provide suitable light-curve data.  Yet clear correlations between \gls{transient} type and host properties have been elusive in past work. For example, while host morphology has been cited as a critical feature, Type Normal Ia \gls{SN}e occur in both star-forming and dead galaxies, whereas Types II P, Ib, and Ic, which arise from different progenitor stars, are all found in star-forming galaxies. Less common 87A-like and Ic-BL supernovae have been observed in dwarf galaxies, whereas \gls{SLSN}-II are detected in both dwarfs and more massive hosts. The role of host galaxy metallicity is similarly murky, historically biased by low metallicity dwarfs that dominate the local volume. The relationships of \gls{transient} types to their host galaxies have not been explored with even a fraction of the data now available nor with a focus on using such links for immediate, i.e., ``night of,” \gls{transient} classification.

%In a pilot study, we have trained and tested a \gls{ML} model to estimate transient likelihoods, while employing only a small range of host galaxy optical/IR magnitudes and colors (Kisley et al. 2022, ApJ, submitted).
Several attempts have been made to classify supernovae from host information alone. \cite{2021GHOST} are able to classify Type Ia and core-collapse supernovae with $\sim60\%$ accuracy. \cite{gomez2020fleet} use a specialized algorithm to find %Type I
\gls{SLSN}e using host and light curve information, achieving high purity for this rare class. A recent pilot study (Kisley, Ko, Qin, Zabludoff, \& Barnard 2022, ApJ, submitted) using the \cite{2022ApJS..259...13Q} database with only a limited feature set and one of many possible \gls{ML} approaches is able to correctly classify 60\% of the LSST alerts expected each year for \emph{five} transient classes, including four major types of \gls{SN}e as well as TDEs. Kisley et al. consider the costs of following up all unclassified transients, finding that one would need to observe only 12-20 of their classified “TDE”s to get one true positive, which is a significant advantage over a random guess, as only <1\% of all the alerts are TDEs. The immediate classification provided, even from this initial work, would then allow spectroscopic confirmation and tracking to be achieved early in the time evolution of the \gls{TDE}, during the super-Eddington phase when the most can be learned about the forming accretion disk and the properties of the supermassive black hole.

Host-galaxy based classifiers need to be updated with additional \gls{ML} analysis tools, host galaxy features, and rarer classes. Incorporating improved class frequency priors, especially from early LSST data,
%broader
%additional multi-survey studies,
ancillary data from other surveys,
and %deliberate
cooperation with the LSST Brokers are all required %before LSST commences.
to realize LSST's full
time-domain science potential.
%We aim to incorporate the expected LSST transient class frequency priors and calculate transient probabilities for galaxies in the LSST \gls{footprint} with existing SDSS, \gls{DES}, and/or \gls{DESI} Legacy Imaging Survey photometry and archival \gls{GALEX} and \gls{WISE} data. We are collaborating with the \gls{ANTARES} event broker to facilitate LSST alert filtering.

\paragraph{Challenges (what makes it hard)}

%Supernova
Explosive transient
training and testing samples are dominated by the hosts of a few
%\gls{transient}
supernova
types (i.e., \gls{SN}e Ia and II account for 85\% of the unambiguously classified events). While it is possible to control for this imbalance in determining the significance of the \gls{ML} results (e.g., Kisley et al.), the deficit of hosts for rarer \gls{transient} types makes isolating their predictive and distinguishing features difficult. More work is needed to add
%additional
hosts of long and short-duration GRBs, \gls{SLSN}e, and subclasses like \gls{SN}e Ia-02 cx, Ibn, and IIL.
%for training and testing.%, but our exploration is still somewhat limited by the absence of certain easily observable features like host optical fluxes.

%As a result, it is necessary to significantly increase the numbers of uncommon \gls{transient} hosts in training and testing samples by
The features of
additional uncommon \gls{transient} hosts
%measuring their features
can be obtained now with
%ground-based
existing facilities. Using 1m-class telescopes, the author of this section and her collaborators have started acquiring SDSS-quality \emph{ugriz} photometry for bright hosts of \gls{transient} types with typically fewer than 30 known occurrences and that lie just outside the SDSS, \gls{Pan-STARRS}, and \gls{DES} footprints. These transients include peculiar \gls{SN}e Ia (e.g., 1991bg-like and Iax) that are likely produced by exploding white dwarfs, but may not be cosmological standard candles, \gls{SN}e Ibn that show signs of interaction with a hydrogen-free circumstellar medium, broad-line \gls{SN}e Ic with ejecta velocities \textasciitilde 3x those of normal \gls{SN}e, TDEs, and short- and long-duration GRBs. With larger telescopes, one can add missing optical features for fainter galaxies near the SDSS, \gls{Pan-STARRS}, and \gls{DES} sensitivity limits that have hosted \gls{transient} types poorly represented in training and testing samples. We encourage similar follow-up campaigns from the community, in preparation for LSST.

For transient pre-classification in surveys like LSST, Kisley et al. show that it is possible to train a successful \gls{ML} model solely on archival optical and \gls{IR} photometric features of known hosts. While such features are the most common in our existing \cite{2022ApJS..259...13Q} database, their connections to harder-to-measure, and thus less available, spectroscopic line strengths and derived quantities like stellar mass, star formation rate, and metallicity are ambiguous. Therefore, this work should be expanded to test classification using a hierarchy of feature subsamples, including  spectroscopic and
%\emph{GALEX} UV
additional photometric features %already measured
from space mission archives and ground-based galaxy surveys, as well as early LSST
%photometric
data.

We also need to employ a broader range of \gls{ML} tools, e.g., \gls{RNADE} or Bayesian Neural Networks, for likelihood estimation. \gls{RNADE} relies on neural networks to infer the density of features for each transient class (as opposed to kernel density estimation). As a result, it may permit modelling of more complex relationships in the data.

%Host galaxy properties will be only one set of inputs, albeit a useful prior set, that can be used for rapid \gls{transient} classification.
%An advantage of
%our approach is that it uses data that we have now.
%As stated above, this one possible starting place uses data we have now;
%Ultimately, our community will develop tools using more complete \gls{LSST} and other %survey data to probabilistically classify transients and direct their follow-ups.

\paragraph{Running on LSST Datasets (for the first 2 years)}

The ML models will improve with training samples that have similar selection to LSST \gls{transient}-host galaxy pairs, i.e., that use early LSST host-transient data, including rarer \gls{transient} types. Better understanding of the behavior and location of variable sources like \gls{AGN}, which early LSST data
%of the sky
will provide, is also critical to exclude them when classifying astrophysical transients. After the start of operations, LSST galaxy photometry can be added back into databases like that of \cite{2022ApJS..259...13Q} to improve the ML model and prior probabilities for a much more complete sample of potential hosts.

\paragraph{Precursor data sets}
The \cite{2022ApJS..259...13Q} transient-host galaxy database covers photometric measurements from the optical to mid-infrared. Spectroscopic measurements and derived physical properties are also available for a smaller subset of hosts. For +36k unique events, the authors have cross-identified \textasciitilde 14k host galaxies using host names, plus +4k using host coordinates. Besides those with known hosts, there are +18k transients with newly identified host candidates. For most hosts, the database contains photometric \gls{flux}, color, luminosity, surface brightness, concentration, and transient spatial offset. For \textasciitilde 6k events, it also includes host spectroscopic line strengths and derived features such as stellar mass and star formation rate. For \textasciitilde 4k events, it contains X-ray measurements from XMM-Newton, Chandra, Swift, and \gls{ROSAT}.

\paragraph{Analysis Workflow}
The transient-host feature database and predictive \gls{ML} model can be applied to the significant fraction of future LSST target galaxies with existing optical and \gls{IR} photometric features. For each potential host, one can estimate its probability across the range of transient classes using the training model and prior probabilities based on expected LSST class rates. By cross-matching deep optical catalogs with the UKIDSS and AllWISE catalogs, we estimate that 50\% of galaxies brighter than $r\sim21.34$ have the $JHK$ and $W1$, $W2$ magnitudes required by the Kisley et al. classifiers. There are, on average, nearly 2600 such galaxies per square degree in a typical high Galactic latitude field. Therefore, in the 18000 square degrees surveyed by LSST, there should be about 46 million of these galaxies. Considering the once-per-century supernova rate in typical galaxies, even the Kisley et al. pilot methodology will provide classifications for hundreds of thousands of transients per year. Furthermore, those transients will be generally at lower redshifts where spectroscopic and multi-wavelength follow-ups are more achievable and rewarding. After the start of LSST operations, LSST galaxy photometry can be used to expand the potential host and host feature samples.

\paragraph{Software Capabilities Needed}
Most host galaxy catalogs (e.g., \citealt{2021GHOST, 2022ApJS..259...13Q}) are rather lightweight, even in FITS format. Rapid/automated access will be critical for follow-up programs for LSST transients due to the enormous number of alerts. \gls{ANTARES} and Las Cumbres Observatory plan to integrate such standard-format catalogs into their event broker systems.

%What tools are best for developing such predictive models? For connecting and communicating effectively with data science experts? For accessing the early \gls{LSST} data and incorporating them into the model? For providing the ability to interrogate the model to obtain insights into the physical drivers of the most successful predictions? What are the best inputs for the \gls{transient} brokers in such cases and what does the community want to know about the potential transients from the brokers?

\pagebreak
\subsubsection{In-Depth Studies of Fast Phenomena} \label{sec:IndepthFast}
%\WOM{No challenges, many Acronyms - check themi, citations as cit and in main.bib please }

\Contributors{Alex Gagliano (\mail{gaglian2@illinois.edu})}
{March 27th, 2022}
\cleanedup{Alex Gagliano}

\paragraph{Abstract}
LSST data will unlock the time-domain sky, with millions of luminous supernovae expected across its decade of operation. A major benefit of scanning the full Southern sky every four nights lies in discovering intrinsically or observationally rare phenomena that have been invisible to us in previous surveys. This includes both rapidly evolving transients for which current samples are small (e.g., \gls{FBOTs}, \gls{FELTs}) and rapid early-time phenomena within \gls{SN} that have been well-characterized at later epochs (e.g., shock breakout, \gls{CSM} interaction, companion interaction). \gls{LSST} will observe these signatures, but will only be able to observe an event in a single band every 2–3 weeks. This photometry alone is much too sparse to probe the physical mechanisms underlying these phenomena. To facilitate scientific discoveries in this unique region of parameter space, we must develop \gls{software} that can identify rapidly-evolving events and prioritize them for high-cadence follow-up with a suite of dedicated resources.

In advance of Rubin first light, effort should be dedicated to exploring the potential pathways to rapid inference of \gls{transient} events with sparse and noisy multiband \gls{LSST} photometry. This will start with real-time classification, and explore complementary information that can be used to improve early classification and follow-up prioritization. At present, synthetic light curves from \gls{SNANA} within large-scale transient simulations like PLAsTiCC \citep{2018_plasticc} provide the baseline dataset for training classifiers for fast events, but additional work must be done to create software flexible enough to capture realistic and scientifically valuable variation among common events. \gls{LSST} data will be unique in the diversity of events they will contain (similar in depth and wavelength coverage to the \gls{Pan-STARRS} Medium Deep Survey; \citealt{2017Huber_MDS}), but significantly larger in scale), and we aim to maximize the scientific yield of these data with complementary software.

\paragraph{Science Objectives}
Accurate \gls{LSST} photometry at early times is critical for robust inference, as is accurate host galaxy association and photometric redshifts.  With this framework in place, LSST data can be used to probe early-time interaction signatures of the supernovae it discovers. These signatures will shed light on a broad range of open questions in supernova science, including the explosion mechanisms driving the majority of \gls{SN}~Ia, signatures of binary companion interaction in SN explosions, the formation of dust in SN ejecta, and the degrees of stripping in stripped-envelope systems. These analyses in the coming years will push us closer to a unified framework for the explosive deaths of stars, and allow us to weave a connecting thread back to the late-stage stellar evolution of their progenitor systems.

\paragraph{Challenges (what makes it hard) }
An ongoing question is the realism of the current generation of \gls{transient} simulations: classifiers trained with \gls{SNANA} data typically underperform on real data for extant surveys, and the reasons for this must be identified and corrected. Alert brokers operating on data from the Zwicky Transient Facility (\gls{ZTF}) perform some fraction of the tasks described in this science case (follow-up scheduling, real-bogus separation or more sophisticated \gls{transient} classification), and a good example of this is the ALeRCE postage stamp classifier; but a public end-to-end framework starting from raw photometry to real-time follow-up for fast phenomena remains elusive. Further, the vast majority of existing classifiers are either validated on idealized synthetic datasets or require full-phase light curves for classification and anomaly detection. These softwares are insufficient for finding the fast, early signatures of transient explosion physics and the rapidly decaying events that we hope to study with LSST.

\paragraph{Running on \gls{LSST} Datasets (for the first 2 years)}
The \gls{LSST} data necessary for enabling this science are the live alert stream, and if the framework is in place early it could be run at first light (although realistically this will be an iterative process in the first year of the survey stream to improve our pipeline, and this may take the form of active learning to adapt our search in real time). The challenge of this work is conducting this analysis with as few points of \gls{LSST} photometry as possible – potentially only one or two observations for fast phenomena. Contextual information can aid in classification at early times, and \gls{LSST} images of the field (which will be available via a postage stamp service within 24 hours of observation) may be valuable for providing these contextual data. To train real-time characterization pipelines in advance of \gls{LSST}, a synthetic dataset should be constructed consisting of millions of \gls{transient} events embedded within LSST-like imaging data. This next-generation simulation should encode the transient correlations with their host galaxies that have been previously identified in literature, so that algorithms may investigate the added value of this information. %This is easy to do with \gls{SNANA}, but generating realistic simulated images with transients painted on will be significantly more difficult.

\paragraph{Precursor data sets}
A variety of existing datasets will be needed to prepare our algorithms for fast transient studies with Rubin. Simulated data from \gls{SNANA} (\gls{transient} and host-galaxy photometry) will be useful for testing the value of contextual information for early classification, but synthetic postage stamps would foster additional development and the \gls{ZTF} alert stream will be critical for validating on real data. The question of enabling follow-up for different science cases requires more than just classification results, and work must also be done in advance to consolidate a large sample of human-labeled “high-priority” events for follow-up to achieve different science goals (e.g., nearby \gls{SN} II with strange early-time light curve signatures, or an \gls{FBOT} fading quickly). This science driver is made more challenging by the fact that most follow-up decisions made by individual science teams are not made public, and using all available follow-up data may only allow us to do as well as (and not better than) current human-driven follow-up efforts.

\paragraph{Analysis Workflow}
In advance of \gls{LSST} first light, the general steps to enable this analysis include:
\begin{itemize}
\item Generation of a large number of \gls{LSST}-like light curves with \gls{SNANA} for both short and long-timescale events. This will include realistic photometric redshifts and observed host-galaxy photometry (globally and at the site of the \gls{transient}).
\item For the longer-timescale events, manipulation of synthetic photometry to include early-time signatures that \gls{LSST} might observe (informed by theory; see references at the end of this section)
\item Fast classification using synthetic properties, with results validated from events discovered within the \gls{ZTF} alert stream
\item Input of raw photometry (event + host) and classification probabilities to an \gls{ML} method for follow-up prioritization, using prior follow-up decisions to identify “interesting” events even if classification cannot be done at these early epochs.

\end{itemize}
\paragraph{Software Capabilities Needed}
Multiple pre-existing softwares will need to be combined/repurposed for this task:
\begin{itemize}
\item Data will come from alert brokers that have already done initial real-bogus separation and initial downselection of candidates.
\item A real-time classifier, probably RNN-based \citep[e.g., RAPID;][]{2019Rapid} that combines event photometry with contextual information.
\item A target selection \gls{software} that combines raw info (e.g., data directly from brokers) with value-added parameters (classifier results plus priors based on what events observers have found interesting in the past)
\item The \gls{pipeline} should be scalable to millions of alerts and fast (\textasciitilde hours or less for all objects detected per night) in order to facilitate meaningful follow-up.
\end{itemize}

\paragraph{References for Further Reading}
Some references for the theory of early-time SN signatures include:
\begin{itemize}
\item \textit{Shock breakout}: \cite{2017SBO}
\item \textit{Companion interaction:} \cite{2010Companion}
\item \textit{\gls{CSM} Shock Cooling:} \cite{2015Piro,2021Margalit, 2021Piro} )
\item \textit{Hydrodynamical Interaction of SN Ejecta with Surrounding \gls{CSM}-Ejecta:} \cite{2018Chandra}
\end{itemize}
Real-time transient classification:
\begin{itemize}
\item Overview of methods: \cite{2021Broccia}
\item \textit{RAPID:} \cite{2019eeu..confE..36M}
\item \textit{GHOST:} \cite{2021GHOST}
\item \textit{SuperRAENN:} \cite{2020Villar}
\item \textit{\gls{ALeRCE} Stamp Classifier:} \cite{2021CarrascoDavis}
\end{itemize}

\pagebreak
\subsubsection{ToO Science (Beyond LIGO \gls{GW} Triggers)} \label{sec:TooBeyondLigo}
\cleanedup{VAV}
\Contributors{V. Ashley Villar (\mail{vav5084@psu.edu}), Igor Andreoni (\mail{andreoni@umd.edu}), Tomas Ahumada (\mail{tahumada@astro.umd.edu}), Suvi Gezari (\mail{sgezari@stsci.edu})}
{March 28, 2022}

\paragraph{Abstract}
Little exploration of a target-of-opportunity (\gls{ToO}) mode for \gls{LSST} has been explored beyond LIGO/Virgo fourth observing run (O4) binary neutron star merger followup. Here we enumerate other high-impact ToO observing science cases for \gls{LSST} and specify the technical challenges which may prohibit these \gls{ToO} observational modes. We note that for many of these cases, the expected cost (i.e., the impact on other scientific goals) is limited, and the scientific gain may be extraordinarily large.

\paragraph{Science Objectives}

We begin by enumerating the potential \gls{ToO} triggers enabled by observatories other than the Vera Rubin Observatory:
\begin{enumerate}
\item Fast Radio Bursts, the enigmatic new observational discovery of \textasciitilde millisecond long bursts of radio emission, seemingly from extragalactic origin. It is still unknown if optical emission may accompany these events. \gls{LSST} is particularly well-matched for FRBs which are non-repeating with low-dispersion measures (i.e., in the nearby universe) and are poorly localized (\textasciitilde degrees uncertainty). Here, a single pointing of \gls{LSST} (with a few visits) may be sufficient to isolate an optical counterpart. Such localizations are expected regularly in upcoming and ongoing radio observatories such as \gls{CHIME} and \gls{SKA}.
\item \gls{BHNS} mergers. These events are substantially more rare than binary neutron star mergers and often have much larger uncertainty regions. It is additionally not necessarily guaranteed (or even known) if such mergers have an electromagnetic counterpart. \gls{BHNS} with small localization region ( \textasciitilde degrees uncertainty) may be an excellent target for \gls{LSST} \gls{ToO}s.
\item Gamma-ray Bursts. Reasonably localized (\textasciitilde degrees uncertainty), nearby gamma-ray bursts would again be well-matched to the field-of-view of \gls{LSST} for a \gls{ToO} campaign. Ongoing and upcoming gamma-ray and X-ray observatories, such as Fermi, Swift and \gls{SVOM} should detect dozens of \gls{GRB}s annually, many of which will likely be well-localized.
\item Finally, we note that neutrino and pulsar timing array hotspots could also be of potential scientific interest.

\end{enumerate}

\paragraph{Challenges (what makes it hard)}
In each of these cases, we must develop ways to cross-match large (degrees) probabilistic areas (over \gls{RA}, Dec and potentially redshift) with galaxy/star catalogs.

Simulations for each of these science cases must be developed, and the potential impact of disruptive ToOs should be explored. However, as in the case for \gls{BNS} \gls{ToO}s (see references), the impact is likely to be minimal, with the opportunity for highly impactful and timely science.

\paragraph{Running on \gls{LSST} Datasets (for the first 2 years)}
 Within the first two years, template images must be available for efficient search of \gls{ToO} targets. In a \gls{ToO} mode, we expect the same alert datastream which will be sufficient for our purposes.

 We additionally note that improved galaxy catalogs from \gls{LSST} (ideally with some photo-z estimate) will greatly our ability to cross-match targets.

\paragraph{Precursor data sets}
 \gls{ZTF} transient alerts. \gls{PS1-MDS} catalogs and \gls{GLADE}\citep{2018MNRAS.479.2374D} galaxy catalogs.

\paragraph{Analysis Workflow}
\begin{itemize}
\item Receive external alert and trigger \gls{LSST} for a \gls{ToO} observing mode
\begin{itemize}
\item Highly specialized criteria must be established for automating (ie. democratizing) these triggers.
\end{itemize}
\item Run specialized filters on the \gls{ToO} alert streams.
\begin{itemize}
\item This can be done via the Brokers, but these specialized filters must be built.
\end{itemize}
\item Cross-match potential candidates with known sources
\begin{itemize}
\item This includes galaxies (especially in the case where redshift probability maps are known) and star catalogs (especially to rule out false positives, such as flaring stars).
\end{itemize}
\end{itemize}

\paragraph{Software Capabilities Needed}
 A \gls{software} which can cross-match large probabilistic areas which use HEALPIx.

\paragraph{References for Further Reading}
\cite{2022AJ....163..209S,andreoni2021target}

\pagebreak
\subsubsection{\gls{TDE} filtering} \label{sec:TDEfiltering}

\Contributors{Andreja Gomboc (\mail{andreja.gomboc@ung.si}), Sjoert van Velzen (\mail{sjoert@strw.leidenuniv.nl})}
{ March 27, 2022 }
\cleanedup{Andreja Gomboc}
\paragraph{Abstract}
\gls{TDEs} are rare transients, occurring (mostly) in galactic centers. Currently about 10 \gls{TDEs} are discovered per year. This number will substantially increase with Rubin \gls{LSST}. Simulations show that we could expect \textasciitilde 10 \gls{TDEs} detected per night (depending on their rate and \gls{LSST} observing strategy).
Two main challenges are: (1) How to identify a \gls{TDE} based solely on \gls{LSST} photometric data? And (2) how to identify a \gls{TDE} before the peak in the light curve?

We propose a dedicated \gls{TDE} filter to run on \gls{LSST} \gls{Alert} stream data on broker(s). It could be developed in stages: 1) extracting nuclear flares (from the centers of galaxies), 2) photometric feature extraction, 3) photometric typing, including \gls{ML}.
Data it would require would be the information included in the \gls{LSST} Alerts: history of an object, full photometric light curve, astrometric data (galaxy cross-match, off-set from galactic center), galaxy \gls{photo-z}, galaxy color/type.
Its output would be a stream of nuclear flares with light curve features, including classification labels or probabilities.

\gls{LSST} with its large \gls{FoV}, depth and image quality has the potential to detect many \gls{TDEs}, enable sample studies, probe \gls{SMBH} mass distribution, emission processes etc.
\paragraph{Science Objectives}
\begin{itemize}
\item To reliably identify a \gls{TDE} in order to enable targeted follow-up observations (including spectroscopy), ideally before the peak in the light curve in order to (together with follow-up observations) help determine better the time and magnitude of the peak, and the decay afterwards.
\item Infer light curve properties that are helpful in determining the mass of the black hole, type of the disrupted star, impact parameter etc.
\item Photometric typing of \gls{TDEs} using only Rubin data. We want to measure the purity and efficiency of a given filtering approach. The simplest approach is cut on a set of features: rise-time, color, color evolution, and fade timescale. More advanced selection will use \gls{ML} (either on these features or on the light curve directly).
\end{itemize}
\paragraph{Challenges (what makes it hard) }
\begin{itemize}
\item Reliable identification of \gls{TDEs} based solely on Rubin photometric data, which may not have ideal time and multi-band coverage (in particular in u-band filter).
\item Measure the purity of our filter using realistic light curves of the \gls{background} population: \gls{SN} and \glspl{AGN} variability.
\item Time constraint is to preferably identify a \gls{TDE} before the peak, i.e. on the order of days to weeks, depending on the time of the first detection.
\item Implementation on one or more brokers. Another possibility is to build a \gls{TDE} \gls{Filter} for each broker separately.
\end{itemize}
\paragraph{Running on \gls{LSST} Datasets (for the first 2 years)}
\begin{itemize}
\item Running on the Rubin Alerts stream. Using available information on galaxies (redshift, type, activity).
\item Requirements: reliable separation of stars and galaxies, cross-match with a galaxy, off-set of the \gls{transient} from the center of a galaxy, \gls{photometric redshift}, galaxy type.
\end{itemize}
\paragraph{Precursor data sets}
\begin{itemize}
\item To develop and test a \gls{TDE} \gls{Filter} an existing sample of observed \gls{TDEs} could be used to simulate Rubin \gls{TDE} alerts with \gls{LSST} simulation suit.

\end{itemize}
\paragraph{Analysis Workflow}
\begin{itemize}
\item Data cleaning (e.g. removing bad data)
\item Identify nuclear flares (an  intermediate data sets)
\item Matching to existing data sets (e.g., \gls{AGN} catalogs)
\item Feature extraction from the light curves (e.g., the empirical light curve model in \citealt{2021ApJ...908....4V})
\end{itemize}
\paragraph{Software Capabilities Needed}
\begin{itemize}
\item We will run in real-time. The \gls{LSST} archive might be needed sometimes to update the catalog matching.
\item Constructing a nuclear flare sample should be possible on all brokers.
\item The light curve feature extraction is more challenging but should also be possible on most brokers. Stress testing will be needed.
\item This filtering needs to be tailored for each broker. We therefore plan to pick one broker (\gls{AMPEL}, most likely) and provide the light curve features of the nuclear flares sample as a new data product (either a new \gls{Avro} stream, or via TNS).
\item We run on the entire stream. After selection “clear” nuclear flares for the feature extraction, we can expect \textasciitilde 1000 new sources per night. However the active source also have to be updated, so we will quickly be working with 1e5 sources at any given time.

\item To visualize the data, we need to be able to pull up the light curve for each nuclear \gls{transient} including the features. This can be done with \gls{TOM} or \gls{TNS}.
\end{itemize}
\paragraph{References for Further Reading}
\cite{Bricman_2020}; and K.\ Bučar Bricman, S.\ van Velzen, M.\ Nicholl, A.\ Gomboc. “Rubin Observatory's Survey Strategy Performance for Tidal Disruption Events”, in prep., ApJ Focus Issue.

\pagebreak
\subsubsection{Understand real photometric classification performance} \label{sec:PhotoPerf}

\Contributors{Catarina Alves (\mail{catarina.alves.ucl.ac.uk}), Fabio Ragosta (\mail{fabioragosta@inaf.it})}
{March 30, 2022}

\paragraph{Abstract}
Traditionally, \gls{SN} used in astrophysical and cosmological studies need to be spectroscopically classified. However, this will be impossible for most events detected by \gls{LSST} due to the limited spectroscopic resources; thus, \gls{LSST} will rely on photometric classification, using the events that will be spectroscopically classified as its training set.
Recent efforts, such as the Photometric \gls{LSST} Astronomical Time-Series Classification Challenge (PLAsTiCC; \citealt{Kessler2019}) and its follow up ELAsTiCC provide simulated \gls{LSST}-like datasets of light curves and ancillary data that can be used to train and evaluate photometric classifiers.
However, simulations cannot fully model the complexity of real data so the performance of the classifiers can be different on real \gls{LSST} data.
In this analysis, we would use the light curves from the spectroscopically confirmed events of \gls{LSST} and assess the real performance of the classifiers. Performing this analysis in the first years of \gls{LSST} allows us to optimize classifiers, including snmachine \citep{2016Lochner, 2022Alves}), RNN-based classifiers \citep{2019Rapid, 2020Villar}, and CNN-based classifiers \citep{Boone_2019, Boone_2021, qu2021scone} for the remainder of the survey.
Moreover, the analysis can give insights if it is needed to further adjust the observing cadence to improve photometric classification of \gls{SN}.

\paragraph{Science Objectives}
This analysis will provide a framework to compare transient light curve classifiers. It will also lay the ground work to improve said classifiers, e.g., for higher Type Ia purity for cosmological studies. We note that this is especially important for classifiers trained with \gls{ZTF} data, which is significantly different from the expected LSST data stream.

\paragraph{Challenges (what makes it hard)}
\begin{itemize}
\item Modify the inputs of the classifiers to be able to train with LSST-like data.
\item Create a tool to allow users to create custom comparison metrics (e.g., purity of Type Ia \gls{SN}) using a combination of classifiers.
\item Run full light curve analysis can be run periodically (e.g. every year, or each time there is a data release).
\end{itemize}

\paragraph{Running on \gls{LSST} Datasets (for the first 2 years)}
The analysis requires light curves from the spectroscopically confirmed events of \gls{LSST}. It is only relevant after 1 year (after complete, difference imaged light curves are available).

\paragraph{Precursor data sets}
ZTF data can be used as a precursor to evaluate real-data classification. In terms of \gls{LSST}-like simulated datasets, we have the Photometric \gls{LSST} Astronomical Time-Series Classification Challenge (PLAsTiCC; \citealt{Kessler2019}) and its follow up ELAsTiCC.

\paragraph{Analysis Workflow}
\begin{itemize}
\item Generate a method to easily compare/contrast publically available classification algorithms.
\item Select classification \gls{algorithm}s and metric to compare algorithms or optimize an ensemble of algorithms.
\item Clean the relevant data and extract features accordingly to the chosen \gls{algorithm}
\item Classify the \gls{LSST} data, and compare classifiers.
\end{itemize}

\paragraph{Software Capabilities Needed}
\begin{itemize}
\item \gls{Alert} Brokers will be essential to identify transients in their early phases. Crosstalk between the software and the brokers will enhance the ability of the software to classify the \gls{transient}. Imperative for the scope – but also to follow up on transients discovered already– will be the ability to reconstruct the history of detection of the source through crossmatch with other catalogs.
\item It is also crucial to have access to catalogues of transients in the data releases to perform complete light curve classification and optmise these classifiers.
\item There are several existing softwares, which require a common framework to load in data, train, and compare results.
\item This work would be applied to archival LSST time series, or ensemble classifiers could be deployed in real time.
\item For functionalities related to visualization, see technical case Interactive Data Visualization at scale \secref{sec:VizScale}
\end{itemize}

\paragraph{References for Further Reading}
Photometric classifier links: \href{https://docs.google.com/spreadsheets/d/1PQZtp9T-NVdHfN7YTZwDMHAFOf7g6gJjLHGwGUsqTdc/edit#gid=0}{literature} review TVS \gls{SN}  from Fabio Ragosta

\pagebreak

\subsection{Extragalactic variable science} \label{sec:evs}
\subsubsection{Augmenting \gls{AGN} Variability} \label{sec:AugmentAGN}
\Contributors{Matthew J. Graham (\mail{mjg@caltech.edu})}
{ Date: March 28, 2022}

\paragraph{Abstract}
 Variability is one of the quintessential properties of \gls{AGN} and different enough from that exhibited by other classes of astrophysical object that it can used to identify them in a wavelength agnostic way. The LSST is expected to observe tens of millions of \gls{AGN} and will thus provide an unprecedented data set for studying population statistics, e.g., tracers of large-scale structure, as well as rare \gls{AGN} phenomena, such as supermassive black hole binaries, changing-look \gls{AGN}, and electromagntic counterparts to compact object mergers in \gls{AGN} accretion disks. However, the timescales of \gls{AGN} variability typically mean that decadal light curves are required for good characterization and identification and we may be waiting until the 2030s for much of this science.

 Bayesian experimental design allows optimizing a set of specified observations to distinguish between different models. In this case, it should be possible to identify sets of additional observations of AGN candidates with external facilities, e.g., \gls{ZTF}, to maximize the identification of AGN (differentiate between competing statistical models) and reduce the timescale to a viable (good enough) data set.

\paragraph{Science Objectives}
\begin{itemize}
\item Identify a high probability variability-selected \gls{AGN} data set with minimal temporal coverage from LSST, i.e., as soon as possible.
\item Create a complete and uncontaminated catalog to produce an independent estimate of the fraction of LIGO mergers produced by \gls{AGN} from the excess spatial correlation of \gls{AGN} and LIGO error volumes.
\end{itemize}

\paragraph{Challenges (what makes it hard)}
\begin{itemize}
\item The characteristic timescales of \gls{AGN} variability typically require decadal baseline time series to constrain in an unbiased fashion \citep{bartos2017multimessenger}.
\item It is unknown what is the minimum amount of data required to differentiate between different statistical/ML models for \gls{AGN} variability to allow high probability identification.
\item It is unknown if multiband data aid in the above or is sparse coverage across six bands a hindrance, i.e., is it better to work in a limited subset of bands?
\item It is unknown what are the best models of \gls{AGN} variability to identify sources as \gls{AGN} with limited data.
\item It is unknown much augmentation from external sources is required to improve \gls{LSST} data.
\end{itemize}

\paragraph{Running on \gls{LSST} Datasets (for the first 2 years)}
\begin{itemize}
\item A subset of AGN can be identified from the alert stream (\gls{ZTF} statistics suggest about 1\% of AGN will show 5 sigma variability).
\item The data release catalogs will form the base for the analysis. The analysis will use all galaxies/stars in the data release.
\item Deep drilling fields will have a higher cadence but it is not clear whether this helps - specific modeling for this data will be required.
\item Other data products, e.g., LIGO probability maps, external catalogs, are essential for this analysis.
\end{itemize}

\paragraph{Precursor data sets}
\begin{itemize}
\item The Catalina Real-Time Transient Survey and \gls{ZTF} data can be used to test aspects of this analysis. It is possible that SDSS Stripe 82 data may also help for full multicolor modeling of LSST. All this data is public.
\item Preseeding possible \gls{AGN} candidates from other surveys may bias identification but this also needs to be tested.
\end{itemize}

\paragraph{Analysis Workflow}
\begin{itemize}
\item Define a base set of \gls{AGN} variability models to compare, e.g., \gls{DRW}, \gls{DHO}, \gls{RNN}.
\item Define a utility function to compare the success of each model.
\item Define a set of possible photometric augmentations.
\item Test with the first year of data and second year of data to determin if there is a substantial improvement in AGN identification.

\end{itemize}

\paragraph{Software Capabilities Needed}
\begin{itemize}
\item This could conceivably run over the full \gls{LSST} data set but some downscaling to just “variable” sources will be probably happen
\item New software infrastructure will be needed to determine best additional observations for subsets of \gls{AGN} candidates.
\item Iterations of fitting \gls{AGN} models with augmented data will also be needed.
\end{itemize}

\paragraph{References for Further Reading}
\cite{sravan2021autonomous}

\pagebreak

\cleanedup{Andjelka}
\subsubsection{Conditional Neural Processes for learning \gls{AGN} light curves} \label{sec:CondNeuralProc}

\Contributors{ Andjelka {B}. Kova{\v c}evi{\'c} (\mail{andjelka@matf.bg.ac.rs}), Dragana Ili{\'c} (\mail{dilic@matf.bg.ac.rs}), Paula S\'anchez-S\'aez (\mail{pasanchezsaez@gmail.com}), Iva {\v C}vorovi{\'c} Hajdinjak, Robert Nikutta  (\mail{robert.nikutta@noirlab.edu}), Nikola Andri{\'c} Mitrovi{\'c}, Mladen Nikolic{\'c} (\mail{nikolic@matf.bg.ac.rs}, Viktor Radovi{\'c} (\mail{rviktor@matf.bg.ac.rs}), Luka {\v C} Popovi{\'c} (\mail{lpopovic@aob.rs})}
{03/26/22}

\paragraph{Abstract}
 The next generation time domain surveys, such as Vera Rubin Observatory Legacy Survey of Space and Time \citep[LSST, see][and references therein]{2019ApJ...873..111I}, will provide observations with different cadences over ten years for millions of active galactic nuclei \citep[AGN]{ 2021arXiv210801683B, 2018arXiv181106542B} in six filters - \textit{ugrizy}.
 The consequences of complex, disturbed environments in the vicinity of a supermassive black hole are not well represented by standard statistical models of optical variability in  \gls{AGN}. Thus, developing new methodologies for investigating and modeling \gls{AGN} light curves is crucial e.g. \citep{Tach20}.
Conditional Neural Processes \citep[CNP,][]{Gar18} are nonlinear function models that forecast stochastic time series based on a finite amount of known data without the use of any object parameters or prior knowledge (kernels).
\cite{2022AN....34310103A} provide an initial  \gls{CNP}  \gls{algorithm} that is specifically designed for learning  AGN light curves for the intended periodicity \gls{pipeline}. It was trained using data from:
\begin{itemize}
\item the All-Sky Automated Survey for Supernova \citep[ASAS-SN;][]{Hol17}, which included 153 \gls{AGN}
\item about 40,000 light curves from Zwicky Transient Facility data release 5 \citep[\gls{ZTF} DR5;][]{2021Sanchez_AGN}.
\end{itemize}

We note that \cite{park2021inferring} provide a Bayesian version of the CNP, along with an attentive layer. The method simultanously performs regression and reconstruction, with the regression step acting as a regularization on the learned latent space.

Preliminary parallelization experiments show that \gls{CNP}  could efficiently handle large amounts of data. These results imply that \gls{CNP}  could be more effective than standard tools in modeling large volumes of AGN data.In addition to the \gls{MLP} encoder, it is necessary to attempt to develop a \gls{1D} convolution neural process and an attention neural process capable of detecting sequential structure.
However, more set stratification testing is required, i.e. light curves in training, validation, and test sets should have a consistent distribution of certain parameters such as time baseline, cadence, gradient, and gaps.

\paragraph{Science Cases Needing this Tool}
\begin{itemize}
\item Estimation of accretion disk and broad line region sizes.
\item Estimation of supermassive black hole masses.
\item Mining periodicity signal  in \gls{AGN} and stellar light curves.
\end{itemize}

\paragraph{Challenges (what makes it hard)}
\begin{itemize}
\item Light curves  show specific phenomena such as flares, possible quasi-periodic oscillations, time gaps, possible irregular cadence due to unpredicted weather conditions,
\item Complexity of light curves varies: monotonic light curves, single peaked light curves, more complex (many peaks and valleys).
\end{itemize}

\paragraph{Running on \gls{LSST} Datasets (for the first 2 years)}
\begin{itemize}
\item We will apply \gls{CNP}  on  the data release light curves. The \gls{software} will focus on yearly data releases.
\item We expect to start modeling  light curves in the first few months.
\item We will use  data from the Deep Drilling Fields  as a training set. However, \gls{algorithm} training will be tested with slower cadences.

\end{itemize}

\paragraph{Precursor data sets}
CNP has been initially tested on simulated datasets, as well as 153 \gls{ASAS-SN} \citep{Hol17} light curves with time baseline about 2,000 days and  40,000 light curves from the Zwicky Transient Facility data release 5 (\gls{ZTF} DR5) with longer time baseline \citep{2021Sanchez_AGN}.

\begin{figure}
\begin{centering}
\includegraphics[width=0.9\textwidth]{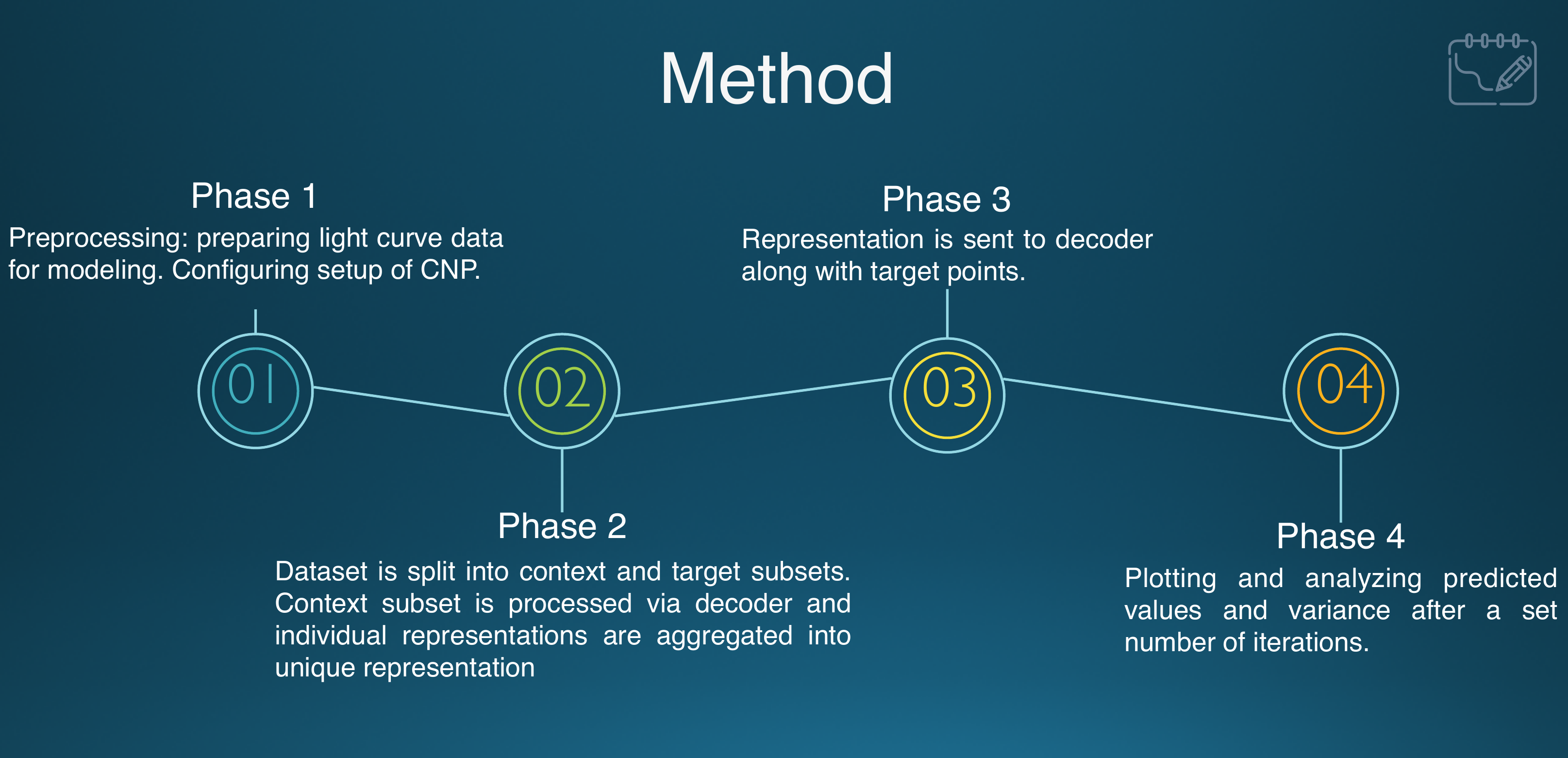}
	\caption{Steps in the conditional neural process method's application. The encoder
and aggregator process context points. This process's output is passed to
the decoder, together with target points, to calculate predictions. The
last unit visualizes both the original and predicted data. \label{fig:CondNeuralAlgo}}
\end{centering}
\end{figure}

\paragraph{Algorithm}
The \gls{algorithm} runs on data-set O, which consists of n points. Context points
, subset $O_N$ with N points,
are used for training the neural network, whilst the target points, subset $T$ with M points, are used for making predictions, i.e. modeling the light curve. The input to the  encoder is a context set, as illustrated  in the Figure \ref{fig:CondNeuralAlgo}  \citep[see details in][]{2022AN....34310103A}.
The encoder is a multi-layer perceptron (\gls{MLP}) neural network which produces as output lower dimensional representations (represented with $h$ in \figref{fig:CondNeuralAlgo})
$r_i = h((x, y)_i)$
for each context point.  The aggregator ($a$) merges  all these representations.

Then these   representations are  combined  with  target points $x_t$ (the unlabeled points). Finally, the decoder (a multi-layer perceptron neural network), calculates predictions for each target value and outputs mean and variance of the predicted distribution.

\paragraph{Software Capabilities Needed}
\begin{itemize}
\item Our neural process architecture is developed in Python and uses PyTorch.  It consists of two multilayer perceptron neural networks and an aggregator. In addition to \gls{MLP} encoder, needed to try to implement \gls{1D} convolution neural process, and attention neural process that will catch sequential structure.
\item To be run on large computing facility based on \gls{GPU}
\item \gls{CNP}  time running  complexity is $\mathcal{O}(N+M)$ where $N$ is the number of known data points and $M$ is the number of unlabeled data points. It is faster than Gaussian Processes which time running complexity is $\mathcal{O}((N+M)^{3})$ \citep{Gar18}. However, we found that it is important set stratification, i.e. light curves in training, validation, and test sets should have a comparable distribution of certain parameters such as time baseline, cadence, gradient and gaps.
Stratification of  training data  is in testing phase and demands  construction of additional algorithms.
\end{itemize}

\paragraph{References for Further Reading}
 Background on deep learning of \gls{AGN} light curves: \citet{Tach20}\\
 Background of \gls{CNP}  learning of \gls{AGN} light curves: \citet{2022AN....34310103A}\\
 Background on Conditional Neural Process: \citet{Gar18}\\

\pagebreak
\subsubsection{Find All the \gls{AGN} ASAP} \label{sec:FindAllAGN}
\Contributors{Weixiang Yu (editor), K.~E.~Saavik Ford, Matthew Graham, Yusra AlSayyad, Colin Burke, James Chan, Neven Caplar, Matt O’Dowd}
{03/30/2022}

\cleanedup{Weixiang Yu}

\paragraph{Abstract}
 The Rubin Observatory \gls{LSST} will discover \textasciitilde 100 million AGN after its 10-year survey. Classifying and cataloging those AGN out of a total of \textasciitilde 40 billion LSST sources without the assistance of spectroscopy will be extremely challenging. This challenge will be further elevated by the short baseline (i.e., temporal coverage) at the early times of LSST and a lack of matching multi-wavelength data. Producing a complete and pure sample of AGN as soon as possible (ASAP) once LSST starts is crucial to the timely follow-up of interesting objects (e.g., \gls{CSQ}), robust association of \gls{MMA} events, and statistical studies of AGN (e.g., to estimate the luminosity function). Here we present the actual challenges for a complete, pure and timely classification of AGN in LSST and how we may/can overcome them to bring LSST to its full potential.

\paragraph{Science Objectives}
\begin{itemize}
\item Find as many AGN as possible and as soon as possible
\item Produce a large sample of highly probable AGN candidates for continuous monitoring and follow-up
\item Find as many lensed quasars as possible and at the same time measure the basic lens parameters
\end{itemize}

\paragraph{Challenges (what makes it hard)}
\begin{itemize}
\item AGN require years-long (even decade-long) baselines to be confidently identified using variability, which LSST will not provide until many years into the survey. Cross-calibrated light curves from precursor surveys (\gls{ZTF}/\gls{DES}/\gls{PTF}/\gls{Pan-STARRS}/\gls{SDSS}), for the sake of extending the baseline, are needed to select as many AGN as possible during the early operations of LSST.
\item Efficient time-series feature extraction and parametric fitting (e.g., \gls{CARMA}) for billions of light curves are needed to classify AGN through variability. Light-curve feature extraction should be run at least following each planned data release.
\item Optical color selection of Type-1 (unobscured) AGN suffers contamination from stars, where multi-wavelength data can help reduce (stellar) false positives. In addition, successful selection of Type-2 (obscured) AGN relies on multi-wavelength photometry. Efficient large-scale cross-matching between LSST and external multi-wavelength catalogs is a huge technical challenge.
\item Deblending of extended AGN (nucleus vs.\ host galaxy) and lensed AGN.

\end{itemize}

\paragraph{Running on LSST Datasets (for the first 2 years)}
\begin{itemize}
\item We will utilize both LSST data release catalogs and light curves of mainly all 5-$\sigma$ sources
\item Re-calibrated light curves from precursor surveys and time series features extracted therefrom.
\item Cross-matched external catalogs (e.g., eROSITA, \gls{WISE}, Gaia, etc.)
\end{itemize}

\paragraph{Precursor data sets}
\begin{itemize}
\item Time-domain: ZTF/PTF/Pan-STARRS/SDSS/\gls{DES}/\gls{HSC}/Gaia
\item Multi-wavelength: eROSITA/\gls{VISTA}/{WISE}/\gls{VLASS}
\end{itemize}

\paragraph{Analysis Workflow}
\begin{itemize}
\item Cross-match LSST sources with external catalogs (including time-series and multi-wavelength information)
\item Compute time-series features (e.g., parametric fit: \gls{CARMA}) on joint light curves from precursor surveys and LSST.
\item Apply minimum quality cuts.
\item Feed LSST catalog data, multi-wavelength photometry, and time-series features into \gls{ML} algorithms for (probabilistic) classification.
\end{itemize}

\paragraph{Software Capabilities Needed}
\begin{itemize}
\item Large-scale cross-matching between LSST and external catalogs (including time-series and multi-wavelength information)
\item Development of efficient algorithms to compute light curve features (e.g., best-fit \gls{CARMA} parameters) for billions of light curves.
\item Algorithms to select lensed quasars (e.g., \gls{CNN} on image postage stamps)
\item Tools to visualize/interact with selected AGN candidates in a high-dimensional parameter space (e.g., color, variability, proper motion, and others)
\end{itemize}

\paragraph{References for Further Reading}
Longer baseline justification: \cite{Kozlowski2017, Kozlowski2021}\\
More Precursor Surveys: \gls{BlackGEM}~\citep{Groot2019}\\
AGN Classification: \cite{Richards2002, MacLeod2011, Butler2011, Stern2012, Peters2015}

\pagebreak
\subsubsection{Connection between short term variability of \gls{AGN} and their long term behavior} \label{sec:ConnectSLAGN}
\Contributors{ Neven Caplar (\mail{ncaplar@princeton.edu}), Colin Burke (\mail{colinjb2@illinois.edu}), K.E. Saavik Ford (\mail{sford@amnh.org)}}
{03/29/2022}

\cleanedup{Neven Caplar}

\paragraph{Abstract}
 It is well established that individual \gls{AGN} which are more variable on short timescales ($\sim$days to tens of days) are also more variable on long timescales ($\sim$years to tens of years). In particular, recent studies have shown deviations from damped random walk at short timescales, with variability being more strongly correlated than expected. It does seem that there is a dependence with mass, with less massive AGN exhibiting a stronger correlation. Are there more connections between short term variability with long term variability, beyond just the amplitude? \gls{LSST} data promises to deliver AGN curves with relatively high cadence and high precision of measurement. This precision is necessary in order to study relatively small changes of flux at short timescales.
 A special class of variable AGN are \gls{CSQ}, i.e., AGN that exhibit large changes of \gls{flux} and/or change in the spectral properties. The physical mechanism behind these changes is unclear, but high cadence multi-color photometry and ‘before and after’ spectroscopy are critical to improving our understanding. Predictive models to anticipate state change events would be extremely valuable.
 We aim to quantify the number of AGN that decrease/increase in luminosity, the number density of sources that change states, as well as their dependence with mass, luminosity and redshift. For changing state AGN it is important to obtain spectra before the change, and follow up spectra (for at least a representative subset of objects). This data will allow us to infer a change in the sound speed and temperature, as well as mass accretion rate as a function of radius from the \gls{SMBH}. We can also constrain the timescales required for the development and/or collapse of the narrow line region, corona, and disk outflow structures (e.g., warm absorbers).

\paragraph{Science Objectives}
\begin{itemize}
\item Create power spectral density/structure functions for complete list of \gls{AGN}, previously found in other surveys
\item Find the best characterization statistic(s) for \gls{AGN} variability
\item Connect \gls{LSST} measurements of variability with long term measurements, and find correlations with long term behavior
\item In particular, determine short-term variability properties for \gls{AGN} that exhibit a single large change (\gls{CSQ})
\item Identify outlier variability behavior – all variability cannot be driven by stochastic gas processes (e.g., microlensing, transits, \gls{SN}, embedded \gls{TDE}s, possible \gls{MMA} sources)
\item Identify rapidly-variable AGNs (intermediate-mass black holes as a probe of seeding mechanisms; Dwarf \gls{AGN} variability for intermediate-mass black hole identification, see \secref{sec:DwarfAGN})
\item Use \gls{CSQ} to constrain \gls{AGN} turn on/turn off process, lifetime of \gls{AGN}, sound speed/temperature/mass accretion rate as a function of radius, physical drivers of state change, influence of feedback on the rest of a galaxy
\end{itemize}

\paragraph{Challenges (what makes it hard)}
 Challenges may include:
\begin{itemize}
\item Large amounts of data (at least 100,000 \gls{AGN} - if considering only spectroscopically confirmed \gls{AGN}) - but really want 50 million (and will have to process more to distinguish galaxies on the basis of variability)
\item Identifying \gls{AGN}. This sample will change over time.
\item The need for complete datasets (50 million). When we want to estimate \gls{SF}$^2$ or equivalent quantities for the whole dataset, how often do we need to run/re-run this?
\item For \gls{ML} classification, the time and number of light curves is again overwhelming.
\item Galaxy contamination, especially for fainter the AGN host galaxy contribution will be significant. This contribution can, in general, be distinguished from variable AGN flux as the galaxy contribution does not vary. But, if we are interested in small changes, small differences in the estimation of galaxy flux (i.e., due to \gls{seeing}, and chromatic effects) can influence the results. Low luminosity sources cannot be followed up with spectra.
\item Low luminosity / dwarf \gls{AGN} host light contamination  for off-nuclear BHs - can we trust those light curves. How do we identify such galaxies with this separation?
\item Do we want to capture \gls{AGN} that exhibit large-changes as they happen, and calculate their short term variability just before the change happened? Is there a way to observe these in almost real time?
\item The need to develop predictive models for large changes? This is an \gls{ML} problem, and whether this is even possible in principle depends on the physical cause of the large changes.
\item For small-amplitude, constantly variable sources (\gls{AGN}s, intermediate mass black holes, off-nuclear \gls{AGN}s), what are the differential image analysis artifacts?
\item For low-luminosity sources, we likely need to determine redshift and mass (and the Eddington ratio) from the variability alone. So we need to figure out how to do that–\gls{ML}.
\item The need to follow-up  higher luminosity sources, requires spectroscopic/photometric followup capacity. Some followup might time frames might be as long as  2 or more weeks, others as short as 1\,hr (e.g., X-ray observations for quasi-periodic oscillation events).
\item Connecting observations to astrophysics requires more theory development and/or \gls{ML}
\end{itemize}

\paragraph{Running on \gls{LSST} Datasets (for the first 2 years)}
\begin{itemize}

\item Primarily will be run on data release catalogs. Several months of data need to be available to create reasonable estimates of short-term variability.
\item Value added catalogs would include descriptions of AGN variability on short scales, (e.g., \gls{SF}$^2$ estimates at different times; see also the discussion in  \citealt{DMTN-118})
\item The initial dataset could be AGNs that are spectroscopically confirmed (e.g., the \gls{SDSS} homogeneous sample that is available from the south)
\item Large analysis could be done on all objects that exhibit AGN-like variability on time scales of $\sim$1\,yr (i.e., after the first data release). This would lead to a \gls{monitoring} catalog for e.g., \gls{CSQ} behavior, and outliers relative to past \gls{SF} behavior
\end{itemize}

\paragraph{Precursor data sets}
\begin{itemize}
\item \gls{ZTF}
\item \gls{WISE} (helps with initial color cuts at the bright end)
\item Gaia
\item All currently known \gls{CSQ}s
\item \gls{DES}
\item Lightcurves generated by Rubin reprocessing of precursor data  described in  Section \ref{sec:FindAllAGN}
\end{itemize}

\paragraph{Analysis Workflow}
\begin{itemize}

\item Compute variability \gls{metric} (e.g., total variance) of every object in the data release
\item Select the objects that exhibit some minimal amount of variability for one sample
\item Select all known \gls{AGN} for a second sample
\item Remove data points that exhibit very large change from previous measurements (in particular if the measurements are in the same night, or only one measurement is very different)
\item Select only light curves for which coverage is ``reasonably’’ homogeneous in time domain (e.g., 10 data points, not separated by more than 2 months)
\item For each color and/or for all colors jointly, run the \gls{PSD} and \gls{SF} determination algorithm
\item Match with known objects, if available, and determine/retrieve long term parameters
\item Look for correlations between variability metrics from the modeling done above and long-term parameters

\end{itemize}

\paragraph{Software Capabilities Needed}
\begin{itemize}
\item Structure function estimation \gls{algorithm}
\item PSD estimation algorithm - e.g., \gls{CARMA} modeling. Perhaps Gaussian model processing based on code by Dan Foreman-Mackey (EzTao: \url{https://ui.adsabs.harvard.edu/abs/2022ascl.soft01001Y/abstract)}? \gls{ML} modeling algorithm trained on existing data?

\end{itemize}

\paragraph{References for Further Reading}

 Extreme variability
 \cite{2022ApJ...925...50R}\\
 Stronger dependence of \gls{PSD} on short time scales
 \cite{2018ApJ...857..141S}\\
 Timescale-Mass relation
 \cite{2021Sci...373..789B}\\
 Spectroscopic follow up of changing state QSO
 \cite{2019ApJ...874....8M}\\
 Changing state QSO from Catalina
 \cite{2020MNRAS.491.4925G}\\
 High redshift changing state QSO
 \cite{2020MNRAS.498.2339R}\\
 Changing state QSO in \gls{WISE}
 \cite{2018ApJ...864...27S}\\

\pagebreak
\subsubsection{Developing machine learning methods for AGN selection and calculating \gls{photometric redshift}} \label{sec:MLAGN}

\Contributors{ Đorđe Savić (\mail{djsavic@aob.rs}), Isidora Jankov (\mail{isidora\_jankov@matf.bg.ac.rs}), Anđelka Kovačević (\mail{andjelka@matf.bg.ac.rs}), Dragana Ilić (\mail{dilic@matf.bg.ac.rs}), Luka Č. Popović (\mail{lpopovic@aob.rs}), Mladen Nikolić (\mail{nikolic.matf@gmail.com}), Aleksandra Ćiprijanović (\mail{aleksand@fnal.gov})}
{3/25/22}

\paragraph{Abstract}
\gls{LSST} will produce catalogs for a vast number of sources, which will usher astronomy into a new era of ``big data.'' Machine learning (ML) deployment will be helpful in developing efficient models for various classification and regression tasks. We are focused on three main problems 1) \gls{AGN} selection, 2) parameterization of \gls{AGN} light curves and 3) estimating photometric redshifts of AGNs and galaxies. Variability will be the cornerstone for separating AGNs from variable stars. The addition of high quality imaging data will be crucial for separating AGNs from regular galaxies, allowing us to train ML classifiers with superb accuracy $>99\%$. Redshift estimates for the vast majority of LSST AGNs will rely on photometric estimates. Our goal is to develop an empirical regression method using all the possible sources of information: colors, fluxes, variability, differential chromatic refraction and multiwavelength data.

\paragraph{Science Objectives}
\begin{itemize}
\item \gls{AGN} selection
\item \gls{AGN} characterization and parametrization of light curves
\item Photo-z estimation
\end{itemize}

\paragraph{Challenges (what makes it hard)}
\begin{itemize}
    \item Identifying the variable sources will require iterative processing of time series data for removing artifacts and to improve \gls{AGN} selection
    \item Pixel level information will be used for separating galaxies with clear morphological traits from the \gls{AGN} which requires a lot of computing power when working with convolutional neural networks
    \item Efficient handling of data quantities greater than that of ongoing surveys
    \item Running User Defined Functions at scale on \gls{LSST} data (across all light curves)
\end{itemize}

\paragraph{Running on LSST Datasets (for the first 2 years)}

\begin{itemize}
    \item We will analyze Data \gls{Release} light curves. Most of the analysis will be based on catalog data
    \item The initial analysis will process all variable point sources with more than 60 epochs of observations
    \item We expect to develop a classifying \gls{algorithm} based on machine learning methods in the first 6 months with further improvements every 6 months.
\end{itemize}

\paragraph{Precursor data sets}
\begin{itemize}
    \item Precursor data are drawn from two main survey fields, an extended \gls{SDSS} Stripe 82 area and the \gls{XMM}-LSS region. The datasets were established by the \gls{AGN} science collaboration which hosted a data challenge in summer 2021 and are hosted publicly on sciserver (\url{https://www.sciserver.org/}). The total amount of objects is of the order of $\sim$450,000.
\end{itemize}

\paragraph{Analysis Workflow}
\begin{enumerate}
\item Data cleaning and identification of artifacts within the data (this will be iterative as we progressively remove/flag bad data from the time-series catalogs)
    \begin{itemize}
    \item Remove poorly calibrated photometric data and sources flagged with suspicious photometry (e.g., on edge of CCD or diffraction spikes) and/or \gls{LSST} \gls{DM} source quality flags.
    \item Remove/flag outlier measurements from a light curve.
    \end{itemize}
\item Data storage and archives
    \begin{itemize}
    \item The processed data at this stage does not require additional storage.
    \item For the \gls{photometric redshift} estimates, the AGN sources will be matched with the multimessenger data for accurate \gls{photometric redshift} measurements.
    \end{itemize}
\item Training and applying machine learning methods
    \begin{itemize}
    \item Select sources with $N > N_\mathrm{threshold}$ epochs of data, and within a specific \gls{SNR} range.
    \item Compute the LC features for all bands.
    \item Training machine learning methods
    \item Classifying all selected sources
    \item Further development on machine learning methods used on the accurately classified AGNs for deeper understanding of \gls{AGN} physics.
    \end{itemize}
\end{enumerate}
\paragraph{Software Capabilities Needed}
\begin{itemize}
    \item Ability to apply selection filters to data (\gls{SQL} query)
    \item Storage of outputs of filtered and classified data (or flags based on filters applied to existing catalogs)
    \item Visualization of distributions of selected sources on the sky and relative to \gls{camera} coordinates
    \item Visualization of postage stamps and light curves for individual sources
    \item Visualization of distribution of properties of sources (e.g. color-color scatter plots and histograms) colored by flags
\end{itemize}

%
% Commented out by mjuric (editor) on Aug 3, 2022, as the provenance of the figure is not clear and it is not referred to in the text. Author: if this is not ideal, please contact the editors for re-instatement. 
%
% \begin{figure}
% \begin{centering}
% \includegraphics[width=0.9\textwidth]{images/DeepEmbeClust.png}
% 	\caption{An example of deep unsupervised clustering based on autoencoder (left branch) and clustering with t-distributed stochastic neighbor embedding (right branch).  \label{fig:DeepEmbeClust}}
% \end{centering}
% \end{figure}

\paragraph{References for Further Reading}
\begin{itemize}
    \item Background on machine learning \citep{10.5555/3379017}
    \item Deep learning in Python \citep{2018ascl.soft06022C}
    \item \gls{AGN} light curves \citep{2011ApJ...733...10R}
\end{itemize}

\pagebreak
\subsubsection{Dwarf \gls{AGN} variability for intermediate-mass black hole identification} \label{sec:DwarfAGN}
\Contributors{ Colin J. Burke (\mail{colinjb2@illinois.edu})}
{ 03/25/2022}

\cleanedup{Colin Burke}

\paragraph{Abstract}
 With a photometric precision of \textasciitilde  percent level or better at g < 22, LSST Rubin has the potential to uncover a population of AGNs in dwarf galaxies (``dwarf AGNs'') hosting \gls{IMBH}s with low luminosities. Owing to their low AGN luminosities and dilution from star-forming host galaxy light, dwarf AGNs have variability amplitudes of a few tenths of a magnitude or less, making them difficult to detect until recently. This proposal is to create a catalog of dwarf AGNs using the first \textasciitilde year of LSST light curves. Dwarf AGNs with \gls{BH} masses smaller than a million solar masses are most variable on days to months timescales, depending on the \gls{BH} mass. Therefore, the full 10 year LSST light curves are not strictly required.

 Recent theoretical and observational results suggest \gls{IMBH}s may lie outside the nucleus of their host galaxies, so difference images will be an invaluable tool toward identifying these so-called “wandering” off-nuclear AGNs using optical variability.
% not included (see use case \S\ref{sec:PinpointWMBH}).
\gls{IMBH} candidates can be identified from their small-amplitude, stochastic, and short variability timescales using the characteristic variability timescale – \gls{BH} mass scaling relation. Stellar mass catalogs can be used to select for AGN-like variability in dwarf galaxies.

\paragraph{Science Objectives}
\begin{itemize}
\item Cataloging \gls{AGN}s in dwarf galaxies and/or \gls{AGN}s with rapid optical variability (e.g., days), like NGC 4395, to identify IMBH candidates
\item Characterize their variability properties and connect with multi-wavelength data
\item Enables studies of \gls{AGN} demographics at low stellar/\gls{BH} masses which traces early \gls{SMBH} seeding scenarios
\end{itemize}

\paragraph{Challenges (what makes it hard)}
\begin{itemize}
\item For small \gls{IMBH}s with 100 - 10,000 solar masses, the variability power is predicted to be strongest on timescales of \textasciitilde hours to days (Burke et al. 2021), so getting \textasciitilde hourly observations for a few days in some fields (e.g., DDFs) would be useful.
\item Host galaxy light contamination dilutes variability amplitude
\item Inferring the variability timescale in large numbers of light curves is computationally intensive
\item Host galaxy stellar mass estimates can be hard if an \gls{AGN} is involved
\end{itemize}

\paragraph{Running on \gls{LSST} Datasets (for the first 2 years)}
\begin{itemize}
\item We will analyze both the alert stream and the data release light curves. Most of the analysis will be based on catalog data (as most low-mass and low-luminosity AGNs are not expected to be variable enough to generate an alert) but we will need to go back to \gls{postage stamp} cutouts and potentially full fields to visualize any problems with the data.
\item The initial analysis will process all variable point sources, identify those with correlated \gls{AGN}-like variability (e.g., using the Ljung-Box portmanteau test or other test with weak priors on the SF), then identify their characteristic variability timescales (e.g., using the EzTao code; \citealt{Yu2022}) or, more likely, piggy-back on \gls{AGN} variability catalogs and select a sub-sample from them
\item Identify opportunities for stellar-mass estimates using multi-wavelength data
\end{itemize}

\paragraph{Precursor data sets}
\begin{itemize}
\item \gls{ZTF} (\citealt{2021ApJ...913..102W}), \gls{DES} \citep{2021arXiv211103079B}, \gls{SDSS}/PTF \citep{2018ApJ...868..152B,2020ApJ...896...10B}
\end{itemize}

\paragraph{Analysis Workflow}
\begin{itemize}
\item Retrieve variability properties (amplitude/timescale)
\item Restrict to dwarf galaxies and/or off-nuclear sources in massive galaxies
\item Match to external catalogs
\end{itemize}

\paragraph{Software Capabilities Needed}
\begin{itemize}
\item EzTao \citep{Yu2022}
\end{itemize}

\paragraph{References for Further Reading}
\begin{itemize}
\item \gls{IMBH} review \citep{2020ARA&A..58..257G}
\item Timescale-Mass relation \citep{2021Sci...373..789B}
\item Variability-Selection Papers \citep{2018ApJ...868..152B,2020ApJ...896...10B,2021arXiv211103079B}
%\item Wandering IMBHs % withdrawn by Suvi (see refs. within \S\ref{sec:PinpointWMBH})
\end{itemize}

\pagebreak
\subsubsection{Mapping SMBH Near Fields with Microlensing} \label{sec:CaustingCrossing}

% Caustic Crossing in Google

\Contributors {Matt O'Dowd (\mail{matthew.odowd@lehman.cuny.edu}), James Chan (\mail{hung-hsu.chan@epfl.ch}), Timo Anguita (\mail{tanguita@gmail.com}), Henry Best (\mail{hbest@gradcenter.cuny.edu}),  Joshua Fagin (\mail{jfagin@gradcenter.cuny.edu})}
{ April, 5, 2022}

\paragraph{Abstract}
 In strongly lensed quasars, the magnification structure in the plane of the source is highly inhomogeneous due to compact bodies in the lensing galaxies, resulting in magnification fluctuations known as microlensing. The rare and brief passage of the quasar central accretion disk across one of these caustic structures results in particularly intense differential magnifications on the scale of the SMBH event horizon. This has the potential to enable tomographic mapping of the structures on that scale. These caustic-crossing events are known to occur but have never been adequately followed up due to their rarity, with average per-lens event frequency <1 per 20 years, and due to the expense of continuous monitoring. High-cadence, multiwavelength monitoring of such an event enable us to measure: 1) the detailed accretion disk structure; 2) the size and geometry of the innermost stable circular orbit and photon sphere, providing a test of general relativity; 3) the geometry and kinematics of the X-ray corona and the X-ray Fe-Kalpha line (also a \gls{GR} test).

 LSST and the Euclid space telescope are poised to discover >200,000 strong gravitational lenses, including >8,000 lensed quasars. In addition, LSST will monitor the time variability of every lens over its 10-year survey, and has the potential to detect the onset of hundreds of high magnification microlensing events capable of resolving the inner accretion disk each year (\textasciitilde \% of the monitored lensed images per annum). The goal would then be to trigger extensive multi-platform follow-up of the most promising events (with, e.g. \gls{JWST} + Chandra/\gls{XMM} - Newton/\gls{XRISM} + ground-based optical/IR integral field spectroscopy) as the caustic feature scans the inner accretion disk. Due to the volume of the data, the complex behavior of light curves in different filters, and the expense of followup, it is important that we develop a specialized analysis pipeline to confidently identify impending caustic crossing events well in advance of the event itself.

\paragraph{Science Objectives}
\begin{itemize}
\item Studying the inner structure of accretion disk via microlensing light curves
\item Predicting the caustic crossing events in lensed quasars through the early LSST data release
\item Triggering the extensive follow-ups to capture the caustic crossing events in different filters
\item Simulating a comprehensive accretion disk model including microlensing, in order to analyse the observed light curves via machine learning techniques

\end{itemize}

\paragraph{Challenges (what makes it hard)}
\begin{itemize}

\item Finding the lenses:  The first objective is to obtain the largest sample of lensed quasars to date based on LSST imaging, outpacing existing catalogs by at least a factor of about \textasciitilde 20 \citep{2010MNRAS.405.2579O}. Currently, there are only \textasciitilde 200 lensed quasars found in various surveys. This objective will provide the unprecedentedly large sample size necessary to carry out the subsequent statistical and scientific analyses.

\item Timely lightcurve extraction: caustic crossing events can last from \textasciitilde days to weeks, with a lead-up that can last months. It’s important that we identify impending events early (several weeks before the event) to begin intermediate, high-cadence monitoring on smaller telescopes that will be able to confidently define and trigger multi-platform followup timed to the passage of the caustic across the SMBH region.

\item Custom deblending:  The two or four images of lensed quasars often have sub-arcsecond separation and so will be highly blended. Very careful deblending is critical both for lens discovery and precise photometry, and so we envisage building our own deblending pipeline for all lensed quasars to update lightcurves \textasciitilde 24 hours of observation. Followup triggers will be based on color changes in single lensed images over the first \textasciitilde few weeks of the microlensing event that leads to a caustic crossing event. Relative photometry at \textasciitilde a few \% between LSST bands is needed to trigger intermediate followup monitoring. An added challenge is that seeing will be different for bands observed on different nights, adding to the importance of careful deblending. Interpolation of lightcurves to enable simulated epoch-matching of photometry in different bands. We note that the flux and relative color changes of an impending caustic-crossing event will not typically generate an alert, necessitating a custom pipeline.

\item Follow-up: this program hinges on extensive, multi-platform followup, including by space-based and other high-demand facilities. This will require a suite of companion proposals. It’s important that we can demonstrate the reliability of LSST monitoring for the success of these proposals.

\end{itemize}

\paragraph{Running on LSST Datasets (for the first 2 years)}
\begin{itemize}

\item We’ll run our caustic crossing trigger watch on calibrated image cutouts within \textasciitilde 24 hours of observation to ensure timely, accurate deblending
\item We’ll run our algorithm on LSST coadd images in ugrizy bands to classify the lensed quasar candidates
\item Using the LSST imaging, new lenses will be modelled in order to proceed microlensing analysis

\end{itemize}

\paragraph{Precursor data sets}
\begin{itemize}
\item COSMOGRAIL: database of observed light curves  \url{https://www.epfl.ch/labs/lastro/scientific-activities/cosmograil/}
\item Lensed quasar database: \url{https://research.ast.cam.ac.uk/lensedquasars/index.html}
\item Gerlumph: containing the LSST light curve and microlensing simulators (\url{https://gerlumph.swin.edu.au})
\item The HSC survey: similar image quality will help build the lens search algorithm (\url{https://hsc.mtk.nao.ac.jp/ssp/survey/})

\end{itemize}

\paragraph{Analysis Workflow}

\begin{figure}
\begin{centering}
\includegraphics[width=0.8\textwidth]{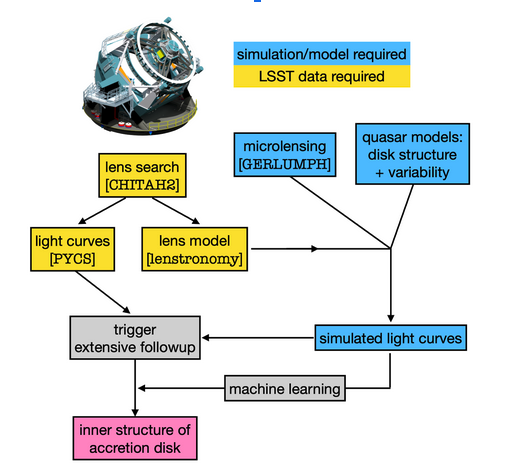}
	\caption{Analysis Workflow - \textbf{Figure attribution pending} \label{fig:CausticAlgo}}
\end{centering}
\end{figure}

See \autoref{fig:CausticAlgo}.

\paragraph{Software Capabilities Needed}
\begin{itemize}
\item PyCS: extraction of light curves in lensed quasars
\item GPU-D: microlensing simulation
\item Lenstronomy: lens modelling

\end{itemize}

\paragraph{References for Further Reading}

\begin{itemize}
\item Lens prediction in LSST \citep{2010MNRAS.405.2579O}
\item Microlensing light curve simulation in LSST \citep{2020MNRAS.495..544N}
\item Light curve extraction \citep{2020A&A...640A.105M}
\item Lens search algorithm \citep{2022A&A...659A.140C}
\item Lens modelling tool \citep{2018PDU....22..189B}
\end{itemize}

\pagebreak

\subsection{Local universe static science} \label{sec:luss}
\subsubsection{Mapping the Accreted and Intrinsic Stellar Populations in the Milky Way} \label{sec:MappingMWStellarPops}
\WOM{acronyms to myacronyms.txt}
\cleanedup{Nicol\'as Garavito-Camargo (May 16)}

\Contributors{Nicol\'as Garavito-Camargo, Adrian Price-Whelan, Alex Riley, Ilija Medan}
{ 2022-03-28}

\paragraph{Abstract}

The Milky Way as a galaxy represents a unique opportunity to constrain the detailed structure and formation history of a galaxy as a way of connecting fine-scale models of galaxy formation to statistical properties of galaxies in the universe. Critical to understanding its formation and evolution is mapping the stellar populations and density structure of the Galaxy as a way of constraining its accretion history, internal structure, and kinematics.

LSST/Rubin will enable mapping the stellar density structure of the Milky Way out to distances never before achieved over large areas of the sky. In particular, the depth of \gls{LSST} single and co-added images over \textasciitilde 18,000 sq. degrees will provide precise stellar photometric measurements for main sequence stars throughout the outer Galactic disk and inner stellar halo, and will enable tracing evolved stars (\gls{HB} and  \gls{RGB} stars) to the expected edge of the Galactic halo. This precise photometry will then allow measuring photometric stellar parameters (e.g., metallicity and luminosity or distance) for billions of stars, a potential intermediate science product of this work. With stellar population parameters and three-dimensional positional information for these samples, we can (1) map the density structures and asymmetries in the outer Galactic disk as it transitions to the inner stellar halo, (2) model the global structure of the inner halo constrain the clustering scales and stellar populations of known stellar streams and substructures, and (3) chart the density profile and structure of the (largely unconstrained) outer Galactic halo (beyond distances of 150 kpc).

%\begin{itemize}
%\item Wide-area sky coverage of more tracers to the outer edge of the Galaxy
%\item Resolving inner halo structures with main sequence stars vs giants/evolved stars (signal to noise/density)
%\item Better coverage toward outer disk midplane (dust extinction prohibitive for existing surveys)
%\item What are the measurables?
%\begin{itemize}
%\item Past: spherically averaged density profile, asymmetric parametrized density model fits
%\item Future, with \gls{LSST}-like numbers of tracers: Non-parametric or flexible models of density structures (e.g., BFE). Photometric metallicities -> chemodynamics
%\item Main sequence stellar populations in known dwarf galaxy streams (Orphan/Chenab, Sagittarius, \ldots)
%\item Density variations in (outer) Galactic disk – quantify degree of density asymmetry, scale height, etc.
%\end{itemize}

%\end{itemize}

\paragraph{Science Objectives}

\begin{itemize}
\item Catalog tracers - (B)HB stars or otherwise - and their distances (10\% accuracy) from the LSST data
\item Characterize their distribution around the Milky Way, using both spherically-averaged density profiles and more flexible models
\item Connect features in the distribution (e.g., breaks in density profile, global/local underdensities) to the Milky Way’s accretion history (e.g., \gls{GSE}, perturbations from \gls{LMC})
\end{itemize}

%{\bf Global note:} intermediate science product = spherically-averaged density profile, to subtract and look at residuals

With photometry from \gls{LSST}, we can probe the region that would transition from the outer disk (R > 20 kpc) to the inner halo (R < 100 kpc). To do this, distances and metallicities of the stellar populations must be known to place the populations at this boundary and use chemistry and kinematics to trace the transition zone \citep{2018MNRAS.478..611B}. When probing this zone we can determine if there is a smooth transition or if these stellar populations are well mixed out to larger radii. Additionally, properties of these populations can be determined as a function of radii to see, for example, if the density structure (i.e. scale height, warp) of the Galactic disk changes at larger radii.

In the above, there are multiple challenges to overcome. To determine photometric metallicities of the sample, a sufficient \gls{calibration} sample, matched with spectroscopic surveys like 4MOST, \gls{GALAH}, \gls{LAMOST}, etc.,  must be identified to relate LSST photometry to metallicity of these various stellar populations. Additionally, in the probing of the outer disk we will need to employ crowded-field deblending techniques to ensure a good level of precision (<10\%) in the resulting photometric metallicity estimates.

To focus on the inner halo population, in an isotropic universe we expect the inner halo to be defined as a spherically-averaged density profile. Previous studies have shown that this is not true due to the various substructures in the halo (i.e. stellar streams, merger debris, perturbations from dwarf galaxies, etc.). that can cause density enhancement and rapid density variations in the density profile of the halo, such as the break in the halo profile \citep{2011MNRAS.416.2903D}.  It has been shown that the main contribution to the inner halo comes from satellite galaxies that have or are in the process of merging with the MW. In particular, the \gls{GSE} and Sagittarius (Sag) are the two main accretion events that have contributed to the stellar halo structure \citep{2020ApJ...901...48N}. \gls{LSST} will discover many stars in the halo that will belong to these two mergers, such stars will enable the characterization of further details of those mergers, time, mass ratios and orbits. In addition, global properties of the halo, such as shape and mass, can be measured with the shape and kinematics of stellar streams.  To identify these substructures, we need to look for over/under-densities (density enhancements) in the inner halo population as compared to a model of a spherically-averaged density profile that has been convolved with the selection function for the survey.

Recently, data from the \gls{Gaia} satellite along with radial velocities and photometric data have allowed to measure the kinematic structure of the outer stellar halo, proving that the MW is out of equilibrium. The LMC just passed its first pericenter around the MW, during this first approach into the galaxy the LMC is perturbing the galaxy (\citealt{GC21,Rozier22}). The phase-space of the inner regions of the MW ($\approx \leq 30 kpc$) has been predicted to be displaced by at least 30 km/s and 50 kpc in the last Gyr \citep{Gomez15}, resulting in an apparent motion of halo tracers with respect to the inner halo. Such `reflex motion' was measured to be 32 km/s (\citealt{2020MNRAS.494L..11P,Erkal21}). While the LMC orbits the MW's halo it induces a DM wake trailing the LMC as a result of dynamical friction \citep{GC19}, and the counterpart stellar wake has been detected in \citet{2021Natur.592..534C}.
Both the reflex motion and the DM wake are important to characterize in detail since many measured quantities of the MW depend on the dynamical state of the halo (e.g., MW mass, halo triaxiality). LSST's large sky coverage with proper motions and photometry will allow to disentangle the signal from both the reflex motion and the wake with that of substructure in the halo. One of the main challenges of measuring the impact of the LMC in the MW's halo is the presence of substructure that can bias the measured values of the reflex motion (Riley et. al. in prep) and the amplitude of the wake \citep{Cunnigham20}. LSST's photometry and proper motions will help alleviate these problems by allowing to characterize substructure in the outer halo of the MW.

With the large sky coverage and depth from LSST the structure of density of the outer halo will be revealed. Key science objectives are: 1) Measure density enhancements caused by substructure, perturbations from the \gls{LMC}, and the smooth halo structure; 2) Determine if the structure of the outer halo similar to that of the inner halo, and if it consists of only substructure and streams or also from mixed structures; 3) See if there is an outer boundary of the outer halo or if the debris fall of continuously. Answering such questions could be addressed with intermediate LSST products such as star catalogs of Horizontal Branch (HB) Stars and RR Lyraes, which have been shown to be a good probe of the density profile of the outer halo \citep{2011MNRAS.416.2903D, 2018ApJ...852..118D} . Measuring the extent and density enhancements in the halo will require catalogs that provide 1) star/quasar separation and 2) calibration of distances and metallicities with 10\% precision 3) HB stars and RR Lyrae catalogs.

\paragraph{Challenges (what makes it hard) }

Technical challenges:  filtering on full catalog of photometry.

Algorithmic:

\begin{itemize}
\item Dealing with probabilistic star/galaxy separation
\item Dealing with probabilistic stellar parameters
\item Constructing appropriate selection functions for density measurements
\item Determining detection limits and color dependencies (vs. magnitude)
\item Determining efficient selection functions
\item Dealing with the effects of extinction by dust
\item Dealing with crowding effects
\item Making use of only positional data (i.e., how to select halo populations without kinematics or precise chemistry)

\end{itemize}
Other points:
\begin{itemize}
\item How often should the full analysis be run? \textasciitilde yearly? (gain depth and photometric precision over time)
\item No time urgency
\end{itemize}

\paragraph{Running on \gls{LSST} Datasets (for the first 2 years)}

\begin{itemize}
    \item Create value-added object catalogs with different stellar populations \gls{BHB}/RRLyr photometry
    \item Determine photometric distances for objects in the catalogs
    \item Cross-match catalogs with previous surveys such as: \gls{Gaia}, DES, SphereX, \gls{SDSS}, \gls{DELVE}, \gls{DESI}, but sky coverage varies by band
    \item Determine the selection function
    \item Analysis could be run every time the new catalogs are published (yearly)
\end{itemize}

\paragraph{Precursor data sets}

\begin{itemize}
\item Existing mock data don’t have substructure / streams
\item Could “train” algorithms on \gls{Gaia}/\gls{DELVE} catalogs, connect to \gls{LSST} on bright end, then extrapolate to fainter magnitudes (fainter than g \textasciitilde  20)
\end{itemize}

\paragraph{Analysis Workflow}

\begin{itemize}
    \item Data cleaning
    \item Star/galaxy separation
    \item Selection of high-quality stars at the catalog level. Filter using color-magnitude selections
    \item Matching to existing data sets
    \item Computing distances using photometry
    \item Establishing density estimation method that accounts for selection function
\end{itemize}

\paragraph{Software Capabilities Needed}

\begin{itemize}
\item Ability to query stars from full co-add source catalog and perform color and magnitude selections
\item Ability to query selection function / detection probabilities in arbitrarily-refined \gls{HEALPix} pixelizations of the \gls{survey footprint}
\item Flexible global density modeling with uncertain distances (marginalizations) with 100s of millions of sources
\item Traditional integration methods will be slow!
\item Flexible mixture density modeling to separate background halo from known structures, like Sagittarius stream
\item Need \textasciitilde few node capability on a compute cluster (handle fairly fast likelihood evaluations on \textasciitilde 100’s of millions of sources)
\item Ability to install custom \gls{software} analysis packages on compute resources
\item Visualization of large areas of the sky of the stellar halo without extragalactic sources.

\end{itemize}

%\paragraph{References for Further Reading}
%[1] \cite{2021A&A...649A...8G}

%[4] Helmi, Amina et al 2018. The merger that led to the formation of the Milky Way's inner stellar halo and thick disk
%[5] Naidu, R et al 2020.  Evidence from the H3 Survey That the Stellar Halo Is Entirely Comprised of Substructure

\pagebreak
\cleanedup{Olsen}
\subsubsection{Local Group Dwarf Galaxies Bound and Unbound} \label{sec:LGdwarf}

\Contributors{Knut Olsen, Adrian Price-Whelan, Alex Riley}
{3/29/22}

\paragraph{Abstract}
The dwarf companion galaxies of the Milky Way, \gls{LG}, and \gls{LV} are critical probes both of the dark matter halos that are the seeds of forming galaxies and of the physical processes that \gls{shape} their formation.
This is both because the bound dwarf galaxies directly trace the accretion history and building blocks of the stellar mass in halos, but also because the unbound dwarf galaxies — that now form \emph{stellar streams} that permeate the Milky Way and M31 halos — enable measurements of the detailed dark matter distribution in the Local Group and a full reconstruction of their assembly histories.
\gls{LSST} will provide an extraordinarily rich dataset for the detection and characterization of faint dwarf galaxies and stellar streams, and will enable heavily automated, statistical searches for these classes of objects, allowing for rigorous comparison to theoretical predictions.
As described in the report from the \gls{LSSTC}-sponsored workshop “Searching for Dwarf Companions in the Milky Way and Beyond” \citep{Bechtol-2017}, detection techniques for \gls{LV} dwarf galaxies will include catalog- and pixel-based searches, as well as the use of RR Lyrae as signposts.
As described in the report “Probing the Fundamental Nature of Dark Matter with the \gls{LSST}” from the \gls{LSST} Dark Matter Group \citep{2019arXiv190201055D}, the detection of new stellar streams and of density variations in already known streams (a signature of dark matter subhalo interactions) will be possible because of the depth to which \gls{LSST} can detect main sequence stars around the Milky Way. For this science case, the power of \gls{LSST} will be its wide, deep, and uniform survey, as well as the complement of the time domain for identifying rare but informative tracers like RR Lyrae stars.

\paragraph{Science Objectives}

\begin{itemize}
\item Identify and characterize the dwarf galaxy population of the Milky Way and nearby galaxies
\item Identify and characterize the population of stellar streams around the Milky Way (resolved stars) and nearby galaxies (integrated light and resolved RGB stars)
\item Constrain the stellar mass functions and density structures of known dwarf galaxy stellar streams around the Milky Way
\item Count the number of streams and measure their properties as a function of radius in the MW
\item Characterize and measure the detection probability of dwarf galaxies in catalog- and pixel-based searches for comparison to theoretical predictions
\item Connect the population of bound dwarf galaxies and satellites to the population of streams: Which dwarf galaxies and satellites have tidal tails or tidal distortions?
\end{itemize}

\paragraph{Challenges (what makes it hard) }

\begin{itemize}
\item All current dwarf galaxy and stellar stream searches have humans in the loop, in order to weed out contaminants and false positives (of which there are many)
\item Systematic searches and statistical analysis will require automation of detection techniques
\item All results will be candidates, requiring spectroscopic information for follow-up. This is expensive and resource-limited %(and technically-limited?)
\item For measuring stellar density variations in streams, requires knowing the selection function (sum of: detection probability, dustmap, crowding issues, …)

\end{itemize}
\paragraph{Running on \gls{LSST} Datasets (for the first 2 years)}

\begin{itemize}
\item Catalogs of photometry, ideally at coadd depth
\begin{itemize}
\item Based on sub-selecting metal-poor, distant sources, so \textasciitilde 100’s of millions
\end{itemize}
\item Time-domain photometry for a subset of (color-selected) sources (RR Lyrae)
\begin{itemize}
\item \textasciitilde 100,000s of sources
\end{itemize}
\item Coadd images for validation of dwarf galaxy candidates (especially distant ones)

\begin{itemize}
\item \textasciitilde 10,000’s of candidates
\end{itemize}
\end{itemize}

\paragraph{Precursor data sets}

\begin{itemize}
\item Dark Energy Survey (as an exemplar in semi-automated dwarf galaxy and stream searches)
\item Gaia data
\item Spectroscopic surveys of streams (S5)
\end{itemize}

\paragraph{Analysis Workflow}

\begin{itemize}
\item Data cleaning (e.g. removing bad data)
\item Derived or intermediate data sets and how these will be stored and accessed
\item Matching to existing data sets
\item The types of analysis techniques or \gls{software} packages that will be applied to the data (with a reference)

\item Select high-quality stars at the catalog level. \gls{Filter} using color-magnitude selections
\item Run \gls{algorithm} for finding candidates and quantifying significance

\begin{itemize}
\item Dwarf galaxies - simple/ugali
\item Streams - see \gls{software} capabilities needed! Existing methods all require humans
\end{itemize}
\end{itemize}

\paragraph{Software Capabilities Needed}
\begin{itemize}
\item Real/bogus candidate discrimination for dwarfs (automated)
\begin{itemize}
\item Algorithmic developments
\end{itemize}
\item Automated stream search \gls{algorithm}
\begin{itemize}
\item Algorithmic developments
\item How to find curved but coherent features that also have distance dependence (on the sky)
\item Would be helpful to have a way of mosaicking large contiguous regions of imaging in an inspectable way, maybe at degraded resolution, for inspection/validation (but that’s still human in the loop)
\end{itemize}
\item Ability to query selection function
\item Ability to inject artificial sources and re-process (more important for dwarf discovery)

\end{itemize}

\paragraph{References for Further Reading}
Dwarfs and Milky Way streams in \gls{LSST}, mainly from a dark matter perspective \citep{2019arXiv190201055D}\\
Dwarf galaxy detections out to 5 Mpc \citep{Mutlu_Pakdil_2021}\\
Streams in external galaxies with integrated light \citep{Pearson_2019}

\pagebreak
\subsubsection{The properties of the faint end of the Main Sequence: the stellar/sub-stellar boundary} \label{sec:FaintEnd}
\cleanedup{Ilija Medan}

\Contributors{ Ilija Medan, Knut Olsen, William O’Mullane, Luis Sarro}{}

\paragraph{Abstract}Low-mass stars are the most abundant stars in the Galaxy \citep{Bochanski_2010} and their long main sequence lifetimes mean they have the potential to trace the entirety of the Galactic star-formation history, and provide clues to understand the structure and evolution of the Milky Way. Additionally, low-mass stars are of great interest for the exoplanet community, as these sources are the best targets for detecting terrestrial planets in habitable zones and knowledge of the chemistry of these stars allows for constraints to be made on the physical properties of the orbiting planets and their formation history.  It is also at this faint end where we expect a large increase in non-stellar objects, like white dwarfs and brown dwarfs. For white dwarfs, \cite{2020ApJ...900..139F} predict 150 million to be observed down to 10-year depth, \textasciitilde 300,000 with 5-sigma parallaxes. So we estimate this dataset to be around ${10}^{9}$ objects or a significant fraction of the entire object catalog. Here we outline the objectives and challenges in utilizing the \gls{LSST} photometry to estimate the physical properties of these faint objects, where we specifically discuss the process that may allow for the estimation of properties like mass, radius, metallicity and age. These properties will allow for great progress in many studies that are fundamental to many areas of astrophysics.

\paragraph{Science Objectives}
Physical properties of low-mass stars (mass, radius, metallicity, age) are fundamental to many areas of astrophysics like Galactic archaeology, studies of the \gls{IMF} and the birth and evolution of star clusters and associations. In this science case we aim to:

\begin{itemize}
\item Identify a set of bona fide late-type sources (stellar and substellar) using \gls{astrometry} and photometry. This is a detection/classification problem where potential contaminants are unresolved galaxies and evolved higher mass stars (giants).
\item Estimate basic stellar parameters from the \gls{SED}: at least the effective temperature ($T_{\mathrm{eff}}$). Identify and tag low gravity (young) and low metallicity sources at least in the stellar regime. An example of this kind of analysis for the \gls{CARMENES} project can be found here.
\item Revisit and assess the validity of existing photometric metallicity relations and recalibrate them if needed. Then, infer photometric metallicities for the sample with a recalibration relationship. This is only possible for the stellar regime.
\begin{itemize}
\item A good starting point is, e.g., \cite{2005ApJ...626..465R}, \cite{2016MNRAS.460.2611S}, \cite{2021AJ....161..234M}.
\end{itemize}
\item A recalibration will require the identification of members in binary systems with bright primaries of well determined metallicities due to difficulties in determining metallicities of low-mass stars from spectra \citep[see][]{1997ARA&A..35..137A}.
\item Characterize the selection function so that inference at the level of population distributions can be achieved. This is crucial for studies of the luminosity and mass functions and their dependence on age and location in the local Galaxy.
\begin{itemize}
\item Potential selection filters appear in the limiting magnitude (survey design) and the classification stage described above.
\end{itemize}
\item The derivation of luminosities, masses and radii using several sets of theoretical models and revisit the possible existence of a radius minimum beyond the stellar/sub-stellar boundary where degeneracy pressure takes over as supporting mechanism \citep[e.g.~][]{2014AJ....147...94D}.
\item Recalibrate mass-luminosity and radius-luminosity relations.
\item Evaluate, where possible, age estimation techniques like gyrochronology \citep[requires time series modeling to infer rotation periods; see e.g.~][]{Popinchalk_2021, Godoy_Rivera_2021}. Evaluate alternative methods like hierarchical models that infer age and mass distributions as hyperpriors.
\item Identifying chemo-kinematic sub-populations based on the astrometric properties (proper motions and parallaxes) and the metallicities estimated above. See, for example, \cite{Hallakoun_2021}.
\item Identify new faint companions to existing binaries \citep[e.g.,~from][]{Hartman_2020, El_Badry_2021} and identify new binary systems using astrometric methods. With these detections, can study the multiplicity rate of stellar systems \citep[compare to e.g.~][]{2010ApJS..190....1R, 2019AJ....157..216W} and of sub-stellar systems.

\end{itemize}
See references for further reading to see prior work.

\paragraph{Challenges (what makes it hard) }
There are multiple challenges with accomplishing the above goals. These include:
\begin{itemize}
\item Understanding selection effects, which is an issue and a general problem for many science cases. Particularly, we must understand the effect selecting these stars from the full set and what the level of either misclassification or over-classification (i.e., being too restrictive in our cuts) has on the overall sample.
\item Generally, understanding how this work will interact with dynamic stellar science.
\item Getting a large enough \gls{calibration} subset for a large range in mass. There can be a challenge in getting a large enough sample that truly covers the mass range of M dwarfs, leading to not well constrained results at the lower mass end.
\item Removing contaminants (i.e., unresolved binaries and active stars) from the \gls{calibration} subset. This is most easily done with parallax measurements (to spot clearly over luminous sources that would have poorly estimated physical parameters from spectra). While this can be done if the pair is in Gaia (which may be a requirement for identifying wide binaries used to calibrate metallicity), this means the same kind of cleaning may not be able to be done once the photometric metallicity relationship starts to be used on field stars that may not have parallax measurements.
\item Photometric metallicity calibrations may not be as accurate for M dwarfs when just using optical data \citep[see][]{2016MNRAS.460.2611S, 2021AJ....161..234M}.
\item Large number of light curves to extract properties from ${10}^{9}$ objects.
\item Need more epochs to constrain this problem (i.e.~identify contaminants and get \gls{astrometry}), so need more years of data. Additionally, overall quality of data will improve later in the mission as we probe to fainter and fainter magnitude limits.
\item Star-Galaxy distinction for these low mass stars will be an issue.
\end{itemize}

\paragraph{Running on \gls{LSST} Datasets (for the first 2 years)}
For this work, we will mainly use the object catalog data to estimate physical parameters and will do so on every data release. For detecting variable stars and estimating ages using rotational properties, we will also use the source data to extra data via light curve analyses. We do note that \gls{LSST} data products will not be entirely sufficient for the first two years and we will need to extrapolate further properties (like proper motion) with external catalogs for the initial releases.

\paragraph{Precursor data sets}
We will utilize multiple precursor datasets throughout the analysis.
For astrometry during the first two years, we will utilize data from Gaia and \gls{DECam} (which will be used to bootstrap proper motions).
Additionally we will need data from large spectroscopic surveys (e.g., \gls{SDSS}, \gls{GALAH}) for the metallicity \gls{calibration} samples.

\paragraph{Analysis Workflow}
In the analysis for this project, we can generally split the methodology for how properties are estimated into those that only need color and those that need color and \gls{astrometry}. We expect that we can classify objects with color alone and also maybe metallicity, though it will not be as precise. Then, we expect that we will also need \gls{astrometry} to determine multiplicity, luminosity, mass and radius, and a more precise estimate of metallicity.

For the above, we expect the steps in the analysis to be the following for all properties:
\begin{enumerate}
\item Make cuts on the object catalog to classify faint objects. Initial cuts will be made with color alone, and more precise classification with early proper motion and parallax measurements.
\item Get the source data for all objects identified in the object catalog. Use light curves built with the source data to:
\begin{itemize}
\item Identify variable stars
\item Identify prime candidates for age \gls{calibration}
\end{itemize}
\item Cross-match the above samples with external catalogs
\item Construct a \gls{calibration} subset for each physical property of interest. The \gls{calibration} subset should evenly span the mass range probed.
\item Calibrate relationships for each physical property. These calibrations will vary for each physical property.
\item Apply the calibrated relationships to the objects initially selected from the object catalog.
\item Repeat and refine these calibrations with subsequent data releases.
\end{enumerate}

\paragraph{Software Capabilities Needed}
For the proposed science case, the main \gls{software} capabilities needed are:
\begin{itemize}
\item The ability to query the \gls{LSST} archive based on a combination of color, proper motion and parallax.
\item The ability to quickly process ${10}^{9}$ light curves.
\item Have the ability to cross-match the data to other large astronomical surveys.
\item Create new tables of physical properties for the sample.

\end{itemize}
We believe that most of the above can be done with existing infrastructure as the above queries can be written in \gls{SQL} and executed with \gls{Qserv}. The exception to this may be with the processing of the large number of light curves for this sample.

\paragraph{References for Further Reading}
\cite{DMTN-118}:  covers Rubin variable properties\\
\cite{Bochanski_2010}: Luminosity and Mass functions of low-mass stars from \gls{SDSS}\\
\cite{2012ApJ...757..112B}: Temperature and radii of KM Dwarfs related to photometry\\
\cite{2019ApJ...879...49C}: example of stellar aging from rotation (i.e.\ variability)\\
\cite{2020ApJ...900..139F}: WDs in \gls{LSST}\\
\cite{Garc_a_Berro_2016}: White Dwarf \gls{LF} \\
\cite{Document-37650} : Brown Dwarf \Gls{cadence} note\\
\cite{2009A&A...497..497G}: temperature of FGK dwarfs\\
\cite{1993AJ....106..773H}: Mass Luminosity relationship GKM dwarfs\\
\cite{2021AJ....161..234M}: photometric metallicities of KM dwarfs\\
\cite{2005ApJ...626..465R}: photometric metallicity FGK dwarfs\\
\cite{2016MNRAS.460.2611S}: Photometric metallicity/temperature of K/M Dwarfs\\

\pagebreak
\subsubsection{ The local \gls{IMF} as inferred from nearby star forming regions and clusters} \label{sec:LocalIMF}
\cleanedup{Luis M. Sarro}

\Contributors{Luis M. Sarro}
{March 28th, 2022}

\paragraph{Abstract}\label{abstract}

  Analysing the local Initial Mass Function and studying potential
  variations amongst various star forming regions requires a careful
  selection of members that is as consistent and homogeneous as possible
  in order to avoid biased determinations of this probability
  distribution. This has been done in the past based on hard cuts in
  proper motions and photometry, and more recently using probabilistic
  models of the distribution of sources (both the targets and the fore-
  and background contaminants) in astro-photometric space (including
  parallaxes when available). {Including the effects of extinction and
  reddening} has proved a difficult, computer intensive task.

  So far, the incomplete and patchy data collected as part of the
  Dynamical Analysis of Nearby Clusters project \citep[DANCe;][]{2013A&A...554A.101B} DANCe project contain $\sim 10^7$
  sources with varying depth, amongst which at most only a few thousands
  are members. These $10^7$ sources represent a lower limit to those expected from Rubin LSST.
  The LSST data set will include parallaxes (if available with sufficient
  accuracy), proper motions, and multi-band photometry, which can be used for membership
  determination using the Miec hierarchical bayesian inference code \citep{2021A&A...649A.159O}.

  Gaia \gls{astrometry} reaches down to $G\sim21$ which is
  insufficient to trace the low mass end of the \gls{IMF}. Existing surveys
  suffer from severe inhomogeneities (both in depth, accuracy and
  completeness) which severely hinder the possibility to infer
  distributions, because the selection functions are often very poorly
  understood.

  Examples and references for this kind of analysis can be found in
\cite{2022NatAs...6...89M}, \cite{2021A&A...654A.122G} and \cite{2021A&A...649A.159O}.

\paragraph{Science Objectives}\label{science-objectives}

%Provide a list of science objectives for the analysis. Points to
%consider:

%\begin{itemize}
%\item What will limit the science that you can achieve (e.g. the size of the sample, the accuracy of the photometry or \gls{astrometry})?
%\item Is there existing work in this area

%\end{itemize}

\begin{itemize}
\item Obtain complete censuses of nearby star forming regions of different ages in the LSST footprint
\item Convert the existing photometry (Rubin LSST plus complementary external archives mainly in the near and mid infrared) into physical quantities (mainly effective temperatures and masses)
\item infer probabilistically the Initial (or present-day) Mass Functions as a function of age for the SFRs in the  Rubin LSST footprint.
\item derive intermediate astrometric catalogs with improved uncertainties by using deep historic archival images, as exemplified by the \gls{TGAS} catalog.
\end{itemize}

\paragraph{Challenges }\label{challenges}
%Describe what are the primary challenges in undertaking the analysis. Points to consider:
%What challenges need to be overcome to undertake this use case. This could be technical challenges (e.g. how to analyze 107 light curves), algorithmic (e.g. current period finding algorithms are slow and don't work well with poorly sampled data), scientific (e.g. a lack of good models to fit to the data), logistic (e.g. getting access to follow-up telescope time).
%Will the quality of the LSST data impact the analysis (e.g. the number of false positives in the alert stream) and what will be needed to overcome any of these limitations (e.g. writing a specialized real-bogus classifier).
%How often the analysis will be run (e.g. will it be rerun for each Rubin data release or periodically with the alert stream).
%Is there an important timing/urgency constraint on your analysis (e.g. finding candidate microlensing events and triggering space-based followup)? If so, quantitatively, what are those time constraints?

\begin{itemize}
\item Prior to the Rubin LSST data releases, the main limitations are i) the unavailability of \gls{astrometry} for faint sources and ii) the  incompletenesses of the photometry. Existing data sets comprise a  collection of data from very different sources and the elicitation of the selection functions has been nearly impossible. As a result, the reliability of the inferred IMF is restricted to the mass ranges were completeness is achieved.
\item The inference problem takes $\sim 10$ hours for a selection of $10^5$ likely candidate member sources in a dedicated computing facility with eight Nvidia
GForce RTX 2080i GPUs. The preselection of these likely candidate members, however, is an undesirable step that leaves out sources with high extinction as well as outlying members, thus rendering the inferred \gls{IMF}s incomplete.

\item In principle, the quality of the \gls{astrometry} will be the major impact factor affecting the results. The analysis would be carried out early in the release schedule and updated with subsequent data releases. New data will imply a better selection of members and a more reliable determination of the \gls{IMF} for the low mass end which, depending on the star forming region and its extinction level, may be significantly undersampled in the first releases. For the bright end of the distribution, historic and archival data can be used to produce early good precision proper motions and parallaxes in a fashion similar to how the \gls{TGAS} catalog was produced \citep{2015A&A...574A.115M}.
\item Extinction maps (either produced as part of the Rubin LSST data releases or external) will be needed as part of the membership analysis. In star forming regions with strong extinction, members can appear significantly displaced from the (isochronal) photometric sequence. Existing codes \citep[Miec;][]{2021A&A...649A.159O} can (probabilistically) identify extincted members using preexisting extinction maps.

\end{itemize}

\paragraph{Running on LSST Datasets (for the first 2 years)}
%What data sets and LSST data products will be analyzed. Points to consider:
%What data sets will you utilize? For example, the alert stream, calibrated images, data release catalogs, the deep drilling fields. How long must the survey be in operation before you will run your analysis (e.g. you need 20 points in a light curve)
%Are the LSST data products sufficient for your analysis or will you need to create value-added catalogs or other derived data products
%What is the size of the data you will use (e.g. the number of light curves you will analyze). Does your science case require analyzing a subset of the population or will you use all galaxies/stars in the data release.

The first milestone for this project arrives with the availability of proper motions. As shown in the literature, proper motions are sufficient for complete (but contaminated) censuses. Parallaxes then allow (when and if available) for the cleaning of field contaminants with proper motions and photometry consistent with that of the \gls{SFR}.

We propose to complement LSST positions with archival data from existing surveys and projects to produce long time baseline estimates of the proper motions that help constrain parallaxes that would be otherwise difficult to estimate with only 2 years data. According to \cite{2019ApJ...873..111I}, proper motion uncertainties of the order of 0.2-1 mas yr$^-1$ (for $r$=21 and 24 respectively) are expected after 10 years of operation. After two years, the uncertainties would be a factor of 11 larger assuming a $t^{-3/2}$ scaling, with $t$ being the time span of the observations. This implies proper motion uncertainties of the order of 11 mas yr$^{-1}$ at $r$=24. For comparison, \cite{2022NatAs...6...89M} using archival data with a time baseline of 20 years in Upper Scorpio achieve average proper motion uncertainties of $\sim$ 1 mas yr$^{-1}$ for $r<24$. Having accurate proper motions derived using archival images can then allow for the determination of parallaxes (especially for nearby star forming regions where the signal is expected to be significant) in early data releases. For reference, parallax uncertainties of the order of 2.9 mas are expected after 10 years of operation which translates into 5.8 mas after two years (assuming a $t^{-1/2}$ scaling of the uncertainties). These uncertainties can be significantly reduced with the availability of accurate and precise proper motions.

\paragraph{Precursor data sets}\label{precursor-data-sets}

%What data can be used today to develop and test these use cases (e.g.
%the ZTF public data set). Is this data public?
%Are there other data sets that need to be assembled/collected 'prelaunch'
%to achieve your science? Do you need a validation set? Will there need to be
%cross-calibration (e.g. of photometry or astrometry) with those precursor data sets
%(e.g. gaining longer time baselines from adding LSST to ZTF to CRTS)?

Currently, the Gaia EDR3 provides the astronomy needed for the analysis only down to $G \sim 21$, which, depending on the youth and distance of the star forming region, translates into a mass lower limit that depends on the \gls{SFR}. The Dynamical Analysis of Nearby Clusters project \citep[DANCe;][]{2013A&A...554A.101B} provides the main stepping stone for testing the procedures. It provides  photometry and proper motions based on historical images of a number of star forming regions and clusters. The data sets include typically tens of millions of sources but the inhomogeneity derived from combining $\sim 10^4-10^5$ images of very heterogeneous instruments, cameras and surveys translates into every complex selection functions and difficult to assess incompleteness.

\paragraph{Analysis Workflow}\label{analysis-workflow}
%Provide a step-by-step description of how you will analyze the data including:
%Data cleaning (e.g. removing bad data)
%Derived or intermediate data sets and how these will be stored and accessed
%Matching to existing data sets
%The types of analysis techniques or software packages that will be applied to the data (with a reference)

\begin{enumerate}
    \item For early data releases, derivation of astrometry using archival archival images for a long time baseline. This is a demanding computation for tens of thousands of images available before the start of operations of the LSST \citep[see for example the analysis in][for Upper Scorpio]{2022NatAs...6...89M}.
    \item Selection in celestial coordinates of subsets of the catalog (one for each \gls{SFR}). This includes proper motions, photometry and parallaxes when and if available.
    \item Addition of complementary photometry if and when available (mainly near and mid infrared). This will require a cross match engine (either through the Rubin Science Platform or through an external service).
    \item Addition of an extinction map, preferably in probabilistic form.
    \item Identification of \gls{SFR} members using Miec or evolutions thereof.
    \item Probabilistic inference of \gls{IMF}s by conversion of the isochronal photometric sequence into masses. Incorporation of Selection function.
\end{enumerate}

\paragraph{Software Capabilities
Needed}\label{software-capabilities-needed}

%Describe the functionality needed in the software to undertake the science use case. Points to consider:
%Will we need to be able to query the LSST archive and what parameters will be used in the query (e.g. what columns in the database will be used)?
%Are you planning to access your data through community alert brokers? Are there software components (e.g. classification algorithms) needed to enhance these brokers.
%Are there new algorithms that will need to be developed for the science use cases or are there existing software packages that will need modifying for the science use case (e.g. to optimize for speed)?
%Is there new software infrastructure needed to run at the scale of the LSST data and estimates of how quickly these analyses should be run? This could include the ability to access other datasets or cross-match to other catalogs?
%How big are the data sets that will be run on and how slow are current approaches?
%What will need to be stored from these analyses (e.g. for the derived data products) and how much data will that be?
%What functionality will be needed to visualize the data, or the results, or to debug the analysis if it fails?

So far, Miec \citep{2021A&A...649A.159O} has only been applied to pre-filtered subsamples due to limitations in the computing power of existing infrastructure and the high computing needs of the software. This results in potentially incomplete censuses. As mentioned above, the existing software takes 10 hours to model a data set of $10^5$ preselected sources in the Pleiades cluster (with negligible extinction) on an infrastucture with eight Nvidia GForce RTX 2080i GPUs. For the regions with high extinction that constitute the objective of this science case, we aim at computation times of several days using a much more powerful infrastucture and data sets of the order of millions.

There is no need for on-the-fly queries to the data base. All data (astrometry and photometry) can be queried prior to the application of Miec.

For the derivation of astrometric quantities using historic archival images, additional computing and storage facilities will need to be allocated. Typically, $10^4-10^5$ wide field images will be added to the LSST images.

If approved, the new astrometry can be made available to the community which would imply additional tables in the catalog.

\paragraph{References for Further
Reading}\label{references-for-further-reading}

\cite{2013A&A...554A.101B}: description of the Dynamical Analysis of Nearby Clusters project that illustrate the possibility to combine LSST images with historic archival images to derive improved astrometric solutions as in \cite{2015A&A...574A.115M}.

\cite{2021A&A...649A.159O}: description of the Miec hierarchical bayesian software to identify Star Forming Regions members and model their distribution in the space of astrometric and photometric measurements.

\cite{2022NatAs...6...89M}: example of the kind of analysis proposed here as applied to the Upper Scorpio and $\rho$ Ophiucus Star Forming Region.

\pagebreak

\subsection{Local universe variable \& \gls{transient} science} \label{sec:luts}
\subsubsection{Identifying symbiotic binaries by their color and variability} \label{sec:SymbioticBinaries}

\begin{figure}
\begin{centering}
\includegraphics[width=0.9\textwidth]{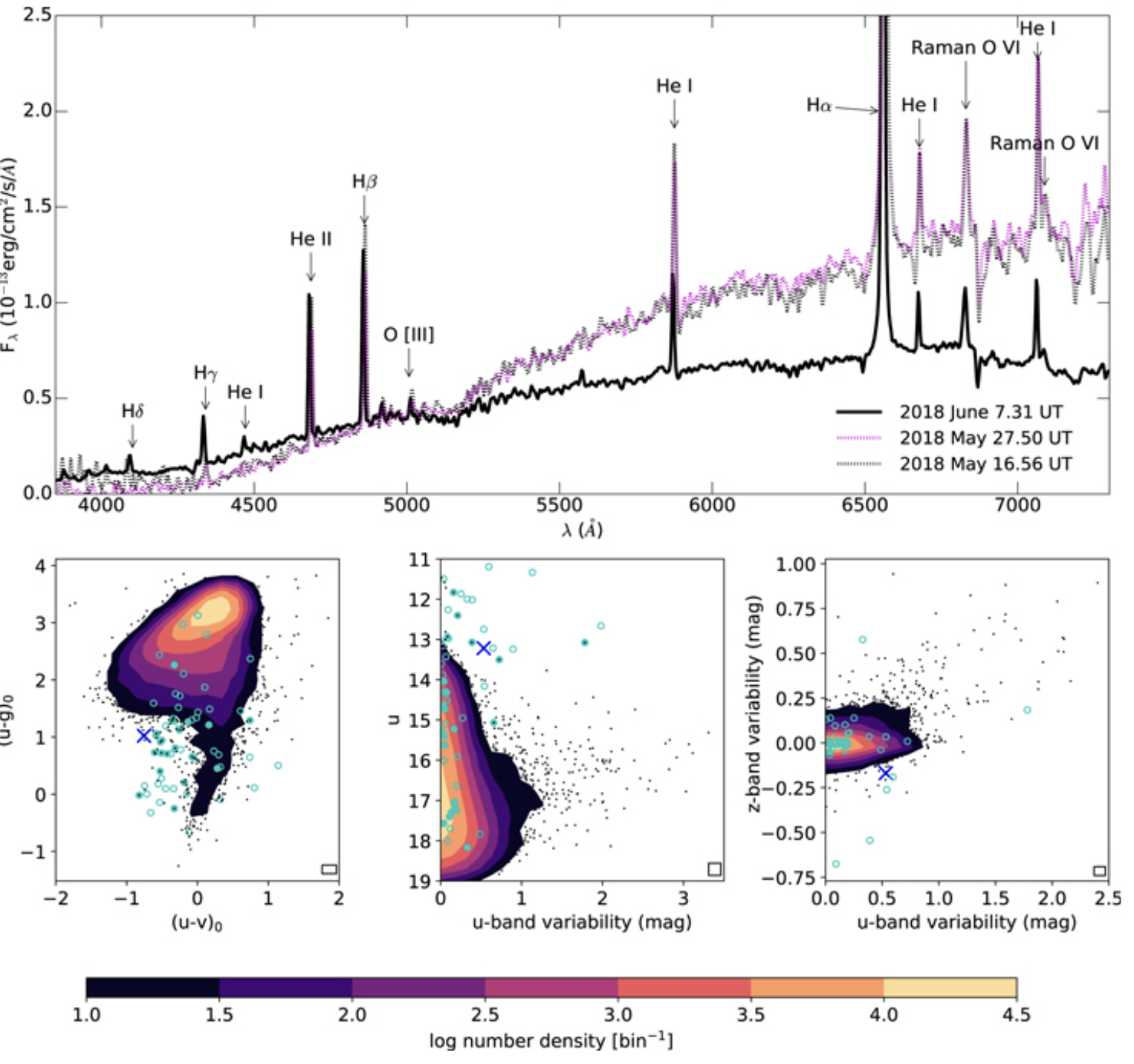}
	\caption{
 Upper panel: Our follow-up spectra of Hen 3-1768 exhibited the labeled emission lines. The first two were obtained with a 279mm Schmidt-Cassegrain telescope equipped with a LISA spectrograph (R$\sim$1500-1900, smoothed here to R-700; dotted lines) on 2018 May 16.56 and 27.50 \gls{UT} and reduced in \gls{ISIS}, the third at \gls{CASLEO} with the REOSC spectrograph (R-700; solid line) on 2018 June 7.31 \gls{UT} and reduced in \gls{IRAF}.
Flux calibrations are extremely preliminary, and absolute \gls{flux} \gls{calibration} was unavailable for the 2018 May 27.50 spectrum.
Lower panels: We selected Hen 3-1768 (blue cross) from a sample of $\sim$210,000 RGs (black points with contours over dense regions), because known symbiotics (turquoise circles, filled for symbiotics in the RG sample and hollow for symbiotics excluded from the RG sample by \gls{IR} or quality cuts) are outliers in spaces defined by these parameters from SkyMapper data: (u-g)0, (u-v)0, average u magnitude, maximum u variability between any two SkyMapper epochs, and z variability between those same two epochs (available for only $\sim$90,000 RGs).
The densest regions of parameter space are replaced by contours delineating the logarithmic number density of RGs (per bin; hollow rectangles).
De-reddening was performed using total Galactic extinctions.
Intervals between epochs range from 3 minutes to 1.5 years; the plotted interval for Hen 3-1768 is 142 days, larger than the sample median of 15 days. (Figure credit: \citealt{2018RNAAS...2..229L})
\label{fig:SymbioticBinaries}}
\end{centering}
\end{figure}

\Contributors{Gerardo Juan Manuel Luna (\mail{gjmluna@iafe.uba.ar})}
{03/22/2022}
\cleanedup{Juan Luna}
\paragraph{Abstract}
 The deep, repeated and multi-filter observations during 10 years of the LSST provide an opportunity to search for the yet-predicted, but not found, population of ${10}^{5}$ symbiotic binaries in our galaxy. Symbiotics are accreting binaries where a white dwarf (or a neutron star) accretes from the wind of a red giant. So far, about 400 symbiotics are known in our Galaxy and the Local Group, but theory predicts about ${10}^{5}$ \citep{1984PASA....5..369A, 2003ASPC..303..539M, 2019ApJS..240...21A}. The luminosity in symbiotics can be powered either by accretion or by nuclear burning on the white dwarf surface. Shell-burning, luminous symbiotics, can be detected by searching for their strong H$\alpha$ emission lines \citep[IPHAS]{2008A&A...480..409C}. Those accretion powered, less luminous symbiotics, display flickering of significant amplitude \citep{2013A&A...559A...6L}. Color-color and color-variability diagrams from surveys such as SkyMapper (\figref{fig:SymbioticBinaries}), \gls{WISE} and \gls{2MASS} have been used to identify regions where symbiotics might stand-out. This proposal aims to use LSST color-color and color-variability amplitude diagrams in order to identify new symbiotics either in our Galaxy or within the Local Group.

\paragraph{Science Objectives}
\begin{itemize}
\item Identify new symbiotic star candidates using color-color diagrams and variability information.
\item Further spectroscopy follow-up to nail-down the true symbiotic nature of the candidates
\end{itemize}

\paragraph{Challenges (what makes it hard)}
\begin{itemize}
\item Identify which combination of LSST filters is most suited to discriminate symbiotics from other objects such as cataclysmic variables, T Tauri, planetary nebulae, \gls{YSO}.
\item Identify variability in the light curves of different filters.
\item Build the control sample of known symbiotic stars observed with the \gls{LSST} filters in order to determine where most symbiotics are located in the color-color diagram. An alternative to obtain this control sample before the start of the \gls{LSST} is to use archival optical spectra and obtain colors in the \gls{LSST} filters system.
\item Implement machine-learning techniques to automatize the identification process and determine the regions on the color-color diagram where most symbiotics would be located \citep{2019MNRAS.483.5077A}.
\end{itemize}

\paragraph{Running on LSST Datasets (for the first 2 years)}
\begin{itemize}
\item We will analyze both the alert stream and the data release light curves, incorporating them to the best color-color and color-variability diagrams.
\item Once a candidate is identified, we will take an optical spectrum in order to verify its symbiotic nature.
\end{itemize}

\paragraph{Precursor data sets}
\begin{itemize}
\item Optical spectra from approximately 100 symbiotic in our own archive plus spectra taken by amateurs and available in the \gls{ARAS} database

\end{itemize}

\paragraph{Analysis Workflow}
\begin{itemize}
\item Remove poorly calibrated photometric data and sources flagged with suspicious photometry (e.g. on edge of \gls{CCD} or diffractions spike).
\item Remove/flag outlier measurements from a light curve
\item Remove/flag extended sources
\item Populate the color-color and color-variability diagrams.
\item Data storage and archives
\begin{itemize}
\item Collect control samples and store in a \gls{DB}
\end{itemize}
\item Variability Identification
\begin{itemize}
\item Characterize variability amplitude (e.g. rms of light curve) and filter based on this amplitude
\end{itemize}
\end{itemize}

\paragraph{Software Capabilities Needed}
\begin{itemize}
\item Ability to apply selection filters to data (\gls{SQL} query)
\item Storage of light curves as objects (time, \gls{passband}, noise, flags for bad points) with annotations (e.g. classifications) that can be queried and cross matched to other data sets
\item Storage of outputs of filtered and classified source candidates
\item Visualization of postage stamps and light curves for individual sources
\end{itemize}

\paragraph{References for Further Reading}
\begin{itemize}

\item \cite{2019MNRAS.483.5077A}

\item \cite{2021PhDT........17L}

\end{itemize}

\pagebreak

\subsubsection{Light Echoes: study the reflection of transients on interstellar medium in the LSST Era} \label{sec:LightEchoes}
\cleanedup{Xiaolong Li}
\Contributors{ Xiaolong Li (\mail{lixl@udel.edu}), Federica Bianco (\mail{fbianco@udel.edu})}
{3/26/2022}

\paragraph{Abstract}
\gls{LEs} are the reflections of astrophysical transients on interstellar dust. \gls{LEs} enable the study of the dust as well as the source transients. But they are rare and extremely difficult to detect because they appear as faint, diffusive features, presenting along with a lot of false positives, such as the star streaks, reflections of telescope, satellites, artifacts and so on. To date, only a few examples (\textasciitilde 100) have been discovered.

{While \gls{LSST} will detect thousands of  transients in the sky and release millions of alerts every night, the Rubin data processing pipeline is designed for point-sources, and it will entirely miss diffuse transient features (in the low SNR regime)}.

The combination of Rubin \gls{LSST}’s cadence, depth, and excellent image quality will have the potential to revolutionize \gls{LEs} studies, pushing it from serendipity discovery to a statistical regime.
But this requires the creation of data processing and analysis software tools specific for this purpose starting at the image-analysis level.
While Rubin's WFD strategy with repeat visits every $\sim3$ days would be excellent for LE studies,  \gls{LEs} appear in dusty regions where the LSST observing strategy may be different than the nominal WFD, potentially requiring imaging follow-up observations to be conducted at other observatories. Spectroscopic follow-up is critical to characterize the source transient (e.g., spectral typing, temperature constraints).
To maximize the science throughout, an end-to-end pipeline is necessary. This pipeline should include modules for the detection of the LEs, built upon \gls{LSST}'s image data, for automated follow-up management (including selection of the best candidates to follow up and integration with telescope observing software), and follow-up spectroscopic analysis in order to characterize the transients and constraint the dust properties.

\paragraph{Science Objectives}

\begin{itemize}
\item Study the evolution of \gls{LEs} over time in the \gls{LSST} image series
\item Analysis of the LE spectra to characterize the transient
\item Reconstruct the relevant dust properties (geometry, density)
\end{itemize}

\paragraph{Challenges}
The end-to-end LE pipeline would include: \begin{itemize}
\item Software to detect \gls{LEs} from \gls{LSST}'s image data within $\sim$days-to-weeks from LE first brightening to enable spectroscopic follow up before the LE vanishes;
\item Software to automate follow-up spectroscopic (and possibly imaging) observations
\item Software for the identification of the source transients through multi-epoch directional analysis and  cross-matching with historical catalogs;
\item Pipelines for the analysis of imaging and spectroscopic data to constrain properties of the reflecting dust and emitting transient.
\end{itemize}
Challenges include
\begin{itemize}
\item Potential need to tune the Rubin image subtraction models to work effectively in the low SNR regime for diffuse sources,
\item Creating a robust \gls{ML}/\gls{AI}-based detection model that can handle the large volume of \gls{LSST} image data and has high recall (as the phenomenon is rare) and high precision (as there are many potentital sources of false positives) is complicated by the small number of \gls{LEs} to train the model on,
\item Science throughput is enhanced by cross-matching and joint analysis of multi-band observations including infrared, X-ray, HI, and CO to study the reflecting interstellar medium properties, with the usual challenges associated to cross-matching that are discussed elsewhere in this document.
\end{itemize}

%--------
\paragraph{Running on LSST data sets (for the first 2 years)}
For the purpose of searching for \gls{LEs}, accurate sky templates in dusty regions of the sky are critical. While the Rubin-DM-pipeline-generated templates may be sufficient for our purposes, the current plan to minimize observations in dusty regions suggests the pipeline may not be specialized to handle these data and the template generation process may need revisions.
We may need to re-build template, especially dust regions.
%The current OpSims try to avoid dust regions, but it is important to allocate a certain amount visits to those region within the first 2 years.
As the Rubin detection pipeline will not discover extended transients at the SNR limit we will need to run a specific detection pipeline for \gls{LEs}. If we are unable to reprocess all \gls{LSST} images we would prioritize regions rich in interstellar dust, which have a higher probabilities of reflecting light of transients.

%{\textbf{TIME DOMAIN ASPECT}}

We'll create LE catalogs that include images postage stamps and transient sources. The postage stamp size will vary depending on the distance, size, and angle of the reflecting dust as \gls{LEs} can get as large as $\sim1$ {\arcmin }

\paragraph{Precursor data sets}
Precursor datasets include ATLAS and DES (and other DECam data). Both datasets are currently being used to create training sets for \gls{LEs} discovery. Images from DECam, while shallower, have comparable quality with \gls{LSST}; a cross analysis between \gls{LSST} and DECam needs to be developed.

\paragraph{Analysis Workflow}
\begin{itemize}
\item Image preprocessing
\item Creating templates;
\item Image subtraction: derive the subtracted images
\item Detection model: input images to detection \gls{pipeline}
\item Prioritize detections for follow up
\item Identify sources through photometric and spectroscopic analysis
\item Cross analysis with other \gls{passband} observations
\item Retrieve relevant interstellar dust structures
\end{itemize}

\paragraph{Software Capabilities Needed}

\begin{itemize}
\item Query \gls{LSST} image data,  including position and image specialization;
\item Access to the interstellar dust map;
\item Able to deploy well-trained detection models;
\item A visualization portal that can easily view \gls{FITS} images,  annotating and classifying;
\item Analysis tools for photometric and spectroscopic data
\end{itemize}

\paragraph{References}

\cite{rest2012light}: Light Echoes of Transients and Variables in the Local Universe.\\
\noindent \cite{patat2005reflections}: Reflections on reflexions I. Light echoes in Type Ia supernovae.\\
\noindent \cite{patat2006reflections}: Reflections on reflexions II. Effects of light echoes on the luminosity and spectra of Type Ia supernovae.\\

\pagebreak
\subsubsection{Compact White Dwarf Binaries in \gls{LSST}} \label{sec:CompactWD}

\Contributors {Alekzander Kosakowski (\mail{alekzander.kosakowski@ttu.edua})}
{2022-05-02}

\cleanedup{Alekzander Kosakowski}

\paragraph{Abstract}
Double-degenerate compact white dwarf binaries will be the dominant source of gravitational wave emission detectable by the upcoming \gls{LISA} mission. On the order of O(${10}^{2}$) systems are predicted to show measurable variations in both gravitational waves and electromagnetic radiation \citep{2017MNRAS.470.1894K}. Large sky surveys such as \gls{LSST} allow for the efficient identification of photometric variables in anticipation of the launch of LISA. Compact white dwarf binaries may show variability due to eclipses, tidal distortions, relativistic beaming, and reflection. These sources of variability are regularly used to characterize the physical properties of the binary, providing clues to its formation and eventual fate.

The Zwicky Transient Facility \citep{2019PASP..131a8002B,2019PASP..131a8003M} has enabled the swift discovery of such binaries in the northern sky. Simple period-finding searches such as Lomb-Scargle \citep{1976Ap&SS..39..447L,1982ApJ...263..835S}, conditional entropy \citep{2013MNRAS.434.2629G}, and Box Least Squares \citep{2002A&A...391..369K} have resulted in many scientifically valuable discoveries \citep[see][]{2020ApJ...905...32B}. We expand this search to the southern sky using \gls{LSST} and \gls{BlackGEM} to create a catalog of photometrically variable, compact, double-degenerate binaries that will merge due to the emission of gravitational waves. Combined with the ELM Survey \citep{2020ApJ...889...49B,2020ApJ...894...53K}, which detects binarity through a photometric and astrometric selection followed by measured radial velocity variability, we aim to create a complete catalog of compact white dwarf binaries that can be used to provide clues to the relatively-poorly understood stages of compact binary evolution, such as common-envelope evolution, and the formation rates of Type Ia supernovae and post-merger helium-rich objects. Having a complete list of compact binaries will provide valuable insight to the expectations of LISA. The combined data sets from various surveys over many years additionally facilitates direct measurements of the effects of gravitational waves (and tidal effects) on the orbits of these compact binaries \citep[see][]{2012ApJ...757L..21H}.

\paragraph{Science Objectives}

The main objective of this project is to create a catalog of photometrically-variable white dwarfs in the southern sky, with strong emphasis on ultra-compact ($P\lesssim60~{\rm min}$) white dwarf binaries, which will act as \gls{LISA} verification sources. Because this is a large sky survey, many other variable white dwarfs will be discovered, classified, and cataloged, including massive rotating white dwarfs, pulsating white dwarfs, and various eclipsing white dwarf binaries.

Follow-up spectroscopy will be performed on candidate binaries to confirm their spectral classification, atmospheric parameters ($T_{\rm eff}$ and $\log{g}$), and the presence of a companion evidenced by radial velocity variations.

\begin{itemize}
\item Crossmatching with other surveys such as Gaia, VST ATLAS, SkyMapper, and \gls{BlackGEM} will be the first step to identify the relatively bright white dwarf binaries. Fainter white dwarfs only visible to \gls{LSST} will be identified through photometry and astrometry cuts to the \gls{LSST} data as it becomes available.
\begin{itemize}
\item Even a simple positional cross-match can be slow when using large surveys such as Gaia. Developing a parallelized cross-matching \gls{algorithm} will be necessary to efficiently identify targets across many surveys.
\end{itemize}
\item Light curve processing may need to be completed on the user's local machine, thus a form of bulk data download is necessary in order to allow users to work off of the Rubin \gls{Science Platform} for more advanced analysis such as light curve modeling.
\begin{itemize}
\item Remotely processing many light curves may be too taxing on the Rubin \gls{Science Platform} servers considering that the resources are shared.
\item This will require a streamlined command-line tool for users to easily query for large numbers of sources at once.
\end{itemize}
\item Use of the \gls{RSP} to run period finding algorithms on a large number of light curves at once is important. Many useful period finding algorithms exist and can be ready for bulk use after minor modifications.
\item Finally, classifying each light curve based on features identified in their light curves (periodicity, amplitude, variability \gls{shape}, eclipse duration, etc) will allow for an organized catalog of white dwarf variables.
\begin{itemize}
\item This objective will require a machine learning classification \gls{algorithm}. Use of existing data from other surveys will help create training sets for classification.
\end{itemize}
\end{itemize}

\paragraph{Challenges (what makes it hard)}
\begin{itemize}
\item Because white dwarfs are inherently faint due to their size, nearby field stars may easily dominate a blended \gls{PSF} in crowded fields. Developing a deblending algorithm capable of handling a large difference in stellar \gls{PSF} profiles while not significantly affecting potentially weak periodic signals is a non-trivial task.
\item Period-finding algorithms on dense frequency grids are computationally expensive, especially when trying to properly handle multi-band data. Developing an efficient \gls{algorithm} capable of identifying multiple types of variability and handling multiple filters simultaneously is not an easy task.
\item The \gls{algorithm} will need to be rerun as more data is taken. Because computational resources are likely to be scarce, determining the most efficient times to rerun the periodogram will also be a challenge.
\item Follow-up time-series spectroscopy is expensive.
\end{itemize}

\paragraph{Running on LSST Datasets (for the first 2 years)}
\begin{itemize}
\item This project will make use of the calibrated light curves and \gls{FITS} images taken over the first two years in combination with other large survey data.
\item \gls{LSST} light curves will be combined with other surveys, such as ZTF and \gls{BlackGEM} to increase temporal sampling and allow for rapid identification.
\item Focused deep drilling fields may provide early detection of scientifically valuable binaries.
\item Given the expected eclipse durations for these double-degenerate binaries ($\sim$$60~{\rm s}$), having greater than $\sim$100 data points across all filters may allow eclipsing binaries to be identified programatically. Sinusoidally varying systems may be identified with fewer data points.
\end{itemize}

\paragraph{Precursor data sets}
\begin{itemize}
\item ZTF’s southern-sky overlap with \gls{LSST} will provide a valuable test-case for our analysis.
\item During the early stages of \gls{LSST}, SkyMapper, VST ATLAS, Gaia DR3, and BlackGEM will provide valuable information that will augment the \gls{LSST} data.
\end{itemize}

\paragraph{Analysis Workflow}
\begin{itemize}
\item "Poor quality" images will removed using the recommended \gls{LSST} data quality flags.
\item For the relatively bright objects still visible in other surveys, the "good quality" \gls{LSST} light curve data will be combined with light curve data from other surveys to produce more complete light curves with longer baselines and better sampling than \gls{LSST} alone would provide.
\item Various periodograms will be run on each light curve to identify different types of variability. Initially, a simple Lomb-Scargle and Box Least Squares periodogram can be run to identify compact ellipsoidal and eclipsing binaries.
\begin{itemize}
\item Light curves that do not show M N-sigma deviant points through \gls{LSST} alerts should not use resources on an expensive Box Least Squares search for eclipses. Determining M and N such that we efficiently identify real eclipses while successfully rejecting false positives is a challenge; too few cuts will result in significantly higher run time and time lost filtering false positives from the final sample, while too many cuts will result in many missed eclipsing binaries. It is important to use forced-photometry for these light curves and not reject truly deep eclipses from the light curves.
\end{itemize}
\item Create phase-folded light curves at the most-probable frequency determined through each periodogram.
\begin{itemize}
\item For Lomb Scargle: record light curve statistics of the phased light curve, including most-probable period, amplitude of variability, and goodness of fit for a single- and double-sine fit.
\item For Box Least Squares: record most-probable period, eclipse depth, eclipse duration, and number of in-eclipse data points per filter.
\end{itemize}
\end{itemize}

\paragraph{Software Capabilities Needed}
Given the volume of data expected to come from the \gls{LSST} program, performing a complete analysis on even subsamples of this data will be computationally expensive. White dwarfs in general display a broad range of photometric variability in terms of shape and period, with variations between milli-magnitudes and magnitudes on periods on the order of single-minutes to days. To perform large-scale periodicity searches on candidate white dwarfs requires an efficient algorithm capable of equally identifying both short- and long-period eclipsing systems (such as Box Least Squares) and sinusoidally-varying systems (such as Lomb-Scargle and Conditional Entropy). However, the scale of this problem may be reduced if only considering compact gravitational wave LISA verification binaries, which have a relatively narrow frequency range.

\begin{itemize}
\item The efficient completion of this project requires a consistent object identifier number across all filters. Having a common multi-filter identifier allows for quick extraction of all epochs per object and allows each object to be assigned complete \gls{LSST} color information which will be used for astrometric and photometric target selection of candidate binaries. Astronomers working on specific objects looking to crossmatch their work with \gls{LSST} will greatly benefit from a bulk light curve search on position that returns a single object ID per object.

\item A sophisticated period-finding algorithm will need to be run in order to identify variability and perform classification. Even restricting our period search to within the expected detection limit on LISA, the volume of data expected requires a highly-parallelized algorithm capable of handling non-regular light curve sampling and heteroskedastic flux errors, which would be run on a state-of-the-art high-performance computing center. A GPU-accelerated program may perform the periodicity search over a dense grid best. However, over the 10-year \gls{LSST} baseline, many of these LISA verification binaries will show measurable period decay, which may negatively affect a standard periodogram's ability to detect orbital periods. As the \gls{LSST} project evolves, the need for a periodogram that detects both period and period decay will increase.

\item The results may be presented to the user through an online GUI. Given the users target selection region, the back-end would collect target information, such as colors, light curves, and periodograms, and present it in a table format. Selecting specific targets on the table would display the calibrated \gls{LSST} light curve, its periodogram, the calibrated light curve phase-folded to the most-probable frequency and its half-frequency, and the object's location on the HR diagram for quick by-eye classification. The calibrated \gls{LSST} light curve will help eliminate flaring systems, such as Cataclysmic Variables. Providing a phase-folded light curve at the half-frequency will help classify ellipsoidal variability and reflection effects.

\item Combining light curves from \gls{LSST} with other southern-sky surveys may provide significantly better sampling, which would improve the likelihood that periodic variables are detected at their true periods and increase detection efficiency. With this in  mind, the \gls{LSST} program would benefit from providing the user the ability to import their own light curve data and augment the \gls{LSST} data for additional periodogram analysis.

\end{itemize}

\paragraph{References for Further Reading}

\cite{2019PASP..131a8002B}, \cite{2020ApJ...889...49B}, 
\cite{2020ApJ...905...32B}, 
\cite{2002A&A...391..369K}, 
\cite{2013MNRAS.434.2629G}, 
\cite{2012ApJ...757L..21H}, 
\cite{2017MNRAS.470.1894K}, 
\cite{2020ApJ...894...53K}, 
\cite{1976Ap&SS..39..447L}, 
\cite{2019PASP..131a8003M}, 
\cite{1982ApJ...263..835S}

\pagebreak
\subsubsection{Analysis of Microlensing events by stars and compact objects} \label{sec:MicroLensingGP}

\Contributors{Rachel Street (\mail{rstreet@lco.global}), and TVS Microlensing Group}
{
2022 March 25
}

\cleanedup{Rachel Street}

\paragraph{Abstract}
The technique of microlensing is routinely used in the Galactic Plane to explore populations that are too faint for direct electromagnetic detection.  Requiring only multiband optical time series photometry of a background source star for detection, rather than of the target lens that passes in front of it, microlensing will provide insights into the population of low mass stars, planets and isolated compact objects in regions across the Milky Way where they are otherwise inaccessible.  Historically, almost all lensing events have been identified in the Galactic Bulge; \gls{LSST} will discover thousands of events per year across the Galactic Plane, allowing us to explore these populations in a greater range of formation and evolutionary contexts.  Lensing by isolated black holes will enable us to constrain the black hole mass function and hence theories of their formation \citep[e.g.,][]{2000ApJ...542..785G}.  This and other complementary science is described in \cite{2018arXiv181203137S} and  \cite{PoleskiMroz2018}. Furthermore, \gls{LSST} will operate contemporaneously with the Roman Space Telescope’s near-infrared survey of the Galactic Bulge, and can provide highly complementary optical data that will not only provide additional model constraints for >1400 bound exoplanetary events, but also constrain the masses of \textasciitilde 250 free-floating planets, and increase the number of planetary anomalies discovered by filling in gaps in the Roman survey \gls{cadence} \citep{2018arXiv181204445S}.

 Models of microlensing are derived from the multi-band lightcurves, but events suffer from multiple degeneracies,  so the analysis of anomalous events in particular entails searching a large and non-linear parameter space for the best fit.  With \gls{LSST} expected to identify thousands of events, but provide relatively low cadence data in most cases, this model fitting process will be computationally  intensive, but highly parallelizable, as each event can be evaluated independently.  For some events where contemporaneous lightcurve data is available from other facilities, the analysis will need to access external data catalogs (e.g., from Roman) and include these data in the model fitting process.

\paragraph{Science Objectives}
\begin{itemize}
\item Evaluate how well microlensing events are identified from the \gls{LSST} alert stream by different brokers and the insights gained by combining their output.   Tests \citep[e.g.,][]{2019A&C....2800298G, 2020RNAAS...4...13M} have shown that classification algorithms are most effective when a baseline lightcurve of $\sim$1\,yr duration is available, with a photometric precision of <1\%.
\item Identify microlensing anomalies from the lightcurve data, ideally while the event is ongoing and the lightcurve data is incomplete.  Prior work exists in this area (see above links, e.g.) but needs implementation in an \gls{LSST} context.
\item Implement parallelized modeling of large numbers of lensing events.  This will need to robustly identify cases where degeneracies or data gaps warrant further evaluation.  Several open-source modeling packages are available (listed on the Microlensing-Source website \citealt{microlensing-source-software}, see \citealt{2017AJ....154..203B, 2019A&C....26...35P, 2010MNRAS.408.2188B}).

\end{itemize}

\paragraph{Challenges (what makes it hard)}
\begin{itemize}
\item The scale of the analysis: infrastructure is needed to manage the modeling of (potentially) hundreds of events at any given time, while keeping track of the status of each one.
\item Identifying events that need additional attention: some events can be accurately modeled in a fully automated manner, but many suffer from model degeneracies, especially for the scientifically most interesting events.
\item Identifying anomalies in real-time, to enable follow-up observations to be triggered where appropriate. Ideally this should take place with as short a latency as possible, within hours.  Modeling of ongoing events will need to take place continuously, with updates whenever new data are acquired.
\item Reliably combining downstream alert streams from different brokers.
\item The \gls{cadence} of LSST lightcurves, depending on the observing strategy.  Combining data from multiple passbands and acquiring follow-up data where necessary can help to address this.

\end{itemize}
\paragraph{Running on \gls{LSST} Datasets (for the first 2 years)}
 This science will utilize the real-time \gls{LSST} alert stream and lightcurve catalogs. For effective false-positive rejection, we anticipate at least 6--12\,months of baseline data will be needed.  It is expected that (approximately) thousands of events will be detected per year \citep{2019ApJ...871..205S}.  The derived data product will be a catalog of the fitted model parameters and derived physical parameters for the sample of lensing events.

\paragraph{Precursor data sets}
 There are various datasets from existing surveys that are currently being analyzed for microlensing events, some of which are public, for example, some \gls{ZTF} data, Gaia data, MOA data.

\paragraph{Analysis Workflow}
\begin{itemize}
\item Query multiple brokers for microlensing detections and combine their assessments to prioritize events.
\item Fit multi-band \gls{LSST} lightcurves together with any other lightcurves available, in real-time.
\item Review combined lightcurves for anomalous deviations; if detected, evaluate for follow-up potential.
\item Identify and evaluate events subject to degeneracies.
\item Analyze the CMD for the region of the event and constrain source parameters.
\item Derive model, and hence physical, event parameters

\end{itemize}
\paragraph{Software Capabilities Needed}
\begin{itemize}
\item Detection and accurate classification of microlensing events in progress by LSST alert brokers, as well as access to the resulting alert data.  A number of algorithms are in place at \gls{ANTARES} and other brokers.
\item Real-time access to LSST lightcurve data on (RA, Dec) query, and previous Data \gls{Release} photometry for the target and surrounding region of sky.
\item Access to Roman lightcurve catalogs on (\gls{RA}, Dec) query, as well as Gaia, VVV and other data catalogs for brighter events.   This should be available from a NASA data archive.
\item Infrastructure to automate and parallelize the modeling of hundreds of events simultaneously and access to a \gls{CPU} cluster facility on which to run the model processes.  GUI to enable user monitoring of the modeling processes
\item Real-time anomaly detection \gls{software}.  Prior algorithms are available but will need revamping for LSST’s context.
\item \gls{TOM} system to manage follow-up observations.  A prototype system is in operation, which could also manage the modeling processes if deployed suitably.
\item Multi-TB storage for lightcurve collections and \gls{TOM}/modeling system database.
\item Webservice to serve GUI for TOM system to collaborators worldwide.
\item Note that the current state of the art for the modeling of a single binary lensing event is \textasciitilde 2\,hr.

\end{itemize}

\paragraph{References}

\cite{2000ApJ...542..785G}: A Natural Formalism for Microlensing

\noindent\cite{2018arXiv181203137S}: The Diverse Science Return from a Wide-Area Survey of the Galactic Plane

\noindent\cite{PoleskiMroz2018}: The First Extragalactic Exoplanets — What We Gain From High \Gls{cadence} Observations of the Small Magellanic Cloud?

\noindent\cite{2018arXiv181204445S}: Unique Science from a Coordinated LSST-WFIRST Survey of the Galactic Bulge

\noindent\cite{2019A&C....2800298G}: A machine learning classifier for microlensing in wide-field surveys

\noindent\cite{2020RNAAS...4...13M}: Gravitational Microlensing Events from the First Year of the Northern Galactic Plane Survey by the Zwicky Transient Facility

\noindent\cite{microlensing-source-software}: Microlensing Source - Publically available software for microlensing modeling, simulation and analysis

\noindent\cite{2017AJ....154..203B}: pyLIMA: An Open-source Package for Microlensing Modeling. I. Presentation of the Software and Analysis of Single-lens Models

\noindent\cite{2019A&C....26...35P}: Modeling microlensing events with MulensModel

\noindent\cite{2010MNRAS.408.2188B}: Microlensing with an advanced contour integration algorithm: Green's theorem to third order, error control, optimal sampling and limb darkening

\noindent\cite{2019ApJ...871..205S}: Predictions for the Detection and Characterization of Galactic Disk Microlensing Events by LSST

\pagebreak
\subsubsection{Young stellar objects and their variability} \label{sec:YSOvariability}
%\WOM{CItations are not in main.bib nor cited.}
\Contributors{Sara (Rosaria) Bonito, (\mail{rosaria.bonito@inaf.it}), Rachel Street, Sabina Ustamujic (\mail{sabina.ustamujic@inaf.it}), Laura Venuti (\mail{lvenuti@seti.org})}
{2022 March 28 - April 1}

\paragraph{Abstract}

\glspl{YSO} show short-term as well as long-term photometric variability related to the physical processes at work in these complex systems and their geometry, e.g. mass accretion from circumstellar disks (which can proceed in a steady, funnel-flow pattern as well as in bursts or eruptive events), flares, rotation, or the presence of dusty warps within the inner disks.
 Monitoring the variability over different timescales is critical to constrain the dynamics of these processes, and studying the color dependence of such variability is essential to disentangle the characteristic signatures of distinct potential mechanisms at play.
Vera C.\ Rubin Observatory \gls{LSST} will be the ideal instrument to allow us to obtain well-sampled multicolor lightcurves of star forming regions (as e.g. Carina, Orion Nebula Cluster, NGC 2264, NGC 6530, NGC 6611) to acquire the first statistically significant data on how young stars vary on both short and long timescales.
Thanks to its depth and duration, the Rubin \gls{LSST} survey will provide us with crucial information to both discover new populations of accreting young stars for classification, and to characterize known objects.
 From the point source catalogs extracted from stacked images, we will be able to achieve the most extensive reconstruction to date of the manifold accretion processes at play in YSOs, and of how they evolve as a function of stellar mass and average cluster age.
 A deep, uniform sky coverage is crucial to ensure statistical representation for young stars of all spectral types in clusters, and to sample the impact of different cluster ages and different environmental conditions (e.g., field crowdedness, presence of massive stars).
By exploring the light curves acquired with LSST at different filters ($u$,$g$,$r$,$i$), we can characterize and classify YSO variability.
 We can discriminate different processes at work in YSOs by examining how the location of individual sources changes with time on color-color and color-magnitude diagrams.
 From the comparison between individual stellar colors and reference photospheric color sequences, we can discern accreting young stars, or \glspl{CTTS}, from non-accreting young stars, or \glspl{WTTS}, while the spectrum of the observed \gls{flux} variations (i.e., amplitude of color variations and slope of photometric variations on color-magnitude diagrams) on different timescales can reveal whether such variations are dominated by accretion events or circumstellar extinction events (e.g., by inner disk warps).
See White Paper by \citet{2018arXiv181203135B}, \Gls{cadence} Note by \citet{bvcadence}, and \citet{2014A&A...570A..82V,2015A&A...581A..66V} for more information on the science case discussed here.

\paragraph{Science Objectives}
\begin{itemize}
\item Variability due to different physical processes in young stellar objects
\item Discrimination of different processes at work in young stars
\item Description of both short-term and long term variability in young stellar objects
\item Classification and characterization of variable stars
\item Spectroscopic follow-up of variable processes as, e.g., EXor objects \citep{2008AJ....135..637H}
\item We need a proper \gls{cadence} to populate the lightcurves to follow short-scale variability
\item We require the bluest filters to describe the accretion process in YSOs
\item Multi-filter observations will allow us to explore color-color and color-magnitude diagrams of YSOs

\end{itemize}

\paragraph{Challenges (what makes it hard)}
\begin{itemize}
\item A proper \gls{cadence} to retrieve different  shapes of lightcurves for different processes at work is needed for the short-term variability
\item Classification of different kind of variability
\item \gls{Alert} stream for variability of EXor objects (long-term variability) is important also for spectroscopic follow-up
\item Early Science exploration of a testbed star forming region (e.g. Carina) is strongly suggested to extend the investigation to other \glspl{SFR} (see \citealt{2018arXiv181203135B}).

\end{itemize}

\paragraph{Running on \gls{LSST} Datasets (for the first 2 years)}
\begin{itemize}

\item Early Science exploration of \glspl{LC} in one selected star forming region (e.g., Carina) has been proposed in \citet{bvcadence} as a micro-survey for the optimization of the survey strategy
\item Similar micro-surveys can extend the investigation to other SFRs (see \citealt{2018arXiv181203135B}): we have identified as ideal a coverage of the LC with 140 points in 1 week for each filter:
\item one point every 30 minutes in a 10 hour/night observation for 7 consecutive nights,
\item motivated by the fact that this sampling will allow us to reveal short-lived phenomena in
\item YSOs with a probability of 5\% (Fig.~1, \citealt{bvcadence})
\item Additional spectroscopic follow-up and multi-band (including also \gls{UV} and X-ray data) will complete the characterization of the sample
\end{itemize}

\paragraph{Precursor data sets}
\begin{itemize}

\item We have simulated the Rubin LSST LCs \gls{cadence} by using \gls{CoRoT} and \gls{K2} data for short-term variability and \gls{ZTF} data for EXor objects (long-term variability)
\item Earlier large-scale surveys covering the same area of the sky to be targeted with LSST in similar filters (e.g., \gls{ZTF}, VPHAS+, \citealt{2014MNRAS.440.2036D}; SkyMapper Southern Sky Survey, \citealt{2018PASA...35...10W}; PanSTARRS, \citealt{2016arXiv161205560C}) will be ingested and will serve both as preliminary tools for the identification of a reference sample of well-known young stellar populations in the region, and to extend the time coverage of the LCs reconstructed for our LSST targets.
\item Photometric cross-calibration between different surveys will be achieved statistically by selecting the population of field stars (not YSOs, which are intrinsically variable) common to all catalogs and spatially located in the same region as our young stellar populations of interest (see, e.g., \citealt{2014A&A...570A..82V}).

\end{itemize}

\paragraph{Analysis Workflow}
\begin{itemize}
\item Preliminary mining of data archives from large-scale surveys available in the literature at similar wavelengths (e.g., VPHAS+, SkyMapper Southern Sky Survey), to inform an initial catalog of known young stellar populations in the areas to be surveyed with \gls{LSST}.
\item Analysis of diagnostic diagrams, such as the distribution of photometric scatter vs. apparent magnitude (e.g., \citealt{2021AJ....162..101V}) and the distribution of \gls{flux} aperture radius vs. apparent magnitude on the CCD frame, to flag and discard saturated data, spurious detections (e.g., cosmic rays), and background-dominated point sources.
\item Identification of young stellar objects in the \gls{LSST} catalog, by inspecting properties such as photometric clustering on color-magnitude diagrams (to locate the sequence traced by cluster stars over the field population), and kinematic or astrometric association in the field. The preliminary catalogs of young stellar populations built from the literature mining step will be used as a guide to interpret the distributions in photometric properties of \gls{LSST} targets. This step will enable discoveries of new young stellar populations, as well as the achievement of a more complete census for already known young stellar populations, thanks to the unprecedented depth of the \gls{LSST} survey (critical, for instance, for studies of the stellar initial mass function).
\item Cross-correlation of different catalogs based upon the RA, Dec coordinates (using tools such as \gls{TOPCAT}; \citealt{2005ASPC..347...29T}) to match the same sources detected with different filters,  at different epochs, or from different surveys, and thereby reconstruct the light curve of each identified young star. The matching radius will be defined based on the astrometric precision of each catalog, and a spatial analysis of field crowdedness (or typical intra-source separation in the region under exam) will be conducted to assess the probability of fortuitous matches resulting from the cross-correlation procedure.
\item Analysis of color-color and color-magnitude diagrams to separate young non-accreting stars (which exhibit colors consistent with photospheric/chromospheric emission) from young accreting stars (which exhibit continuum excess emission compared to the photospheric level as a result of the emission from the accretion shocks; revealed particularly at short wavelengths, yielding bluer colors compared to non-accreting young stars in the same population; e.g., \citealt{2014A&A...570A..82V}). The same analysis will be repeated for all single-epoch catalogs, and the frequency of cases that switch between accreting/non-accreting classifications at different epochs will provide critical information on potential \gls{transient} accretion states during the later stages of protoplanetary disk evolution (e.g., \citealt{2013ApJ...762..100C}).
\item Implementation of multiwavelength variability indicators \citep[e.g.,][]{1996PASP..108..851S} to identify young stellar objects that exhibit significant variability above the noise level traced by field stars.
\item Implementation of previously tested techniques, such as studying the color slopes associated with distinct young stellar variables \citep[e.g.,][]{2015A&A...581A..66V,2022AJ....163..263H}, and assessing the degree of periodicity and asymmetry in the measured \gls{flux} variations \citep[e.g.,][]{2014AJ....147...82C,2018AJ....156...71C,2022AJ....163..263H}, to discern the dominant physical drivers of the observed variability behaviors (e.g., geometric modulation by surface spots at a given temperature; stable magnetospheric accretion activity vs. episodic accretion; circumstellar occultation).
\item Analysis of the wavelength-dependent amplitudes of variability measured as a function of epoch difference (from hours, to days, to years), to parse the dominant timescales of variability for each photometric behavior \citep[e.g.,][]{2020MNRAS.491.5035S}, and the intensity of \gls{flux} variations (as a fraction of the typical luminosity state of the object) that are triggered over those timescales.
\item Identification of the most reliable predictors of young stellar status and variability behavior (i.e., specific color signatures, specific time intervals over which distinct variability processes can be most easily distinguished) and development of automated routines to extend the same classification analysis from the testbed case (microsurvey of the Carina star-forming region) to other regions encompassed by the \gls{LSST} survey. This step will be crucial to assess the time evolution of the physical processes that govern the star-disk interaction over the protoplanetary disk lifetimes.

\end{itemize}

\paragraph{Software Capabilities Needed}
\begin{itemize}
\item $ugri$ bands for LC, color-color, and color-magnitude diagram analysis
\item \gls{Alert} Brokers for long-term variability and spectroscopic follow-up is important for EXor objects
\item We plan to develop a \gls{software} for the classification of YSO variability at all the time-scales
\item Interactive visualization will be used
\item We plan to use \gls{3D} models and \gls{3D} rendering to reproduce the configurations that lead to the observed LCs. The use of \gls{3D} models will allow to explore different line of sights and to understand how geometric effects affect the observed LCs

\end{itemize}

\pagebreak
\subsubsection{Long Period M dwarf Variability} \label{sec:LongPeriosMD}

\Contributors{ Mark Popinchalk (\mail{popinchalkmark@gmail.com})}
{3/30/22}

\cleanedup{Mark Popinchalk}

\paragraph{Abstract}
M Dwarfs are the most common type of star, but are difficult to observe due to their intrinsic low luminosities. They are also the most long-lived stars, making them important for understanding the oldest limits of gyrochronology. Previous M dwarf studies include the MEarth sample \citep{Newton_2018_mearthsouth}, which looked at a volume limited sample of 25\,pc and successfully recovered rotations for objects down to \gls{Gaia} $\mathrm{G} \approx 17$\,mag. These M dwarfs have a roughly bimodal distribution of periods, with peaks at $\sim$1 and $\sim$100\,d and relative sparsity in between. This points to a potential sudden transition from rapid to long periods, but the age at which this occurs is poorly constrained. The  cause of the spin down is also unknown. Furthermore, not all M dwarfs are the same, as there is a transition from partially convective to fully convective interiors around the M3 spectral type. This is thought to have implications for angular momentum evolution, as descriptions used for core and envelope angular momentum transfer in more massive stars is not applicable to fully convective objects. Current space based missions like
\gls{TESS} struggle to 1) measure periods >28 days, and 2) recover periods from objects fainter than \gls{Gaia} $\mathrm{G} = 17$\,mag. The depth and duration of \gls{LSST} will be a powerful tool for increasing the number of M dwarfs with rotation periods $\gtrsim$30\,d.

\begin{figure}
\begin{centering}
\includegraphics[width=0.9\textwidth]{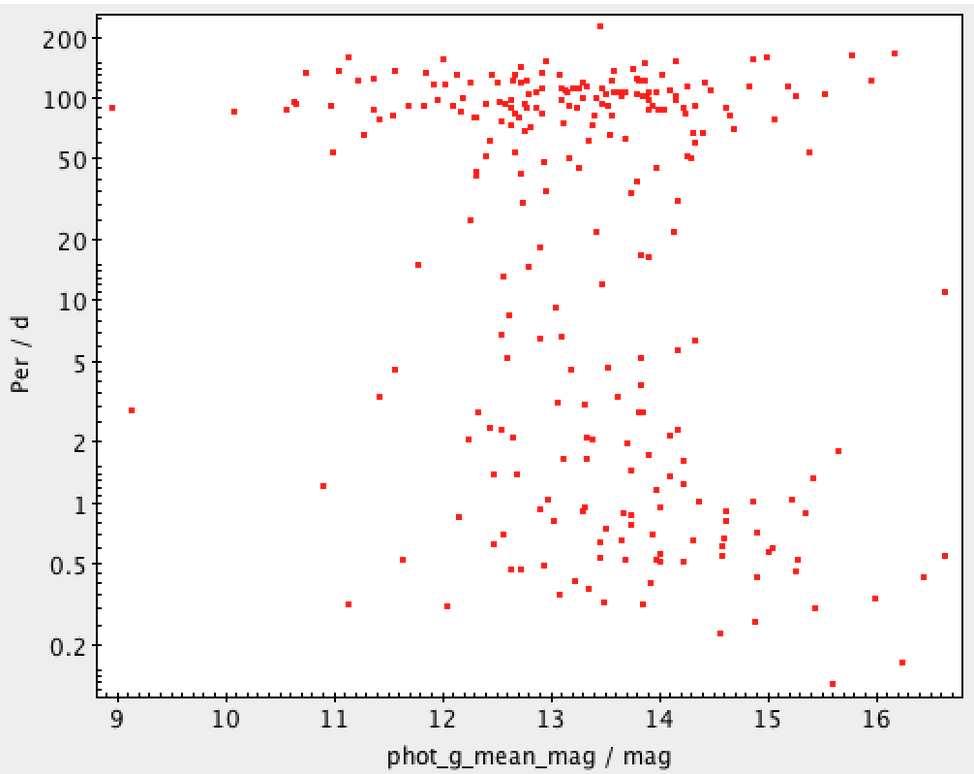}
	\caption{Period vs Gaia G mag for objects in the MEarth sample \citep{Newton_2018_mearthsouth}. Notice the bimodality of the rotation periods between \textasciitilde 1 and \textasciitilde 100 days. The brightness limit is based on the volume limit of the sample.  \label{fig:MearthPG}}
\end{centering}
\end{figure}

\paragraph{Science Objectives}
\begin{itemize}

\item Identify M dwarfs in LSST.
\item Measure the rotation periods for a large sample of M dwarfs.
\item Probe the distribution of long-period rotation in M dwarfs, and understand the importance of the transition from partially to fully convective interiors.
\item Constraining the evolution of angular momentum in M dwarfs.
\item Investigating color dependence in M dwarf variability.

\end{itemize}

\paragraph{Challenges (what makes it hard)}
\begin{itemize}
\item Because the LSST magnitude depth will discover so many new M Dwarfs, even a subset of potential stars will be massive. Defining and prioritizing appropriate targets in the \gls{LSST} field will be necessary.
\item The bimodality of M dwarf rotation periods means that many targets will likely be rapid rotators with $\sim$1\,day periods. These may be challenging to identify in light curves from the first two years of \gls{LSST} depending on the survey strategy due to the irregular observations and inter-night gap being greater than a day. This challenge will lessen with additional \gls{LSST} data releases.
\item Blending will make precise measurements difficult in crowded fields.
\end{itemize}

\paragraph{Running on LSST Datasets (for the first 2 years)}

\begin{itemize}
\item Starting with a smaller data subset would allow for trouble shooting of the survey strategy. One such data set could be drawn from MLSDSS-GaiaDR2 sample \citep{kiman_2019_mlsdss} which has $<$ 20000 M Dwarf sources that have been matched with \gls{Gaia} DR2 within the \gls{LSST} field of view, most of which are fainter than \gls{Gaia} $\mathrm{G} = 17$\,mag.
\item The Wide Fast Deep survey will provide a sufficient cadence for identifying long rotation periods ($> 100$\,d).
\item Additional follow-up of suspected short period objects can speed up characterization.
\end{itemize}

\paragraph{Precursor data sets}

\begin{itemize}
\item The MEarth South data set provides a small sample ($<$ 100) of known long rotation periods. While these sources will saturate in \gls{LSST} observations, they provide an important reference for comparison with the \gls{LSST}-discovered sample.
\item Other large long-term surveys that overlap with the \gls{LSST} field can be used to extend light curve coverage back and provide addtional characterization (e.g., \gls{ZTF}, \gls{TESS}, \gls{Pan-STARRS}).
\item MLSDSS-GaiaDR2  can serve as an initial sample.
\end{itemize}

\paragraph{Analysis Workflow}

\begin{itemize}
\item Identify and characterize M Dwarfs from color-color and color-magnitude diagrams. Older field M dwarfs are more likely to have longer rotation periods \citep[e.g.,][]{2020_angus_gal_kin, lu_2021_gyro-kin, Popinchalk_2021}, and so priority can be placed on older objects that should appear fainter and bluer compared to the main sequence \citep[see][]{kiman_2019_mlsdss}.
\item Various periodograms will be run on each light curve to identify different types of variability. These will be re-run as new data points are added with each visit.
\item Create phase-folded light curves determined from the best period based on each periodogram. Record amplitude of variability, most-probable period.
\item A period-sensitivity analysis algorithm will need to be run to flag periods that are under explored due to irregular observation cadence.
\end{itemize}

\paragraph{Software Capabilities Needed}

\begin{itemize}
    \item Interactive color-color and color-magnitude diagrams.
    \item Period detection algorithms that update as data points from subsequent visits are added.
    \item Interactive visualization of phase-folded light curves, with the capability to display user-supplied periods.
    \item An algorithm that flags periods that may be due to cadence windowing effects.
\end{itemize}

\paragraph{References for Further Reading}
 \citet{Newton_2018_mearthsouth}, 
 \citet{kiman_2019_mlsdss}, 
 \citet{2020_angus_gal_kin}, 
 \citet{Popinchalk_2021}

\pagebreak
\subsubsection{Identifying Substellar Companions to White Dwarfs} \label{sec:SubstellarCompanions}

\Contributors{Alekzander Kosakowski}
{2022-05-03}

\cleanedup{Alekzander Kosakowski}

\paragraph{Abstract}
Planets at distances of $\sim$$1~{\rm AU}$ will be engulfed by, and potentially merge with, their host star during red giant evolution.  More distant planets may be destroyed due to the effects of tides from their host star. However, massive planets and brown dwarfs may survive these events and end up on a compact orbit ($P\sim1~{\rm h}$) with an evolved star \citep[see][]{2021ApJ...913..118R}. Additionally, more distant companions may avoid common-envelope evolution entirely, but still migrate to shorter periods ($P\sim1~{\rm d}$), potentially due to scattering interactions as the orbits of other planets become unstable during their host star's evolution.

Main sequence stars with masses $M\lesssim8~{\rm M_\odot}$ will evolve into white dwarfs, potentially leaving behind remnants of their planetary systems. In rare cases, these remnants can be observed as debris disks \citep{2020ApJ...897..171V,2021ApJ...917...41V,2021MNRAS.504.2707G} or even as evidence for intact planetary bodies \citep{2020Natur.585..363V, 2021Natur.598..272B}.

Planetary and sub-stellar companions to white dwarfs will cause rapid, deep eclipses in astronomical survey data with eclipse depths likely below the noise floor of each individual observation. Thus, the deep multi-band observations of \gls{LSST} provide an ideal platform for identifying these deeply-eclipsing systems in the southern sky. Combined with \gls{ZTF} in the north, and overlap with \gls{BlackGEM} in the south, this project will provide insight to the fate of planetary systems around white dwarfs, potentially including formation rates of second-generation planets around evolved stars.

\paragraph{Science Objectives}
This project aims to create a catalog of deeply-eclipsed white dwarfs with low-mass stellar and sub-stellar companions.

\paragraph{Challenges (what makes it hard)}
\begin{itemize}
\item White dwarfs eclipsed by sub-stellar companions are likely to show eclipses with depths below the noise-floor of a single observation, with eclipse depths approaching $100\%$.
\begin{itemize}
\item Stacked images may provide estimates on the true mid-eclipse depth, but detecting enough mid-eclipse data points for sufficient image stacking is unlikely given the eclipse duration relative to the expected orbital periods.
\end{itemize}
\item These systems are relatively rare; we can’t simply ignore the Galactic plane and expect to obtain acceptable results. While the deep eclipses should still be visible in blended photometry, developing a proper deblending algorithm is still a large problem for this project.
\item The expected eclipse duration of these systems is on the order of $1\sim10~{\rm min}$, with orbital periods that may be longer than 1 day. Obtaining enough data to identify and constrain the periods of these systems through eclipses may require many months to years of data collection with LSST alone.
\begin{itemize}
\item Combining data across many surveys will help mitigate this issue, but requires an efficient multi-survey cross-matching algorithm and support for importing data from non-public surveys and single-night user photometry.
\end{itemize}
\item Obtaining follow-up spectroscopy for radial velocity measurements towards estimating companion masses requires a lot of telescope time, which may not be easy to obtain without a well-constrained orbital period from precise eclipsing light curve data.
\end{itemize}

\paragraph{Running on LSST Datasets (for the first 2 years)}
While the LSST alerts system will provide valuable information towards identifying these deeply-eclipsing systems, many epochs will be required to determine orbital periods.

Until accurate periods can be determined, follow-up photometry observations will be unfeasible. Despite this, the validity of the LSST alerts can be easily verified by examining the fits images for a completely disappearing star. Follow-up spectroscopy will be completed on verified deeply-eclipsing white dwarfs. Systems which show infrared excess in their SED must contain low-mass stellar companions, providing a simple filter when targeting sub-stellar companions.

Because these systems can show nearly $100\%$ eclipse depths, a sophisticated forced-photometry pipeline is important for this project. Photometry extraction with varying aperture sizes will be required for minimizing the noise level within the eclipses, specifically the ingress and egress where the flux is still measurable in a single exposure.

\paragraph{Precursor data sets}
The ZTF coverage in the north, and especially its overlap with LSST in the south, provides a valuable prototype for this project. Methods applied to the existing ZTF data can be immediately applied to the LSST data.

While less deep in terms of limiting magnitude, BlackGEM will augment the LSST dataset, both in time-domain astronomy and mapping the southern-sky in multiple filters. Having access to both BlackGEM photometry and LSST photometry will significantly speed up the detection of the relatively bright objects detectable in both surveys.

\paragraph{Analysis Workflow}
Measuring precise orbital periods from eclipsing systems requires many epochs of data. However, the LSST alerts combined with reference images for each field allows progress to be made on this project while waiting on sufficient data to be collected for orbital periods.
\begin{itemize}
\item First, a catalog of white dwarf candidates will need to be created using LSST colors. \gls{Gaia} DR3 parallax can only be used down to $G\approx20\sim21~{\rm mag}$, but may still help create a color-region for selecting faint LSST white dwarfs.
\item LSST alerts will be monitored closely for triggers from this LSST white dwarf catalog. Objects that show $\approx100\%$ flux dips will be cataloged for potential follow-up observations.
\item Follow-up spectroscopy will be obtained on objects without infrared excess in their SEDs. Companion masses will be estimated from radial velocity shifts measured in the white dwarf's spectrum.
\end{itemize}

After many months of operation, LSST will have enough data to perform a proper period search on these cataloged deeply-eclipsing white dwarfs. Analyzing the light curve data will roughly follow the same steps as other programs:

\begin{itemize}
\item Filter "poor-quality" images
\item Deblend the photometry of affected objects in crowded fields.
\item Perform forced photometry each white dwarf, using various aperture sizes to obtain the best signal-to-noise ratio per image.
\item Run a modified box least squares periodogram on each system, searching for eclipses in candidates that have produced a certain number of alert triggers.
\item Place limits on the true eclipse depth by creating co-added images using only mid-eclipse epochs.
\end{itemize}

\paragraph{Software Capabilities Needed}
This project has similar software requirements as presented in section \ref{sec:CompactWD} for identifying compact white dwarf binaries.
\begin{itemize}
\item This project will require a highly-optimized, modified box least squares algorithm to identify variability on both short ($P\sim1~{\rm h}$) and long ($P\gtrsim1~{\rm d}$) timescales. Thus, support for GPU acceleration is required to handle a wide and dense frequency grid.
\item The ability to create user-defined sub-catalogs based on LSST alerts and photometry and astrometry from LSST and other sources, such as \gls{Gaia}.
\item A custom alert system capable of handling user-defined filters. Alerts for only objects present in user-defined catalogs will be sent to the user. In addition to basic alert information, alerts should record cumulative alert number and time since previous alert per object.
\end{itemize}

\paragraph{References for Further Reading}
\cite{2021Natur.598..272B}, 
\cite{2021ApJ...913..118R}, 
\cite{2021ApJ...919L..26V}, 
\cite{2020ApJ...897..171V}, 
\cite{2021ApJ...917...41V}, 
\cite{2020Natur.585..363V}

\pagebreak
\subsubsection{RR Lyrae Catalogs} \label{sec:RRLyrae}

\Contributors{Andy Connolly (\mail{ajc@astro.washington.edu})}

\cleanedup{Andy Connolly}

\paragraph{Abstract} With a single visit depth of $\mathrm{r} \approx 24.7$\,mag and
$\sim$1000 repeated observations over a 10-year period, \gls{LSST}
provides an opportunity to measure the distributions of \gls{RRL}
within the Galactic disk and halo. The tight correlation between
period, luminosity, and metallicity of \gls{RRL} enables the calculation of
accurate distances ($\sim$3\%) out to $>$100\,kpc within the first 2\,years of \gls{LSST}. This proposal is to create a catalog of \gls{RRL} using the first
2\,years of \gls{LSST} data, to measure their metallicities and estimate the
3D metallicity distribution within the Galactic disk and halo. To
accomplish this, \gls{RRL} must be identified within the variable sources
detected by \gls{LSST} (either the alert stream or batch processing of the
static data) and their periods (and any additional modulations to the
light curves) measured.  \gls{RRL} have a typical period of 0.2--1.1\,d and
amplitudes of variation of $|sim$0.5--1\,mag.

Discriminating between \gls{RRL} subclasses will determine the accuracy of
the period and metallicity estimates. \gls{RRL} are divided into three
subclasses: \gls{RRab} which have a fundamental radial-mode pulsation with
skewed, non-sinusoidal light curves with large amplitude; \gls{RRc} which have a radial first overtone mode with sinusoidal variation
but at a smaller amplitude than \gls{RRab}; and \gls{RRd} which simultaneously pulsate in
both modes. \gls{Blazhko} \gls{RRL} show modulations in the light curve amplitude
of $\sim$0.1\,mag over weeks to months. \gls{Blazhko} \gls{RRL} account
for 20-30\% of the total \gls{RRL} population.

\paragraph{Science Objectives}
\begin{itemize}
\item Cataloging \gls{RRL} from the \gls{LSST} data and measuring their distances to $\lesssim 3$\% accuracy.
\item Characterizing \gls{RRL} distances and metallicities requires accurately measuring the periods of \gls{RRL}  from their light curves and identifying those \gls{RRL} showing modulations due to the \gls{Blazhko} effect.
\item Estimating of the metallicity of the \gls{RRL}  from a Fourier analysis of the phased light curves.
\item Identifying binary \gls{RRL}  within the catalog to estimate the masses of the pulsators.
\end{itemize}

\paragraph{Challenges (what makes it hard)}
\begin{itemize}
\item Identifying the variable sources will require iterative processing of time-series data to remove artifacts within the data and to improve the model/period fitting
\item Data will continue to be updated as new observations of \gls{RRL} are obtained.
\item Running \gls{UDF} at scale on \gls{LSST} data (across all light curves)
\item Sampling of \gls{RRL}  will be sparse, and the observations will be noisy. ``For a distance modulus up to 19, for more than half of the survey footprint more than half of the light curves can be fit correctly using time intervals of 30 days. For a distance modulus of 21, we have to move to a time interval of 50 days to get a correct fit for 10\%.'' \cite{Hernitschek_2021}
\item Period finding (multiband) can be slow
\item Aliasing of periods from sparse sampling will be common in the early data
\item Identification of those  \gls{RRL} affected by Blazhko modulation
\end{itemize}

\paragraph{Running on \gls{LSST} Datasets (for the first 2 years)}
\begin{itemize}
\item We will analyze both the alert stream and the data release light curves. Most of the analysis will be based on catalog data but we will need to go back to postage stamp cutouts and potentially full fields to visualize any problems with the data.
\item The initial analysis will process all variable point sources and then once  \gls{RRL} have been identified we will fit periods and metallicities to a subset of the data.
\item We expect to detect $\sim$10$^8$ variable stars in the first 2 years of Rubin with $\sim$160 observations spread over $ugrizy$. The first analysis can be taken after 6 months of normal survey operations and will be reanalyzed every 3--6 months.
\item We will use data from the Deep Drilling Fields to validate the analysis (due to its better sampling) and then the Wide Fast Deep data for the analysis.
\end{itemize}

\paragraph{Precursor data sets}
Precursor data from Gaia and \gls{ZTF} (bright but well sampled) and \gls{Pan-STARRS} (multiband)

\paragraph{Analysis Workflow}

Data cleaning and identification of artifacts within the data (this will be iterative as we progressively remove/flag bad data from the time-series catalogs):
\begin{itemize}
\item Remove poorly calibrated photometric data and sources flagged with suspicious photometry (e.g., source on edge of the \gls{CCD} or a diffraction spike).
\item Filtering undertaken using \gls{LSST} \gls{DM} source quality flags.
\item Remove/flag outlier measurements from a light curve.
\item Remove/flag extended sources.
\item Visualize light curves and sequence of images to explore images and light curves for ``bad" \gls{RRL}.
\end{itemize}

Data storage and archives:
\begin{itemize}
\item Collect training samples and store in a \gls{DB} (known  \gls{RRL} stars from \gls{Pan-STARRS}, Gaia, \gls{ZTF})
\item Collect training samples for metallicity measurements of known  \gls{RRL}
\item Build representation of light curve (multiband) with time, passband, noise, flags for bad points from either the Level 1 database or Alert Stream
\item Cross match to \gls{WISE} (or other \gls{IR} data) to improve  \gls{RRL} color selection
\end{itemize}

Variability Identification and RR Lyrae selection:
\begin{itemize}
\item Select sources with given number of epochs of data, and \gls{SNR} (magnitude range)
\item Characterize variability amplitude (e.g., \gls{RMS} of light curve) and filter based on this amplitude
\item Measure a multiband structure function \citep{Hernitschek_2016}
\item Separate RR Lyrae based on structure function (characteristic time, variability) and IR color (probably using a \gls{CNN} approach). Expect 85\% purity and 80\% completeness
\end{itemize}

Period finding and metallicity estimation:
\begin{itemize}
\item Fit Lomb-Scargle multiband period finder and return periodogram and peaks in periodogram
\item Apply prior to periods (0.2--1.1\,d) to remove aliased periods
\item For low \gls{SNR} or poorly sampled light curves we can also fit  \gls{RRL} templates to estimate periods
\item Estimate Fourier series for phased  \gls{RRL} light curves and fit  metallicity measurements using $\phi_{31}$
\end{itemize}

\paragraph{Software Capabilities Needed}

\begin{itemize}
\item Ability to apply selection filters to data (\gls{SQL} query)
\item Ability to run Lomb-Scargle fitter across all variable sources (e.g., as a \gls{UDF} using  Apache Spark)
\item Storage of light curves as objects (time, passband, noise, flags for bad points) with annotations (e.g., classifications)  that can be queried and cross matched to other data sets
\item Storage of outputs of filtered and classified data (or flags based on filters applied to existing catalogs)
\item Visualization of distribution of properties of sources (e.g., color-color scatter plots and histograms) colored by flags
\item Visualization of distributions of selected sources on the sky and relative to camera coordinates
\item Visualization of postage stamps and light curves for individual sources
\end{itemize}

\paragraph{References for Further Reading}
\cite{Sesar_2017}, 
\cite{Hernitschek_2016}

\pagebreak
\cleanedup{davenportj}
\subsubsection{Exceptional Variability: New Astrophysics \& Technosignatures} \label{sec:ExceptionalVari}

\Contributors{ James R. A. Davenport (\mail{jrad@uw.edu})}
{3/28/22}

\paragraph{Abstract}
 Rubin provides an unmatched ability to carry out searches for truly exceptional forms of stellar variability, which are sure to challenge our understanding of stellar evolution. These may take the form of e.g. dramatic outbursts or dimming from stars that were thought to be ``non-variable'', subtle changes in stellar properties over the \gls{LSST} baseline, or unexplained variability on a variety of timescales (e.g., ``Boyajian’s Star'', \citealt{boyajian2015}). These behaviors may not be remarkable as compared to the vast array of light curve morphologies that Rubin will observe for stars, but when placed in context (e.g. compared to similar stars on the H-R Diagram) they would reveal themselves as outliers.

 Perhaps the most extreme form of exceptional variability would come from a technosiganture -- a signal or byproduct of extraterrestrial technological activity. These signals may be subtle in amplitude, but highly unusual in timing, for example. The most unusual behavior from otherwise ``boring'' stars must be scrutinized under this lens, which will motivate extensive follow-up study.

\paragraph{Science Objectives}
\begin{itemize}
\item Identifying new forms of stellar variability on all timescales
\item Identifying exceptional or unique star systems compared to other similar stars
\item Making robust parameter space constraints from technosignature searches with Rubin data
\end{itemize}

\paragraph{Challenges (what makes it hard)}
\begin{itemize}
\item Outlier detection in the behavior of stars. Sparse sampling makes the distribution of possible ``normal'' behavior for stars very large.
\item Developing scalable, appropriate technosignature algorithms. Identifying those from the e.g. Radio astronomy community that are applicable to optical surveys.
\item Outlier detection with $<100$ epochs is challenging.
\end{itemize}

\paragraph{Running on LSST Datasets (for the first 2 years)}
\begin{itemize}
\item The alert stream will be useful for \gls{monitoring} stars along the ``SETI Ellipsoi'' \citep{lemarchand1994} or other favorable directions for unusual behavior, e.g. the Earth Transit Zone \citep{heller2016}.
\item The deep drilling fields will provide the best empirical models of what ``normal'' behavior for stars is.
\item First challenge is to find exemplars, things that stand out as being patently unusual with $<100$ epochs.
\item Final challenge is to ``Classify Everything'', and whatever remains is therefore exceptional.
\end{itemize}

\paragraph{Precursor data sets}
\begin{itemize}
\item \gls{ZTF} possibly the best precursor dataset given the sampling and baseline, but many systematics to consider in identifying stellar outliers.
\item Having a robust set of templates for every ``type'' of star would be ideal, akin to the ``One of Everything'' \citep{lacki2020} catalog. This could be drawn from high-cadence data from e.g. \gls{TESS} \citep{tess}.
\end{itemize}

\paragraph{Analysis Workflow}
\begin{enumerate}
\item Gather large sample of stars, estimate rough stellar parameters, place in context on the color--magnitude diagram (CMD)
\begin{itemize}
\item Better yet: use the color-color-color-color-magnitude diagram, and include upper limits for colors at the extreme red or blue end.
\item include extinction corrections and their uncertainties in this space.
\end{itemize}
\item Dynamically segment the CMD into small regions of self-similar stars
\item Compute a large number of features for every light curve, including variability metrics, Lomb-Scargle timescales, etc.
\item Actively hunt for outliers in each CMD bin. What are the most unusual stars relative to their siblings?
\begin{itemize}
\item We are building towards the goal of, for $\sim$10B stars, being able to immediately say: What kind of star is this? What is the limit of ``normal'' behavior for a star like this?
\end{itemize}
\item Coordinate alert \gls{monitoring} with other targeting approaches (e.g. technosignature methods)
\end{enumerate}

\paragraph{Software Capabilities Needed}
\begin{itemize}
\item Ability to create a reliable Color-Magnitude Diagram, including extinction corrections, for $\sim$10B stars without precise distances from e.g. \gls{Gaia}.
\item Ability to quickly cross-match \& place a star from e.g. the \gls{Alert} Stream into a CMD bin
\item Ability to generate multi-band variability features (e.g. general light curve stats, Lomb-Scargle periods, fast Gaussian Process predictions)
\item Ability to cross match catalogs at scale, for alerts and known stars
\item Ability to quickly compute 2-D and 3-D separation between objects across the sky
\item Ability to do feature selection, outlier detection, and basic machine learning at scale

\end{itemize}

\paragraph{References for Further Reading}
 SETI Ellipsoid: \cite{lemarchand1994}\\
 Earth Transit Zone: \cite{heller2016}\\
 SETI w/ surveys: \cite{djorgovski2000}\\
 ``Boyajian’s Star'': \cite{boyajian2015}\\
 ``One of Everything'': \cite{lacki2020}\\
 Systematic Serendipity: \cite{giles2019}\\
 SETI in the Spatio-Temporal Survey Domain: \cite{davenport2019a}\\

\pagebreak

\subsection{Solar system science} \label{sec:solar}
\subsubsection{Non-Tracklet Discovery for Small Body Populations} \label{sec:NonTracklet}

\Contributors{Joachim Moeyens (\mail{moeyensj@uw.edu})}

\cleanedup{Joachim Moeyens}

\paragraph{Abstract}
The Vera C. Rubin Observatory’s Legacy of Survey of Space and Time (\gls{LSST}) will discover over 5 million new Solar System small bodies over the course of its 10-year survey. The Solar System pipelines that will enable these discoveries will rely on observing “tracklets”: two-dimensional sky-plane motion vectors that constrain the position and rate of motion of moving objects \citep{Kubica2007, Denneau2013, Jones2018, Holman2018}.
A \gls{tracklet} requires at least two observations to be made in a single night typically within 90 minutes. Three such tracklets are required over the course of a 15-day linking window to identify the presence of a moving object.
The requirement to observe tracklets has two immediate consequences: 1) to successfully discover minor planets tracklets must be observed which imposes a strong constraint on cadence, 2) any minor planets that are not observed in at least three tracklets over a 15-day window will not be discovered by Rubin Observatory pipelines. For example, \gls{NEO}s may move too quickly for a \gls{tracklet} to be formed, and at the extreme end, very distant objects may not exhibit sufficient discernible motion within 90 minutes for two observations to be identified as separate.
Tracklet-less Heliocentric Orbit Recovery \citep{2021AJ....162..143M} is a small body discovery \gls{algorithm} capable of discovering Solar System small bodies without the need to use tracklets.
The \gls{algorithm} currently focuses on discovering Main Belt asteroids and trans-Neptunian objects, with future extensions planned to tackle the \gls{NEO} population. While capable of discovering minor planets without tracklets, THOR requires significantly more computational power than traditional tracklet-based algorithms.
The Asteroid Institute, a program of the B612 Foundation, has started work to deliver the \gls{THOR} \gls{algorithm} as part of their \gls{ADAM}. \gls{ADAM} is a scalable, cloud-based astrodynamics platform designed to enable large-scale analyses with small body science in mind.
The aim of the discovery service is to provide the community with a tool to search for asteroids in any astronomical dataset, and ultimately, to find the asteroids that may have been missed in the \gls{LSST} dataset.

\paragraph{Science Objectives}
\begin{itemize}
\item Filter the LSST alert stream to contain only unattributed observations. Additionally, filter out false positive detections using tools such as a real-bogus filter
\item Create an automated test orbit selection \gls{algorithm} to maximize discovery space. This selection \gls{algorithm} should scale with data properties such as density of observations and the phase space density of observed small body populations
\item Run discovery \gls{pipeline} and validate results
\item Submit discovery candidate observations and observations of known objects to the \gls{MPC}
\end{itemize}

\paragraph{Challenges (what makes it hard)}
\begin{itemize}
\item \gls{LSST}’s depth will push the computational power needed to successfully identify small bodies
\item Collaboration/communication between the discovery service and the internal Rubin pipelines: we do not want to compete for discoveries but instead we want to deliver complimentary analyses so that we maximize discovery potential. There are a number of questions in defining this interaction. Do we ignore tracklets in the alert stream? Is there a way to get access to the observations that were combined into tracklets? Do we include a \gls{tracklet}-builder as part of the discovery service?
\item The false-positive density in the alert stream (and in difference images) may have a significant impact on the processing time, particularly orbit determination.
\item Depending on the linking window size (15 days or more), processing should occur nightly as new observations are made. This would require being able to query for the current night’s observations and the previous 14 nights of observations quickly.
\item Longer baseline discovery searches should be encouraged to maximize the discovery potential for very distant objects (effectively increasing the linking window to \textasciitilde 30 days or longer), doing so will add additional processing time and the need for querying for larger volumes of observations.
\item To avoid pipeline bottlenecks, the discovery \gls{algorithm} hosted in the cloud should return results before the next night’s observations.
\item \gls{THOR} needs to be extended to the \gls{NEO} population since initial development focused on easier-to-link populations. The on-sky motion of NEOs may increase the computation cost of \gls{THOR} by an additional factor of 10.
\item Orbit determination techniques can decrease the computational cost but still need development: the current technique falls back to on-sky coordinates and ignores much of the benefits gained by linearizing the linking problem relative to the motion of test orbits.
\end{itemize}

\paragraph{Running on LSST Datasets (for the first 2 years)}
\begin{itemize}
\item LSST \gls{Alert} Stream (or difference image sources) initially
\item Possibly the \gls{LSST} observation catalogs for longer-baseline discovery searches
\end{itemize}

\paragraph{Precursor data sets}
Current datasets used for development:
\begin{itemize}
\item \gls{THOR} was initially developed and tested on two weeks of \gls{ZTF} observations (specifically, the alert stream hosted at \gls{UW})
\item The discovery service is being actively developed with the goal of processing the NOIRLab \gls{Source} Catalog (DR2) \citep{Nidever2021}
\end{itemize}
Future datasets to search and use for development:
\begin{itemize}
\item \gls{Pan-STARRS} is a possible dataset to mine for more discoveries
\item The simulated \gls{LSST} dataset is a good candidate for testing to \gls{LSST} scale
\end{itemize}

\paragraph{Analysis Workflow}
\begin{itemize}
\item Nightly \gls{Operations}
\begin{itemize}
\item Trigger discovery search on new nightly observations
\item Gather previous 15 nights' worth of observations
\item Filter alert stream to discard non-moving object sources and observations of known moving objects, filter out as many false positive observations as possible
\item Calculate optimal test orbits and submit discovery job to cloud-based infrastructure
\item In a test scenario, observations of known moving objects should not be discarded so that algorithmic completeness can be calculated
\end{itemize}
\item >Monthly \gls{Operations}
\begin{itemize}
\item Trigger discovery search on monthly threshold of observations
\item Gather previous 30 or more nights' worth of observations
\item Query the \gls{LSST} observations catalog for unassociated observations that span \textasciitilde months, apply filters to remove non-moving object sources and observations of known moving objects
\item Calculate optimal test orbits and submit discovery job to cloud-based infrastructure
\end{itemize}
\item Discovery search launches in the \gls{cloud} (for both nightly and monthly operating modes) and produces a list of candidate discoveries (their constituent observations and orbits) and observations of known objects that may not have been correctly identified as such
\item Discovery candidate observations and observations of known objects are submitted to the \gls{MPC}
\end{itemize}

\paragraph{Software Capabilities Needed}
\begin{itemize}
\item Fast/reliable querying of the \gls{LSST} alert stream
\item Fast/reliable querying of the \gls{LSST} observation catalogs on monthly cadences
\item Filters to remove static and known sources from the \gls{LSST} alert stream, tools to filter out or minimize the presence false positive detections in the alert stream
\item Identification of tracklets in the alert stream (if \gls{LSST} tracklet building can be accomplished near real-time)
\item \gls{THOR} discovery performance extended to the \gls{NEO} population with enhancements for orbit determination and test orbit selection
\item Continued development of \gls{cloud}-based infrastructure underlying \gls{ADAM}
\item Task-queue system with autoscaler for \gls{THOR} discovery jobs (parallelized by test orbit or chunks of sky or both)
\item Visualization tools to track in-progress discovery searches
\item Speed enhancements in \gls{THOR} and cloud-based infrastructure to reduce computational cost with cost-benefit analysis of cores/machines vs processing time to handle \gls{LSST} data volume

\end{itemize}

\pagebreak
\subsubsection{Characterizing Populations of Active Small Bodies} \label{sec:SmallBodyActivity}

\cleanedup{Colin Orion Chandler, Agata Rozek, and Henry Hsieh}

\Contributors{Orion Chandler (\mail{orion@nau.edu}), Henry H. Hsieh (\mail{hhsieh@psi.edu}), Agata Ro\.{z}ek (\mail{a.rozek@ed.ac.uk})}
{2022 April 1}

\paragraph{Abstract}

Small solar system objects can exhibit activity, or visible mass loss, due to a variety of mechanisms. The most common of those mechanisms is the sublimation of volatile ices, which drives the activity of the vast majority of known comets. While typically associated with comets from the outer solar system, sublimation-driven activity has also been observed on a small number of objects in the main asteroid belt (known as main-belt comets),
presenting intriguing new opportunities for studying the origin of terrestrial water.
%\citep[now known as main-belt comets;][]{2006Sci...312..561H,2017A&ARv..25....5S},
%counter to intuition that such objects are too close to the Sun to still have sufficient near-surface ice to drive activity.
%As only a relatively small number of main-belt comets
%have been discovered to date, this newly recognized population remains poorly understood.
%of these objects, which are known as main-belt comets \citep[review by][]{2017A&ARv..25....5S},
Recently, comet-like activity attributed to mechanisms such as impacts or rotational destabilization has also been detected.
%for a small number of otherwise asteroid-like objects, which are now known as \glspl{active asteroid} \citep[review by][]{2015aste.book..221J}.
Activity produced by these mechanisms is unpredictable and \gls{transient}, lasting anywhere from several months to just a few days. If identified promptly, these active events present rare opportunities to perform observational studies of processes %in the solar system
that are largely only studied using theoretical or computational models, or in laboratory settings that can only approximate certain aspects of those processes in the natural world.

With its unprecedented imaging sensitivity and sky coverage, \gls{LSST} has the potential to revolutionize active solar system object science by detecting activity that has been too weak or too short-lived to be reliably detected by other surveys, and providing regular deep monitoring of known active objects
%(e.g., for long-term evolution studies or detection of interesting events like cometary outbursts)
that cannot currently be done for large numbers of objects.
LSST will be able to discover cometary activity at much larger distances than current surveys, enabling long-term studies of incoming dynamically new comets %from the outer solar system
as they pass through different regions of the solar system.
It should also greatly increase the number of impact- or rotationally-driven active events that are detected in time to conduct real-time observations,
%, enabling us to better understand the physics of individual events as well as the overall frequency and diversity of such events.
%Finally, the survey should
increase the number of known members of populations of rare objects like main-belt comets and active Centaurs, and potentially uncover active objects in small-body populations in which no active objects are currently known (e.g., the Jupiter Trojans).

Achieving these myriad goals, however, will require \gls{LSST} to have automated activity detection algorithms that can promptly search all moving objects detected in a night for wide range of activity morphologies, and trigger observational follow-up for confirmation or detailed activity characterization.  Other challenges associated with activity detection and characterization include the potential need for larger image cutouts than are currently planned to be provided as part of LSST alert packets for all solar system object detections, and the need for a system that can track the outcomes of non-pipeline analyses (e.g., human vetting or follow-up observation results) and link them to the appropriate solar system objects in the LSST or broker databases.

\paragraph{Science Objectives}
%One of the primary outstanding questions in studying active objects is \textit{what triggers and drives activity?} % rephrased 5/12/22 COC, AOK?
Active object science with LSST consists of two major components: the discovery of new active objects and the characterization and long-term monitoring of known active objects.  LSST has the potential to revolutionize active object science in both of these areas, but substantial software challenges will need to be met for the survey to achieve this potential.

The \gls{LSST} will discover a large number of active bodies from myriad known populations, including \glspl{active asteroid}, active Centaurs, and active \glspl{NEO}. Given its unprecedented sensitivity, the survey may also discover activity in other small body populations that are not currently known to contain active bodies, such as the Jupiter Trojans and \glspl{TNO}. \gls{LSST} will be especially well-positioned to identify long-period comets and interstellar comets at large heliocentric distances, which will enable study of dynamically new bodies as they enter the inner solar system for the first time, and identify candidate targets for the \gls{ESA} Comet Interceptor mission. By exploring the dynamical and physical parameter spaces of all known and newly found active bodies, we will be able to better understand the conditions required to produce activity. To achieve that goal, we will quantify activity occurrence rates in various populations as functions of heliocentric distance and orbital position, and also take into account other factors such as activity strength, recurrence, and dust dynamics.

For all objects, we will measure activity properties (e.g., dust production rates, tail morphologies) and track these properties as functions of time to enable diagnosis and understanding of underlying activity mechanisms. We will also conduct forward and backward dynamical integrations to trace the origins and future dynamical evolution of active bodies. These dynamical insights will give context to the presence of active objects within each dynamical population and inform sublimation modeling  \citep[see][]{2020ApJ...892L..38C} that can enable estimates of which molecules are most likely responsible for observed sublimation-driven activity.

These objectives should be achievable by \gls{LSST} if \gls{software} can be successfully deployed that detects activity (or any form of mass loss) for small solar system bodies over a wide range of heliocentric distances, brightnesses, morphologies, and time periods, and is also able to conduct uniformly-defined quantitative analyses on known active objects with extended morphologies (e.g., for detecting events like cometary outbursts, which are characterized by rapid increases in brightnesses of cometary comae). Current surveys are limited by relatively shallow single-visit image depths in addition to the inherent challenges of activity detection in survey data and uniform analysis of extended objects that can vary widely in morphology. At present, identifying activity requires a mix of automated flagging procedures and human vetting. The \gls{LSST} single-visit image depths will surpass those of other ground-based surveys, yet significant challenges with automated flagging and human vetting will remain, and these may be exacerbated by the high-volume data flow \gls{LSST} is expected to produce.

\paragraph{Challenges (what makes it hard)}
The full suite of activity detection software to be used for \gls{LSST} data analysis will ideally include a variety of automated activity detection techniques in order to account for a range of activity morphologies. Examples of different morphologies we would like to be able to detect include coma strong enough to affect an object’s \gls{PSF} \gls{FWHM}, low-level activity too faint to affect an object’s \gls{PSF} \gls{FWHM}, completely unresolved activity that only affects an object’s photometry, circularly symmetric activity, and asymmetric or highly directed activity. Automated tools exist for flagging the potential presence of some of these types of activity morphologies, but not all, and human vetting is still required to confirm the presence of activity in essentially all cases. For \gls{LSST}, new \gls{software} will need to be developed to automate searches for activity using specific approaches that currently require some degree of human intervention, while other code will need to be enhanced to work at \gls{LSST} scales, both in terms of data processing speed and method of ranking and prioritizing results for any needed follow-up.

Software will also need to be developed to perform uniformly-defined quantitative analyses of known active objects in such a way that derived quantities can be used to quantify an active object's evolution over time and make comparisons to other active objects.  Such characterization tasks can be relatively simple, such as measuring the amount of flux within a radius of a certain angular or physical distance from the nucleus, or substantially more complex, such as characterizing the morphology of an active object's comae (e.g., identifying whether one or more tails are present or not, and potentially quantifying the lengths, directions, and shapes of those tails).

In terms of scale, all activity detection algorithms will need to be applied to all solar system object detections made each night (expected to be on the order of 1 million).  This approach is needed to avoid selection biases, and is also necessary for many activity detection algorithms themselves due to their need to establish templates for inactive moving objects to which potentially active outliers can be compared.  In contrast, however, advanced activity charaterization tasks will only need to be applied to detections of objects that are already known to be active, which should comprise a far smaller data set.

Further increasing the computational demands of active object analysis, many activity detection and characterization algorithms will require access to image data \citep[see Appendix B11, pg 30, of ][]{2019arXiv190611346H}. Notably, some algorithms may require larger image cutouts than will be available from the \gls{LSST} alert stream, though exact minimum viable cutout sizes for each algorithm are yet to be determined.
%Another challenge is that the non-sidereal motion of solar system objects complicates comparisons with sidereally-tracked \gls{background} stellar templates. % moved here from above ¶, merged into this ¶ 5/11/22 COC
Crucially, it will be essential to process the image data rapidly in order to keep pace with image acquisition and to enable follow-up observations for confirmation and characterization.

Human vetting and even follow-up observations will be required in many cases to confirm activity detection results from LSST, especially early in the survey while the application of detection algorithms to LSST data and the algorithms themselves are being refined based on their early performance. This adds subjectivity and external elements to \gls{pipeline}-centered search efforts. A mechanism for keeping systematic records of any human vetting and follow-up observation results will need to be developed in order to facilitate on-the-fly \gls{algorithm} evaluation and improvements, as well as later debiasing efforts.

Mechanisms for taking past activity scores of objects into account when searching for activity would be very useful to have if possible, given that multiple recent marginal detections of activity can collectively comprise a stronger detection of activity than any of those detections alone.  This will require activity detection code to have the ability to link past data for a given object to current data, and incorporate those data into activity evaluation procedures. %To date there is no centralized repository or service that tracks activity occurrence. % COC added this last sentence 5/11/2022

We anticipate eventually making use of machine learning techniques to improve activity detection algorithms, and potentially remove the need for human vetting, but these techniques are not currently in widespread use for this purpose. Additionally, some \gls{ML} techniques, such as neural networks, require large labelled training data sets for training and testing. Consequently, \gls{software} and a training data set will need to be developed for this purpose, further underscoring the need for careful tracking of all activity detection-related actions and decisions, automated or otherwise. %AR added sentence on training data set 5/12/2022

The ultimate quality of \gls{LSST} difference imaging is currently uncertain, but it is already expected that it will not be particularly good early in the survey when sufficient data for developing high-quality sky templates for static sources have not yet been obtained. As such, given the very high probability of blending of solar system objects with \gls{background} sources (due to \gls{LSST}’s unprecedented image depth), procedures will need to be developed to distinguish spurious activity detections from image artifacts.

\paragraph{Running on \gls{LSST} Datasets (for the first 2 years)}
Alert stream data, particularly ``postage stamp'' image cutouts and photometry, will be used to identify and (to some extent) characterize active object candidates and their activity. However, additional custom cutout image data may be required for specific activity detection and characterization algorithms and techniques.

Most image-based activity detection algorithms will, in principle, be able to run on survey data on Day 1 (if the \gls{software} is ready) as they primarily use other sources in the same field for comparison in order to identify active objects.  It should be possible to conduct many tasks related to characterization of known active objects from Day 1 as well, since they generally only require photometric analysis of single-visit images as long as the positions of those objects in the images can be identified. The lack of a complete set of template images for difference imaging analysis early in the survey is expected to limit transient detection during this period, and will present a challenge from Day 1 of the survey. Nonetheless, some mitigation can be achieved by targeting known solar system objects with exceptionally small ephemeris uncertainties until \gls{DIA} is fully operational.  Activity characterization efforts may also be hindered if nearby or blended background sources affect photometry measurements (especially for objects that are particularly extended), but some mitigation of this problem can be achieved by simply ignoring images where objects are close to known background sources or are in high-density star fields in general, or by adding functionality to activity characterization software to flag outlying data that could be due to background contamination.

Photometric searches for activity will require temporal baselines of adequate duration in order for the anticipated solar system object photometric behavior to occur, thereby enabling outlier detection. Thus this process will be less efficient early on in the survey, although such searches may still be possible using photometric data obtained for known asteroids by other past and concurrent surveys.

Some \gls{LSST} data products will be useful for this work, such as \gls{PSF} moments for \gls{PSF} comparisons and photometry and phase functions for photometric detection of activity. Nonethelesss, significant development of new \gls{software} tools for producing additional data products will be required to address, for example, other activity morphologies like unusual tails, or very faint and irregular comae. % such as? 5/11/22 COC response added 5/12/2022 AR; looks great, thanks! 5/12/22 COC
Also needed is infrastructure to achieve broader science objectives. Examples include human vetting in a highly systematic and documented fashion, prioritizing and triggering observational follow-up at facilities temporally and geographically appropriate given target orbital positions. % added this last bit and broke up the run-on sentences 5/12/22 COC

While clear activity may be detectable in a single image, multiple images will be required to satisfy the \gls{MPC}’s comet classification requirements. As of this writing, the \gls{MPC} policy states that unnumbered objects require multiple nights with multiple images acquired within 1-2 days.

\paragraph{Precursor data sets}
At present there is no single data set or service comparable to what the \gls{LSST} and its brokers will supply. Data sets with comparable image depths include public archives of \gls{DECam} data (hosted at \gls{NSF}'s \gls{NOIRLab} AstroArchive\footnote{\url{https://astroarchive.noirlab.edu}}), MegaPrime data (hosted at the \gls{CADC} archives\footnote{\url{https://www.cadc-ccda.hia-iha.nrc-cnrc.gc.ca}}), and SuprimeCam and \gls{HSC} data (if the data are reduced). The \gls{ZTF} \gls{Alert} stream provides a comparable alert stream data product that includes preliminary analyses (e.g., extendedness) as well as template subtracted data. An alert broker simulating time domain events at similar depth to the \gls{LSST} can be constructed to utilize publicly available \gls{DECam} data. The data selected would necessarily have been acquired over a period greater than one month, such as \gls{DES} and/or \gls{DECaLS}. Extending \gls{Broker} analyses to include \gls{PSF} and extendedness would provide a similar service as to what is needed from the LSST data stream for this science case. A vetted dataset of active objects within this dataset would be needed to adequately test activity detection and characterization techniques.

\begin{figure}
\begin{centering}
\includegraphics[width=0.9\textwidth]{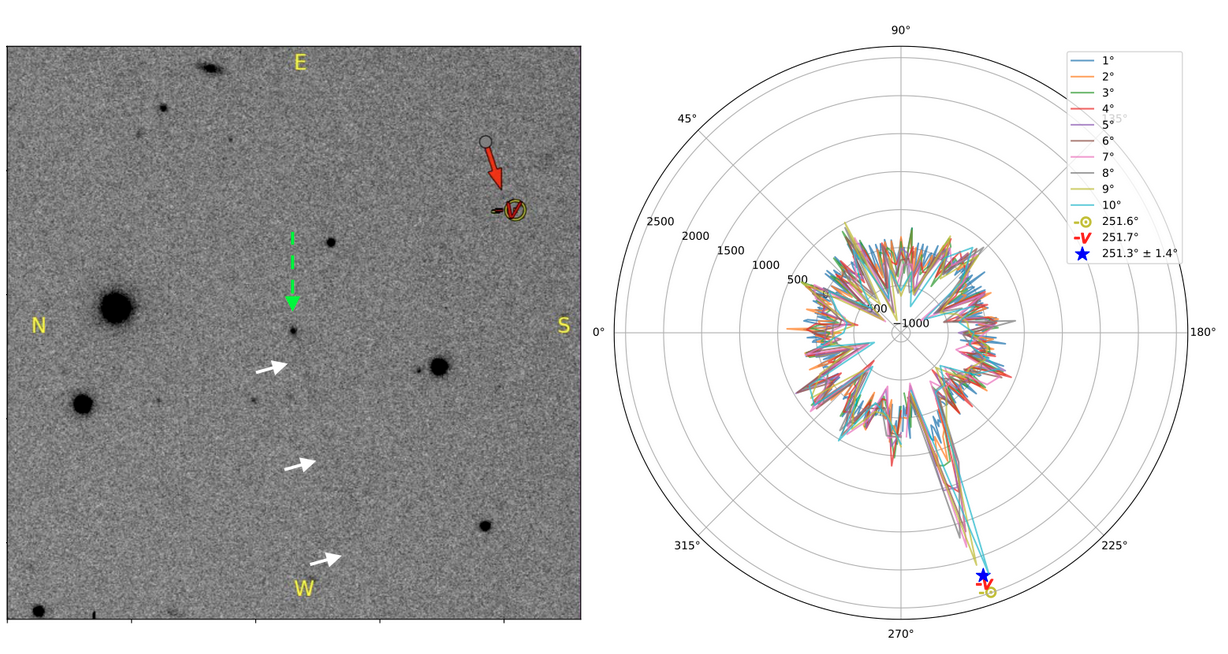}
	\caption{Left: \gls{active asteroid} (248370)~2005~QN$_{173}$ (green dashed arrow) with a tail (white arrows).  Right: the application of a ``wedge photometry'' tool designed to detect tails by measuring counts within variable width bins (wedges of an annulus). The wedge measurement of the tail angle is in close agreement with the anti-Solar and anti-motion vectors computed by \gls{JPL} Horizons. From \cite{2021ApJ...922L...8C}.}
\label{fig:Chandler21}
\end{centering}
\end{figure}

\begin{figure}
    \centering
    \begin{tabular}{|c|}
        \hline % I'm not wed to all the lines but it was touch to realize the top was a pair and the bottom was a pair without. 5/11/22 COC
         \includegraphics[width=0.85\textwidth]{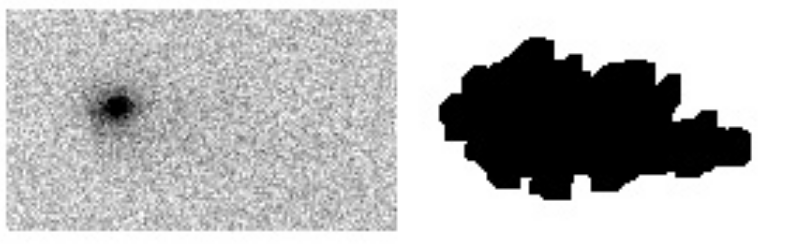} \\
         \hline
         \includegraphics[width=0.85\textwidth]{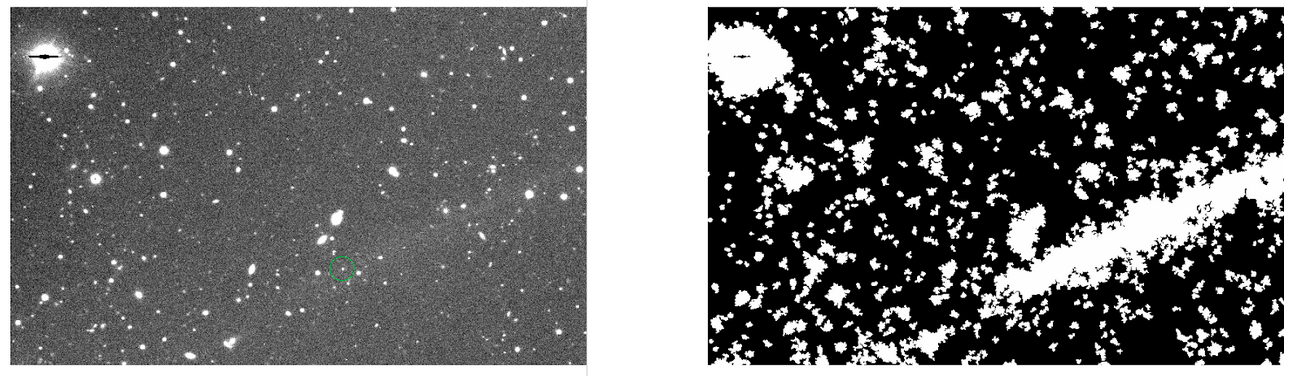}\\
         \hline
    \end{tabular}
    \caption{Example results from the tool NoiseChisel \citep{2015ApJS..220....1A,2019arXiv190911230A}.
    Top: Comet 358P/\gls{Pan-STARRS} with a tail and coma that are difficult to identify in the original image (left), but the resulting NoiseChisel output (right) provides clear evidence of activity. Image courtesy Mohammad Akhlaghi and Henry Hsieh.
    Bottom: The original image (left) of comet 67P/Churyumov–Gerasimenko (green circle) with a faint tail extending roughly ENE. The NoiseChisel output (right) provides much more contrast, important for manual analysis by humans and automated searches by computers. Image courtesy Agata Ro\.{z}ek.
    }
    \label{fig:NoiseChisel}
\end{figure}

\paragraph{Analysis Workflow}
\begin{enumerate}
\item Identify known and new moving object in nightly data
\item Retrieve sufficient surrounding image data as required for activity detection and characterization tools
%(e.g., wedge photometry, NoiseChisel; see Figures~\ref{fig:Chandler21} and \ref{fig:NoiseChisel})
\item Apply various activity detection algorithms (e.g., photometric enhancement analysis, PSF analysis, multi-aperture photometry analysis, wedge photometry, NoiseChisel; see Figures~\ref{fig:Chandler21} and \ref{fig:NoiseChisel}).  As all of these analyses will be performed on individual detections, this work should be highly parallelizable.
\item Apply various activity characterization algorithms (e.g., $Af\rho$ calculations, surface brightness profile characterization, automated basic dust modeling analysis, temporal image subtraction for outburst detection) and identify any unusual changes in those parameters that could indicate the onset of events such as cometary outbursts.  As all of these analyses will be performed on individual detections, this work should be highly parallelizable.
%\item Perform \gls{PSF} measurements, incorporating \gls{deblend}ing techniques as needed when \gls{Template}-subtracted data is not yet available
%\item Measure \gls{extendedness} (if not provided by \gls{Alert} stream)
%\item Compare measured photometry with magnitude(s) predicted by %\gls{JPL} Horizons (etc.) or
%internally defined phase function fits to previous LSST detections, applying broadband filter offsets as needed
%\item Perform wedge photometry on all sources and compare to \gls{JPL} Horizons computed anti-Solar and anti-motion vectors (potential orientations of tails)
%\item Employ NoiseChisel to identify faint activity in images where sufficient \gls{background} image data are available
%\item Apply additional techniques (e.g., image matching via \gls{ML})
%\item Extend all analyses to incorporate full time span of image data available
\item Compute combined weighted activity confidence metrics for new active object candidates
\item Retrieve activity confidence metrics for previous detections of new active object candidates and increase priority for objects considered to be active candidates for multiple recent detections, even if activity metrics for individual detections have low confidence levels
\item Prioritize activity candidates for additional investigation to confirm and characterize activity via:
\begin{enumerate}
\item Observational follow-up
\item Searches for and analysis of archival data
\item \gls{Citizen Science} classification and activity vetting
\end{enumerate}
\item Prioritize cometary outburst candidates for observational follow-up to confirm outburst detections and analyze confirmed outburst events
\item Record results of confirmation analyses for new activity candidates and cometary outburst candidates, and link those results to the appropriate detections in LSST or broker databases
\item Incorporate results from initial activity detection analyses and confirmation analyses into \gls{ML}-based tools
\item More advanced activity detection in the future might involve shifting (and possibly rotating) and stacking of multiple detections of the same object, or even of multiple objects (e.g., that share certain physical or dynamical characteristics, such as belonging to the same collisional family), to search for ultra-faint activity, so infrastructure for facilitating this would be desirable, but this is considered lower priority at the moment relative to single-visit activity detection and characterization efforts.
\end{enumerate}
% COC note: mention where alert stream / broker should be handling each step

Our overarching goals will be to make use of various activity detection tools on all \gls{LSST} solar system object detections and return parameters related to the activity confidence levels of those detections for each algorithm, and apply activity characterization tools on known active object detections to quantitatively characterize activity evolution to enable both quantitative studies of long-term activity evolution and identification of unusual changes in that evolution (e.g., outbursts). Activity and outburst detection parameters may include detection-level confidence parameters and object-level confidence parameters (e.g., parameters indicating whether an object consistently shows indications of activity or an outburst over multiple recent detections).
%, meaning that an object with many marginal activity likelihood parameter values over several nights could rise in priority to match objects with stronger activity likelihood parameter values).
Activity or outburst likelihood parameters should be automatically associated with their corresponding solar system objects in LSST or broker databases, while results of any additional follow-up analyses to confirm the presence of activity or an outburst should be recorded and also linked to the appropriate objects in available databases.
%Prioritize active object candidates for human vetting (at least initially)
%Prioritize active object candidates for observational confirmation and send to observation management software
%Prioritize high-confidence active object detections for observational follow-up for detailed characterization and send to observation management software
%Logging of both affirmative and negative results of human vetting steps to enable construction of training sets for machine learning and also quantifying detection rates for debiasing analyses

\paragraph{Software Capabilities Needed}

An alert broker will be needed for data integration, such as object matching with external catalog data, photometry, and activity likelihood parameters. We may make use of training sets developed using vetting by professional astronomers, follow-up observations, and \gls{Citizen Science} projects (e.g., \textit{Active Asteroids}\footnote{\url{http://activeasteroids.net}}) to train \gls{ML}-based image pattern recognition algorithms.
%(i.e., to look for tails and comae).
We expect the computational overhead to be negligible for broker-provided analyses, however the computational cost for additional analysis tools, especially \gls{ML}-based approaches, is unknown.

Image cutouts of each activity candidate will need to be stored for myriad reasons, including adherence to applicable \gls{MPC} reporting mandates. Storage requirements have yet to be determined.

We do not anticipate the need for custom visualization tools. Widely adopted  visualization tools like Matplotlib\footnote{\url{https://matplotlib.org}} and Bokeh\footnote{\url{https://bokeh.org}} should suffice for static and dynamic/interactive data visualization, respectively.

% Most functionality for these tasks will likely make the most sense being part of a community alert broker, given that activity detections will likely be desired as part of secondary alerts sent out following initial baseline \gls{LSST} alerts

Software implementations of some activity search algorithms do exist, such as those outlined in Appendix \gls{B}.11 of \cite{2019arXiv190611346H}. However, most need to be optimized to run at \gls{LSST} scale and some will need to be developed from scratch. To our knowledge, no \gls{software} yet exists to perform the type of logging that will be needed to record the outcomes of human vetting and follow-up observations to confirm activity detections for the purposes of later debiasing analyses, especially at the scale of the \gls{LSST} survey.

% \begin{figure}
%     \begin{centering}
%     \includegraphics[width=0.5\textwidth]{images/NoiseChisel358P }
% 	\caption{NoiseChisel analysis of comet 358P/\gls{Pan-STARRS}, courtesy Mohammad Akhlaghi \& Henry Hsieh.}
%     \label{fig:NoiseChisel358P}
%     \end{centering}
% \end{figure}
% \begin{figure}

% \begin{centering}
%     \includegraphics[width=0.9\textwidth]{images/NoiseChisel67P}
% 	\caption{NoiseChisel analysis of 67P/Churyumov–Gerasimenko, courtesy Agata Rożek.} % love this!
%     \label{fig:NoiseChisel67P}
%     \end{centering}
% \end{figure}

We emphasize that cutout images larger than the postage stamps that will be automatically generated for all transient detections will be required for some activity detection and characterization analyses, meaning that custom extraction of image data, potentially for all solar system object detections every night, will be needed.
%For example, NoiseChisel  \citep{2015ApJS..220....1A,2019arXiv190911230A} %(Akhlaghi \& Ichikawa 2015, Akhlaghi 2019)
%requires a large image area in order to effectively account for background noise.
%Figure \ref{fig:NoiseChisel} shows example NoiseChisel results from two different images of active objects.

\paragraph{References for Further Reading}

\textbf{Science Background:}\\
\cite{2020ApJ...892L..38C}, ``Cometary Activity Discovered on a Distant Centaur: A Nonaqueous Sublimation Mechanism'', ApJL, 892, L38\\
\cite{2006Sci...312..561H}, ``A Population of Comets in the Main Asteroid Belt'', Science, 312, 561 \\
\cite{2015aste.book..221J}, ``The Active Asteroids'', Asteroids IV, 221-241\\
\cite{2017A&ARv..25....5S}, ``The Main Belt Comets and Ice in the Solar System'', Astron. Astrophys. Rev., 25, 5 \\

\noindent\textbf{Software/Algorithm Background:}\\
\cite{2015ApJS..220....1A}, ``Noise Based Detection and Segmentation of Nebulous Objects'', ApJ, 220, 1 \\
\cite{2019arXiv190911230A}, ``Carving out the low surface brightness universe with NoiseChisel'', The Realm of the Low-Surface-Brightness Universe, Proc. \gls{IAU} Symposium No. 355 \\
\cite{2018PASP..130k4502C}, ``SAFARI: Searching Asteroids for Activity Revealing Indicators'', \gls{PASP}, 130, 114502 \\
\cite{2021ApJ...922L...8C}, ``Recurrent Activity from Active Asteroid (248370) 2005 QN173: A Main-belt Comet'', ApJL, 922, L8\\
\cite{2009Icar..201..714G}, ``Searching for Main-belt Comets Using the Canada-France-Hawaii Telescope Legacy Survey'', Icarus, 201, 714 \\
\cite{2009A&A...505.1297H}, ``The Hawaii Trails Project: Comet-hunting in the Main Asteroid Belt'',  A\&A, 505, 1297-1310 \\
\cite{2015IAUGA..2251973H}, ``The Main-belt Comets: The Pan-STARRS1 Perspective'', Icarus, 248, 289-312 \\
\cite{2019arXiv190611346H}, ``Maximizing \gls{LSST} Solar System Science: Approaches, Software Tools, and Infrastructure Needs'', arXiv:1906.11346 \\
\cite{2011Icar..215..534S}, ``Limits on the Size and Orbit Distribution of Main Belt Comets'', Icarus, 215, 534 \\
\cite{2013MNRAS.433.3115W}, ``Main-belt Comets in the Palomar Transient Factory survey - I. The Search for Extendedness'', \gls{MNRAS}, 433, 3115 \\

\pagebreak
\subsubsection{Constraining the Number Density and Mass of the Galactic Interstellar Small Body Reservoir} \label{sec:InterstellarEjecta}
\Contributors{W.~Garrett Levine (\mail{garrett.levine@yale.edu})}
{last edited 5/3/22 (began 3/29/22)}
\cleanedup{W. Garrett Levine}

\paragraph{Abstract}
We estimate the mass of small bodies per unit stellar mass that is ejected from planetary systems into the interstellar medium as rogue objects. In particular, we combine the first two years of Rubin/LSST data with the results from n-body simulations of exoplanetary systems. This initial LSST dataset is especially helpful because the expanded field-of-view means that the \gls{ISO} detection rate should be highest during the initial survey stages. Estimating the reservoir of ISOs is a question of detectability and requires a rigorous debiasing of the population of interlopers traversing the Solar System discovered by LSST. Although this calculation is most dependent on orbital parameters and asteroid size, small body \gls{shape} and composition could be concerns as well. For this science case, we leverage the Rubin Science Platform and real LSST images with synthetic injected sources to construct forward models of ISO detections by the survey and fit these to the results of our orbital simulations. Through this study, we estimate the prevalence of dynamical instabilities like that of the Nice Model and put preliminary constraints on the occurrence and multiplicity of giant planets with Safronov numbers larger than unity. Because this calculation is based solely on ISO occurrence rates, this work probes the architectures of extrasolar systems independently from either dedicated searches for exoplanets or observations of circumstellar disks.

\paragraph{Science Objectives}
The purpose of this science case is to develop an estimate of the Galaxy-wide aggregate interstellar small body ejecta. This objective can be broadly divided into two steps: modeling the \gls{ISO} detection efficiency of Rubin/LSST, and conducting numerical simulations of dynamical instabilities in exoplanetary systems. The first task will result in a robust estimate of the \gls{ISO} number density with well-constrained uncertainty, while the second one will couple observational results with theoretical astrophysics to interpret the LSST data in the context of planetary formation. These numerical simulations could constrain the size, composition, number, and age distributions of ISOs through forward modeling with the Solar System science collaboration’s post processing \gls{pipeline} to model detections.

The nature and composition of \gls{ISO}s will be constrained through concentrated follow-up efforts; this question is outside of the scope of this study aside from any detectability effects on the types of ISOs which will be identified. For example, the typical Galactic kinematics of \gls{ISO}s will depend on their compositions and characteristic ages.

\paragraph{Challenges (what makes it hard)}
Many of the fundamental challenges for this topic are more generally addressed by other science use cases. For example, constructing comprehensive selection functions to quantify the small body detection efficiency will be critical to debiasing the census of \gls{LSST} detections. However, the parameter space for hyperbolic orbits is significantly larger than for the set of elliptical orbits that are bound to the Sun. Additional selection functions may need to be computed to estimate the occurence and size distribution of interstellar ejecta. In addition, objects on hyperbolic orbits are subject to substantial apparent motion in the sky; this effect could tangibly modify the small body detection efficiency for constant phase function and brightness.

The known population of \gls{ISO}s currently consists of only two members. `Oumuamua and Borisov were markedly different objects with starkly contrasting orbital properties, so these objects could represent a bimodal distribution of \gls{ISO}s, two points on a spectrum, or even outliers.  Since \gls{LSST} may expand this catalog by an order-of-magnitude, this science case also encompasses inherent “unknown unknowns.” ISOs will be only a small fraction of \gls{LSST}’s total small body detections, so each identification will be critical to constraining the galactic number density. Therefore, identifying faint objects is especially important for this science case.

Finally, the nature of `Oumuamua-like ISOs is unknown. The Pan-STARRS survey showed that these objects may comprise a substantial fraction of interstellar small bodies, so \gls{LSST} pipelines must be capable of efficiently detecting objects with highly variable lightcurves (via shift-stacking and other co-additive methods). With only two members of the interstellar small body population, this science use case relates to ``unknown unknowns" since more classes of interlopers may exist. Interstellar objects on highly eccentric ($e \gg 1$) orbits are expected to be more difficult to detect than objects on trajectories that somewhat resemble those of long-period comets. `Oumuamua's eccentricity was e $\sim$ 1.2, and Borisov's eccentricity was e $\sim$ 3.5.

\paragraph{Running on LSST Datasets (for the first 2 years)}
The wide-fast-deep strategy by Rubin/LSST will be ideal for detecting large numbers of small bodies. Especially because orbits of ISOs are hyperbolic, the initial stages of the LSST survey should rapidly yield a number of these objects. When it first surveys the sky, LSST's deep limiting magnitude compared to other wide-field Southern Hemisphere observatories may result in the immediate detection of objects that are passing through this increased volume. New objects would be detected continuously as the small bodies enter the survey volume during the ten-year survey. Therefore, the greatest marginal increase in the scientific understanding of the \gls{ISO} population will occur during the first few years of LSST operations. Population-level studies on the number of interstellar small bodies could be conducted through data releases and would not be subject to immediate time constraints. However, each individual the interstellar objects themselves will demand immediate follow-up for compositional and dynamical characterization.

This science case does not require additional science products from LSST, although it will need for selection functions and detection algorithms to be quantified for an extended parameter space compared to Solar System small bodies. In addition, an optimized post-processing \gls{pipeline} will be important to efficiently evaluate population-level ramifications of LSST findings. Although the composition of \gls{ISO}s should not affect detectability as much as these objects’ orbits, it will be important to understand this complication.

\paragraph{Precursor data sets}
Similar algorithms could be validated using Pan-STARRS small body detection data, although that research might require its own equivalent of the in-development Solar System small body post-processing \gls{pipeline}. Because of the dramatic effect which LSST will have on the catalog of \gls{ISO}s, no precursor dataset can parallel the forthcoming data. Therefore, it is possible that efforts are best spent on developing well-understood simulated LSST datasets of \gls{ISO}s.

\paragraph{Analysis Workflow}
Orbital parameters of identified \gls{ISO}s and small body selection functions corresponding to their orbital properties will be obtained from LSST data sources. Selection functions should decouple orbital data from the object’s position on the sky and position on the the Rubin Observatory detector from the small body’s inferred physical attributes. These data will be coupled with forward models that inject synthetic sources into survey images to determine the detectability of \gls{ISO}s in real LSST frames. From this research, the interstellar object number density can be quantified for the Solar neighborhood.

Under the assumption that the region surrounding our Solar System is representative of the broader galactic interstellar object reservoir, one can extrapolate to the entire Milky Way. Simulations of small body ejecta from young exoplanetary systems can be run at any time before, during, or after the LSST survey and can be done on computing systems outside of the collaboration.

\paragraph{Software Capabilities Needed}
Much of the \gls{software} capability to complete this science case will build upon general \gls{software} for small body selection functions. Moreover, this science use will directly benefit from the development of both tracklet-based and non-tracklet small body identification algorithms. In addition, readily available orbital data from LSST, potentially in conjunction with the Minor Planet Center, would be helpful for population-level research.

\paragraph{References for Further Reading}
\cite{do2018oumuamua}, \cite{portegieszwart2018oumuamua}, \cite{engelhardt2017iso}

\pagebreak
\subsubsection{Multiwavelength studies of Solar System moons and asteroids} \label{sec:Multifrequency}
\Contributors{Ilhuiyolitzin Villicana Pedraza}
{March 27, 2022}
\cleanedup{Mario Juric}

\paragraph{Abstract}
Multiwavelengths studies have been important in characterizing new discoveries across astrophysics \citep[e.g.][]{2017Galax...5....3V}. We expect the similar to be the case in the Solar System.
%The most common studies show complete maps of one region and other important parameters. Existed a lot of observations toward the Planets of the Solar System but we do not have enough information for the moons and asteroids of the Solar System with interest in astrobiology.
We will analyze the dataset of $\sim$1 billion measurements from the Rubin Observatory's catalogs, combine them with light curves and spectra and multiwavelength information coming from instruments such as \gls{ALMA}, \gls{VLA}, and \gls{JWST}.

\paragraph{Science Objectives}
The aim of our work is to support the analysis and observations of small bodies of the Solar System using the Rubin Observatory, and complement these with observations in the radio and the infrared. Using the Rubin Observatory we can cover a large portion of the bodies in the Solar System, using \gls{ALMA} and \gls{VLA} we can obtain important molecular spectroscopic information \citep{2017IAUS..321..305V} and using the \gls{JWST} we obtain the \gls{IR} data.

\paragraph{Challenges (what makes it hard)}
Obtaining new observations from highly-oversubscribed telescopes is always difficult. We can also search information from archival data, catalogs, and the literature. Getting access to follow-up telescope time will be a challenge.

\paragraph{Running on LSST Datasets (for the first 2 years)}
Data release catalogs \gls{LSST} from data products will be analyzed. For the light curve we need 30 points for every object. We are planning to create catalogs for the \gls{LSST} as products. We estimate that the light curves to be analyzed are 10. We can use the Asteroid Light curve Photometry Database\footnote{\url{http://alcdef.org/}} to supplement the light-curve information.

\paragraph{Analysis Workflow}
First we need compare the public catalogs for asteroids to avoid duplication. We will use these catalogs to compare with our observations to match the existing data sets. The non-LSST information will be analyzed using tools such as DS9 for imaging and \gls{IRAF} for spectroscopy. For \gls{ALMA} and \gls{VLA} data we will use \gls{CASA}. These external data will need to be cross-matched to the LSST sample\footnote{An example of cross-matching with present-day tools can be found at \url{https://whitaker.physics.uconn.edu/wp-content/uploads/sites/2038/2017/02/PythonTutorial_Ashas.pdf}}.

% To compare catalogs  we have some information of the code
% 1. Save or charge 2 catalogs to compare.\\
% \begin{verbatim}
%   catalog_path =“desktop/catalog1/”
%   catalog_path =“desktop/catalog2/”
% \end{verbatim}
%  2. In Python:\\
% \begin{verbatim}
% from astropy.table import Table, Column Join
% from astropy.table import Column
% from numpy import *
% import matplotlib.pyplot as plt
% #matplotlib inline
% import numpy as np
% from astropy.io import ascii
% \end{verbatim}
% Follow the intructions from the Ashas paper \url{https://whitaker.physics.uconn.edu/wp-content/uploads/sites/2038/2017/02/PythonTutorial_Ashas.pdf}

\paragraph{Software Capabilities Needed}

We will use Rubin Data Management and LINCC software for the LSST. For \gls{ALMA} and \gls{VLA} we will use \gls{CASA}. We will access other datasets or cross-match to other catalogs like \url{https://alcdef.org/} for asteroids.

\pagebreak
\subsubsection{Small Bodies in Rubin/LSST Data for Population-Level Studies} \label{sec:SmallBodyPops}
\Contributors{W. Garrett Levine (\mail{garrett.levine@yale.edu}),
Henry Hsieh (\mail{hhsieh@psi.edu})}
{updated 5/3/22 (began 3/24/22)}
\cleanedup{W. Garrett Levine and Henry Hsieh}

\paragraph{Abstract}
Conducting population-level studies of small bodies can illuminate the ancient Solar System’s dynamical history. The orbital evolution of the giant planets sculpts the statistical distributions of small bodies, making these objects a viable tracer of the Solar System's assembly. We discuss a modeling pipeline through which users can impose physical parameters on synthetic sets of small bodies and simulate the resulting observations from Rubin/LSST. In addition, this pipeline can be used to simultaneously fit the physical parameters of small bodies to brightness data in \gls{LSST}. Generally, the brightnesses of small Solar System objects depend on multiple physical parameters including heliocentric and geocentric distances, phase angles (i.e., the Sun-object-observer angle), rotational phase, viewing aspect angle (i.e., orientation of the object from the point of view of the observer), and filter. Phase functions expressing the brightness dependence of an object on phase angle and rotational lightcurves expressing the brightness dependence of an object on rotational phase are typically solved on their own, but our novel approach solves these parameters simultaneously. Because this pipeline is efficient, it keeps pace with the nightly alert stream and can be readily applied to simulate the small body yields and characteristics from \gls{LSST}.

\paragraph{Science Objectives}
Rubin/LSST will detect millions of Solar System small bodies with a diversity of orbital and physical properties. Determining the composition of these objects is necessary to advance models of planet formation and to constrain the dynamical history of the ancient Solar System. Broadly, we can divide this science case into two deliverables. First, it will be necessary to develop a module that computes best-fit properties (size, shape, composition, phase function, and rotation period) from input data from LSST (a series of times of observation, \gls{RA}, dec, apparent motion, and calibrated brightnesses in multiple filters). For efficiency purposes, physical and orbital properties could be decoupled.

To conduct population-level studies with this tool, it will be necessary to finalize development of the Solar System small body post-processing \gls{pipeline} to enable timely simulations of realistic Rubin/LSST datasets. Modeling small body populations via injection/recovery testing will be crucial. The aforementioned joint-fitting capabilities for color, rotation, and phase function will be important. Below, we provide a few examples of specific questions that may be addressed by this more general science case.

\begin{enumerate}
    \item Connecting the elongation of monoliths (<100m objects) to the collisional physics which generated these bodies.
    \item Estimating the population of `Oumuamua-like (elongated) interstellar objects.
    \item Comparing the intraclass and interclass characteristics of collisional families.
    \item Identifying outliers in groups of asteroids with similar orbits.
    \item Characterizing trends in shapes or rotation periods as a function of orbital elements and sizes
\end{enumerate}

\paragraph{Challenges (what makes it hard)}
Currently, a holistic model of asteroid brightness that accounts for the relevant physical parameters has yet to be integrated with the Rubin/LSST post-processing \gls{pipeline}. Common routines that fit the parameters of small bodies often fit each attribute independently, which may lead to internal inconsistencies for the outputs. Joint-fitting routines can be algorithmically expensive, so computational efficiency in the Solar System post-processing module will be critical to the success of this science case.

Solar System objects may fall on different parts of the detector over different nights and possibly within the same night, especially for fast-moving and nearby small bodies. Therefore, well-constrained pixel-by-pixel detection efficiencies and selection functions will be important for this science case. Small bodies will move relative to \gls{background} stars – accurate photometry for these objects will rely on high-quality difference image processing, or will otherwise need to be disentangled from any imperfectly subtracted \gls{background} sources or image artifacts. Finally, the detectability of objects with extreme large-amplitude lightcurves (e.g., binary systems or objects with highly elongated shapes) that are close to the LSST detection limit may be highly dependent on the rotational phase at which they happen to be at the time of any given observation.

\paragraph{Running on LSST Datasets (for the first 2 years)}
In order to identify scientifically valuable objects that require immediate follow-up, preliminary fits for physical parameters should be performed using photometry from the nightly alert stream once a certain amount of data judged to be sufficient for deriving those parameters with meaningful precision and reliability, where the threshold for data set size per object has yet to be determined.  As more data are collected, these fits could be updated at regular intervals, where the ideal incremental increase in data set size needed to trigger re-fitting has similarly yet to be determined.  Finally, given that some parameters may be expected to change over time due to changes in viewing aspect angle (which can cause an object's average projected cross-section on the sky to change, changing its apparent average size), resurfacing due to close encounters with other bodies or otherwise undetected collisions that can change an object's observed average color, or radiative forces that can accelerate or decelerate an object's rotation, the capability to perform physical parameter fits on Solar System objects using subsets of the full LSST data set would also be very desirable.
Inactive (or marginally active) objects could be evaluated from small postage stamps, but active small bodies may require larger cutouts. More detailed fitting and population-level studies could be conducted through annual data releases.

\paragraph{Precursor data sets}
Algorithms to predict asteroid brightness from a set of physical parameters could be tested on \gls{Pan-STARRS} data along with selective follow-up of some small bodies to confirm the validity of results. When \gls{LSST} is running, objects that have previously appeared in \gls{Pan-STARRS} data could be cross-matched to assign longer baselines. Cross-calibration of the photometry would be necessary for executing this idea.

\paragraph{Analysis Workflow}
%Newly-identified objects in the nightly alert stream should be analyzed via the algorithm for joint-fitting of physical attributes. The most interesting and potentially valuable objects should be flagged for follow-up, and these targets should be disseminated to the Solar System \gls{Science Collaboration}. This part of the science case will be related to the anomaly fitting technical case for timeseries analysis. Any objects with sufficient uncertainties on their physical parameters should also be analyzed and updated through data in the alert stream, and the lightcurves of all small bodies should be updated.

Once a certain minimum amount of data (to be determined) has been acquired for a given Solar System object, those data should be processed using the algorithm(s) for simultaneous fitting of physical properties (primarily color, phase function, rotation period, and axis ratio). The most interesting and potentially scientifically valuable objects should be flagged for follow-up observations, and these targets should be disseminated to the Solar System \gls{Science Collaboration} or to designated \glspl{TOM}. In addition to storing the results of physical parameter fitting procedures in the LSST or broker databases, it will also be essential to record the precise data set used to derive those parameters to inform algorithms for triggering updated fitting analyses when the available data set for an object has increased by a sufficient amount that re-fitting is likely to produce meaningfully more precise results. Mechanisms should also be put into place to regularly perform fitting analyses only using subsets of the total LSST data set (e.g., all data in a given year or a given observational apparition, or every $N$ consecutive detections, for example) to enable relatively unbiased searches for changes in physical parameters over time.

More computationally-expensive \glspl{algorithm} could be run on annual data releases. Any population-level analysis involving the post-processing pipeline could also be run on the data releases, since this work would not be time-critical.

\paragraph{Software Capabilities Needed}
A model of small body observables (brightness, variability, color) must be linked to physical properties (size, shape, rotation). This code should be integrated into the post-processing module for Solar System objects. Since each small body is independent, this problem can be considered “embarrassingly parallel.” %A period-finding algorithm must be written that’s optimized for \gls{LSST} cadence. Small body studies often use \gls{PDM}, but it may be possible to implement a method that has uses outside of small bodies.
Various algorithms have been developed for simultaneous fitting of some physical parameters (see References for Further Reading), but more evaluation is needed to determine which (if any) of these algorithms are best-suited for achieving our scientific goals and how much further development (if any) is needed to be able to operate them at LSST scales.  It would also be highly desirable to have automated mechanisms to continuously monitor and evaluate the data available for every Solar System object in the LSST database to determine when both initial physical parameter fitting and updated fitting analyses should be triggered, where appropriate triggering criteria are likely to be somewhat more sophisticated than simply the number of available data points (e.g., in particular, also taking into account phase angle coverage).

\paragraph{References for Further Reading}

\cite{kaasalainen2001inversion} ``Optimization Methods for Asteroid Lightcurve Inversion. I. Shape Determination'', Icarus, 153, 24\\
\cite{2022AJ....163...29L} ``Characterizing Sparse Asteroid Light Curves with Gaussian Processes'', Astron.\ J., 163, 29\\
\cite{lujewitt2019lightcurve} ``Dependence of Light Curves on Phase Angle and Asteroid Shape'', Astron.\ J., 158, 220\\
\cite{thirouin2016manos} ``The Mission Accessible Near-Earth Objects Survey (MANOS): First Photometric Results'', Astron.\ J., 152, 163\\
\cite{2015AJ....150...75W} ``Asteroid Light Curves from the Palomar Transient Factory Survey: Rotation Periods and Phase Functions from Sparse Photometry'', Astron.\ J., 150, 75

\pagebreak
\subsubsection{Shift-and-Stack for faint object detection} \label{sec:ShiftStack}
\cleanedup{Mario Juric}

\Contributors{Mario Juric, Steven Stetzler, David Trilling, Andy Connolly, Hayden Smotherman}
{3/28/22}

\paragraph{Abstract}
A foundational goal of the \gls{LSST} is to map the Solar System small body populations that provide key windows into understanding of its formation and evolution. This is especially true of the populations of the Outer Solar System -- objects at the orbit of Neptune and beyond (further than 30~AU). LSST, on its own, will detect individual KBOs to $r$\textasciitilde 24.5.
But advanced shift-and-stack algorithms \citep[e.g., such as][]{Whidden_2019} would enable significantly deeper searches. This would allow for a census of the outer Solar System (OSS) to deeper magnitudes. For example, stacking just 10 epochs would reach $\sim$25.5 magnitude, yielding 4-8x more TNO discoveries than the single-epoch baseline, enabling rapid identification and follow-up of unusual distant Solar System objects in $\gtrsim 5 x$ greater volume of space \citep{https://doi.org/10.48550/arxiv.1901.08549}.
Stacking O(36) exposure ($ \sim$~half a year) would reach \textasciitilde 26.5, potentially yielding as many as $10^6$ KBOs (extrapolating the early results of the DEEP survey; Trilling et al., in prep). These increases would enhance the science cases discussed in the \cite{Schwamb_2018} whitepaper, including probing Neptune's past migration history as well as discovering hypothesized planet(s) beyond the orbit of Neptune (or at least placing significant constraints on their existence).

\paragraph{Science Objectives}
\begin{itemize}
\item Exploratory analysis of TNO populations across the entire sky to 25.5+ magnitude.
\item Improve the characterization of OSS populations (size and color distributions at the small-size end).
\item Increasing the number of tracers of dynamical populations in the OSS by 5-10x.
\item Increased likelihood for discovery of distant or unusual objects
\item Prior art: the work in progress on the DEEP survey (w. DECam), papers by \cite{Whidden_2019} and \cite{Smotherman_2021}.
\end{itemize}

\paragraph{Challenges (what makes it hard)}
\begin{itemize}
\item Need access to (all) images (eventually) instead of catalog. Access patterns similar to building coadds, but requires all pixels in a \textasciitilde 3deg radius of a typical boresight. Significantly more expensive to produce the “shift and stack coadds,” though the output dataset is small, \textasciitilde O(TBs).
\item Fast searches require access to accelerated computing hardware (GPU).
\item Software is required to interface with a GPU both efficiently and with ease for a user.
\item {\bf Significant} algorithmic and implementation improvements needed:
\begin{itemize}
\item Run time scales with the number of images searched. In the first year of LSST, we expect 80 images available to search. Based on a 24.5mag 5-sigma detection threshold on a single exposure, 7 images are needed for a 25.5mag search, 40 images for 26.5mag search, and 80 images allows for a search to 26.9mag. These numbers scale the run time by 10-100x.
\item Run time also scales with the time baseline searched, due to the search parameter space expanding. With present-day algorithms, the number of trajectories searched scales with the square of the time baseline. A rough calculation implies \textasciitilde 500 GPU-days to perform a search on 80 images over a 3 month baseline. This becomes \textasciitilde 63 days with 10 images, and \textasciitilde 7 days with 10 images over a 1 month baseline.
\item Current implementation needs to fit the required pixels into limited GPU memory. This imposes a constraint on the quantity (time baseline)x(sky area searched) since an increased time baseline increases the number of images and the sky area searched increases the number of pixels per image used. Assuming a stack of 80 images, cutting out a (1/16)~deg$^2$ region of the sky requires \textasciitilde 36 GB of memory, reaching the memory limits of current GPUs.
\item Finally, the run time scales with the area of sky searched. The calculation above shows we can process (1/16)~deg$^2$ of the sky with one GPU. Taking the size of the ecliptic to be $360{\rm deg} \times 10{\rm deg} = 3600{\rm deg}^2$, this implies that -- assuming just present day codes were used without algorithmic and implementation improvements -- a full-sky search would take 57,600 GPU-years to complete.
\end{itemize}
\item The need for repeated runs – analysis would be run \textasciitilde quarterly-yearly, eventually over 10yr dataset.
\end{itemize}

\paragraph{Running on LSST Datasets (for the first 2 years)}
\begin{itemize}
\item Either raw/calibrated/difference LSST image dataset (both WFD and deep-drilling). The analysis can start as soon as a \textasciitilde month of data is collected, but will likely be done on a quarterly or an annual basis (depending on the speed of the algorithm).
\item The input dataset is the amount of single-epoch data LSST will collect over the stacking window (e.g., 3 months or 1 year).
\item The output data size is small O(few TB).
\end{itemize}

\paragraph{Precursor data sets}
\begin{itemize}
\item These searches are already being performed with data from the \gls{DECam}: \gls{HITS} survey, \gls{DEEP} survey, and the deep drilling fields of the \gls{DECAT} survey (a direct precursor to \gls{LSST} \gls{DDF}s)
\end{itemize}

\paragraph{Analysis Workflow}
\begin{itemize}
\item For a given (set of) test orbit(s), the overlapping set of LSST visits will be determined. This will be \textasciitilde 30-90 nights worth of visits.
\item For each LSST visit, a difference image exposure (data/mask/variance planes) will be obtained from the LSST archive and loaded into GPU memory. Images will be additionally masked to mask variable stars from the images.
\item The GPU search will be performed, producing a set of candidate trajectories for moving objects \cite{Whidden_2019}.
\item Postage stamps of the calibrated images will be queried from the LSST archive for each visit along a candidate trajectory. Forced photometry will be performed on the cutout image using LSST pipeline code.
\item Postage stamps are validated using both human vetting and machine learning methods (a CNN trained with simulated postage stamps and previously discovered real objects).
\item Validated detections are fit with an orbit to produce a discovery.
\end{itemize}

\paragraph{Software Capabilities Needed}
\begin{itemize}
\item Ability to query the LSST archive for calibrated exposures, difference images, and cutouts.
\item New software infrastructure will be required to run this analysis, likely requiring many (10s-100s) machines with 1 or more GPU(s).
\item Derived data products will include test orbits corresponding to likely stacked detections. %, \textasciitilde \% of initial test orbits. These are filtered and/or deduplicated, reducing data volume by \textasciitilde \%. Filtering of test orbits and validating of filtered orbits will likely require cutouts of images.
\end{itemize}

\paragraph{References for Further Reading}
\begin{itemize}
\item Shift-and-stack with GPUs: “Fast Algorithms for Slow Moving Asteroids: Constraints on the Distribution of Kuiper Belt Objects”, \cite{Whidden_2019}.
\item Extra depth from shift-and-stack with LSST data and KBO discovery estimates: “Enabling Deep All-Sky Searches of Outer Solar System Objects”, \cite{https://doi.org/10.48550/arxiv.1901.08549}.

\end{itemize}

\pagebreak

\subsection{Cosmology} \label{sec:cosmology}
\subsubsection{Weak lensing cosmology analysis / cosmic shear} \label{sec:CosmicShear}
\cleanedup{Rachel Mandelbaum}

\Contributors{ Rachel Mandelbaum (\mail{rmandelb@andrew.cmu.edu}), Arun Kannawadi (\mail{arunkannawadi@astro.princeton.edu}), Andresa Campos (\mail{acampos@cmu.edu})}
{ March 27, 2022}

\paragraph{Abstract}
We present the first cosmological analysis of the weak lensing shear-shear correlation functions, which trace the evolution of large-scale structure from intermediate to low redshift.  With the first year of LSST data, we can already constrain the amplitude of matter fluctuations to greater precision than precursor surveys such as \gls{KiDS}, \gls{DES}, and \gls{HSC}, reaching \textasciitilde 1\% precision when considering statistical and systematic uncertainty.  This improved result is due to the increased area of LSST, and the fact that the first year of data is comparable in depth to DES and deeper than KiDS.  This work will rely on analysis of the catalogs produced by the LSST \gls{Science Pipelines} down to detections with \textasciitilde 10$\sigma$ significance.  This result is sufficiently precise to permit a robust test of a \gls{LCDM} by comparing with the amplitude of fluctuations from the \gls{CMB}, which current analyses suggest may be higher than that from weak lensing (implying some uncontrolled systematic or a failure of the LCDM model).  Together with the CMB data, we can go beyond the LCDM model and provide competitive constraints on the equation of state of dark energy.

\paragraph{Science Objectives}

Milestones to this result include defining samples that pass basic null tests, validating that the shear estimates meet the needs for the science, and estimating the ensemble redshift distribution N(z) for the samples selected based on photo-z.  For the last step, we do not necessarily need very precise photo-z (though that would improve the localization of the selected redshift bins and improve the constraint on w) but we do need a very precise understanding of the N(z) for the ensemble.  At this stage, we anticipate that photo-z calibration and the impact of blending on shear and photo-z are likely to be the key limited systematics.

\paragraph{Challenges (what makes it hard)}

Computational challenges related to image analysis, shear calibration, and cosmological likelihood analysis tend to compete with building sufficient models for systematics as the biggest challenge.

 Broadly, there are two kinds of systematics: modeling systematics that are mostly astrophysical in nature and systematics in our data. Understanding astrophysical systematics that arise from our lack of understanding of baryonic physics and galaxy formation relies on information beyond the \gls{LSST} data to mitigate their impact on  cosmological parameter constraints. While some amount of self-calibration may be possible, we need to improve external (beyond \gls{LSST}) measurements and simulations that give us direct constraints on baryonic feedback, intrinsic alignment etc.

Key systematics of both types are listed below:

\begin{itemize}
\item \gls{IA}:
 At \gls{LSST} precision,  modeling the intrinsic alignments of galaxy shapes with the large-scale density field is necessary. They are one of the major drivers of the overall uncertainty in current cosmic-shear analyses. We will need tighter priors from external measurements and/or from hydrodynamical simulations. Self-calibration of IA parameters from the observations is also a possible approach.
\item Blending:
 At LSST depths, a substantial fraction of the galaxies are blended. \citet{2021arXiv211207659N} estimates 12\% of LSST galaxies would constitute undetected blends, i.e., two overlapping galaxies  In general, they could have vastly different redshifts which necessitates a deblending algorithm that can apportion the \gls{flux} to the different galaxies. This is currently done using the  SCARLET algorithm \citep{2018A&C....24..129M}, which uses multiband images to model out the neighboring galaxies.
Synthetic source injection tools that are developed jointly between \gls{DM} and DESC can be used to estimate the sensitivity of the measurement to the amount of blending present. On longer time scales, pixel-level joint processing of Euclid+Rubin images to create Euclid+Rubin DDPs will provide an even better handle on deblending, especially undetected blends.

\item \gls{PSF} estimation:
 \citet{2022arXiv220507892Z} estimates the bias in the cosmological parameters from errors in higher-order moments. They find that errors from the current version of the state-of-the-art algorithm Piff (now default in Rubin \gls{Science Pipelines}) are comparable with PSFEx and that they introduce biases that are $0.2\sigma$ in the structure growth parameter $S_8$ in LSST Y1 analysis.
\item Shear \gls{calibration}:
 Even in the absence of any blending, source detection algorithms may preferentially detect sources that are sensitive to ellipticities/shape of the source. This introduces a detection or selection bias in the sample \citep{2019A&A...624A..92K}. This could be corrected for by introducing a detection step in the shear measurement, which is essentially what metadetection achieves \citep{2020ApJ...902..138S}.
 The requirement on the shear multiplicative bias is at the level of $10^{-3}$. While shear measurement algorithms such as metadetection reach this level in simulations with a number of realistic effects (E. Sheldon, internal communication), this could go beyond the requirements if various uncorrected systematic effects add up coherently.
Can we get shear measurement codes to the stage where they do not need any image simulations for \gls{calibration} (only validation)? It would be an ever-increasing burden to calibrate shear if \gls{calibration} simulations are a requirement.
\item Photo-z \gls{calibration}:
 Blending impacts the ability to precisely estimate ensemble redshift distributions, which can ultimately cause a bias in the cosmological parameters, as shown in \citet{2021arXiv211207659N}. Although we expect to have a different deblending methodology from the one in the above paper, it shows the importance of being able to perform deblending accurately.
%\begin{itemize}
%\item Do we have to worry about sample bias in the spec sample given that the LSST area will be so big?
%\end{itemize}

\item Sociological systematic:
 True blinding is difficult to achieve since the Rubin data products are available to the entire data rights community. \gls{DESC} products can be blinded, but that would necessarily imply that Rubin data products cannot be used off-the-shelf for cosmological analysis by \gls{DESC}. %The more the Rubin products are science ready, the less effective blinding challenges can get.

\end{itemize}
 These analyses will be done with a subset of the \gls{LSST} data releases. The analyses will be mostly constrained by the major validation effort that is required before getting the results published, which has meant that most precursor surveys do not carry out cosmological shear measurements with each new data release.

\paragraph{Existing Tools}
\begin{itemize}
\item The following tools already exist today for this analysis
\begin{itemize}
\item RAIL+qp for getting a p(z) for each object
\item Metadetection \gls{algorithm} (descwl\_coadd, mdet-shear-sims… )
\item Code from Pedersen et al. on self-calibrating \gls{IA}
\item Synthetic \gls{Source} Injection (DM/DESC)
\item Theoretical modelling (e.g., CCL, HMCode/Halofit), correlation functions (e.g., TreeCorr, NaMaster, TXPipe), Covariances,  Inference (e.g., CosmoSIS)

\end{itemize}

\item The following functionality is still missing from these tools:
\begin{itemize}
\item Storing the p(z) for each object from different colors could be memory-intensive. We need either an uber-fast \gls{algorithm} that can take in colors and reference catalogs and spit out p(z) on-demand, or find an effective compression technique to store them.
\item Current \gls{SSI} \gls{software} is not equipped to inject sources onto coadds, since the ability to inject a source and additional noise that is consistent with the source is not yet present. Developing this ability could save a lot of processing time.
\end{itemize}

\end{itemize}

\paragraph{Running on \gls{LSST} Datasets (for the first 2 years)}
 We will use the following data products:
\begin{itemize}

\item Data release catalogs for the \gls{WFD} survey - we will use essentially all galaxies down to some relatively low significance (\textasciitilde 10-sigma).
\item Some \gls{calibration} and systematics tests will use the DDFs.
\item External spectroscopic catalogs to validate photometric redshifts.
\item We will need some image simulations (beyond DESC DC2) that mock-up the cell-based coaddition to validate the shears, and cross-matching and cross-correlation against spectroscopic surveys to validate the \gls{photo-z} and ensemble N(z) estimates.
\item Euclid+Rubin DDPs for deblending (\citealt{2019arXiv190108586S} predict mean shrinkage in ellipticity error bars by more than a factor of 2 when \gls{LSST} is combined with Euclid)
\end{itemize}

\paragraph{Precursor data sets}
\begin{itemize}
\item DES, \gls{HSC}, and KiDS all have public data releases that could be used for precursor analyses.  In some cases, the catalogs are most accessible while reanalyzing the images would be a substantial challenge despite data being made public.
\item Reference spectroscopic samples (deep ones for direct \gls{calibration} and wide ones from e.g. DESI for cross-correlation) will be important for the N(z) validation.

\end{itemize}

\paragraph{Analysis Workflow}

Assuming the existence of catalogs with galaxy \gls{shape} or shear measurements and with photometric redshifts, the analysis workflow is as follows:
\begin{enumerate}
    \item Apply selection criteria to the catalogs following some pre-defined process to mitigate selection biases (potentially using the metadetection technique).
    \item Apply any shear \gls{calibration} or other steps.
    \item Divide the sample into tomographic (redshift) bins based on the \gls{photometric redshift} estimates.
    \item For each tomographic bin, apply a method for inferring the ensemble redshift distribution, N(z), based on the photometric redshifts and/or other information such as clustering with a reference spectroscopic sample.
    \item Measure various null tests designed to detect the impact of systematic biases; depending on the results, may need to iterate on the above steps.
    \item Measure shear correlation functions using pairs of tomographic bins $i$ and $j$.
    \item Produce a covariance matrix using some method potentially based on numerical integration of analytic expressions or based on mock catalogs.
    \item Carry out a likelihood analysis for the cosmological parameters based on the measured data vectors and covariance matrices, including models for astrophysical and other systematics.
\end{enumerate}

\paragraph{Software Capabilities Needed}

Needed capabilities include the following:

\begin{itemize}
\item Survey property maps (e.g., skyproj with healsparse\footnote{\url{https://github.com/LSSTDESC/healsparse}})
\item Rubin calibrated exposures and coadds
\item Estimators for ensemble redshift distributions
\end{itemize}

\paragraph{ References}

Precursor survey analyses:
\begin{itemize}
    \item \gls{KiDS}-1000: \citet{2021A&A...645A.104A}
    \item \gls{DES}-Y3: \citet{2022PhRvD.105b3514A}
    \item \gls{HSC}-Y1: \citet{2020PASJ...72...16H}
\end{itemize}

\pagebreak
\subsubsection{Probabilistic Type Ia supernova cosmology analysis} \label{sec:Type1aCosmo}
\Contributors{Alex Malz (\mail{aimalz@nyu.edu}), Rachel Mandelbaum (\mail{rmandelb@andrew.cmu.edu})}
{March 29 2022}

\paragraph{Abstract}
 Our scientific objective is to constrain the expansion history and cosmological parameters using photometric Type Ia supernovae from the first \textasciitilde year of LSST data, in combination with precursor distance ladder measurements.  In particular, we would like to go beyond the traditional analysis methods, by self-consistently incorporating probabilistic information as follows:
\begin{itemize}
\item Classification probabilities (e.g., with \gls{BEAMS})
\item Host identification probabilities (e.g., with \gls{zBEAMS})
\item Redshift probabilities -- e.g., with \gls{SCIPPR}
\item Selection function
\item Debiasing on the (probabilistic) absolute magnitude.
\end{itemize}

 The probabilistic approach will permit the use of a larger sample in early LSST data compared to a non-probabilistic approach that requires secure follow-up in early LSST data.

\paragraph{Science Objectives}
 The key steps in the analysis:
\begin{itemize}
\item Alert broker provides classification probabilities and redshift posteriors for possible hosts for potential supernovae.
\begin{itemize}
\item If using early (first year) data, it will be challenging to understand the selection functions which come with light curves correctly classified as Type Ia SNe.
\item We run our own classification using additional points in the light curve after the initial detection of the supernovae.
\end{itemize}
\item We apply additional selection criteria to identify ones that can be used for science, and potentially get new host probabilities and redshift posteriors (but still need the originals because they go into the selection function).
\item Light curve fitting with population-level debiasing function.
\item Hierarchical inference of cosmological parameters factoring in probabilistic information for hosts, classification, and redshift (though none of the existing algorithms currently work at scale).

\end{itemize}

\paragraph{Challenges (what makes it hard)}
 The key technical limitation is that most of the steps in this analysis do not run at scale, while it’s not clear how to do selection function quantification from the level of alerts, so algorithm development needs a lot of thought.

Early light curve fits will be particularly impacted by sparsity.  Lack of redshifts from Rubin in the first year will be an issue - the brokers will get them from different sources and they will definitely differ from DESC photo-z.

\paragraph{Running on LSST Datasets (for the first 2 years)}
The analysis will use:
\begin{itemize}
\item The alert stream.
\item DESC value-added photo-z catalogs.
\end{itemize}

\paragraph{Precursor data sets}
PLAsTiCC and ELAsTiCC data challenges can be used as a test-bed, but photo-z are not self-consistently generated with the light curves, especially for PLAsTiCC.  ELAsTiCC results could potentially be used to determine the selection function (validate a method for doing so), but the sample may not be large enough. A preliminary study using Pan-STARRS Medium Deep Survey data has already been conducted (see \citealt{jones2018measuring}).

\paragraph{Analysis Workflow}
Running the analysis requires us to run these steps:
\begin{itemize}
\item Classify lightcurves using additional points after the initial detection.
\item Quantify selection function - limited by basic algorithmic questions.
\item Probabilistic lightcurve and absolute magnitude fitting.
\item Use redshift uncertainties and classification probabilities for cosmology (\gls{BEAMS} - does not work at scale), host identification probabilities (\gls{zBEAMS} - simple assumptions, does not work at scale), redshift probabilities (scippr - open source, does not work at scale)
\item Hierarchical inference at scale, combining all of the above components in a self-consistent manner.
\end{itemize}

\paragraph{Software Capabilities Needed}
The needed software capabilities boil down to integrating and scaling up existing probabilistic approaches.
There are also visualization needs for probabilistic data products connected to intuitive traditional plots, which could be implemented in the RSP.

\paragraph{References for Further Reading}
\begin{itemize}
% \item DES photometric supernova analysis (did not use probabilistic methods): \citet{2021arXiv211110382V} and \citet{2022arXiv220111142M}
\item Bayesian Estimation with multiple Species (BEAMS): \citet{PhysRevD.75.103508} \item \gls{zBEAMS} (BEAMS plus host identification uncertainty)- \citet{2017JCAP...10..036R}
\item Supernova Cosmology Inference with Probabilistic Photometric Redshifts (SCIPPR): open source code for incorporating probabilistic classification and redshift to cosmology inference, does not yet work at scale - \citet{2018AAS...23124503P}; \url{https://github.com/aimalz/scippr}
\item Photometric LSST Astronomical Time-series Classification Challenge (PLAsTiCC): \citet{2018_plasticc,2020arXiv201212392H}
\end{itemize}

\pagebreak
\subsubsection{Optimal spectroscopic follow-up algorithms for Type Ia supernova cosmology} \label{sec:OptimalSpecCosmo}
\cleanedup{AIM}

\Contributors{Alex Malz (\mail{aimalz@nyu.edu}), Emille E. O. Ishida (\mail{emille.ishida@clermont.in2p3.fr}), Bruno Quint}
{6 April 2022}

\paragraph{Abstract}

The \gls{RESSPECT} project is a COIN-DESC joint endeavour which aims to enable the construction of optimized training samples for \gls{LSST} photometric \gls{SN} cosmology. It takes into account a realistic description of the astronomical data environment and available follow-up resources. Early \gls{LSST} data can be used to validate the ReSSpect pipeline, ensuring the system is able to scale to \gls{LSST} requirements and helping define which categories of transients have the potential to boost performance of machine learning classifiers -- if included in the training sample. The system requires daily updates on its data sets with every newly observed photometric point and the entire pipeline should be daily run in order to identify the optimal candidates for the subsequent night. As a consequence, significant computing resources are required to daily process all updated data which survives selection cuts. By construction, half of the objects selected for follow-up  will be composed of SNIa and the other half will contain transients which are easily mistaken with SNIa by \gls{ML} classifiers. The final queried spectroscopic sample, when used as a training sample, will enable optimum results from any supervised learning classifier - thus ensuring exploitation of purely photometric \gls{LSST} light curves for \gls{SN} cosmology.

\paragraph{Science Objectives}
Cosmological analysis with \gls{LSST}’s photometric supernova sample will be essential because the spectroscopically confirmed (and even spectroscopically confirmable) sample will not be sufficiently large to improve upon the current status quo. In order to fully take advantadge of the information contained in \gls{LSST} data, it is necessary to optimize the exploitation of purely photometric light curves.

Photometric supernova cosmology requires classification of lightcurves, primarily done by machine learning algorithms conditioned on a training set. Currently, the only possible available option is to train on simulations or on the available spectroscopic samples. Although encouraging results have been obtained  by using simulations as training, this method inevitably biases the results towards the already known -- and modelled -- light curve features present in templates. Alternatively, currently available spectroscopic samples are highly biased towards SNIa-like events, which leads to sub-optimal performance from \gls{ML} classifiers. From the algorithm point of view, an ideal training sample  should enclose examples not only of the class of interest, but also on all possible objects which can be mistaken by it.
Our final goal is to enable purely photometric supernova cosmology by using \gls{RESSPECT} to construct informative training samples for \gls{ML} classifiers.
 In parallel, the construction of this highly informative sample will also enable further development in the modelling of astrophysical transients - in particular \gls{SN} events which lie in the boundary of the Ia class.

\paragraph{Challenges (what makes it hard)}
\begin{itemize}
\item Feature extraction: the \gls{pipeline} requires that all incoming data are represented as a homogeneous rectangular matrix. It currently transforms all data into features using a parametric fit. This procedure needs to be repeated every time a new photometric point is observed. This process is slow and scales fast with the number of objects which needs to be processed at each night.
\item Retraining: The active learning loop requires the model to be retrained at each night after the data has been updated. Moreover, in order to use the cosmology \gls{metric} to disentangle equally valuable batches of potential candidates, the training needs to be repeated multiple times for each batch.  This procedure needs to be done before the subsequent observation night. As a consequence, the classifier inside the AL loop needs to be cheap and straight forward to train.
\item Cosmology fit: the cosmology \gls{metric} calculation is quite slow and is based on a simplistic model because anything more sophisticated would be even slower. Nevertheless, it still requires a lot of computational time to enable the cost calculation for a series of multiple batches.
\item Updated substream of alerts: after the retraining is done, the updated \gls{ML} model needs to be used to filter the original data stream (provided via one of the community brokers). This will require interaction between the pipeline and the broker interface, since the entire process should be automatized.
\item Communication with follow-up facilities: once the list of candidates is identified, we need to automatically distribute the list to the available spectroscopic follow-up facilities. And once the labels are obtained they need to be included in the training. Ideally these processes would be automatic.
\item Timing constraint: results cannot be put to use right away. What we learn from doing this will build the training set we use for photometric SN cosmology in early \gls{LSST} data and  inform follow-up strategies after year 1.
\end{itemize}

\paragraph{Running on \gls{LSST} Datasets (for the first 2 years)}
 The analysis will use:
\begin{itemize}
\item The filtered alert stream (removing artifacts and already known transients identified via cross-match with publicly available catalogs)
\item Optionally utilize \gls{DESC} value-added photo-z catalogs
\item Initial training requires a minimum number of spectroscopically classified objects (5 Ias and 5 non-Ia would be enough)
\item \gls{Validation}: requires set of complete light curves which could potentially be used for cosmology (pass quality cuts, pending classification)
\item Spectroscopic classifications for appointed candidates: candidates would be known at each night and classification will be acquired using external follow-up resources and publicly available catalogs
\end{itemize}

\paragraph{Precursor data sets}
Our initial tests show that it is not advisable to use precursor data sets for training. Starting from scratch with \gls{LSST} data allows the construction of a tailored data set and avoid bias in classification results.

\paragraph{Analysis Workflow}

\begin{figure}
\begin{centering}
\includegraphics[width=\textwidth]{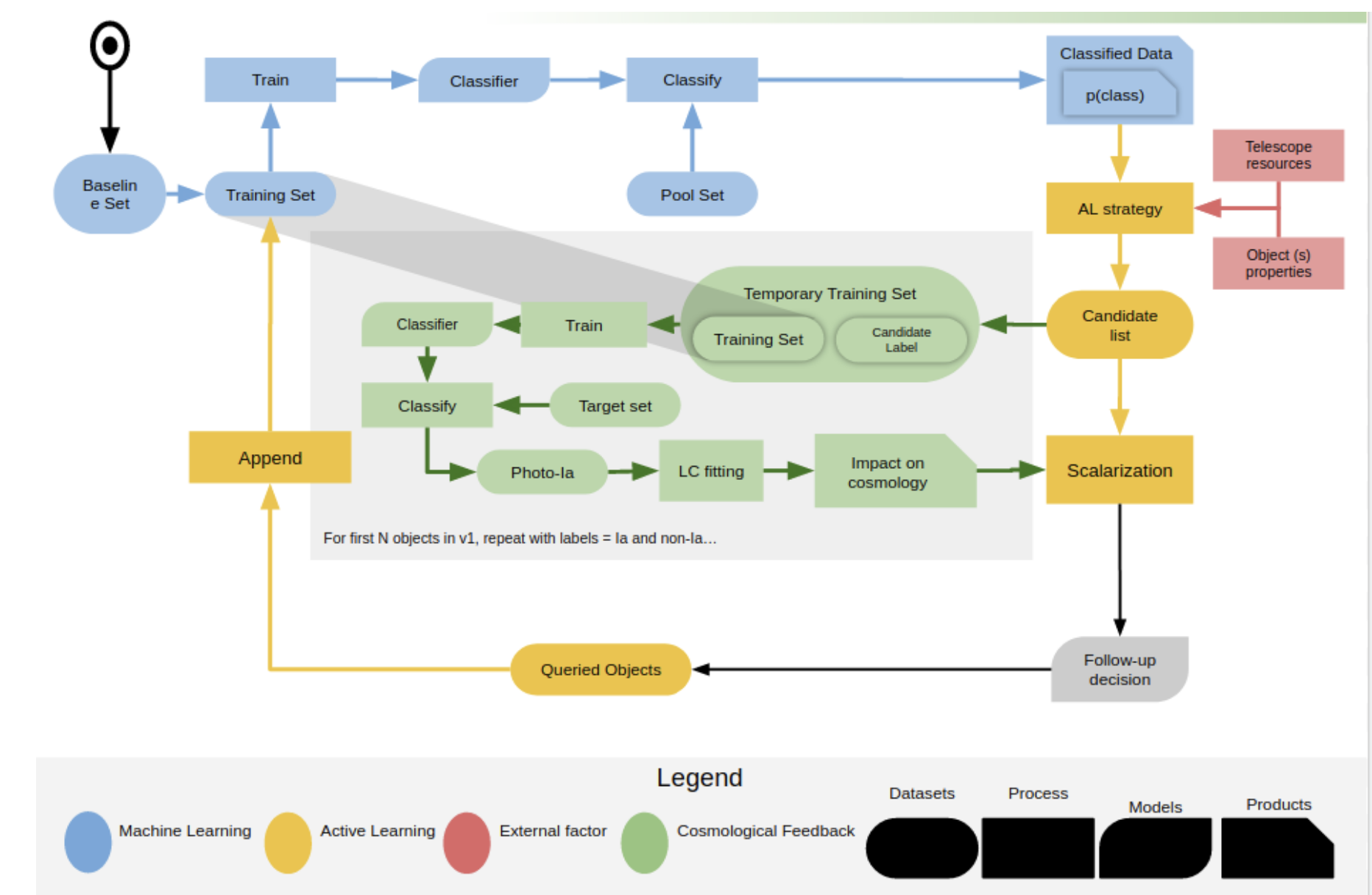}
	\caption{ Workflow (figure courtesy of Bruno Quint)
\label{fig:OptimalSpecCosmo}}
\end{centering}
\end{figure}

The key steps in the analysis (see also \autoref{fig:OptimalSpecCosmo}):
\begin{itemize}
\item \gls{Construction} full light curve (validation) and small spectroscopic confirmed (initial training) samples from alerts
\item Apply feature extraction to the validation and initial training samples
\item At each day (active learning loop):
\begin{itemize}
\item Apply feature extraction to the pool and training sample including newly observed epochs
\item Train classifier
\item Classify test and pool sets (yields classification probabilities)
\item Identify candidates for spectroscopic follow-up
\item For each candidate, for each possible class:
\begin{itemize}
\item Add the candidate with presumed label to the training sample
\item Train the classifier
\item Classify the test sample
\item Select photometrically classified \gls{SN} Ia
\item Perform SALT2 fit (parametric \gls{SN} lightcurve model) on photometric \gls{SN} Ia sample
\item Add bias correction
\item Infer cosmological parameters (currently using a simplified Bayesian model due to computational expense)
\item Quantify cosmological impact for the new candidate in training set
\end{itemize}
\item Combine cosmology \gls{metric} of candidates into batches
\item Evaluate observation cost of each batch
\item Choose batch of candidates optimizing the overall \gls{metric}
\item Stream candidates to spectroscopic follow-up facilities
\item Add new available labels (candidates from previous nights) to training set
\end{itemize}
\end{itemize}

\paragraph{Software Capabilities Needed}
 We have all the moving parts in terms of the software described above, and the \gls{pipeline} works end-to-end on catalog data.
 We need an interface to the real alert stream as well as the spectroscopic follow-up facilities.
 We need software engineering to scale-up the computation (specifically the classifier retraining and cosmology \gls{metric} evaluation) and the computing resources to run the pipeline nightly.

\paragraph{References for Further Reading}
\begin{itemize}
\item Basic active learning \gls{algorithm}: \citet{2019MNRAS.483....2I}
\item \gls{RESSPECT} pipeline:  \url{https://github.com/COINtoolbox/RESSPECT}/
\item Cosmology \gls{metric} validation code: \url{https://github.com/emilleishida/resspect_metric}
\item Details on the active learning strategy and cost calculation: \citet{2020arXiv201005941K}
\item Active learning application to real alerts data: \citet{2021arXiv211111438L}

\end{itemize}

\pagebreak
\subsubsection{Cross-correlation between LSST and \gls{CMB} probes of gas physics} \label{sec:CMBprobes}
\cleanedup{Giulio Fabbian}
\Contributors{ Giulio Fabbian (\mail{FabbianG@cardiff.ac.uk}), Rachel Mandelbaum (\mail{rmandelb@andrew.cmu.edu})}{}

\paragraph{Abstract}
During its journey towards us, the \gls{CMB} interacts with the large-scale structures of the universe (e.g. galaxy clusters, filaments, voids) as they form. This happens mainly through gravitational lensing of its photons and their inverse-Compton scattering with electrons having large thermal or bulk velocities (\gls{tSZ} and \gls{kSZ}). Thus, the \gls{CMB} is also a powerful tracer of the matter distribution in the universe. The matter distribution can also be probed surveying the distribution of galaxies at different redshifts and frequencies. The inhomogeneity of the matter distribution in the universe affects galaxy observations and \gls{CMB} similarly, but, crucially, not exactly in the same way so that a significant amount of information can be extracted thus from their joint analysis and through their cross-correlations. Cross-correlations will help to break degeneracies between observables and to isolate different physical effects otherwise indistinguishable.
The \gls{SO}  will play a major role in this endeavor, providing a deep CMB polarization-sensitive survey covering about half the sky and the southern hemisphere starting in \textasciitilde 2023.

Multi-frequency observations of the CMB can be used to extract maps of the \gls{tSZ} effect through its characteristic spectral distortion imprinted in the CMB, and to separate it from galactic foregrounds and extragalactic emission in the \gls{IR} band. The \gls{kSZ} signature is observed in maps of the CMB temperature anisotropies that cannot be easily isolated without the help of an external tracer given that its \gls{SED} is the same as the one of CMB anisotropies. Alongside \gls{SO} and on the same timescale, the next generation of galaxy surveys, the European Space Agency’s Euclid mission and the Vera C.~Rubin  Observatory \gls{LSST}, will map the mass distribution in the recent universe over more than half the sky with complementary experimental strategies. Given their full overlap, SO, LSST and Euclid will provide ideal data sets for joint studies and cross-correlation analyses.

\paragraph{Science Objectives}
\begin{itemize}
\item The cross-correlation between \gls{LSST} galaxy density map and tSZ Compton $y$-map will give us insight on the bias-weighted pressure profile and energy content of electrons in halos.
\item The cross-correlation between galaxy density of LSST and the square of the \gls{CMB} temperature map (essentially a probe of the $\langle TT\delta\rangle$ bispectrum) is sensitive to both the electron density and velocity field (and hence a good probe of baryonic feedback processes).
\item A good knowledge of the galaxy sample can be used to investigate the variation of both the physical processes above as a function of redshift and galaxy environment or mass.
\item The analysis will use data products provided by \gls{CMB} experiments and LSST catalogs with potentially augmented properties to allow for systematic tests of the sample selection and or investigation of variations as a function of environment properties.
\item The accuracy of the   characterization of the galaxy sample's ensemble redshift distribution (in a statistical sense) and extragalactic and galactic foregrounds affecting \gls{CMB} data sets (e.g. modulation of sample selection function by extinction of galaxy dust)  will be the major potential systematic factor. Additive systematics affecting only the \gls{CMB} or the LSST data should not have a large impact in cross-correlation studies.
\item Demonstrating analyses are currently being carried out with current generation of CMB data sets and publicly available surveys (that so far have been mainly optical spectroscopic surveys or \gls{IR} photometric data, such as catalogs built from \gls{WISE} data).
\end{itemize}

\paragraph{Challenges (what makes it hard)}
\begin{itemize}
\item Scientific challenge: The need for joint simulations to test the analysis methods.
\item Technical challenge: Some cross-correlation \gls{software} used for current surveys might not scale well to LSST data volume.
\item Data quality:
\begin{itemize}
\item Practical limit in resolution is imposed by \gls{CMB} survey. \gls{ACT}/\gls{SO} give \textasciitilde arcmin resolution $y$ and $T$ maps on potentially the full LSST footprint, SPT-3G will have lower noise but will cover a smaller sky area, while Planck has lower angular resolution and higher noise but potentially larger sky coverage. The details of the best data set to be used depends on the choice of the observing strategy and timing of the \gls{CMB} experiments, i.e., which sky area will be covered (and delivered) first.
\item The need to have a reasonably well-understood redshift distribution for the galaxy sample used to define the density field could be a practical issue when defining the \gls{LSST} galaxy sample.
\end{itemize}
\item Timing issue: There is not a major urgency to this analysis, though carrying it out earlier rather than later might inform strategy and analysis plans for \gls{CMB-S4}.
\end{itemize}

\paragraph{Running on LSST Datasets (for the first 2 years)}

 We’ll use the photometric galaxy sample derived from object catalogs from 1 year of data (DR2).  It will be important to have a set of well-characterized \gls{photo-z} to define tomographic source samples for which we will  infer ensemble redshift distributions.  As long as the sample is reasonably homogeneous and inhomogeneities are well understood, the analysis can proceed.

 We’ll want a value-added galaxy catalog with additional characterization of the galaxy SEDs, local density, selection function, etc.

 We’ll also need a \gls{CMB} dataset with sufficient resolution to carry out this analysis, such as \gls{ACT}, \gls{SPT}, \gls{SO}, etc.

\paragraph{Precursor data sets}
 Photometric catalogs from \gls{KiDS}, \gls{DES}, and/or \gls{HSC} can be used to develop this use case though they cover a smaller area than LSST, so \gls{DES} might be the only viable option.  Overlap with a CMB dataset is a key practical consideration to define the details of the analysis. Joint simulations including realistic evolution of the dark matter and baryon distribution on both the galaxy sample as well as on the CMB probes  will likely be important for carrying out this science with LSST and SO, as these probes are expected to be measured with high precision and statistical significance.

\paragraph{Analysis Workflow}
\begin{itemize}
\item Select galaxy catalog and split it in different mass or tomographic bins according to \gls{LSST} redshift estimates.
\item Make galaxy overdensity maps and store them.
\item Compute and store cross-correlation statistics (usually angular power spectra) between galaxy overdensity maps and external \gls{CMB} temperature and/or tSZ maps.
\item Compute or estimate covariance from jackknife resampling or simulation or analytical approximations.
\item Perform robustness or consistency/null tests using data splits constructed with subsets of the galaxy catalog or \gls{CMB} products built from subset of the data  depending on which data splits are available from the \gls{CMB} side.
\item Inference of relevant parameters.
\end{itemize}

\paragraph{Software Capabilities Needed}
\begin{itemize}

\item Query the \gls{LSST} archive and make galaxy catalogs with various cuts and desired properties, including division into tomographic samples (distinct redshift bins).
\item Map-making \gls{software}, to construct maps of projected quantities such as galaxy overdensity at a variety of resolutions
\item Ensemble redshift distribution inference for photometric samples.
\item Cross-power spectrum calculation that takes maps as inputs.
\item Theory predictions of 2-point statistics  (through \gls{CCL} or similar \gls{software})
\item \gls{MCMC} sampler or likelihood-free inference code [which would require fast simulations / new \gls{software} development]
\end{itemize}

 In some cases new methods or optimized \gls{software} are needed but they are common needs for other cosmology science cases (e.g., 3x2pt analysis).  Joint simulations would be more specific to this LSST+CMB science case.\\
 Maps and power spectra points and covariances will be stored.

\paragraph{References for Further Reading}
 Example of prototype joint tSZ and kSZ analysis: \citep{2021PhRvD.103f3513S} (\gls{ACT} and Planck \gls{CMB} + \gls{BOSS} \gls{CMASS}).\\
 kSZ from Planck + \gls{DESI} (with spectroscopic data): \cite{2022MNRAS.510.5916C}.\\
 kSZ from ACT + \gls{KiDS}: \cite{2021arXiv211002228S}.\\
 tSZ from ACT and Planck + \gls{KiDS}: \cite{2022A&A...660A..27T}.\\
 SPT and Planck y-maps \citep{2020MNRAS.491.5464K}.\\
 Projected field kSZ (without spectroscopy): estimator and its application on data \citep{2016PhRvL.117e1301H,2021PhRvD.104d3518K}.\\
 tSZ - galaxy density cross-correlation \citep{2020MNRAS.491.5464K}.

\pagebreak
\subsubsection{Self-consistent cosmological parameter constraints from galaxy clustering and galaxy-galaxy lensing using the \gls{DESI} Y1 \gls{LRG} sample} \label{sec:Constrain}
\cleanedup{Kate Storey-Fisher}
\Contributors{Kate Storey-Fisher (\mail{k.sf@nyu.edu}), Francois Lanusse, Sam Schmidt, Arun Kannawadi}
{3/28/2022}

\paragraph{Abstract}
We present the results of a galaxy clustering and galaxy-galaxy lensing analysis using the Rubin \gls{LSST} photometric sample and the \gls{DESI} spectroscopic sample.
\gls{DESI} is the current largest spectroscopic redshift survey, and is still ongoing.
The two surveys have an overlap region of at least \textasciitilde 4000 deg$^2$, and could increase to as much as 6000 deg$^2$ depending on footprint decisions \citep{2022ApJS..259...58L}.
We use the \gls{DESI} \gls{LRG} sample, which contains \textasciitilde 8 million galaxies from 0.3 < z < 1,  as the lenses.
We model galaxy bias, baryonic feedback, and intrinsic alignments, and the high precision of \gls{DESI} allows us to break degeneracies to better handle these systematics.
We constrain $\Omega_m$ to $X \pm Y$ and $S_8$ to $XX \pm YY$.
This lensing analysis from combining \gls{LSST} and \gls{DESI} is an important cross-check of the \gls{LSST} cosmic shear results; we find that [they give consistent constraints].

\paragraph{Science Objectives}
To perform this analysis, will require Rubin’s \gls{photo-z} catalog and shape catalog for the \gls{LSST} sample.
We will also need the \gls{DESI} spectroscopic galaxy catalog, of which there are already releases.
The catalogs required for this joint analysis are also necessary individually for each survey’s analysis, so this analysis would leverage the existing work already performed to improve constraints at little additional cost.

Our analysis will be limited by the size of the survey overlap region; the choice of extending the \gls{LSST} footprint north further into the DESI footprint would allow for significant increased analysis power.
To understand these limitations and make forecasts for the improvement possible with this joint analysis, we could perform a precursor analysis with other data sets of photometric sources and spectroscopic lenses such as \gls{KiDS} and \gls{GAMA}, or \gls{HSC} and \gls{SDSS}-III/IV.

\paragraph{Challenges (what makes it hard)}

This joint analysis will require address a number of challenges, including:
\begin{itemize}
\item Understanding of clustering sample selection functions, but this should be handled by \gls{DESI}. Would result in catalogs of random points illustrating the coverage as a \gls{DESI}-provided data product.
\item We will rerun the analysis for each Rubin data release; the challenge is to make the analysis easily re-runnable.
\item The photo-z quality for Y1-Y2 \gls{LSST} data will be lower than Y10, and the early LSST overlap with \gls{DESI} may be less uniform in image quality (and the high airmass may be prioritized by the scheduler). Essentially, the depth maps may be more complex and have more structure early in the survey at the edge of the footprint.
\item Most/all LSST source photo-z tomographic bins assume calibration with a \gls{DESI}-like (probably \gls{DESI}) sample; may need to worry about covariance when also using \gls{DESI} for cosmology measurements.
\end{itemize}

\paragraph{Running on LSST Datasets (for the first 2 years)}
We will require the following data sets and pay attention to relevant considerations as follows:
\begin{itemize}
\item Shape catalog: Metacal should be available from the DM \gls{pipeline}, Metadetect may be available (otherwise would be a value added catalog from \gls{DESC}).
\item As the \gls{LSST}/\gls{DESI} overlap area is at higher airmass than the average \gls{LSST} airmass, we will have to recalibrate the images in this region. As this region is \textasciitilde 1/5 of the entire \gls{LSST} survey area, this will result in \textasciitilde $\sqrt{5}$\textasciitilde 2.2 times larger error bars. (Note that we will have to put in additional effort for the recalibration, which requires rechecking and potentially correcting for  the multiplicative and additive biases of the weak lensing shears, but we will not have to fully re-do the shape measurements.). As g-g lensing is more forgivable in terms on calibration parameters, this may be tolerable. So a subset of the image simulations may be sufficient.
\item Estimate of number of galaxies that can serve as source galaxies for DESI LRGs  in early LSST data: 70 million. We obtain this from considering that in early \gls{LSST} data, an estimated source galaxy number density is 10~arcmin$^{-2}$, of which approximately half are above the LRG redshift \citep{2018arXiv180901669T}.  Given the expected 4000 deg$^2$ overlap region, this gives approximately 70 million sources.
\item Photoz for the \gls{shape} catalog: Tomographic sample selection based on Rubin-provided photo-z, calibrated by cluster-z with \gls{DESI}.
\end{itemize}

\paragraph{Precursor data sets}
\begin{itemize}
\item For the spectroscopic galaxy sample, we will be able to use early \gls{DESI} data as a precursor. We could also use \gls{GAMA}, a spectroscopic survey, or \gls{SDSS}-III/IV.
\item For the galaxy \gls{shape} data set, we could use the \gls{UNIONS} survey, \gls{HSC}, or \gls{KiDS}.
\end{itemize}

\paragraph{Analysis Workflow}
We lay out the following workflow as a possible route to completing this analysis:
\begin{itemize}
\item Perform an optimized tomographic binning of \gls{shape} sample, likely using Rubin-provided photo-z estimates.
\item Use \gls{DESI} \gls{ELG}s \& quasars for \gls{photo-z} calibration.
\item Obtain the \gls{DESI} \gls{LRG} catalog for the lens sample.
\item Using the window functions of both surveys, create samples in the overlap region of the two surveys.
\item Compute correlation functions (galaxy-galaxy lensing, as well as the auto-correlation) on these samples.
\item Implement null tests. This might include, for example, checking that high-redshift lenses with low-redshift sources produce no signal.
\item Estimate a covariance matrix using theory predictions or empirical estimators (jackknife etc.), accounting for correlations between the different quantities being measured.
\item Carry out the cosmological parameter likelihood analysis given the data vectors, covariances, and models for systematic errors.
\end{itemize}

\paragraph{Software Capabilities Needed}
\begin{itemize}
\item We will require software to compute the correlation functions, such as TreeCorr\footnote{\url{https://github.com/rmjarvis/TreeCorr}} or a similar correlation function package. While we are working with large data quantities, we expect current techniques and computing power will be sufficient.
\item We will need to store these correlation functions as intermediate data products, but this will not require a large amount of space.
\end{itemize}

\paragraph{References for Further Reading}
\begin{itemize}
\item \citet{https://doi.org/10.48550/arxiv.2105.13545}: Dark Energy Survey Year 3 Results: Constraints on cosmological parameters and galaxy bias models from galaxy clustering and galaxy-galaxy lensing using the redMaGiC sample
\item \citet{Zhou_2020}: Preliminary target selection of \gls{DESI} \gls{LRG} sample
\item \citet{lsstdesi}: Maximizing the Joint Science Return of \gls{LSST} and \gls{DESI}
\end{itemize}

\pagebreak
\subsubsection{Weak lensing cosmology analysis / 3x2pt} \label{sec:WLCosmo}
%\WOM{No workflow}
\Contributors{Andresa Campos (\mail{acampos@cmu.edu}), Francois Lanusse (\mail{francois.lanussse@cea.fr})}
{March 29, 2022}

\paragraph{Abstract}
We present the first cosmological analysis of the weak lensing shear-shear correlation functions in combination with galaxy clustering and galaxy-galaxy lensing, which trace the evolution of large-scale structure from intermediate to low redshift.  With the first year of \gls{LSST} data, we can already constrain the amplitude of matter fluctuations to greater precision than precursor surveys such as KiDS, \gls{DES}, and HSC, reaching \textasciitilde 1-\% precision when considering statistical and systematic uncertainty.  This improved result is due to the increased area of LSST, and the fact that the first year of data is comparable in depth to \gls{DES} and deeper than KiDS.  This work will rely on analysis of the catalogs produced by the LSST Science Pipelines down to detections with \textasciitilde 10-sigma significance.  This result is sufficiently precise to permit a robust test of LCDM by comparing with the amplitude from the CMB, which current analyses suggest may be higher than that from weak lensing (implying some uncontrolled systematic or a failure of the LCDM model).  Together with CMB data, we can go beyond the LCDM model and provide competitive constraints on the equation of state of dark energy.

\paragraph{Science Objectives}

 Milestones to this result include defining samples that pass basic null tests, validating that the shear estimates meet the needs for the science, and estimating N(z) for the samples selected based on \gls{photo-z}.  For the last step, we do not necessarily need very precise \gls{photo-z} (though that would improve the localization of the selected bins and improve the constraint on w) but we do need a very precise understanding of the N(z) for the ensemble.  At this stage, we anticipate that \gls{photo-z} calibration and the impact of blending on shear and \gls{photo-z} are likely to be the key limited systematics.

\paragraph{Challenges (what makes it hard)}
 Systematics! This science case directly inherits from the cosmic-shear systematics, we refer to \secref{sec:CosmicShear} for details, and here we document the additional galaxy-clustering challenges.
\begin{itemize}

\item  Galaxy bias modeling:
 accurate modeling for galaxy bias used for galaxy-galaxy and galaxy clustering modeling will be needed. Ongoing \gls{DESC} bias challenge is investigating various nonlinear bias models to reach Y1 and Y10 requirements.

\item Baryonic Effects: feedback from stars and supermassive black holes (AGN) redistributes gas. The effects of galaxy formation are degenerate with cosmology/
neutrino effects, therefore ignoring baryons at small scales leads to biases (see \citet{Schneider_2020}). It is necessary to model the baryonic effect and marginalize of it.

\item Deprojection of systematics maps:
         to clean potential contamination from varying survey and observing conditions variability, a step of mode deprojection from a set of systematics maps can be applied (see e.g. \citet{10.1093/mnras/stz093}).
\item Blinding: a blinding scheme that encompasses multiple probes is a challenge on its on, as discussed in \citet{10.1093/mnras/staa965}. Added to that, there is the additional question on how effective the blinding of \gls{DESC} data products will be, given that Rubin products will be available.

\end{itemize}

\paragraph{Existing Tools}
\begin{itemize}
\item What tools exist today to undertake these analyses
\begin{itemize}
\item \gls{RAIL}+qp for getting p(z) for each object
\item Metadetection \gls{algorithm} (descwl\_coadd, mdet-shear-sims… )
\item Code from Pedersen et al. on self-calibrating \gls{IA}
\item Synthetic \gls{Source} Injection (DM/DESC)
\item Systematics maps building (e.g. Supreme)
\item Theoretical modelling (e.g., \gls{CCL}, HMCode/Halofit), correlation functions (e.g., TreeCorr, NaMaster, TXPipe), Covariances (TJPCosmo),  Inference (e.g., CosmoSIS)
\end{itemize}

\item What functionality is missing from these tools and frameworks that would need to be developed, or is there some issue with their application to the dataset at \gls{LSST} scale?
\begin{itemize}
\item Toolset is fairly complete for Y1
\end{itemize}
\item If new tools were built what components from existing frameworks would be critical to keep? (e.g. what parts of existing tools work well)
\begin{itemize}
\item \gls{SACC} file format for interchangeable data files.
\end{itemize}
\end{itemize}

\paragraph{Running on \gls{LSST} Datasets (for the first 2 years)}
 We will use the following data products:
\begin{itemize}

\item Data release catalogs for the \gls{WFD} survey - we’ll use essentially all galaxies down to some relatively low significance (\textasciitilde 10-sigma).
\item Some \gls{calibration} and systematics tests will use the DDFs.
\item External spectroscopic catalogs to validate/calibrate photometric redshifts.
\item We will need some image simulations (non-DC2) that mock-up the cell-based coaddition to validate the shears, and cross-matching and cross-correlation against spectroscopic surveys to validate the \gls{photo-z} and ensemble N(z) estimates.
\item Euclid+Rubin DDPs for deblending (Schuhmann et al., 2019 predict mean shrinkage in ellipticity errorbars by more than a factor of 2 with \gls{LSST} is combined with Euclid)
\end{itemize}

\paragraph{Precursor data sets}
\begin{itemize}
\item \gls{DES}, \gls{HSC}, and KiDS all have public data releases that could be used for precursor analyses.  In some cases, the catalogs are most accessible while reanalyzing the images would be a substantial challenge despite data being made public.
\item Reference spectroscopic samples (deep ones for direct calibration and wide ones from e.g. \gls{DESI} for cross-correlation) will be important for the N(z) validation.

\end{itemize}

\paragraph{Analysis Workflow}

This science case directly inherits
from the cosmic-shear,  galaxy clustering and
galaxy-galaxy lensing workflow. We refer to \secref{sec:CosmicShear} and \secref{sec:Constrain} for details on each probe. Here we provide the general steps and document the additional items related to the combined 3x2pt analyses: 

\begin{itemize}
    \item Apply selection criteria to the source and lens catalogs.
    \item Apply the optimal redshift estimation method to each catalog, perform tomography and estimate the N(z) distribution.
    \item Measure each of the three two-point correlations: cosmic shear, galaxy clustering and galaxy-galaxy lensing.
    \item Compute the joint covariance matrix of the three probes.
    \item Model and mitigate systematics.
    \item Perform likelihood analyses and cosmological parameter estimation.
    \item Test for internal consistency of probes before unblinding.

\end{itemize}

\paragraph{Software Capabilities Needed}
\begin{itemize}
\item Survey property maps (e.g., skyproj w/ healsparse, Supreme)
\item Calexps and coadds
\item N(z)s estimator
\item Sampler
\end{itemize}

\paragraph{ References}

Precursor survey analyses:
\begin{itemize}
    \item \gls{DES}-Y3: \citet{PhysRevD.105.023520}
    \item \gls{KiDS}-1000: \citet{2021A&A...646A.140H}
\end{itemize}

\section{Technical areas in detail} \label{sec:techdetails}

\subsection{Introduction} \label{sec:tdintro}
After the science breakouts we identified six main cross-cutting technical areas, and developed use cases within them, with a goal of understanding the needs to support a broad range of Rubin science (\autoref{tab:scitech}). Each of these areas is given a subsection here.

% Generated file 
In this section:\
\begin{longtable}{l l p{0.6\textwidth} r}\hline
\hyperref[sec:tCrossMatch]{\ref{sec:tCrossMatch} }  &  & \hyperref[sec:tCrossMatch]{ Cross Matching}   & \hyperref[sec:tCrossMatch]{ \pageref{sec:tCrossMatch} }  \\
\hyperref[sec:tSelect]{\ref{sec:tSelect} }  &  & \hyperref[sec:tSelect]{ Selection Functions}   & \hyperref[sec:tSelect]{ \pageref{sec:tSelect} }  \\
\hyperref[sec:TimeSeries]{\ref{sec:TimeSeries} }  &  & \hyperref[sec:TimeSeries]{ Time Series}   & \hyperref[sec:TimeSeries]{ \pageref{sec:TimeSeries} }  \\
\hyperref[sec:ImageReproc]{\ref{sec:ImageReproc} }  &  & \hyperref[sec:ImageReproc]{ Image Reprocessing}   & \hyperref[sec:ImageReproc]{ \pageref{sec:ImageReproc} }  \\
\hyperref[sec:ImageAnalysis]{\ref{sec:ImageAnalysis} }  &  & \hyperref[sec:ImageAnalysis]{ Image Analysis}   & \hyperref[sec:ImageAnalysis]{ \pageref{sec:ImageAnalysis} }  \\
\hyperref[sec:Photoz]{\ref{sec:Photoz} }  &  & \hyperref[sec:Photoz]{ Photometric Redshifts }   & \hyperref[sec:Photoz]{ \pageref{sec:Photoz} }  \\
\hyperref[sec:othertech]{\ref{sec:othertech} }  &  & \hyperref[sec:othertech]{ Other technical use cases}   & \hyperref[sec:othertech]{ \pageref{sec:othertech} }  \\
 & \hyperref[sec:JointCal]{\ref{sec:JointCal} }  & \hyperref[sec:JointCal]{ Joint Calibration of precursor surveys for longer-baseline Light Curve Generation}   & \hyperref[sec:JointCal]{ \pageref{sec:JointCal} }  \\
 & \hyperref[sec:MultiStreamTransients]{\ref{sec:MultiStreamTransients} }  & \hyperref[sec:MultiStreamTransients]{ Multi-Stream Transient Detection and Characterisation}   & \hyperref[sec:MultiStreamTransients]{ \pageref{sec:MultiStreamTransients} }  \\
 & \hyperref[sec:VizScale]{\ref{sec:VizScale} }  & \hyperref[sec:VizScale]{ Interactive Data Visualization at scale}   & \hyperref[sec:VizScale]{ \pageref{sec:VizScale} }  \\
\hline \end{longtable}

\subsection{Cross Matching} \label{sec:tCrossMatch}
%\WOM{check secrefs science cases, rough on workflow and citations}
\Contributors{Viviana Acquaviva, Igor Andreoni, Leanne Guy, Saavik Ford, Nico Garavito-Camargo, Mario Juric (editor), Ilhuiyolitzin Villicana-Pedraza, Jeremy Kubica, Samuel Wyatt, Weixiang Yu, Alex Riley}
{2022-03-30}
\cleanedup{Mario Juric}

\subsubsection{Abstract}
A significant fraction of science use cases presented at the Workshop require the ability to (generally positionally) cross-correlate the detections in the LSST catalog with one or more other catalogs – an operation commonly known as “cross-matching”. This capability would enable enrichment of LSST data with information taken in other wavelengths, at other times, different resolutions, or of generally different characteristics. This capability is needed in two regimes: a) real-time – low-latency matching of O(10k) sources to O(10) catalogs each holding O(1Bn) objects (to support adding information to alert streams from other catalogs), and b) offline processing – the ability to match O(10Bn) x O(1Bn) object catalogs, followed by joining data from both catalogs (e.g., full time series of observations, multi-wavelength studies) for analysis at scale. This capability should be easy to use for the end-user. For example, it may be provided at the community broker level for real-time cases, or accessible as simple Python calls or SQL-like statements callable from Python notebooks for the offline-level.

We note that cross-matching is just a first step in the process, nearly always followed by bringing in additional data from catalogs being cross-matched. It may be better to think of it as {\em (distributed) joining of (large) tables on a spatial (user-defined) index, followed by (potentially heavy) computation on the result of the join}. The cross-match capability may therefore need to be a part of a larger scalable analytics system.

\subsubsection{Science Cases Needing this Tool }

{\bf Solar system:}
\begin{itemize}
\item Multifrequency study of the Solar System moons and hazardous asteroids (\secref{sec:Multifrequency})
\end{itemize}

{\bf Local universe static}
\begin{itemize}
\item The properties of the faint end of the Main Sequence: the stellar/sub-stellar boundary. (\secref{sec:FaintEnd})
\item Mapping the Accreted and Intrinsic Stellar Populations in the Milky Way (\secref{sec:MappingMWStellarPops})
\item Local Group Dwarf Galaxies, bound and unbound (\secref{sec:LGdwarf})
\item The local \gls{IMF} as inferred from nearby star forming regions and clusters (\secref{sec:LocalIMF})
\end{itemize}

{\bf Local universe variable \& transient:}
\begin{itemize}
\item All science-cases involving enriching discovered transients or variability with non-LSST data.
\end{itemize}

{\bf Extragalactic static:}
\begin{itemize}
\item Estimation of galaxy physical parameters with SED fitting (\secref{sec:GalPPSED})
\end{itemize}

{\bf Extragalactic variable:}
\begin{itemize}
\item Find All the \gls{AGN} ASAP (\secref{sec:FindAllAGN})
\item Connection between short term variability of \gls{AGN} and their long term behavior (\secref{sec:ConnectSLAGN})
\item Augmenting \gls{AGN} variability (\secref{sec:AugmentAGN})
\end{itemize}

{\bf  Extragalactic Transient:}
\begin{itemize}
\item Real time transient host association
\item Immediate classification of astrophysical transients (\secref{sec:ImmediateClassification})
\item Non-localized alert alert crossmatching
\end{itemize}

\subsubsection{Requirements for the software}

 We identify two rough clusters of use-cases, with differing scalability and performance requirements:
 \begin{enumerate}
\item {\bf Real-time cross-matching:} The ability to return multiple (say, N<=10) neighbors within \textasciitilde 5”, with latency at alert scale (seconds), and the numbers of objects being cross-matched similar to the numbers of alerts. The system well need to cross-match to multiple (arbitrary) catalogs with a choice of distance metrics (e.g., on-sky, real-space). Brokers may enrich alerts w. this information, therefore needing close collaboration with broker teams wishing to operate this functionality.

\item {\bf Large-scale static-sky cross-matching:} The ability to do full catalog cross matches for statistical analysis. Matching to a filtered subset will be an important feature -- for example, a query such as ``for these 10M objects, find 3 nearest $g-r < 0.5$ neighbors''. This system would need to:
\begin{itemize}
\item Scale to O(10Bn) x O(1Bn) catalog cross-matches and repeat on \textasciitilde monthly scales.
\item Allow matching as close to interactive as possible (e.g., with smaller catalogs) to enable exploratory science.
\item Have the ability to do faster cross-matching on smaller subsamples of these objects, for testing purposes.
\end{itemize}
\end{enumerate}

Aspects that are common to both use cases include:
\begin{itemize}
\item Cross-matching is just a first step in the process, nearly always followed by bringing in additional data from catalogs being cross-matched. It may be better to think of this as (distributed) joining of (large) tables on a spatial (user-defined) index – a joint table of 100M objects is not very useful if the data on those objects (e.g., time series) can’t be easily pulled/fused/processed together.
\item Desirability of probabilistic cross-matching, which takes into account finite extent or a large uncertainty ellipse for  some objects. This may be emulated by the user if a basic cross-matching capability where N nearest neighbors are returned is provided For example, the user can take N nearest cross matches, compute some probability function that each of them is the “true” match, and pick the most probable one (or keep the full distribution).
\item N-way cross-matching (cross matching to multiple catalogs)
\item Reporting non-detections is important. E.g., for Solar System, transients/variability, etc. Related to (reliable) selection functions.
\end{itemize}

Challenges:
\begin{itemize}
\item The low latency and partial scalability for the real-time (alerts) use case.
\item The scalability for the online case, where large (likely distributed) catalogs are being joined. For example cross-matching the LSST (located at SLAC) and the DES catalog (located at NCSA), joining the time-series, and passing them on to a Python function to compute some metric of interest on the joined light curve.
\item The join predicate which may be more complex than simple spatial join (e.g., ``find N closest matches in a user-defined subset of a catalog'').
\end{itemize}

\subsubsection{Running on LSST and other Datasets}

As discussed above, this component is expected to be needed both for real-time and batch processing use cases.

% \begin{itemize}
% \item Alert streams
% \item Static catalogs
% \item Arbitrary external catalogs
% \item HEALPix maps
% \end{itemize}

\subsubsection{Existing Tools}
\begin{itemize}
\item Codes such as the Apache Spark based AXS, for scalable distributed joining/cross-matching of extremely large datasets.
\item Codes such as catsHTM and similar for fast in-memory crossmatching
\item Online cross-match services like CDS XMatch and many others
\item HEALPix Alchemy for non-localized events such as GW/Neutrino Alerts. Has the capability to crossmatch catalogs, observational footprints, and all-sky images within a healpix map.
\item Cross-match services being developed as parts of community brokers
\end{itemize}

Needed enhancements:
\begin{itemize}
\item Scalability – many of these tools assume O(<1k-1M) object operation
\item Need arbitrary catalog cross-matching, w/o pre-computed join tables and with user-defined predicates
\item N-way cross-match – the ability to join many catalogs
\item Distributed joins – many of these tools focus on just computing cross-match, but science use cases need to bring together and work on the data as the next steps (e.g., light curves, spectra, sometimes even images).
\item User-friendly distributed operation – the user should be able to use these tools with similar ease to using a RDBMS (Relational DataBase Management System) today. For example, the user should be able to write a declarative statement about the result they're trying to achienve, and have the execution – query optimization, scaling, potential work distribution, fault tolerance – be handled transparently.
\end{itemize}

\subsubsection{Computational Workflow}

Computational workflows are somewhat science dependent, and many have been discussed in specific science use cases. Here we just give two representative examples:\\

\noindent Example from the stellar/sub-stellar boundary science case:
\begin{itemize}
\item Open a notebook and query some sources by making cuts (in color) on the object catalog data to identify candidate low-mass stars.
\item Identify candidates with early proper motion and parallax measurements.
\item Get sources data (light curves) and/or take source measurements for each selected object.
\item Run routines to identify variability.
\item Cross-match with calibration subsets (\gls{Gaia}, spectroscopic surveys, etc.)
\end{itemize}

An example from the SED fitting science case:
\begin{itemize}
\item Open a notebook and query some sources by making cuts for example in redshift or luminosity.
\item Run some object classification algorithm to identify which sources are galaxies.
\item Cross match those galaxies with sources from other multi-wavelength catalogs to obtain a “wider” \gls{SED}.
\item Do science with the cross matched data (validation of LSST-results, \gls{SED} fitting on multi-wavelength \gls{SED}, comparison with simulations, and others).
\end{itemize}

\subsubsection{References for Further Reading}
\begin{itemize}
\item AXS: Astronomy eXtensions for Spark -- Apache Spark based distributed cross-matching system \citep{zecevic2019}
\item HEALPix Alchemy \citep{2022AJ....163..209S}
\end{itemize}

\pagebreak
\subsection{Selection Functions} \label{sec:tSelect}
\Contributors{Yusra AlSayyad, Katelyn Breivik, Giulio Fabbian, Matt Holman, Adrian Price-Whelan, Kate Storey-Fisher}
{2022-03-29}

\subsubsection{Abstract}
Selection functions are core components of any modeling procedure that aims to quantify the population statistics or density distribution of sources or objects. A selection function for a given modeling method may contain things like the detection efficiency of sources with a given brightness or shape, the classification accuracy of sources, the cadence of observations, the Milky Way and intergalactic dust distribution, or the crowdedness (in source counts) of a field. While the LSST Data Management (DM) group plans to provide the core data products, the detection efficiency of a theoretical point source per position and epoch and each coadd, each specific science case, classification, or detection algorithm will need a specialized selection function which depends on the science question or model being studied. What we are therefore currently missing in the community are worked examples of how to construct and use selection functions of varied complexity for different use cases.

\subsubsection{Science Cases Needing this Tool}
Any science cases that want to learn population statistics or source density distributions.

\subsubsection{Requirements for the software}
We recommend producing 3–5 science demonstrations that utilize the DM data products and metadata to construct selection functions that are used in illustrative science examples. As possible example use cases, we recommend:
\begin{itemize}
\item A selection function combining the depth of coadd images over the sky with a dustmap, to evaluate a model for the spherically-averaged distribution of stars in the Milky Way’s stellar halo.
\item Tools and worked examples on how to convert from the pixelized/rasterized representation of detection maps that DM will provide to vector/polygon representations for users who need them.
\item Worked examples for generating simulations either at the catalog-level or realized images for injection into the Rubin images via the Pipelines’ synthetic source injection framework.
\item A selection function that incorporates cadence or time of observation information as a function of sky position to assess the detectability of RR Lyrae stars to measure the period and magnitude distributions of RR Lyrae stars. Similar for variability-selected \gls{AGN} populations.
\item A selection function for detecting asteroids of a given shape in single images to compare the numbers of asteroids with different shape characteristics.
\end{itemize}

\subsubsection{Running on LSST and other Datasets}
This requires LSST (coadd and single-epoch) image masks, property maps and both coadd and time-series catalog information. Specific tutorials will pull in case-specific info such as dustmaps and variability models.

\subsubsection{Existing Tools}
The core data products are planned to be produced by the DM team, but what is missing are demonstrations of how to construct selection functions of different types/forms.

\subsubsection{Computational Workflow}
n/a

\subsubsection{References for Further Reading}
\cite{2022MNRAS.510.4626B}, \cite{2021AJ....162..142R}

\pagebreak

\subsection{Time Series} \label{sec:TimeSeries}
There were multiple technical cases for time series.

\subsubsection{Parametric Fitting} \label{sec:TSparametricFit}

\Contributors{Catarina S. Alves (\url{mailto:catarina.alves.18@ucl.ac.uk}), Matthew Graham, Andrew Bradshaw, Andrew Connolly, Garrett Levine, David Trilling, Fabio Ragosta, Tomas Ahumada, Jing Lu, Alex Gagilano, Neven Caplar, Illija Medan}
{ edited May 4, 2022 (begun March 30, 2022)}

%\cleanedup{Catarina Alves}

\paragraph{Abstract}
Many analyses involve fitting a prespecified model to data where the model parameters have semantic content, for example, they represent physical quantities. Given the numerous and diverse objects that Rubin LSST will observe with unprecedented precision and time coverage, including solar system objects, stars of all stripes, transients, and variable galaxy images, the models which are fit to the observations must similarly be flexible while also including physical information about each object. Commonly used tools in parametric time series analysis should be automatically computed (or trained) on all objects on a regular basis, along with specific subsets being analyzed with targeted tools as needed; all provided through a unified interface with a common data structure. Additionally, model selection criteria such as AIC or Bayes factors shall also be pre-computed to enable comparison between models, along with uncertainties and ranges of validity for model parameters. Providing the data alongside informative statistics in a networked and unified interface should help maximize the potential of LSST.

\paragraph{Science Cases Needing this Tool}
 Parametric fitting to time series is common to many areas of astronomy, including:
\begin{itemize}
\item cosmology (in particular, SN Ia cosmology to produce distance estimates)
\item \gls{AGN} to model stochastic variability
\item stars to model particular periodic structure in light curves, e.g., RR Lyrae
\item exoplanets to model background stellar activity
\end{itemize}

\paragraph{Requirements for the software}
 The primary requirements for the software are that it should:
\begin{itemize}
\item scale efficiently to arbitrary data sizes
\item be flexible across different model structures and fitting constraints, e.g., optimizer choice, loss functions
\item be able to provide uncertainties (posterior distributions) on both model parameters and predicted values
\item run as fast as possible
\end{itemize}
 In most cases, the input will be time series, either individual or batches, and these will be irregularly and sparsely sampled, gappy, and heteroskedastic. The time series may also be univariate (single filter) or multivariate (ugrizy). It may be that this is best supported through a common time series data model.

 A lot of parameter fitting codes exist for specific science cases (see below) so rather than implementing new versions of these, the software requirements is to run these in an appropriate environment. Modification of existing codes might be needed to implement the uncertainty requirement and to improve the speed and scalability of the code.

\paragraph{Running on LSST and other Datasets}
 This technical case will use time-series, whose needed duration depends on the specific use case. The datasets could come both from calibrated images and data release catalogs. The alert stream data may present challenegs for parametric fits because the alerts use different templates across the years, which could result in artificial discontinuities.

 SN Ia cosmology is a science case associated with parametric fits. There are numerous precursor datasets from previous surveys. More recently, we have data from the Zwicky Transient Facility (ZTF) and LSST-like simulated datasets such as the Photometric LSST Astronomical Time-series Classification Challenge \citep[PLAsTiCC;][]{2018_plasticc,Kessler2019} and its update, ELAsTiCC. The analysis of these precursor datasets will lead to publications, in particular when applied to real data. A comparison of template fitting algorithms on the same datasets, including the advantages and limitations of each methodology, would also lead to a new publication.

\paragraph{Existing Tools}
\begin{itemize}
\item Tools (SN Ia light curve fitting + SED templates + AGN): SALT2 \citep{guy2007salt2, 2021MNRAS.504.4111T}, SALT3 \citep{kenworthy2021salt3}, SUGAR \citep{2020A&A...636A..46L}, ParSNIP \citep{Boone_2021}, SNooPy \citep{burns2010carnegie}, BayeSED \citep{han2012decoding, han2014bayesed, han2018comprehensive}; EzTao \citep{Yu2022}
\item There is no easy way to compare between the above tools. Moreover, the SALT light curve fitters are too slow for LSST use.
\item It is crucial to work with the community to understand how to implement the unified interface with a common data structure to maintain the ease to use of the existing tools.
\end{itemize}

\paragraph{Computational Workflow}
 Pick your optimizer, pick your loss function. Compress data from photometry as a function of time into phase and amplitude.

\paragraph{References for Further Reading}
 Links to tools:\\
\begin{itemize}
\item \textit{SALT2} \citep{guy2007salt2, 2021MNRAS.504.4111T}\\
\item \textit{SALT3} \url{https://saltshaker.readthedocs.io/} \citep{kenworthy2021salt3}\\
\item \textit{SUGAR} \citep{2020A&A...636A..46L}\\
\item \textit{ParSNIP} \citep{Boone_2021}\\
\item \textit{SNooPy} \url{https://csp.obs.carnegiescience.edu/data/snpy} \citep{burns2010carnegie}\\
\item \textit{BayeSED} \url{https://bayesed.readthedocs.io/en/v2.0/index.html#} \citep{han2012decoding, han2014bayesed, han2018comprehensive}\\
\item \textit{EzTao} \url{https://github.com/ywx649999311/EzTao} \citep{Yu2022}
\end{itemize}

\pagebreak
\subsubsection{Tools to Facilitate Anomaly Detection and Characterization in LSST Time-Series Data} \label{sec:AnomalyDetect}

\cleanedup{Alex Gagliano}

\Contributors{Alex Gagliano (\mail{gaglian2@illinois.edu}), W.\ Garrett Levine, Neven Caplar,
Ashish Mahabal, Catarina S.~Alves, Matthew Graham, Andrew Bradshaw, Andrew Connolly, David Trilling, Fabio Ragosta, Tomas Ahumada, Jing Lu, Ilija Medan}
{edited 05/03/22 (began 3/30/2022)}

\paragraph{Abstract}
Rubin will generate time series photometry for tens of billions of sources in multiple filters to unprecedented depth over its ten-year survey. These data will contain a dizzying breadth of persistent variable and non-variable phenomena, and some of these will be observed for the first time. %, and a tiny fraction of alerts (still numbering hundreds of millions).
Here we primarily concentrate on the data released through LSST alerts and the anomalies lurking in the alert stream. To optimize the use of the Rubin Observatory as a discovery machine for rare and high-priority events, infrastructure must be developed to efficiently identify anomalies among massive datasets in a timely manner. This will require synergy between state-of-the-art machine learning tools for anomaly detection and visualization techniques for interactive and low-latency high-level analysis. These proposed tools should be sufficiently scalable and fast enough to enable prioritization and follow-up of rapidly-evolving events before they dim (such as, early SN interaction, cometary outbursts or breakup, rapidly changing \gls{AGN}, microlensing events/\gls{TDEs}/kilonovae/other unknown phenomena and extreme cases of known types).

\paragraph{Science Cases Needing this Tool}
Detecting anomalies in the alert stream will be essential for identifying:
\begin{itemize}
\item Comets experiencing volatile outbursts or breakup
\item Intrinsically-rare known classes of transients (kilonovae/\gls{TDEs}/\gls{FBOTs})
\item Mapping \gls{SMBH} Near Fields with Microlensing
\item Connection between short term variability of \gls{AGN} and their long term behavior
\item Extreme cases of known types of variables
\end{itemize}
In addition to elucidating the physics powering known signals, anomalies will also represent entirely new classes of phenomena probing poorly-explored regimes in event brightness and timescale. The science enabled by this infrastructure cannot be fully known in advance but will have significant impact, ensuring the legacy of LSST as a pioneer in transient discovery for decades to come.

\paragraph{Requirements for the \gls{software}}
Anomalies in the event stream will manifest themselves in difference images and require low latency access. Because of the value of rapid follow-up, anomaly detection algorithms must keep pace with the alert stream and should be able to operate on as few bits of information as possible. Provided that anomaly detection is sufficiently fast and high-level processing is done by Rubin, the continuous alert stream could include a flag specifically for these events. Deriving meaningful information from anomalies will require flexible frameworks that can be re-trained and improved with rapid feedback processes, as many of the events we aim to discover will be “unknown unknowns” before Rubin comes online. Software that can accurately characterize astronomical events in real-time remain woefully absent from the literature, and this presents a barrier to survey readiness.

The survey's sensitivity to unusual variable phenomena is limited by the template generation and subsequent image differencing conducted by the survey. Where possible, the template images and strategy for generating them should be reported, and the raw images should be provided through the data releases in case users wish to construct custom templates for their science case. The Rubin \gls{Science Platform} could also allow for users to construct custom difference imaging pipelines close to the data to avoid duplication of data.

While high-redshift events will be contained to small postage stamps, active Solar System objects will require on-demand custom image cutouts (in some cases, as large as 10’x10’) for full characterization. Even for stationary events, contextual information such as that provided by the host galaxy of an explosive transient is valuable for further characterizing a source at early epochs. Events that are not identified through the alert stream or for which larger cutouts are needed could be retrieved through the data releases, although this 24-hour latency would present an additional obstacle to rapid discovery. To manage bandwidth, a queue system or resource allocation framework among the scientific stakeholders could be implemented.

\paragraph{Running on LSST and other Datasets}
Anomaly detection will leverage the alert stream and its associated postage stamps. Algorithms to find anomalies are expected to function more effectively with long baselines, but it would be ideal to identify anomalies in short light curves. This use case would be especially relevant for newly discovered Solar System small bodies approaching perihelia. Data releases for anomalous but persistent variable sources would be valuable for archival studies of these populations. Finally, rapid catalog cross-matching will be useful to determine the multi-wavelength properties of LSST-discovered sources and further investigate their underlying physics.

It is essential to identify the relevant \gls{software} systems and train them with simulated samples in advance of LSST first light. On the hunt for extra-galactic transients spanning a broad range in progenitor physics, the data and models produced by the Photometric LSST Astronomical Time-Series Classification Challenge (PLAsTiCC) challenge have been state-of-the-art. These models have been improved (and the data made more realistic) in the Extended LSST Astronomical Time-Series Classification Challenge (ELAsTiCC), but data are still limited to the classes for which realistic rest-frame \gls{SED} models are available.

\paragraph{Existing Tools}
Central to the question of anomalous behavior is the ``distance" between an event in question and a larger population. An event with a large distance from the population is considered anomalous. These distance measures could be calculated in data space (comparing light curves between events), parametric feature space (characterizing e.g., the period and amplitude of each light curve and comparing these), or non-parametric feature space (reduced-dimensionality embeddings of the original data). To generate feature representations of ingested data, methods often require close integration with feature extraction (period-finding) and dimensionality reduction tools (tSNE, UMAP, \gls{PCA}). These tools can also be used to provide high-level summaries of large datasets, with which the astronomer may be able to identify anomalies by eye.

Methods differ in their focus on characterizing anomalies directly, or indirectly by constraining the properties of the “normal” larger population. The latter is more common, although novel methods are in active development. Isolation forests are an example of direct anomaly identification, and they are used by SNAD \citep{2021Malanchev} to identify unusual supernovae and \citet{2021Sanchez_AGN} to find unusual \gls{AGN} (in the \gls{AGN} case, a variational autoencoder is first used to generate a set of non-parametric features). A wealth of machine learning tools, however, focus on the bulk properties of the population (which can inform objects at the boundaries of classification), and the vast majority of these rely on decision trees (including boosted decision trees and random forests) for classifying common events. An event with low classification probability reported by these methods may be an anomaly.

A non-exhaustive list of open-source codes for photometric classification of supernovae include:
\begin{itemize}
\item \textit{Avocado:} A gradient-boosted decision tree classifier for anomaly detection \citep{Boone_2019}.
\item \textit{ParSNIP:} A deep neural network that calculates latent parameters of light curves with an encoder \citep{Boone_2021}.
\item \textit{RAPID:} A recurrent neural network for real-time classification \citep{2019Rapid}.
\item \textit{SuperRAENN:} A recurrent autoencoder coupled to a random forest algorithm for classification \citep{2020Villar}.
\item \textit{snmachine:} A framework for light curve feature extraction and classification using multiple techniques \citep{2016Lochner, 2022Alves}.
\item \textit{superNNova:} A Bayesian recurrent neural network for real-time classification \citep{M_ller_2019}.
\end{itemize}
Density-based methods are also in development (e.g., DBscan\footnote{\url{https://scikit-learn.org/stable/modules/generated/sklearn.cluster.DBSCAN.html}}; Local outlier factor\footnote{\url{https://scikit-learn.org/stable/auto_examples/neighbors/plot_lof_outlier_detection.html}}).

Because anomalies are defined relative to a population, there must also exist an interface near the data (ideally embedded within the Rubin \gls{Science Platform}) for rapidly interacting with a large sample of alerts. A gap exists in current tools for early/real-time classification, and this needs to be resolved prior to LSST first light as it is central to enabling follow-up before an anomaly has ended.
Paths to resolving major current barriers include:
\begin{itemize}
\item Introducing a broad diversity of unknown signals, parameterized by phenomenology, into simulated data spanning multiple orders of magnitudes in brightness and timescale.
\item Training and validating softwares that can rapidly recover these signals from simulated data streams with noisy and incomplete data (a few e.g., <5 photometric points and small postage stamps).
\item Building tools that are highly scalable to millions of objects with runtimes of \textasciitilde minutes or less
\item Constructing \gls{software} that is adaptable to new datasets and rapid re-training (active learning)
\end{itemize}

In the direction of flexible detection methods, Astronomaly \citep{Lochner_2021} has recently proposed a general framework in which active learning is used to leverage user-identified outliers and inform automated anomaly detection. Active learning in astronomy remains underdeveloped, but will likely play a powerful role in facilitating rapid anomaly identification and follow-up in the first few years of \gls{LSST}.

\paragraph{Computational Workflow}
The raw alert stream data (postage stamps and immediate photometry) will be accessed through community alert brokers.% In a resource-limited setup, known objects could be prioritized for data retrieval and follow-up; %by their probability of being volatile-rich.
%however, the science benefit from identifying active objects in unexpected classes would be a strong consideration.
The brokers and/or Rubin \gls{Science Platform} should provide well-calibrated historical light curves (including non-detections via forced photometry) across all diaSources at the location of the alert. Photometry for Solar System objects, which will be moving relative to the background stars, will need to track the objects themselves.

Real-time anomaly detection with LSST will likely be conducted as follows. Anomaly detection algorithms should first rapidly analyze newly acquired photometry within the context of prior light curve data. Global features such as period, parametric fit parameters, and dimensionality-reduction components should then be calculated for each event. Local features computed across windowed components of a light curve (e.g., outbursts, changing states) should also be computed to  help identify anomalous behavior within an otherwise common event.

In addition, anomaly identification routines must qualitatively and quantitatively estimate the similarity between global and local features of single events and for these events in comparison to all events across the \gls{LSST} catalog. This could occur via machine-learning tools to flag high-distance objects/phenomena (the quantitative approach), or visualization tools for assessing the spread between these scales (the qualitative approach). Ideally, this work will be done in parallel.

Finally, the anomaly detection infrastructure should be integrated into the LSST \gls{monitoring} routines for individual working groups. These groups will need to query enlarged postage stamps and acquire high cadence Target of Opportunity photometry for extremely high-value objects. Supernovae will also need to be classified via spectroscopic follow-up.% The proposed general framework for anomaly detection with LSST data and software is as follows:  %This coordination demands rapid catalog cross-matching to augment LSST data where possible.

\paragraph{References for Further Reading}
\begin{itemize}
\item \cite{2019Rapid}
\item \cite{2021Aleo_Miner}
\item \cite{Lochner_2021}
\item \cite{2021Malanchev}
\item \cite{2021Martinez}
\end{itemize}

\pagebreak
\subsubsection{Feature extraction and statistical representation (including period finding)} \label{sec:FeatureExtract}

%\jeremy{I've moved some of the text around to pull it out of bullet format. I have tried to preserve as much of the original phrasing as possible}

%Authors: Alekzander Kosakowski (\url{mailto:alekzander.kosakowski@ttu.edu}), Fabio Ragosta (\url{mailto:fabio.ragosta@inaf.it}), David Trilling (\url{mailto:david.trilling@nau.edu})
%+ all participants in the time-series breakout session (see the slide who has everyone assigned to each breakout session; I cannot find it)

%In-person participants listed as planning to attend the Time-Series technical breakout:
%\begin{itemize}
%\item Matthew Graham, Neven
%\item Ashish, Alekzander, Mark, Juan, %Tomislav
%\item Andy Connolly, Viviana Acquaviva
%\item Fabio Ragosta, Catarina Alves, Tomas %Ahumada, Jing Lu
%\item Ilija Medan
%\item Garrett, David
%\end{itemize}
%\url{https://docs.google.com/presentation/d%/15hJQoklRJtZ7WuRiLXAhKKY8BiI4vNNMqA7PGl9-L%ZA/edit#slide=id.g11f92a62f75_15_0}
%
%
% Date: 30 March 2022

\Contributors{Alekzander Kosakowski (\mail{alekzander.kosakowski@ttu.edu}), Fabio Ragosta (\mail{fabio.ragosta@inaf.it}), David Trilling (\mail{david.trilling@nau.edu}), Matthew Graham, Neven Caplar, Ashish Mahabal, Mark Popinchalk, Juan Luna, Tomislav Jurkic, Andy Connolly, Viviana Acquaviva, Catarina S.~Alves, Tomas Ahumada, Jing Lu, Ilija Medan, Garrett Levine}
{30 March 2022}

 {\bf Summary:}\\
 Many well-built and useful tools are available, but running many of them efficiently is difficult.
 We considered the design of a decision tree \gls{algorithm} to efficiently classify variables and transients in (nearly) real-time. The full \gls{algorithm} only runs on objects that produce a specific number of alerts and deeper analyses are only triggered based on probabilistic outputs from previous steps.

\paragraph{Abstract}
 Variable sky astrophysics is a vast field requiring many period-finding and feature-identifying tools. Thus, to efficiently identify all sources of variability in (nearly) real-time for classification and catalog-creation, a complex multistage \gls{algorithm} is required. Currently, users must make use of many other tools to handle these different types of variability, resulting in running many similar analyses on the same data set to tease out different features (see \cite{Ivezic2014, VanderPlas_2018,2020arXiv201005941K}, and references therein).

 Here we propose to create a hierarchical classification algorithm designed to handle the entirety of the variable-sky database generated by \gls{LSST}, in combination with other southern-sky surveys. The goal of this tool is to quickly and automatically classify transients and variables based on features in a multi-band light curve (shape, period, filter-specific amplitude and decay, etc) and the available color, magnitude, parallax, angular diameter information from \gls{LSST}. The proposed tool will handle multiple forms of variability simultaneously (eclipses vs pulsations vs non-periodic) in an efficient manner. Classification based on the light curve and \gls{LSST}-specific photometric/astrometric data will be built-in and presented in the output as probabilities of each object being a specific type of variable.

 The proposed \gls{algorithm} will trigger advanced data analysis on objects that generate more than N alerts across multiple surveys (requiring crosstalk with multiple brokers). Based on the type of alert relative to the rest of the object’s data, a specific set of algorithms would be triggered to identify potential periods or the presence of features to within some probability. Further analysis would be limited to cases that show specific probabilities greater than some threshold to reduce overall computation time and prevent running unrelated analyses on all data.

\paragraph{Science Cases Needing this Tool}
 All variable-sky astrophysics science cases will make use of this tool for initial identification leading to targeted follow-up. This tool aims to provide a first guess -- which will be quantified as a probability to be labeled as a certain phenomena -- on the classification of the event based on the analysis of the extracted features.

 List of science cases include:
\begin{itemize}

\item Asteroid lightcurves
\item Eclipsing binaries
\item Interacting binaries
\item Novae
\item Kilonovae
\item Supernovae
\item \gls{AGN}
\item \gls{TDE}
\item Stochastic transients (e.g. stellar activity)
\end{itemize}

\paragraph{Requirements for the \gls{software}}

 In general the input data will be a catalog photometry (in multiple bands). The algorithm would need to simultaneously access the calibrated light curve data and individual stellar color and astrometry information. Connections with Gaia \gls{DR3} will be useful until LSST can provide its own precise parallax measurements. The advanced analyses will be run on set times with increasing time-gaps between runs. (Month1, Month2, Month4, Month6, Month8, Month12, etc)

 There may be cases where alert data is needed (e.g. follow up of anomalous \gls{transient}s). Thus, a well defined data model will give the opportunity to consider a common hyper-plane to compare and discriminate between different transients.

  A single multi-column output table will be saved for each object identified in LSST. The columns can be simple summary statistics using 32-bit floats or Booleans to represent the data in as little space as possible. Intermediate steps used for creating this table will be deleted. Since different science cases may have very different feature needs, it is difficult to write a general description of the exact features will be required. The user will work with the final table for generating their own user-specific output visualization using \gls{ADQL} or some other search function.

 Core requirements of the tool include:
\begin{itemize}
\item Ability to scale to high volumes of data. In principle, nearly every \gls{LSST} source will need to be searched for periodic (or time-varying signatures). The algorithm will be expensive to run.
\item Approaches need to be able to produce results for irregularly spaced data and/or data gaps.
\item LSST will need to communicate and crossmatch with other surveys, including non-public databases uploaded by the user to their collaboration’s \gls{RSP} server.
\item To accurately calculate periods for periodic variables over many years, precise timing measurements must be used.
\end{itemize}

Additional difficulties include:
\begin{itemize}
\item Assigning proper weights to each subclass of variable or \gls{transient} is a difficult task. If the thresholds are too strict, then the resulting table is borderline useless, missing many real detections. If the thresholds are too relaxed, then the runtime will increase drastically and the output table will be filled with false positives.
\item We need access to a large dataset able to provide estimates to these weights for all possible class (and potentially subclass) or variable or \gls{transient}.
\end{itemize}

\paragraph{Running on LSST and other Datasets}

 Some \gls{LSST} science cases will be adequately served by the standard Lomb-Scargle periodogram approach. Examples include, but are not limited to, variable stars and asteroid lightcurves, though we note that there are special problems associated with moving targets.

 Some \gls{LSST} science cases will NOT be adequately served by Lomb-Scargle and will require different approaches. These include eclipsing systems like stellar binaries or irregularly varying astrophysical sources \citet[e.g. AGN, “Boyajian’s Star”, TDE, see \ref{sec:ExceptionalVari} for other examples and for examples of analysis on unevenly spaced data in time domain see ][ and reference therein]{Naul2018}.

\paragraph{Existing Tools}
 There are many existing tools available for period-finding and feature-extraction. A short list of existing tools that this group is aware of:

\begin{itemize}
\item Celerite \citet[][]{Mackey2017}: python library for fast and scalable Gaussian Process regression in one dimension;
\item Astropy/scipy/Sci-kit learn\citet[][]{olivier_grisel_2022_6563718} : python packages that allow users to access open source projects for specific scientific needs;
\item FastAPI\footnote{\url{{https://fastapi.tiangolo.com/}}}: a modern, high performance, web framework for building APIs;
\item Spark\footnote{\url{https://github.com/apache/spark}} is a unified analytics engine for large-scale data processing.
\item Dask \citet[][]{rocklin2015dask} is a flexible library for parallel computing in Python.
\end{itemize}

 All these tools are used for a very diverse environment of transients and variable characterization in a very specific way, due to the fact that each of these tools need its own data model to be ingested to work on the reference science case. The main downside of what has been described above is the lack of homogeneity in methods for feature extraction.
 Nevertheless the impressive amount of data we will need to handle with \gls{LSST} will not easily be tackled with this ensemble of tools because they are not obviously scalable.

\paragraph{Computational Workflow}

 Because of the extreme computational requirements of running many period finding approaches on every potentially varying source, one challenge will be to identify when there is a variable source that is not being well characterized by Lomb-Scargle. That is, before deploying additional algorithms, we need to identify sources that are (i) varying and (ii) not sinusoidal. How to do this? What is the scale (number) of sources that fall into this category? This could probably be estimated (i.e., how many eclipsing binaries will be in the \gls{LSST} catalog, how many multi-periodic systems will be identified, etc.). This would then allow an estimate of the amount of compute resources needed to carry out this “second level” processing.

Workflow requirments include:
\begin{itemize}
\item Access the entire \gls{LSST} data archive for light curves, colors, parallaxes, etc
\item Access to the alert brokers would be needed to provide intel to follow ups, the \gls{software} would be run on sources detected in the alert n (we considered n< 4) times. The cross-talk between the \gls{software} and the brokers will be imperative for early classifications.
\item Pegasus enables scientists to construct workflows in abstract terms without worrying about the details of the underlying execution environment or the particulars of the low-level specifications required by the middleware (Condor, Globus, or Amazon \gls{EC2}). Pegasus also bridges the current \gls{Cyber Infrastructure} by effectively coordinating multiple distributed resources.

\item The output will be a simple multi-column table. Preferably with classification columns appended to the standard color/magnitude/parallax columns to save disk space.

\end{itemize}

\paragraph{References for Further Reading}
\cite{2011ApJ...733...10R},
\cite{Narayan_2018},
\cite{2019Rapid},
\cite{Forster_2020}

\pagebreak

\subsection{Image Reprocessing} \label{sec:ImageReproc}

\Contributors{Francois Lanusse, Joachim Moeyens, Steven Stetzler,
Suvi Gezari, Clare Saunders, Matt O’Dowd,
Charlotte Olsen, Alma Gonzalez, Gabriele Riccio, William O’Mullane}
{29 March 2022}

\subsubsection{Abstract}
There will be a need for reprocessing of images or image cutouts for a variety of science cases as listed below.  In some cases this is making custom cut outs for specific types of objects, potentially reprocessing the cutouts and stacking them in a new coadded image.
This is distinct from reprocessing full images ala data release processing. In some cases custom processing of the single frame cutouts is requested, this may require changing the background subtraction, extraction of photometry,
%The shifted stacking is possibly a second use case taking some patch of sky and applying custom GPU base shifting code to search for faint objects. This may then come back as more standard cutout/forced photometry on request

\subsubsection{Science Cases Needing this Tool}
Cosmology:
\begin{itemize}
\item Strong lensing scene modeling, SN scene modeling, (potentially) Cluster lensing specific deblending and shape measurement pipeline
\end{itemize}
Extragalactic (Static):
\begin{itemize}
\item \gls{SED} fitting (\secref{sec:GalPPSED}): LSST’s extensive deep imaging of the southern sky will allow for synergies with multiple other legacy surveys which have different wavelength coverage. With six bands, the amount of galaxy properties which can be recovered through SED fitting is limited to photometric redshifts, colors, and stellar mass. Adding additional supplementary IR bands from surveys such as VISTA, Spitzer, and WISE will allow us to constrain more galaxy properties as well as reconstruct star formation histories with sufficient bands. For this reason it will be necessary to reprocess LSST images to the same \gls{PSF} of images from crossmatched galaxy catalogs. There is a need to develop and/or implement algorithms that will allow images to be convolved to match pixel scales of \gls{IR} sources. Some of this work is being done and tested with HSC, but making a user friendly interface will greatly serve the community. It should require little when using on the fly calculations, and the only data products that may be necessary for LSST to store would be catalogs of fluxes for crossmatched galaxies from various other surveys.
\item Low surface brightness dwarf galaxy (candidate) catalogs out to 100 Mpc (\secref{sec:LSBdwarfG}). A good fraction of nearby dwarf galaxies are very low surface brightness objects, on which the source extraction algorithm may not perform well. At the image level one would need to obtain calibrated cutouts for this sample together with flags from the deblending process and re-fit the photometry with models that are optimized for \gls{LSB} dwarf galaxies. This would produce a new catalog with matched photometry and colors.
\item Lens discovery where we would like to identify the optimal image coadd if different from the default pipeline coadd.
\end{itemize}
Extragalactic (Transient Science):
\begin{itemize}
\item Deep custom stacking of pre-event and post-event epochs to search for pre-cursor eruptions and measure late-time evolution, and to create weekly/monthly stacks and do difference imaging to probe faint and slow transients
\item Deep custom stacks of \gls{AGN} in a high state vs low-state, to isolate the position of the AGN relative to the host galaxy nucleus wandering/recoiling MBHs,
%(\secref{sec:PinpointWMBH}) withdrawn
IMBHs in dwarf galaxies (\secref{sec:DwarfAGN})
\item Custom deblending of small-separation gravitationally lensed quasars for lens finding (\secref{sec:CaustingCrossing}), microlensing studies, dark matter substructure, and time-delay cosmography
\end{itemize}
Solar System:
\begin{itemize}
\item Active Objects I/II (\secref{sec:SmallBodyActivity}): Determining the presence of activity (sublimation, outgassing, collisional breakups) of Solar System small body populations requires processing custom-sized and custom-shaped nightly cutouts to look for evidence of activity. This would require querying for, at times, greater than >O(100x100) pixel cutouts of difference images for each observed Solar System object (100k-100m in a night). Active object cutouts should also support custom-stacking for pairs of nightly observations, and on the larger-scale, support stacking over \textasciitilde month-long cadences. This would include specifying orientation such that any tails and/or motion can be aligned in the stacked images.
\item Searching for Faint Objects (\secref{sec:ShiftStack}): There is a large population of slow moving solar system objects (KBO/TNO/Planet 9) that LSST will observe below the 5 sigma source detection limit of a single exposure. Objects from these faint populations can be recovered using efficient shift-and-stack algorithms accelerated with \gls{GPU}s (KBMOD) on difference images produced by LSST. This use-case requires access to either user-defined sized cutouts of difference images or access to individual visit focal planes if user-defined cutouts cannot be supplied. To validate moving object discoveries in the difference images, we would perform forced photometry on cutouts of the calibrated exposures centered on the discovered objects locations. A shift-and-stack search that is both deep and complete would require access to images across the ecliptic (\textasciitilde 3600 square degrees) and across a full year of LSST data, with these searches being run quarterly or yearly. These searches would significantly benefit from having public/group access to the images at a supercomputing center or in the cloud.
\end{itemize}

\subsubsection{Requirements for the software}
The given science cases require a few different areas of functionality:
\begin{itemize}
\item Several science use cases will require custom-sized and custom-shaped cutouts of both calibrated exposures and difference images, and these will also be required for some of the functions described below. This can be provided by an image cutout service that allows users to access cutouts of a chosen size for visit and coadd-level images.
\item Extragalactic transients and variable objects required custom image stacking. This would most likely be able to use the existing coaddition algorithms, but would require custom combinations of visits into the coadds. For example, transients would benefit from stacking all the images before or after the lifetime of the transient, or AGN images could be separated into bright and faint time periods.
\begin{itemize}
\item Existing tools: the existing coaddition algorithms in the LSST data processing can be used to do this. Also, Zuds is a tool from ZTF that performs custom stacking.
\item Needed tools: Custom stacking could be made more simple for users with a tool that would allow date ranges to be specified for the input images.
\end{itemize}
\item Alternatively, some nearby objects will require more specialized stacking: to find nearby Solar System objects, images need to be shifted and stacked following the likely trajectory of an object. An additional allowance to specify the orientation of each image in the stack is desired so that the tails of active objects can be aligned in the stack.
\begin{itemize}
\item Existing tools: KBMOD: GPU-based shift-and-stack code to search for Solar System objects
\item Needed tools: KBMOD still needs significant development and infrastructure support to work at the scale of LSST, custom-sized cutouts of difference and calibrated exposures
\end{itemize}
\item Custom deblending will be necessary for analyzing gravitational lenses, where software specific to lensing must be used. Other particular use cases such as low-brightness dwarf galaxies will want to spot-check the standard deblending.
\begin{itemize}
\item Existing tools: scarlet, BEAST, GaaP , lens modeling tools (e.g. lensStronomy, molets, Muscadet)
\end{itemize}
\item Custom photometry will be required for science use cases where it is necessary to combine LSST measurements with external datasets by reconvolving the data.
\item Scene modeling on the calibrated visits will be required for Type Ia Supernova cosmology and for strong lensing.
\begin{itemize}
\item Existing tools: Tractor and scarlet, others tailored to specific projects
\item Needed tools: Something that takes advantage of existing LSST data products, such as the per-visit calibration and WCS information
\end{itemize}
\end{itemize}

\subsubsection{Running on LSST and other Datasets}
This service would run on any image cut outs from the Rubin image cutout service. In principle this could be run on any available multiepoch image dataset with a cut out service.

\subsubsection{Computational Workflow}
There may be more than one use case here but all start with image cutout of some sort. The image cutout service should be a good starting point.
\begin{enumerate}
\item Select an object or set  of objects make appropriate sized cutouts - these may be long polygons for solar system objects
\item For each cutout apply custom processing if required (reconvolve for matched catalog)
\item For each object stack the images
\begin{enumerate}
\item If this is time batched make the stacks according to the given time frames (implies perhaps a single object in step 1)
\item If this is a shifted stack (epoch centered) the images should already be centered on the object and can be stacked ignoring positon.
\item Otherwise standard stacking/dithering should be applied.
\end{enumerate}

\end{enumerate}

 Shifted stack:
\begin{enumerate}
\item For a patch of sky get cutouts from the difference images at all epochs
%\item Currently using butler - would like to be near the images
\item Pass this set of images ot the custom GPU code for stack shifting
\item For candidate objects go back and get cut outs and forced photometry from the locations provided.
\end{enumerate}

\subsubsection{References for Further Reading}

 Functionality Required for the technical Case:
\begin{itemize}
\item Custom coadds (for extragalactic transients), using different time periods (i.e. before and after lifetime of transient; bright and faint times of AGN) - these would use custom times, but not custom stacking algorithms
\item cross-matching : reconvolving LSST photometry to combine it with photometry from external datasets (co-processing) would be needed for coadds and at visit level
\item Gravitationally lensed quasars: want to do own deblending, some may be urgent, so need to be done in real-time.
\item Solar System active objects: look at the cutouts and see if there is an extended body for all Solar System objects observed in a night. May need >100x100 pixel cutouts. Pull out cutouts over the history of a moving object.
\item Solar System faint objects- custom coadds:  shift and stack along the trajectory of Kuiper belt objects - can get to 26th magnitude using all images from first year. Might need arbitrary polygons. Searches are done using GPUs
\item \gls{LSB} dwarf galaxy catalog out to 100 Mpc: want to repeat photometry using models that are optimized for dwarf galaxies. Want the deblending information
\item Lens discovery - are coadded images optimal for lens discovery? May need custom coadds in order to see whether regular coadds are sufficient.
\item Type1a supernovae - reprocess with scene-modeling get highly calibrated measurement for point source.
\item Static local universe - go back to images to validate whether stars and galaxies have been accurately distinguished.
\end{itemize}

 Existing Software Needed:
\begin{itemize}
\item Image cutout service (which is already provided by \gls{RSP})
\item 100K to 1M cutouts per night - bigger than alert cutout since they are extended (solar system)
\item There are scene-modeling codes - Tractor, Scarlet,
\item KBMOD for Solar System shift and stack
\item Deblending tools for lens  modeling (lenstronomy, Muscadet)
\item BEAST, GaaP (coming soon to DM)
\item Need algorithm for reprocess LSST images  to match the pixel scale/PSF of complementary IR surveys (e.g., Tractor)
\item Need a means for users to query other catalogs to pull matched images/IDs for crossmatched galaxies
\item Zuds - custom stacks for transients (ZTF)
\item Monthly and weekly stacks for long lasting fainter transients
\item Galaxy morphology fitting (e.g., GalFit)
%\item Diffuse galaxy with bright AGN .. needs reprocessing ..
\end{itemize}

 New data products:
\begin{itemize}
\item New fluxes for SN lightcurves – on the order of 1M data points
\item 10k  per year SN-1a - 100K objects, postage stamps and possibly new coadds% (if reuse PSF just this for images)
\item 8k lensed quasars need light curves;  \textasciitilde 200K  lensed galaxies need cutout images
\item 100K- 1M Solar System cutouts per night (custom shapes)
%\item Trajectories and postage stamps - terrabytes for all of it
%\item Lens cutouts potentially .. - could potentially stack lens over survey (continually add to it)
\item Euclid, Roman etc co-processing with LSST data to search for lenses (petabytes of data)
%\item Uncertainty on intermediate products ..

\end{itemize}

\pagebreak
\cleanedup{Olsen}
% with exception of decision on keeping citizen science section and adding references at end
\subsection{Image Analysis} \label{sec:ImageAnalysis}

\Contributors{
Federica Bianco (\mail{fbianco@nyu.edu}),
James Chan (\mail{hung-hsu.chan@epfl.ch}),
Colin Orion Chandler (\mail{orion@nau.edu}),
Henry Hsieh (\mail{hhsieh@psi.edu}),
Arun Kannawadi (\mail{arunkannawadi@astro.princeton.edu}),
Ilin Lazar (\mail{i.lazar@herts.ac.uk}),
Yao-Yuan Mao (\mail{yymao.astro@gmail.com}),
Knut Olsen (\mail{knut.olsen@noirlab.edu} (facilitator)),
Tyler A Pritchard (\mail{TylerAPritchard@gmail.com}),
J. Antonio Vazquez-Mata (\mail{jvazquez@astro.unam.mx})
}
{03/30/22}

\subsubsection{Abstract}
While much of the science that will be done with \gls{LSST} will rely entirely on the catalogs delivered by the science pipelines, a significant number of science use cases will also require analysis of the image data.  In this section, we summarize the discussions of the kinds of \gls{software} tools that are likely needed to enable image-based analyses of a broad range of science use cases.  In this breakout, we only considered analysis that would be done at the level of objects; large-scale image analysis is covered in the section on image reprocessing.

We first briefly summarize the ways in which the science use cases will use the image data, then enumerate the \gls{software} capabilities required by one or more of these use cases.

\subsubsection{Science Cases Needing this Tool}
\begin{itemize}
\item Lensed quasars: \secref{sec:CaustingCrossing}
\item Galaxy morphology: \secref{sec:GalMorphML}
\item Transient host galaxies: \secref{sec:ImmediateClassification}
\item Slow transients: The Rubin Observatory LSST Alert \gls{pipeline}'s difference imaging \gls{pipeline} will efficiently capture transient and variable events with rapid rises in brightness.  However, one weakness to this methodology is the reliance on templates that may contain flux from sources that vary slowly over timescales comparable to that between template image acquisition. Slowly evolving events such as high-redshift ($z>2$) \gls{SLSN}e and Type IIn supernovae with timescales of a year or more can be difficult to detect in these classical nightly image subtraction surveys with fixed templates, and can require special processing to detect including image subtraction with custom templates or bespoke ML-algorithms. Custom image analysis on galaxy cutouts with the tools described below will allow for the expansion of our ability to detect these events and enable them to be discovered more promptly and enabling follow-up studies.  %KO contacted Tyler for input
\item Detection and analysis of active small solar system objects, \secref{sec:SmallBodyActivity}: The detection and analysis of visible mass loss from comets and asteroids due to various processes (e.g., sublimation, impacts, rotational disruptions) are a high priority for \gls{LSST} solar system science.  This science case includes the detection of previously unknown activity, characterization of morphology evolution of known active objects, and the detection of anomalous brightening of known active objects (i.e, cometary outbursts).
\item Distances for $<100$ Mpc dwarf galaxies  \secref{sec:ImmediateClassification}: Use ML-based methods (e.g., \gls{CNN}) to estimate distance for very nearby ($< 100$ Mpc), low surface brightness dwarf galaxies. The algorithm will run on postage stamps, with a small training data set based on a subset of objects with known distances/redshifts. Existing photo-z algorithms are not optimized for this very nearby region, and the lack of training data also poses a special challenge.
\item \gls{WL} artifacts/debugging: Failures and outliers in the measured shape of seemingly normal sources may be caused due to a neighboring source, which can be easily identified from image cutouts. Examples of this include: i) a bright star in the vicinity, elevating the local background ii) an extended arm of a spiral galaxy or merger components of an irregular galaxy with a dominant bulge component (so that it is not evident from the catalogs) overlapping a faint source of interest. A quick look at the image can differentiate an algorithmic failure of the measurement from imperfect deblending.

\item Search for resolved dwarf galaxies, \secref{sec:LGdwarf}: A complete census of dwarf galaxies within the Local Volume will rely on both catalog analysis of resolved stars and analysis of the associated images (e.g. \citealt{Carlin21}).  The image cutouts will need to be up to \textasciitilde 10 \gls{arcmin} in size to contain the candidates and their context.  The goal of the search will be to remove humans from the loop to the extent possible, and/or use the citizen science community to help.

\end{itemize}

\subsubsection{Requirements for the \gls{software}}
Because image-based analysis is a very broad technical subject, we first listed the \gls{software} capabilities needed, identified the science cases needing each capability, and defined the data products that each capability would need as input to the analysis.  Later in this section, we will describe the requirements of the capabilities in more detail.

\begin{itemize}
    \item {\bf Capability:} Image cutouts.  While also discussed in the Image Reprocessing breakout, all image analyses discussed presume the ability to make image cutouts of any input image over scales ranging from $\sim$10 arcsec to $\sim$10 \gls{arcmin}.\\
{\bf Needed by:} All. \\
{\bf Input data products:} Coadds, single \gls{epoch} images. \\
    \item {\bf Capability:} Ability to perform custom deblending and reblending of the objects detected in an image. \\
    {\bf Needed by:} Lensed quasars, \gls{transient} host galaxies, distances to dwarf galaxies within 100 Mpc, weak lensing artifacts and debugging.\\
{\bf Input data products:} Coadds, clipped coadds.\\
    \item {\bf Capability:} Ability to link an image cutout to archival data (both catalogs and images) from external sources. \\
    {\bf Needed by:} All. \\
{\bf Input data products:} Images, catalogs.\\
    \item {\bf Capability:} Ability to conduct analyses on any band or combination thereof.\\
    {\bf Needed by:} All except the comet activity use case.\\
{\bf Input data products:} Coadds, single \gls{epoch} images.\\
    \item {\bf Capability:} Ability to conduct image analysis in real time.\\
    {\bf Needed by:} Transient host galaxies, comet activity.\\
{\bf Input data products:} Single-epoch images.\\
    \item {\bf Capability:} Ability to associate image cutouts with measured light curves.\\
    {\bf Needed by:} Lensed quasars, \gls{transient} host galaxies, slow transients.\\
{\bf Input data products:} Coadds, single-epoch images, catalogs, alert stream.\\
    \item {\bf Capability:} AI-aided image analysis and time variable clustering analysis. The idea behind this capability is to use machine learning and AI techniques to help in the construction of difference images, their analysis, or to replace traditional \gls{DIA} entirely. \\
    {\bf Needed by:} Lensed quasars, slow transients.\\
{\bf Input data products:} Coadds, single-epoch images.\\
    \item {\bf Capability:} Machine learning-based clustering of static images.\\
    {\bf Needed by:} Galaxy morphology, comet activity, distances to dwarf galaxies within 100 Mpc, search for resolved dwarf galaxies.\\
{\bf Input data products:} Coadds.\\
    \item {\bf Capability:} Ability to handle diverse morphology found in images.\\
    {\bf Needed by:} Galaxy morphology, comet activity, distances to dwarf galaxies within 100 Mpc, search for resolved dwarf galaxies.\\\
{\bf Input data products:} Coadds, single-epoch images.\\
    \item {\bf Capability:} Ability to conduct visual inspection of images.\\
    {\bf Needed by:} Galaxy morphology, comet activity, distances to dwarf galaxies within 100 Mpc, weak lensing artifacts and debugging, search for resolved dwarf galaxies.\\\
{\bf Input data products:} Coadds, single-epoch images.\\
    \item {\bf Capability:} Ability to use images in citizen science programs, which is being delivered as part of Rubin Construction (and thus not discussed further).\\
    {\bf Needed by:} Galaxy morphology, comet activity, distances to dwarf galaxies within 100 Mpc, weak lensing artifacts and debugging, search for resolved dwarf galaxies.\\
{\bf Input data products:} Coadds, single-epoch images.\\
    \item {\bf Capability:} Ability to create synthetic image cutouts or inject objects into empirical image cutouts.\\
    {\bf Needed by:} All.\\
{\bf Input data products:} Coadds, single-epoch images.\\

\end{itemize}

\paragraph{Image cutouts}

%KO I think this Outline section is unnecessary
%Outline:
%\begin{enumerate}
%\item Size of postage stamps for different use cases
%\item APIs (take RA/DEC, postage stamp size OR objectID and dynamically determine the size based on their footprint
%\item Return a data hyper-cube that allows efficient subset querying of stamps of the same object across multiple single-exposures, coadds over bands etc.
%\end{enumerate}

 The image cutout tool that returns \gls{postage stamp}(s) must be able to take in as input at least one of the following:
\begin{itemize}

\item Positions (RA, \gls{DEC}) on the sky and a size of the image, either in arcsec or in number of pixels. Depending on the science use case, we expect the size of the postage stamps to vary from 10 arcsec x 10 arcsec to 1 deg x 1 deg
\item Unique identifier of a source/object from the Data \gls{Release} catalogs and return postage stamps with size determined by the footprint, along with a buffer optionally. %(new feature? This requires integrating TAP service with image cutout)
\item The postage stamps should contain sufficient \gls{metadata} information to produce color-composite images from stamps that will identify objects such as lensed quasars.
\item Single-visit images should be queriable by metadata information, e.g., observations of (RA, DEC) between ‘datetime1’ and ‘datetime2’ or seeing size $< 0.5$ arcsec, etc. This can be used to look for \gls{transient}/variable objects specifically. (new feature?)
\end{itemize}

 The returned stamps can be single visit images, or any of the coadds produced by DM. The returned images must be viewable on a portal on the browser for some quick visual inspection and allow for bulk downloads, for creating \gls{ML} datasets. The portal must allow for an interactive visualization, i.e., zooming in, finding pixel values, variance estimates and mask bits for each pixel etc.

 Existing tools, such as that provided by the RSP, have most of the above capabilities, but some work may be needed for high-level capabilities such as integrating the cutout service with \gls{TAP} query that can translate ADQL queries to image cutouts through the catalogs.

\paragraph{Deblending / reblending}
High-level requirements for the deblending/rebending tool are:
\begin{itemize}
\item Being able to specify postage stamps to run on.
\item Being able to run on both coadd postage stamps and clipped coadds.
\item Being able to customize the deblending criteria. Should have a well-defined \gls{API} so that users can supply their custom deblending criteria / algorithms easily.
\item Being able to customize the photometry fitting procedure. Users can customize how to refit the photometry and produce a new catalog (for the all postage stamps that it has run on).
\item Generate some sort of scores based on user-supplied metrics.
\item This will mostly be used for static science, so it will not be run very often. It might need to be run on a large number of postage stamps, but that should still be a small fraction of the overall sky coverage (e.g., $< 2$\%). For alternative deblending/reblending methods that need to be run on a large fraction of sky, it should be part of the image reprocessing.
\item The data output (catalog) needs to be stored for a long term for science use.
\end{itemize}

Existing tools:
\begin{itemize}
\item Scarlet \citep{2018A&C....24..129M}, Source Extractor \citep{2010ascl.soft10064B}, \gls{SDSS} Deblender, Tractor \citep{2016ascl.soft04008L}, GalFit.
\end{itemize}

\paragraph{Link to archival data}
For external archive data to be used in conjunction with \gls{LSST} data, the science use cases require:
\begin{itemize}
\item \gls{Object}-based and coordinate based external archive queries
\begin{itemize}
\item Moving objects:
\begin{itemize}
\item something like the \gls{CADC} SSOIS\footnote{\url{https://www.cadc-ccda.hia-iha.nrc-cnrc.gc.ca/en/ssois/}} or PDS CATCH\footnote{\url{https://catch.astro.umd.edu/}}
\item Faster yet: already linked to objects (e.g., \gls{ZTF} alert stream)
\end{itemize}
\item Maintain tables of crossmatches and archive-specific \gls{metadata} in order to facilitate productive external archive searches
\item Augment table data with potentially missing metadata (e.g., depth, \gls{seeing}, …)
\end{itemize}
\item Data products would include images and catalog objects
\item Programmatic access important
\item \gls{SIA} interface for images
\item \gls{TAP} interface for catalogs
\item Build on Astroquery as a general interface
\item Local or Cloud storage for data products
\item Ability to store products in multiple file formats
\end{itemize}

\paragraph{Real-time analysis}
This is a general \gls{software} requirement/consideration rather than a specific \gls{software} tool.  Example science use cases that require  \gls{RTA} include the following:
\begin{itemize}
\item Detection of active solar system objects (\gls{RTA} needed to enable follow-up
\item Comet outburst detection (\gls{RTA} needed to enable follow-up)
\item Characterization of \gls{transient} host galaxies

\end{itemize}
 Requirements for \gls{software} performing real-time analysis include the following:
\begin{itemize}
\item Automation of image analysis tasks for specific science cases (e.g., active solar system object detection, comet outburst detection, characterization of active solar system object morphology, analysis of \gls{transient} host galaxies)
\item Rapid definition of \gls{postage stamp} requirements (e.g., exposure, position, cutout size)
\item Prompt retrieval of postage stamps from single-epoch data via image cutout tool
\item Same-day completion of data analysis tasks for all data acquired in a single night at a minimum given the need to keep pace with overall data acquisition rate
\item For some applications, near-real-time analysis may be desirable to enable extremely rapid follow-up for highly variable targets
\item Job management infrastructure to manage massively parallel data processing to achieve required processing speeds

\end{itemize}

%\paragraph{Associate w/ LCs requirements}\\

\paragraph{ML-based “DIA” / variable clustering}
\begin{itemize}
\item \gls{DIA} is an expensive and critical step for Time Domain Astronomy. \gls{DIA} models will be tested throughout Rubin commissioning to select the most effective method. Critical with Rubin will be the need to limiting false positives and accurately perform Real/Bogus due to the enormous survey data volume (expected 10M alerts per night).
\item Traditional \gls{DIA} will rely on the construction of a template which Rubin will update annually (with each data release). All LSST historical data will be reprocessed with new template in each data release. \gls{DIA} needs to be efficient, with the current computational bottle neck being the PSF matching operations that align the properties of the images that compose the template and those of the template with those of each science image.
\item We expect that a traditional DIA analysis will be the basis for the alert generation throughout the survey lifetime. A typical DIA workflow entails Template creation, PSF matching, Image Subtraction, Transient Detection-Real Bogus (typically done with feature based methods like Random Forests), Photometry, Transient classification (typically performed on multiple data points). Yet we expect that the community will want to customize the \gls{transient} discovery process to, for example, increase sensitivity to specific classes of transients by designing purpose-specific templates and/or modifying the traditional DIA workflow. This workflow can be modified in several AI-aided ways. An AI model could be developed for progressively more complex tasks thus replacing more of the \gls{transient} detection and characterization infrastructure. A simple minimal task would include AI-aided detection and/or real bogus which requires a binary classifier (a many to one NN classification), more complex tasks could include \gls{transient} classification (a many to many NN classification) or photometry (regression), or any combination of the tasks outlined here. Example tasks include:

\begin{itemize}
\item Comparing the science image with a historical (LSST but potentially also precursor surveys) collection of images from the sky position (i.e. the input of a neural network would be a data cube of historical images and a single science image, the output would be a \gls{transient} detection/real bogus classification/transient class classification). This is expected in particular to increase effectiveness in the detection of slow evolving transients where an average template
\item AI-generation of templates: this entails teaching a neural network to perform all the computationally expensive steps of template creation including warping, normalizing, and \gls{PSF} matching of the template and science images
\item Bypassing the PSF matching and image subtraction steps by comparing the template with the science image (steps in this direction have demonstrated the feasibility of this approach \citep{Acero-Cuellar22} in the Real Bogus step, but the detection step and \gls{transient} classification without DIA are yet to be explored…).
\item Direct \gls{transient} classification from \gls{DIA} images, template+DIA images (providing critical context for the \gls{transient}’s origin), template + science image alone, or a collection of images from the same sky position (no template), which is currently being explored (e.g. ALeRCE\footnote{\url{http://alerce.science/}}).
\end{itemize}
\item Need for efficient image subtraction with a variety of templates or template-agnostic image subtraction
\begin{itemize}
\item For example, slowly evolving objects (High-Z \gls{SLSN}e, IIn), sub-threshold events (High-Z transients), variable events with data-in-templates (\gls{AGN}), non-point source/diffuse emission (e.g. light echoes), periodic and semi-periodic variabilities with time scales that are on times scales comparable with the cadence of images selected for templates \citep{Hambleton20}
\end{itemize}
\item Requirements
\begin{itemize}
\item Depending on methodology, AI-aided \gls{DIA} can use a template image and science image to most closely mirror today's \gls{DIA} workflow, or a time-series of single visit images to search for events
\item Robust access to multi-band single visit images as well as yearly/deep stacks depending on choice of AI-aided methodology - postage stamps may be sufficient for this so long as they are large enough to enable matching (large enough to characterize the \gls{PSF}) or \gls{PSF} modeling information is provided
\begin{itemize}
\item One may choose to do this on a variety of scales depending on the science use case - Deep-Drilling Fields for exceptionally long-lived events, a known galaxy list (that can be exceptionally large, e.g. 2<z<5) for subthreshold events, the entire field for diffuse/light-echo type emission
\end{itemize}
\item Efficient access to \gls{ML}-tools, computational cores, and rapid retrieval on the terabyte up to approaching petabyte scale data
\item Significant additional storage for synthetic or historical training data
\item Ability to re-process as tools improve
\end{itemize}
\item Challenges
\begin{itemize}
\item Prototyping of many models similar to this methodology exist in the literature, but have never been run at anything approaching scale.  To do this, we need:
\begin{itemize}
\item Training - for a truly robust AI-aided \gls{DIA} algorithm one must have sufficient training data to capture the known on-sky distribution of objects.  This will require simulation, precursor, and on-sky data to work in concert.
\item Data-scaling - at the largest scales, this requires robust processing across the entire sky of data.  While the simplest examples of this are trivial (e.g. a small number of point sources from a known list of locations) it quickly scales to terabytes of data volume and velocity
\end{itemize}
\end{itemize}

%\subsubsection{Existing Tools}

\end{itemize}

\paragraph{ML-based static clustering}

 Morphology is a fundamental parameter, essential for the full spectrum of extra-galactic \gls{LSST} science. \gls{LSST} offers an unparalleled combination of depth, area and statistics, with \textasciitilde 20 billion galaxies expected from its 18,000 deg$^2$ footprint with a point-source depth of $r_\text{AB}$\textasciitilde 27.5 mag. A rich literature exists on measuring morphologies in surveys, from visual inspection, using systems like \gls{GZ}, to automated methods, either via simple measures (S\'ersic/CAS etc.) or sophisticated supervised or unsupervised machine-learning (ML) techniques. However, the unprecedented size of \gls{LSST} requires a radically different approach. Visual inspection, even using \gls{GZ}, will be prohibitively time-consuming. Furthermore, since the morphological detail in galaxies will increase as \gls{LSST} becomes deeper, morphological catalogs will be needed at multiple depths (e.g. from every data release). This calls for classification/clustering techniques which are able to handle large amounts of high cadence survey data (petabyte to exabyte scales) in an efficient and accurate way.

\begin{itemize}
\item Need for  supervised \gls{ML} to do morphological classification of galaxies and detect particular structures, using previous labeled data.
\item Need for unsupervised \gls{ML} in particular for large scale galaxy morphology classification may provide an advantage since it does not need training sets (i.e., no need for labeled data)

%\end{itemize}

%\subsubsection{Science Cases Needing this Tool}
%\begin{itemize}

\item Photometric redshift estimation
\begin{itemize}
    \item Weak/Strong lensing studies
    \item Studies of Galaxy evolution as a function of environment, redshift and other properties from a statistical perspective (which \gls{LSST} can now enable)

\end{itemize}

 \item Requirements:
\begin{itemize}
\item Need for combined calibrated images in all possible bands to carry out classification.
\item This can be done either combining images in the \gls{RSP} during the training process or having previously combined cutouts in png format. Meanwhile the first one will require computer power to accelerate the process, the second one will require additional 2 Tb of storage for every 500 millions of 50x50 pix images.
\item Being able to access/propose for \gls{CPU} time easily and efficiently (for example 100 cores for each LSST yearly release); maybe even have a portion of the LSST \gls{CPU} capabilities reserved for ML based applications on a yearly basis
\item Need an efficient architecture within the \gls{RSP} to be able to import training data in large scales from other surveys
\item Need for an efficient data transmission to local IDACs with \gls{GPU} facilities.
\end{itemize}

 The classification process is expected to be repeated at every data release to generate morphological catalogs.

\item Existing tools:
 External tools to develop \gls{ML} algorithms have been developed and optimized to do efficient calculation. Tensorflow, Keras, PyTorch are the most accessible and friendly libraries to work with.

 The most popular technique used for galaxy classification is \gls{CNN}s \cite[e.g.,][]{Huertas-Company15,Cheng20,Dai2018} using training data mainly from the Zooniverse. Other techniques are random forest classifiers, Support Vector Machines \cite[e.g.,][]{Goulding2018} or unsupervised algorithms \cite[e.g.,][]{Cheng20}. The downside is that most of these techniques require large amounts of training data and may not be able to operate efficiently at \gls{LSST} scales.

\end{itemize}

%\paragraph{Visual inspection and Citizen science}
% What is \gls{Citizen Science}? \\
% \gls{Citizen Science} is the involvement of the public in scientific research. \\
% It can be also associated with the projects between experts and non-experts. \\
% There is already a sophisticated tool existing, Zooniverse, which we make use of. \\
% An interface is required to link Rubin and Zooniverse, so that Rubin does not need to %scratch a new one. (But maybe it is also possible.)
%\WOM {This is one of th \gls{EPO} use cases and a link to Zooniverse is in the making .. please talk to Lauren Corlies; KO: perhaps we should remove this section? }

%Requirement:

%\begin{itemize}
%\item The cutouts should be generated efficiently from Rubin.
%\item Users are able to query and download the images from Rubin. Perhaps these images can direct to Zooniverse immediately.
%\begin{itemize}
%\item Ability to do some type of semi-automated vetting (e.g., exclude images with no objects in them, or too many objects for a given science case)
%\end{itemize}
%\item A tool for users to describe:
%\begin{itemize}
%\item Tutorial
%\item Field Guide
%\item Workflow
%\item Subject Sets
%\item About information (purpose, science team, etc.)
%\item \gls{NASA} Partner status
%\item Analysis tools
%\end{itemize}
%\end{itemize}

% We probably also need translation tools, such as data analysis, or training set for \gls{ML}.

% (Colin’s project: \url{http://activeasteroids.net})
% e.g.  \url{https://www.zooniverse.org/projects/krojas26/experts-visual-inspection-experiment/classify}

\paragraph{General Visual Inspection}
\begin{itemize}
\item Ability to display a grid or list of relevant images.
\item Ability to pull from different sources (e.g., Pan-STARRS1, \gls{SDSS})
\item Mouse over information: RA, Dec, x, y, \gls{UT} observing date, filter?, ..?
\item Ability to flag images and export results
\item Ability to change angular \gls{FOV}
\item Ability to rotate, flip images
\item Ensure uniform orientation (N up, E left)
\end{itemize}

Examples:

\begin{itemize}
\item \url{https://yymao.github.io/decals-image-list-tool/}
\end{itemize}

\paragraph{Testing synthetic images}
\begin{itemize}
\item This is an overall infrastructure functionality, rather than a specific \gls{software}.
\item Being able to process user-provided synthetic images with both the \gls{LSST} Science Pipeline and all the custom tools above.
\item Should accept both fully simulated images  and real images with synthetic source injections.
\item Provide access to the produced catalogs and other data products in the same way as the real data products.
\item This will mostly be used for testing, verification, and validation. It may be run more often during the development stage. However, the data products may still need to be stored for long terms as they may be needed for testing bias, completeness for science use.
\end{itemize}

\paragraph{Diverse Morphologies}
\begin{itemize}
\item Ability to account for a range of different possible morphologies for extended objects for various purposes
\item Ability to fit user-provided models to image cutout data
\item Example relevant science cases:
\begin{itemize}
\item detection  of different types of small solar system object activity (e.g., circularly symmetric coma, faint dust trails)
\item Recalculation of \gls{astrometry} of highly active comets fitting comet-like profiles

\end{itemize}
\end{itemize}

\subsubsection{Running on \gls{LSST} and other Datasets}
As discussed above, the image analysis capabilities will make use of several Rubin LSST data products, including coadds, clipped coadds, single \gls{epoch} images, catalogs, and the alert stream.  There also was a general need for access to external archival data.  A few image analysis technical cases also had specific needs:

\paragraph{ML-based “DIA” / variable clustering}
\begin{itemize}
\item Existing archives of Survey Data.  Time-series and static, in similar filters as the main survey
\item Simulated static Sky-data from the LSST with source injection for \gls{transient} and variable events
\item \gls{LSST} commissioning data
\item On-Sky LSST Single \gls{epoch} and stacked data
\end{itemize}

\paragraph{ML-based static clustering)}
\begin{itemize}

\item The \gls{LSST} Data Previews, the \gls{HSC} survey (\gls{LSST} precursor), \gls{HST} datasets and the \gls{DESI} Legacy Surveys  will be beneficial to act as training sets
\item The \gls{algorithm} can be run as soon as the first data release is online if unsupervised machine is used and if training is done on precursor or other data

\end{itemize}

\subsubsection{Existing Tools}
Existing tools vary by image analysis capability, but include:
\begin{itemize}
    \item Rubin \gls{Science Platform} (image cutouts)
    \item Scarlet, Source Extractor, \gls{SDSS} Deblender, Tractor, Galfit (Deblending/reblending)
    \item External data archives
    \item Tensorflow, Keras, and PyTorch for \gls{ML} applications
    \item The Rubin \gls{EPO} Citizen Science project for the citizen science use case
\end{itemize}
\subsubsection{Computational Workflows}
Detailed computational workflows were provided for the two of the image analysis capabilities.
\paragraph{ML-based “DIA” / variable clustering}

\begin{itemize}
\item A minimal example for this technology is developing an AI-aided detection and/or real bogus methodology for the Rubin Observatory \gls{LSST}, and it could follow the following development process:
\item Training for AI-Aided \gls{DIA}
\item Prior to \gls{LSST} operations, precursor surveys and simulated data (such as that created by DESC for the Data Preview) could be used to optimize both the neural network structure (building off of extant prototypes) and training methodology
\item With the release of wide-field \gls{LSST} commissioning data at levels similar to both individual (and potentially 10-year) survey depth, transfer learning methodology could be used to evaluate success of the algorithms on precursor/simulated data when applied to on-sky survey-like data.
\item As additional on-sky data from Rubin observatory cameras becomes available, models can be continuously improved with adaptive learning techniques to improve quality
\item On-sky AI-Aided \gls{DIA}
\item The user would identify a list of sky-area to for the \gls{DIA} - depending on the scale of the process this could be a HEALPix map, or a list of galaxy or point source positions with postage stamp size (that may come from catalog cuts or external sources)
\item The user would identify a ‘template image’ to use, including those created by themselves, \gls{LSST} simulated images from precursor surveys, or the \gls{LSST} project
\item The user would identify a AI-aided \gls{DIA} model to use for their science project.  In the optimistic case this could be a single model that has been robustly trained across the sky.  In a more pessimistic case this could be something like a consensus NN infrastructure that the user trains on a subset of objects that are similar to their science needs
\item The user then creates a (potentially embarrassingly parallel) processing \gls{pipeline} on the RSP, LINCC, or local compute that takes template images, survey images/cutouts, and runs the AI-aided model to receive a list of detected events.  This would then be a jumping off point into a further scientific study revolving around the objects of interest…
\end{itemize}

\paragraph{ML-based static clustering}
\begin{itemize}

\item We will need to query galaxy cutouts and catalog properties from the \gls{LSST} archive. Example catalog properties needed: ra, dec, photometric redshift, stellar mass, SFR, colors, object radius.
\item The classification/clustering will be done with algorithms provided by Scikit Learn and suitable processing parallelization will be used
\item  First training will be done using external catalogs and the \gls{DP0}.2 images through the RSP.
\item Once the first \gls{DR} comes out, the algorithms will be run on these images and morphological catalogs will be generated. About 1Tb of storage will be needed to save these catalogs.

\item If 100 cores are used the training timescales may last for a couple of days to a week depending on the depth of the data and sky cover age.
\item If GPUs at IDACs are used, the training timescale could be reduced by a factor of 2. However, It will be very important to estimate what is cheaper and more efficient, using more cores in CPUs at the \gls{RSP} or moving data to IDACs with GPU facilities.

\end{itemize}

\pagebreak

\subsection{Photometric Redshifts } \label{sec:Photoz}
There were multiple technical cases for \gls{photo-z}. The first of these is about how to effectively represent and store photometric redshift information; the other two relate to how statistical uncertainty is quantified and the impact of photometric redshift quality on scientific applications.  Essential background on what Rubin Observatory will provide regarding photometric redshifts, and the anticipated connection with the scientific community in this area, is given in \url{https://dmtn-049.lsst.io/}.

\subsubsection{Photo-z p(z) representation and storage} \label{sec:PhotozRep}
\Contributors{Sam Schmidt, Julia Gschwend, Alex Malz, Raphael Shirley, Ashley Villar}
{March 29, 2022}
\cleanedup{AIM}

\paragraph{Abstract}
Current-generation surveys (\gls{DES}, \gls{HSC}, and others) now commonly supply one-dimensional (1D) \gls{photo-z} \glspl{PDF}, typically stored as evalulations on a grid or as samples, though existing software tools such as qp (see reference below) are being developed to explore more efficient parameterizations of PDFs for next-generation data sets, bearing in mind the limited resources for storing and serving PDFs.  However, dedicated effort will be required to minimize the inefficiencies of lossy format conversions and the computational expense of repeating estimation procedures to achieve the most appropriate format for {\it each} science case expected to use the PDFs. It is also important to provide documentation on how \gls{photo-z} PDFs are represented and how they can be``decompressed" to a traditional grid PDF or other format, if necessary, for a particular use case.

\paragraph{Science Cases Needing this Tool}
The storage of \gls{photo-z} impacts all science cases where redshift is needed beyond a simple ``point estimate''.

Storing a 1D PDF or a 2-dimensional $p(z, \alpha)$ (where $\alpha$ could be star formation rate, stellar mass, etc.) requires that decisions be made as to the format in which the data will be stored, be it on a grid, a specific set of quantile values, a mixture model fit to samples, etc.  Efficient storage and the ability to actually {\it use} the resulting \gls{photo-z} PDFs across all science cases is an essential need.  That is to say, the most appropriate storage format for a PDF may be different from what a science-case-specific code has used for its input in the past, and methods should be provided to transform between representations where necessary. In some cases, adapting the metrics/analysis on the user-side to utilize the efficient storage format directly may actually improve workflow performance: for example if the \gls{CDF} is used but never the \gls{PDF}, then quantiles would reduce computational expense even if past use cases assumed a gridded input format or drawing samples from the PDF.  Some thought will need to go into each science case to decide on an optimal parameterization balancing storage limitations against loss of information to which downstream analysis is sensitive. In the end, the storage method must cover all science use cases.

\paragraph{Requirements for the \gls{software}}
Many of the requirements could be fulfilled by the qp\footnote{available at \url{https://github.com/LSSTDESC/qp}} software package, which aims to be a general, extendable tool for transforming between multiple PDF representations, and includes metric computations. Multidimensional photo-z estimates, where additional quantities are jointly fit, are obviously more computationally intensive and require a more complex solution for storage.  The DM PhotoZ table is expected to have \textasciitilde 200 columns, and even with additional database storage that might be available from the LIneA Brazil in-kind contribution, multidimensional PDFs must be able to be represented by a relatively modest number of parameters if they are to be implemented on the full \gls{LSST} dataset.

For science that will involve a relatively small sub-sample (compared to \textasciitilde billions of \gls{LSST} detected objects, say up to millions of objects), it may be simpler and computationally less expensive to compute multidimensional distributions “on the fly” for that subset as-needed rather than attempt to store them on disk, particularly if the specific multidimensional parameterization does not lend itself to compression to a small set of parameters.
While PDFs are necessary for many science cases, some still use single point estimates of redshift or low-dimensional summary statistics (e.g. moments), and thus these will also be included in any data releases; some discussion of uncertainty quantification for such point estimates is included in \secref{sec:Uncertainty} of this white paper.

\paragraph{Running on LSST and other Datasets}
Both Rubin \gls{photo-z} outputs and additional \gls{photo-z} estimates are very likely to include \gls{PDF} representations, as are all annual \gls{LSST} release catalogs and DDFs for which photo-z’s are computed.  Any multiband precursor dataset could also be used to generate PDFs for experimentation.

Photo-z PDFs could be produced for any new release of photometry and any update in prior information, such as additional spectroscopy for training sets or new SED templates. The provided data products would have to include provenance information indicating the version of \gls{software} as well as the version of prior information and input data used to create it.  However, computational and storage expense for \gls{photo-z} estimates is nontrivial, so some thought will need to go in to how often release catalogs are generated.

\paragraph{Existing Tools}
qp is an existing \gls{software} package to transform between several PDF representations (interpolated grid, histogram, samples, quantiles, parametric mixture model, etc.) and to evaluate metrics of 1D PDFs. qp can be used to identify the optimal parameterization for a given scientific use case by allowing users to experiment with the number of parameters and format, ensuring sufficient information is preserved in the approximation/decompression steps to achieve the desired science goals.

qp deals only with \gls{1D} PDFs currently, though there has been discussion of expanding to multidimensional, but computational and storage efficiency would be an issue.  qp is still under active development, and several of the storage methods may still be somewhat slow, particularly for transformation of large datasets from one format to another, and so additional code optimization may be necessary.  Any extensions should be compatible with existing qp code, or entail refactoring with an eye toward backward compatibility, as there is already an active user base.

\paragraph{Computational Workflow}

Considering a photo-z workflow consuming data and resources directly from the DAC, it might use \gls{Butler} to access the catalog data. It should be able to read individual columns from the Parquet files. Examples of common photo-z inputs are magnitudes (or fluxes) and respective measurement errors, but for machine learning methods the list of inputs can be fairly diverse (e.g. to include shape parameters, concentration, Sersic index, and others).
For the official \gls{photo-z} tables, the ones to be included in the object catalog for the data release, the query would just select columns with no selections at rows level. For other science-driven \gls{photo-z} runs to be done by users, it is expected to have pre-processing/cleaning or filtering based on e.g., signal-to-noise, quality flags, region selections (e.g. cone search), etc.

% Are you planning to access your data through community alert brokers? Are there \gls{software} components (e.g. classification algorithms) needed to enhance these brokers?\\
% A: (I think only if they were in the DM \gls{photo-z} table)

The \gls{photo-z} algorithms to be part of the data releases are to be defined after discussions at the Photo-z Validation Cooperative\footnote{\url{https://community.lsst.org/t/rubin-commissioning-and-the-photo-z-validation-cooperative/6310}}.
There is already a shortlist of algorithms to be tested available in dmtn-049, based on the letters of recommendations collected by the Photo-z Coordination Group. As it is now, DESC has a workflow that contains the estimation of photo-zs included as one of its steps via the {\it Redshift Assessment Infrastructure Layers} (RAIL) \gls{software} package (see link in references). In this case, the output PDFs might be already set to be represented as qp objects. As RAIL is publicly developed code, other cases outside of DESC can use (and extend) RAIL (by adding additional \gls{photo-z} algorithms of interest to the code base, such as othose being developed for the Brazilian IDAC), and therefore take advantage of each algorithm's native output formats. Alternatively, any other \gls{photo-z} pipeline can also import qp at the end of the workflow and provide the parametrized PDFs as qp objects as well.
Besides the main photo-z tables provided by DM, alternative photo-z tables will be provided by international contributors via the in-kind contribution program (e.g. LIneA in Brazil) as federated datasets, using the IDAC's infrastructure. Scalability tests are being done currently at LIneA using Parsl as workflow manager and Lephare and \gls{CMNN}\footnote{ \url{https://github.com/dirac-institute/CMNN_Photoz_Estimator}
} as examples of photo-z codes, using DESC's Cosmo DC2 mocks and DES data as precursor datasets. According to current results, it is estimated to take \textasciitilde 5 days to run \gls{CMNN} for the whole \gls{LSST} DR1, generating \textasciitilde 3 TB of outputs, without any post-processing (formatting/compressing) on the results. Although this estimate already fulfills the minimum requirements regarding computing speed (the order of milliseconds per object, as mentioned in \url{https://dmtn-049.lsst.io/}), even before planned optimization efforts begin, that estimate of computational time and storage footprint points to the necessity of adopting some strategy to transfer data in chunks in advance as soon as it is available, so it can be processed and made available as federated datasets, in the required output format, respecting the data releases timeline.
As some subset of users will likely wish to run ``custom" \gls{photo-z} analyses on specific data subsets, it is worth considering that we should provide an ``easy to use" \gls{photo-z} pipeline to the community, with different algorithms available, providing formatted outputs, by default in the same formats as those official data products provided by DM with which people will already be familiar. In this case the \gls{photo-z} code would be run by users on smaller subsets to address particular science cases (where for these cases the storage of results would either be local on the users' own workspaces, or potentially on a shared resource). The easiest way to accomplish this would be to embed RAIL in the RSP and any IDAC infrastructures under consideration, however these options require development to distribute parallel jobs in different environments.

%Describe whether an analysis will need to be distributed across multiple cores or machines (e.g. can the analysis be undertaken as an embarrassingly parallel application or does it require message passing, is the analysis iterative requiring many passes of the data).
Parallelism, computation, storage and visualization: Many photo-z estimation codes are constructed in such a way that the analysis can be considered as embarrassingly parallel and also there are no constraints in spatial location of objects in the catalog. The partitioning can be optimized based on the machines'  memory availability.

%Do you understand the memory to number of cores ratio needed for the analysis ?\\
There are no universal memory-per-core requirements, as these will be sensitive to choice of photo-z code and \gls{configuration} adopted.
%
%Describe how the outputs will be stored or visualized\\
Regarding visualization, qp includes some visualization tools but a more flexible and platform-independent solution would be desirable.

\paragraph{References for Further Reading} The qp \gls{software} package for PDF storage can be found at \url{https://github.com/LSSTDESC/qp}. For \gls{1D} PDFs,  \cite{malz2018approximating} is a good resource\footnote{Note that uses an older version of qp with additional though inefficient features.}.
The LSSTDESC RAIL \gls{software} package can be found at \url{https://github.com/LSSTDESC/RAIL}.
A Roadmap to Photometric Redshifts for the \gls{LSST} Object Catalog is available at \cite{DMTN-049}.

\pagebreak
\subsubsection{Photo-z: uncertainty quantification} \label{sec:Uncertainty}
\cleanedup{VAV}
\cleanedup{AIM}

%\WOM{I Did not put secrefs to the science cases}
\Contributors{Rachel Mandelbaum, Alex Malz, Raphael Shirley, Ashley Villar}
{3/29/22}

\paragraph{Abstract}
Most science cases that require a photo-z need some way to conveniently and quickly characterize the uncertainty in the photo-z.  A p(z) (whether it is a posterior, a likelihood or some other distribution) can be many bytes of information; therefore, summary statistics describing these distributions are essential.  For many science cases, we would like to answer questions such as:
\begin{itemize}
\item Is there a single ``peak” in the p(z), or is it multimodal?
\item What is the ``best-fit” redshift or the peak redshifts in case of multimodality?
\item What is the spread (second and higher moments)?
\item Is there any statistical property (those listed or another, such as percentiles) of the p(z) that correlates with galaxy type?
\end{itemize}
 Existing photo-z codes do not have a standardized way of reporting these user-defined summary statistics.  We note one solution, RAIL\footnote{\url{https://github.com/LSSTDESC/RAIL}}, which standardizes photo-z outputs such as p(z) and can compute various point-estimated quantities from the p(z) using qp\footnote{\url{https://github.com/LSSTDESC/qp}} as a back-end. RAIL does not necessarily calculate the quantities specified above, but it allows for easy implementation of such calculations. %utilizing qp, which I don't think needs to be specified here? -- I added it back because it's mentioned later in the section.

We also note that ``uncertainty" is a somewhat ambiguous term; what precisely are we quantifying? In this technical case, we only consider quantification of uncertainty \textit{intrinsic} to photo-z estimates, whether the cause is epistemic (e.g. nonrepresentative spectroscopic training data or incomplete template library) or aleatoric (e.g. insufficiently complex estimation algorithm).   Although a subtle difference, this is distinct from the question of the \textit{accuracy} of the redshift estimates compared to reality.%  As such, this use case does not require reference spectroscopic data nor an SED template library.

\paragraph{Science Cases Needing this Tool}
We enumerate a subset of science cases requiring photo-z uncertainty quantification:
\begin{itemize}
\item Extragalactic variable classification (\gls{AGN}): \secref{sec:FindAllAGN}
\item Extragalactic transient classification: \secref{sec:ImmediateClassification}
\item Cosmology (specifically, identification and characterization of tomographic samples might utilize uncertainty estimates): \secref{sec:cosmology}
\item Cosmology: cluster finders might use individual galaxy photo-z uncertainties when estimating cluster redshift and distinguishing cluster members from non-members sharing the line of sight: \secref{sec:cosmology}
\item Many static galaxy cases. For example, measuring the luminosity function of galaxies (especially as a function of redshift).: \secref{sec:GalMorphML}
\end{itemize}

\paragraph{Requirements for the software}
\begin{itemize}
\item Timing:
\begin{itemize}
\item Photo-zs must be produced for all of the galaxies (\textasciitilde 5 billion) at least as often as every data release.
\item A nightly (re)calculation using multiwavelength data is needed for the alert broker-filtered extragalactic alerts in a given night.
\item Estimates must be updated as external information becomes available (e.g., as the spectroscopy training set expands, or with the addition of new \gls{SED}s to template libraries).
\end{itemize}
\item Due to the number of galaxies, the software must be parallelizable.
\item The memory requirements are (relatively) low. Lower-dimensional summary statistics must be much smaller than the p(z) itself (whose possible parameterizations are discussed in Section~\ref{sec:PhotozRep}), and there is no significant temporary/intermediate storage need.
\item If storage parameterizations differ from usage parameterizations, software functionality for conversions must be provided.
\end{itemize}

Finally, we note that we can build these p(z) reduction tools today, or extend RAIL/qp to make them, and test on precursor datasets.

\paragraph{Running on LSST and other Datasets}

We require the following:

\begin{itemize}
\item As stated above, we expect to be provided with p(z) information and summary statistics for all galaxies in the LSST object catalogs as well as p(z) for nearby galaxies in the nightly alert stream\footnote{We note the p(z) information is considered proprietary and thus will not be in the public alert packet. Only the IDs for nearby Objects from the most recent DR will be included in the alert packet. Data-rights holders can use those to IDs query LSST databases for full p(z) information (via brokers or directly).}.
\item In the first year of data (when the ``official" p(z) will not be released), one could rely on a forced photometry measurement
\item Ancillary data is required, in the form of spectroscopic training sets and SED template libraries; multiple versions thereof must be available for characterization of photo-z estimators.
\item Before LSST, \gls{HSC} or \gls{DES} are helpful validation datasets.
\end{itemize}

\paragraph{Existing Tools}
\begin{itemize}
\item Most photo-z codes provide some characterization of uncertainty, however standardization of outputs is still missing.
\item qp supports many formulations of uncertainty (e.g. moments) and can calculate point estimates from p(z); it could be extended to the specific ones requested here.
\item  Through the ELAsTiCC data challenge, brokers are being provided with p(z) quantiles in the alerts for the purpose of classification, along with a tool for converting the quantiles into other parameterizations (e.g. evaluations on a grid).
\end{itemize}

\paragraph{Computational Workflow}
\begin{itemize}
\item Uncertainty quantification will involve accessing the photo-z PDFs for each individual extragalactic object in the object catalog, and doing some operations to the photo-z PDFs.
\item This can be done in an embarrassingly parallel fashion.
\item Results can be stored in a table.  Visualization would typically involve exploring their correlation with object properties (magnitudes, colors, etc.)
\end{itemize}

\paragraph{References for Further Reading}
\begin{itemize}

\item \cite{tanaka2018photometric}
\item \cite{malz2018approximating}
\item RAIL for estimating p(z) and stress-testing estimators: \url{https://github.com/LSSTDESC/RAIL}

\item qp for manipulating 1D PDFs: \url{https://github.com/LSSTDESC/qp}
\end{itemize}

\pagebreak
\subsubsection{Photometric redshifts: Science driven metrics} \label{sec:PhotozMetrics}
\Contributors{Alex Malz (\mail{aimalz@nyu.edu}), Colin Burke (\mail{colinjb2@illinois.edu}), Raphael Shirley (\mail{r.a.b.shirley@soton.ac.uk}), Andresa Campos (\mail{andresar@andrew.cmu.edu}), Christa Gall (\mail{christa.gall@nbi.ku.dk})}
{3/2/22}

\cleanedup{Colin Burke}
\cleanedup{AIM}

\paragraph{Abstract}
 We propose a photometric redshift \gls{metric} infrastructure targeted towards specific science cases that move beyond point estimate-based metrics. Derived population statistics (e.g. \gls{SMF} or luminosity function in a given redshift bin) should ideally have a performance \gls{metric} associated with the standard LSST photo-z data products, including p(z). We propose each science collaboration with an interest in photo-z submit metrics which can be applied to the p(z) and to point estimate outputs of general photo-z estimators. This will permit public data challenges in addition to the development of new algorithms involving additional band measurements, imaging, positions, and other ancillary data.

\begin{itemize}
\item We want to move beyond simple, presumed Gaussian, errors for quantifying \gls{photo-z} point estimates against spec-z measurements; instead, we want to apply metrics targeted to specific science cases (e.g. failure fractions for specific populations such as \gls{AGN}, catastrophic outlier fractions as a function of magnitude and color, and galaxy population statistics for large samples of \gls{photo-z}) that are sensitive to the quality of the photo-z uncertainty characterization.
\item A set of available metrics will demonstrate the applicability of the provided \gls{photo-z} information to a given science use case and be available for development and comparison of algorithms and input data options (e.g. additional bands, imaging, positions).
\end{itemize}

\paragraph{Science Cases Needing this Tool}
\begin{itemize}
\item Broad cosmological and extragalactic use cases including time-domain, \gls{transient}, \gls{AGN}, and galaxy science.
\item Single object studies, selection based on \gls{photo-z} derived values, and population studies all require targeted metrics.
\item Various cosmological probes that incorporate \gls{photo-z} information in different analysis stages, like weak lensing and galaxy clustering.
\end{itemize}

\paragraph{Requirements for the \gls{software}}
 The most important \gls{metric} theme that came out of the discussion was a notion of information; the accuracy of estimated uncertainties is not equivalent to the precision of an estimate. ``True uncertainties" derived from a forward model of synthetic data are essential for comparisons to estimated uncertainties that must guide algorithm development to achieve the goal of accurate uncertainty characterization; such true uncertainties are due to limitations of the data itself rather than the model and are thus dependent on the galaxy population of interest. Generating true uncertainties for mock data requires software to model the space of redshift and data (whether the data is just LSST photometry or also images, positions, and other sources of photometry).

 Such a model is only useful for metric evaluation if it includes ``realistic complexity" to which the metric's corresponding science case is sensitive, including selection effects and imperfect prior information, such as training sets and SED template libraries. The modeling should thus be flexible enough to include emission lines for \gls{AGN} and other specialized galaxy populations to ensure at least one photo-z estimator performs well on such galaxies. Rather than being strictly based on extant data, the model should be extensible potential systematic differences of the types we should expect to discover with LSST (such as galaxy subpopulations in thus far unpopulated regions of color space).

 For metrics that are only meaningful for specific subpopulations (such as \gls{AGN} and transient classes), the \gls{software} should be connected to subsampling or cross-matching \gls{software}. That same connection would also enable the integration of data from outside sources, such as Roman or Euclid photometry, in running an estimator on subsets of galaxies. The \gls{software} should then be able to save additional versions of the photo-z PDFs for those galaxies; not all galaxies in the catalog will have the same number of estimated photo-z data products, a complication the catalog's format must accommodate.

 Metrics needed:
\begin{itemize}
\item For transient and variable source classes (e.g. AGN), quality of match to external catalog of time-domain object identification and classification
\item For each subpopulation of interest, flag for confidence in method based on estimation algorithm's general performance on that subpopulation (e.g. \gls{AGN} will require trustworthy uncertainty but only evaluated on Y1 \gls{AGN}, by strength relative to host).
\item Ensemble metrics relative to a higher-fidelity (spectroscopic) sample.
\item Quantification of information content between multiple photometric redshift estimators (or same estimator with different priors, or additional bands, etc.)
\item Connection to external pipeline for science \gls{metric}, e.g. cosmology 3x2pt

\end{itemize}

However, ideally it should be possible to accommodate other key science-driven metrics.

\paragraph{Running on LSST and other Datasets}
\begin{itemize}
\item Spectroscopic training sets and/or SED template libraries will be necessary as prior information for photo-z estimators
\item Initially, photo-z estimation will be performed using only the Rubin photometric catalogs, though it is likely that ancillary data (see below) will become more important as the survey progresses.
\item Ancillary data where available, including imaging and/or more bands from other instrument(s) and/or positions to assess how additional bands contribute to \gls{photo-z} accuracy.
\item Deblended galaxy model parameters (e.g., S\'ersic parameters) for every galaxy (at scale for \gls{photo-z}) rather than image cutouts
\end{itemize}

\paragraph{Existing Tools}
 \gls{RAIL} includes an extensible emulation suite for obtaining true posteriors conditioned on photometry to compare to estimates.
 Further development would be necessary to interface with cross-matching/subsampling, to extend emulation to additional forms of data (imaging, more bands, positions, etc.), and to ensure realistic complexity for rare subpopulations.
 While the first of these may be straightforward, the latter two would require significant software development.

\gls{RAIL} also includes an extensible framework for metrics, to which science case-specific metrics would have to be added.

\paragraph{Computational Workflow}
\begin{itemize}
\item Forward modeling of data (photometry and ancillary) with corresponding true PDFs to compare to estimates
\item Training (or otherwise informing, for template-fitters) of estimators
\item Estimation of photo-z PDFs on synthetic forward-modeled data
\item Evaluation and comparison of metrics of point estimates and PDFs, including on specific subpopulations individually
\item Evaluation and comparison of metrics computed across sky using random (or binned, such as by \gls{AGN} strength relative to host) samples of objects
\end{itemize}

\paragraph{References for Further Reading}
\begin{itemize}
\item Using image information / \gls{CNN}: \citet{2019A&A...621A..26P}
\item \gls{AGN} photo-z: \citet{2019MNRAS.489..663B}
\item Galaxies photoz use case: \url{https://community.lsst.org/t/lor-the-galaxies-science-collaboration-photo-z-use-case/5887}
\item \gls{AGN} roadmap: \url{https://agn.science.lsst.org/sites/default/files/LSST_AGN_SC_Roadmap_v1p0.pdf}
\item RAIL: \url{https://github.com/LSSTDESC/RAIL}
\end{itemize}

\pagebreak

\subsection{Other technical use cases} \label{sec:othertech}
A few independently developed use cases were also submitted which did not fit in the broad areas outlined in \secref{sec:tdintro}.

\subsubsection{Joint Calibration of precursor surveys for longer-baseline Light Curve Generation} \label{sec:JointCal}

\Contributors{Weixiang Yu (editor; \mail{wy73@drexel.edu}), Colin J. Burke (\mail{colinjb2@illinois.edu}), K.E. Saavik Ford (\mail{sford@amnh.org})}
{03/25/22}

\cleanedup{Weixiang Yu}

\paragraph{Abstract}
Rubin \gls{LSST} is unprecedented in its unique combination of depth, spatial coverage, and time-domain capability. However, many science cases could not take full advantage of LSST’s  time-domain capability until later into the survey, because some classes of objects (e.g., \gls{AGN}, Mira, etc.) exhibit intrinsic long-term variability and a years-long (even decade-long) baseline is needed to classify and characterize them. Thus, we propose to jump start LSST variable sciences through joint calibration of current/past time-domain surveys with LSST and producing/serving forced-photometry light curves for variable LSST sources. Those re-calibrated archival light curves will not only enable early variable science for LSST sources, but also ensure a timely classification of variable sources into different classes (e.g., AGN vs. variable stars). A more complete \& less contaminated AGN catalog is critical for \gls{MMA} and \gls{CSQ} searches.

\paragraph{Science Cases Needing this Tool}
\begin{itemize}
\item Early classification of variable sources in LSST
\item Early variable science with LSST
\item Statistical studies of certain classes of objects where a long baseline is needed to produce a complete and pure sample.
\item Long-term structure function determination / AGN (non)-stationarity (\gls{Stripe 82})
\item Rapid identification of a cleaner/more complete sample of AGN for searching for MMA counterparts, changing-state’ quasars (CSQs) means identification of AGN-driven fraction of \gls{LIGO} sources, disk structure and astrophysics of turn on/off of accretion disks.
\end{itemize}

\paragraph{Requirements for the software}
\begin{itemize}
\item Careful cross-calibration of data from multiple surveys conducted with different hardware and at various sites.
\item Generate/store forced-photometry light curves from re-calibrated archival data at the locations of variable LSST sources
\item Extract time-series features from those light curves and store them in a database
\item Preferably run this analysis a few times throughout the 10-year LSST survey.
\item Visualization tools for serving those light curves to the end users through the RSP will be useful.
\end{itemize}

\paragraph{Running on LSST and other Datasets}
\begin{itemize}
\item The proposed work will only utilize data release catalogs. Ideally, the first run should be carried out once the variable nature of LSST sources can be reliably determined (primarily for saving computing resources).
\item We can test the pipeline now on precursor data: \gls{HSC} + \gls{BlackGEM}/\gls{DECam}/\gls{ZTF}/\gls{PTF}/\gls{Pan-STARRS}/\gls{SDSS}.
\item Because commissioning fields will be chosen for overlap with precursor surveys, this procedure will be possible on Data Preview 2 (the LSSTCam commissioning data release)
\item The data products resulting from running such a pipeline on precursor data sets will enable many new investigations that can benefit from a longer light curve baseline (certainly for AGN variability science).
\end{itemize}

\paragraph{Existing Tools}
\begin{itemize}
\item Light curves from different surveys can be merged by mean/median. However, without careful cross-calibration and color-term correction, the resulting light curves are not reliable, especially for sources exhibiting complex variability properties (non-stationarity on short time scales).
\end{itemize}

\paragraph{Computational Workflow}
\begin{itemize}
\item Cross-calibrate archival data from precursor surveys with LSST.
\item Run forced-photometry on re-calibrated archival single-epoch images at the locations of variable LSST sources, color-correct the photometry, and store those light curves on disk for later retrieval.
\item Extract time-series features from those light curves and store them in a database.
\end{itemize}

\paragraph{References for Further Reading}
Longer baseline justification: \cite{Kozlowski2017, Kozlowski2021}\\
Photometric Calibration of PTF: \cite{Ofek2012}\\
Photometric Calibration of ZTF: \url{https://irsa.ipac.caltech.edu/data/ZTF/docs/ztf_extended_cautionary_notes.pdf}\\
Photometric Calibration of Pan-STARRS: \cite{Schlafly2012}\\
Photometric Calibration of \gls{DES} and Rubin LSST: \cite{Burke2018}

\pagebreak
\subsubsection{Multi-Stream Transient Detection and Characterisation} \label{sec:MultiStreamTransients}
\WOM{Missing workflow and refs - seems in between tech and science}
\cleanedup{Tyler Pritchard}

\Contributors{Tyler Pritchard (\mail{TylerAPritchard@gmail.com}), Alex Gagliano,  Samuel Wyatt, Igor Andreoni, Tomas Ahumada, Catarina Alves, Christa Gall, Suvi Gezari, Jing Lu, Fabio Ragosta, Clare Saunders, Adam Scott, Ashley Villar, Sam Wyatt, Ann Zabludoff}
{03/28/22}

\paragraph{Abstract}
\gls{LSST} will discover millions of variable and transient events per night; the characterization of most of these objects will occur on longer timescales.  With a likely single filter detection and inter-night revisit timescales of \textasciitilde3--5 days, it will take multiple nights to characterize any individual \gls{transient} with only \gls{LSST} data.  Individual groups will have their own follow-up resources; the optimal allocation of these facilities remains an open problem. One tool that could enable more prompt characterization, and help groups efficiently target their follow-up, is the correlation of multiple streams of public time-domain information on a single platform that would allow humans and/or algorithms to select targets and assign resources.

For example, one vision of this could be a broker-like interface with shared data from \gls{LSST}, \gls{ZTF}, \gls{ASAS-SN}, and Gaia. A more comprehensive version  could include publicly reported spectra that are correlated with known alerts (such as those provided by the \gls{TNS}).   This would result in products such as early estimated colors, if for example an \gls{LSST} detection in a single filter is supplemented by detections in \gls{ZTF}, \gls{ASAS-SN}, or Gaia.   Or, rather than different filters, multiple observations in a similar filter between LSST and another survey would result in an estimated magnitude difference between the two observations, and therefore brightness rate-of-change, before LSST or any single survey would be able to provide it.    This could be useful across a range of science cases including the detection of very young or fast evolving explosive transients, X-ray binaries transitioning into an outburst, or changing look \gls{AGN}.

Current limitations are primarily driven by the data volume and  velocity of any single stream, as well as interface issues with a number of (theoretically at least) public data. One current solution that members of the \gls{transient} community currently use is to create their own jupyter notebooks that connect to multiple data servers. This solution faces several technical issues including: queue-rate limits, difficulty of creating queries across different formats that are sufficiently cross-matched, local storage and compute, and a significant amount of custom pre-processing. This is difficult to scale today, and will become increasingly difficult without a unified platform as \gls{LSST} comes online. Similar work is being done today in curated transient surveys, with less efficient tools or data releases, such as  the  Young Supernovae Experiment (the special \gls{ZTF} follow-up of the \gls{TESS} survey region with \gls{Pan-STARRS}),the Global Supernova Project (an LCO follow-up of \gls{ZTF} detections with an open/free-to-join community), Gaia alerts and third party spectroscopy made available on, e.g, the \gls{TNS}. This need will only grow in the future as more overlapping surveys are planned -including the search for transients in DESI and the Roman Observatory, future proposed rapid-transient DECam surveys in the era of the Rubin Observatory \gls{LSST}.

This shares overlap with the multiple-instrument algorithms, but is focused on derived-products such as light curves and \gls{photo-z}’s (and potentially additional \gls{metadata} such as classification outputs).

\paragraph{Science Cases Needing this Tool}

Example science cases that would benefit from this tool include:
\begin{itemize}
\item Anomaly Detection \& Follow-up  -- things can be anomalous along multiple axis, and follow-up is often required as early as possible
\item X-ray Binary Outbursts -- the \gls{shape} and timing of the outburst provides information about the mass and size of the binary system.
\item Fast Evolving Transients \& Early Supernovae -- \gls{LSST} will provide a deluge of newly discovered transients. Identifying the select few that merit follow-up observations will prove challenging, especially as the information from \gls{LSST} will be limited. With a latency of several days between repeated observations in the same filter, it will be difficult to acquire follow-up early in the evolution of newly discovered \gls{LSST} transients.
\item Changing-look \gls{AGN} -- these systems evolve between observational \gls{AGN} types on rapid timescales.
\item Quantifying sample purity, contamination, completeness for e.g., SN\,Ia cosmology.
\end{itemize}

We also note that this tool would be useful for ``archival'' studies, particularly the development of multiple-instrument \gls{ML} algorithms.

\paragraph{Requirements for the \gls{software}}
The requirements for such a tool share many similarities with that of an \gls{LSST} broker, but emphasize the combination of multiple streams and the need for a rapid response.
\begin{itemize}
    \item The ability to combine multiple kafka or \gls{LSST}-alerts streams from publicly available sources (including multi-wavelength or multi-messenger surveys). This could be across the entire survey footprint for multiple streams, or constrained to a well-defined overlap between two streams for a more narrow use case.
    \item Ideally, the ability for \textit{relatively small} surveys or observing programs to easily contribute data, potentially for something like known events, post-detection.
    \item The ability to search for and filter off of cross-matched alert sources - either from each individual stream or in the combined data-set.
    \item The ability to cross-match or query other catalog or annotated information to enable prompt decision making.
    \item Rapid publication of the cross-matched streams for greatest impact.
    \item The biggest challenge is primarily the volume and velocity of data -- while the \gls{LSST} alert stream will dominate the volume of the data, other sources will expand the scope of what is stored significantly. If this can be done in a timely manner, it will help inform the community develop plans for follow-up.
    \item While more challenging, it would be good for the service to additionally annotate data by providing host galaxy cross-matching from catalogs (e.g., Glade or GWGC3) or spectra that get publicly reported.
\end{itemize}
\paragraph{Running on \gls{LSST} and other Datasets}
\begin{itemize}
\item This would be focused on the LSST \gls{Alert}-stream and could begin immediately upon survey start.
\item The LSST \gls{Alert} stream data can then be combined with both extant and future public time-domain data  (e.g., \gls{ZTF}, \gls{ASAS-SN}, Gaia, future DECam Surveys, Roman, other public data).
\item Scaled down test surveys are productive and being conducted now (e.g., the Young Supernovae Experiment), and in a more limited sense with other surveys incorporating ZTF data into their transient detection/characterisation methods (e.g., DLT40, \gls{ATLAS}).
\end{itemize}
\paragraph{Existing Tools}
The tools needed largerly exist as the functionality overlaps with the needs of an \gls{LSST} \gls{Alert} broker, including:
\begin{itemize}
    \item Big Data storage (e.g., postgresql/NoSQL/cloud buckets)
    \item Methods to rapidly and efficiently stream data (e.g., Kafka, pub-sub systems, and shared \gls{cloud} buckets)
    \item \gls{HTTP} \gls{API} access with potential web visualization interface
\end{itemize}
What's missing is the expansion and publication of this to multiple combined data streams, such as:
\begin{itemize}
    \item Methods to efficiently cross-match positional data between different alert streams \gls{ZTF}, \gls{ASAS-SN}, and others (e.g., postgris \gls{W3C} spatial tools \& ToPCAT), AXS (Astronomy Extensions for Spark).
    \item For events with large positional uncertainties and time-critical follow-up, methods to publish observation details for community coordination such as \url{treasuremap.space}.
    \item Methods for users to filter off of stream data including personal jupyter notebooks doing small scale processing on curated streams, Target Opportunity Managers (\gls{TOM}), and broker filter interfaces.
\end{itemize}
\paragraph{Computational Workflow}
This will vary depending upon the chosen methodology, number of surveys or data streams, scientific interest, and wavelength(s),  but one workflow for a minimal implementation for a specific science case could be:
\begin{itemize}
    \item Identify a target list.  For extra-galactic transients, for example, this could be the \gls{TNS}, for variable stars the AAVSO Variable Star Index, for AGN a joint Gaia-WISE curated list, or a \gls{LIGO} or IceCube alert stream.
    \item Query photometry/alert servers for data upon the publication of a new event (e.g., \gls{TNS}, \gls{LSST}/\gls{ZTF} brokers, \gls{ASAS-SN} Sky Patrol, Gaia.
    \item In the case of events with open data but no published photometry stream (e.g., Swift, Fermi), the stream could be annotated to indicate where observations exist even if detections/limits are not available.
    \item (optional) A feedback mechanism for vetted groups without a dedicated photometry stream to provide data on events. This could be potentially supplementary (e.g. spectra, a redshift, classifications or periods) or additional photometry.
    \item A re-publication (on-line or via api) of the \gls{Alert} stream with the additional observation or annotated list of observations).
\end{itemize}
A larger, more science agnostic case could be:
\begin{itemize}
    \item Take a curated Kafka stream from an \gls{LSST} broker focusing on non-\gls{SSO} transient events and a second stream from a second survey (potentially in a defined sub-survey region, e.g., \gls{DDF} or overlap with a DECam Survey)
    \item Spatially cross-match the two streams
    \item (Optional) Add additional streams if available
    \item (Optional) Add annotated information including spectra, epochs of observations by other telescopes/wavelengths (potentially archival/long baseline), classifications, periods, and other information.
    \item Publish a joint stream with the combined data to an end user, either through a broker, \gls{TOM}, or pub-sub channel.
\end{itemize}
This could be scaled up depending on overlapping observation width, survey period, and number of streams.
\paragraph{References for Further Reading}
Brokers, Tools, Sources mentioned above:
\href{alerce.online}{Alerce}
\href{https://antares.noao.edu/}{Antares}
\href{https://fink-broker.org/}{Fink},
\href{https://lasair.roe.ac.uk/}{Lasair},
\href{https://www.wis-tns.org/}{Transient Name Server},
\href{https://www.aavso.org/vsx/}{AAVSO Variable Star Index},
\href{http://gsaweb.ast.cam.ac.uk/alerts/home}{Gaia Alerts},
\href{https://yse.ucsc.edu/}{Young Supernova Experiment},
\href{https://asas-sn.osu.edu/}{ASAS-SN SkyPatrol},
\href{https://supernova.exchange/public/}{Global Supernovae Project Supernova Exchange},
\href{treasuremap.space}{Treasure Map},
\href{https://lco.global/tomtoolkit/}{TOM Toolkit},
\href{https://github.com/astronomy-commons/axs}{Astronomy Extensions for Spark}
\\

\pagebreak
\subsubsection{Interactive Data Visualization at scale} \label{sec:VizScale}

\Contributors{Leanne Guy (\mail{leanne.guy@lsst.org}), Ilija Medan (\mail{imedan1@gsu.edu}), Jing Lu (\mail{jl16x@my.fsu.edu}), Tomislav Jurkic (\mail{tjurkic@phy.uniri.hr}), Juan Luna (\mail{jmluna@iafe.uba.ar}), Tomas Ahumada (\mail{tahumada@astro.umd.edu}), Rosaria (Sara) Bonito (\mail{rosaria.bonito@inaf.it}), Sabina Ustamujic (\mail{sabina.ustamujic@inaf.it}), Markus Hundertmark (\mail{markus.hundertmark@uni-heidelberg.de}), Yiannis Tsapras (\mail{ytsapras@ari.uni-heidelberg.de}), Matthew Graham, Neven Caplar, Ashish Mahabal, Mark Popinchalk, Viviana Acquaviva, Catarina S.~Alves, Garrett Levine}
{2022-03-30}

\paragraph{Abstract}

 At the end of the 10-year LSST, the size of the final Object Catalog is expected to be approximately 15PB in size and contain approximately 40 billion Objects. Developing a framework for automated interactive visualisation of the LSST dataset is key to discovery.  The classic technique of subsetting the data down to a manageable size that will fit into memory, by defining cuts on various object attributes and then visualising the resultant data, limits discovery potential by reducing the dataset based on assumptions about the data. Creating visualisations based on the full dataset will allow scientists to explore the full LSST discovery space interactively. All LSST science domains can benefit from visualisation that enables interactive exploration, subsetting, drilldown, and brushing and linking between plots. In 2016, the Gaia \gls{DR1} density map of over 1 billion sources in the Milky Way was named “One of the five coolest things on Earth this week” by General Electric. Many industry-standard tools exist already that we can take advantage of for creating powerful visualisations of peta-scale datasets, e.g Holoviz (\url{https://holoviz.org/}).
 LSST can benefit enormously by building upon these frameworks and tools.

\paragraph{Science Cases Needing this Tool}
All science domains and use cases can benefit from powerful interactive visualisations. Some notable examples include:
\begin{itemize}
\item Visualizing the near real-time light curves generated from custom cutouts of images. This will be useful for a variety of extended objects, e.g. galaxies, solar system objects, young stellar objects, as well as \gls{transient} events such as gravitational microlensing events, SN, and GRBs in particular in crowded fields.
\item Studying accretion; rapid variability on timescales from minutes to hours or days  is a hallmark of accreting systems, from compact binaries to \gls{AGN}s and young stellar objects (YSOs). The time scale and amplitude of the variability provide information about flares, rotation, and the accretion process including its rate and power. Being able to obtain quick on-the-fly measurements of the variability of a source, such as light curve variance, would enable rapid identification. The nature of the variable sources could then be further narrowed-down with a (linked) visualisation of their location on color-color or color magnitude diagrams.
\item Anomaly detection for young/unknown sources.
\item Density maps.
\item Investigating the attributes of points in linked plots in order to \gls{drill down} to images, which could be retrieved dynamically via the cutout server, in order to perform further analyses (e.g. finding and relating host galaxies to transients and to visualize them in the LSST images).
\item Providing multi-level linked selection of aggregated data to look for correlations between parameters,  e.g  computed scientific parameters or latent parameters following dimension reduction. An example of the progression of levels could be “Sample subsets -> correlation heat map of parameters -> individual x-y plot”, where these can be combined with \gls{drill down} to zoom in on the individual target.
\item On-the-fly calculation of basic statistics or other quantities from selected areas on plots, e.g computing the mean and median of a selected a region on spatial distribution plot.
\item \gls{3D} rendering of select Objects, such as YSOs, whose light curves exhibit varying shapes, e.g. a dip due to warp disks or modulation due to rotation or variability associated with accretion/ejection processes, using external tools such as paraview for data analysis or the sketchfab platform to share models interactively. The \gls{3D} rendering will allow us to explore different line-of-sights and to understand how geometric effects change the observed light curves. The TVSSC is starting a program on \gls{3D} visualization.
\item Mapping a \gls{WCS} onto a plot.
\item Producing interactive dashboards that enable exploration of the parameter space of a given class of object. For example, tweaking \gls{SN} Ia parameters to fit light-curves.
\item Providing an interface to interactively perform period finding and phase folding for light curves. For example, a slider bar for changing the period that would adaptively phase fold the light curve to that period.
\item Provide easy to use Observing Program Management Systems to assist with managing and tracking active observing programs. The TVSSC has solicited an in-kind contribution for this purpose.
\end{itemize}

\paragraph{Requirements for the \gls{software}}
 There are many excellent open-source tools on the market for producing stunning visualizations. We should not think about writing a visualization tool, rather we should adopt an existing tool or a framework and write software to provide an interface between these industry standard tools and the LSST data products. Producing visualizations of large datasets may require additional \gls{CPU} or GPU or the use of a tool like Dask to compute the quantities to visualize. This approach provides a highly optimized rendering pipeline that makes it practical to work with extremely large datasets even on standard hardware, while exploiting distributed and GPU systems when available.

 There is no reason why we cannot start to build these tools today. The biggest impediment is likely to be that people are not used to this manner of working and will require training. This requires something of a change of culture.

 Requirements:
\begin{itemize}
\item Rapidly create visualizations of different classifications of objects, e.g stars vs. galaxies.
\item Link plots that can present different representations of the same data.
\item Brushing and linking between different visualizations of the same data points.
\item Speed for rapid visualization.
\item Good annotation of data across multiple surveys to allow for plotting/visualisation of data on an object from all of these surveys.
\item Drilldown capabilities (images/spectra from other surveys by cross-matching)
\item Integration with external tools (when they can be installed on the science platform), e.g creating \gls{3D} models.
\item Able to integrate new \gls{software} that can calculate needed derived quantities (e.g. statistics on data, representation in healpix) relevant to science use cases
\item Adding domain specific knowledge (i.e. overlaying WCS) onto industry specific \gls{software}
\item Ability to transform data from \gls{LSST} tables/parquet files into a data format is needed by the visualization tool.
\item Ability to automatically and accurately render billions of objects (e.g Datashader) rapidly and flexibly including annotations and interactive capabilities.
\item Need to produce static images from final plots to include in papers
\item Any third-party \gls{software} that we use should be open-source.
\item Rendering billion point datasets could require large processing power or \gls{GPU}
\item Traditionally people will store a pdf/png of the final plot. This means that to recreate the plot (say in order to change colours or point size), all the computations need to be redone to recreate the plot. This can be time consuming. It may be  preferable to store the aggregated data resulting from the computations that go into making the plot. This way the plot can easily be recreated to change plot attributes.
\item Ability to align lightcurve data of a single target from different sources on a visualization.
\item Ability to interactively mask datapoints
\item Include options/capabilities for those that are e.g. visually impaired.
\begin{itemize}
\item Developing \gls{3D} printed kits to include visually impaired scientists and to be used for scientific dissemination.
\item Data sonification
\end{itemize}
\end{itemize}

\paragraph{Running on \gls{LSST} and other Datasets}
Scientists will need to visualize the following data products:
\begin{itemize}
\item Object catalog and \gls{ForcedSource} catalogs to visualize time series and associate object attributes
\item Full visit images or cutouts of the \gls{LSST} images obtained via the cutout server
\item \gls{LSST} data combined with data from various external catalogs in a single visualization. Note that data access might require the use of queries over TAP and of the services of a cross-match service.
\item Precursor and simulated datasets such as  DES DRX, HSC PDRX, DESC DC2. can serve to develop and validate visualization tools in preparation for LSST data and possibly lead to publications if advanced visualization of these datasets leads to new discovery. The \gls{RSP} provides an ideal environment to run on and develop visualizations.
\item Objects with selections based on \gls{HEALPix} maps
\end{itemize}

\paragraph{Existing Tools}
Existing tools for processing/visualising data include:
\begin{itemize}
\item lightkurve (inherits partially from Astropy’s TImeSeries)
\item The Holoviz suite of tools including Bokeh, Holoviews and datashader and panel.
\item Altair to link plots
\item Treasuremap
\item Plotly for dashboards
\item Paraview\footnote{\url{https://www.paraview.org}} data analysis and visualization application. It can be run on supercomputers to analyse extremely large datasets also with Python\footnote{\url{https://www.paraview.org/python}}
\end{itemize}
Many of these tools are used in industry to visualize large datasets and are expected to work at LSST scale. Holoviz has been demonstrated on the DESC \gls{DC2} dataset via Rubin DP0. What is needed is to provide the interface layer to the LSST data products.

\paragraph{Computational Workflow}
\begin{itemize}
\item Input data to a visualization framework would be either a) directly from \gls{parquet} files provided by the project, b) from results of ADQL queries of LSST databases, or c) images retrieved either via the Butler or cutout server. Additionally, it would be useful to work with community alert brokers to visualize data in alert packets.
\item Steps include: Data preparation, including extracting data/parameters from processed images, parametric fitting, classification, dimension reduction, joining with results of cross-matching,
\item We would expect to be able to use the project provided batch and reprocessing tools for any reprocessing.
\item Creating density maps from the full catalog data  will require distributed processing frameworks such as Apache spark or Dask to aggregate, bin or compute other derived quantities for visualization.
\item Outputs could be stored in user databases to avoid recomputation if computation cost is high.

\end{itemize}

\paragraph{References for Further Reading}
\begin{itemize}
\item Astropy \citep{2013A&A...558A..33A, 2018AJ....156..123A}
\item Holoviz\footnote{\url{https://discourse.holoviz.org/}}
\item \gls{Firefly}\footnote{\url{https://github.com/Caltech-IPAC/firefly}}
\item Matplotlib  \citep{matplotlib}
\item Lightkurve \citep{2018ascl.soft12013L}
\item Altair\footnote{\url{https://altair-viz.github.io/index.html}}
\item TreasureMap \citep{2020ApJ...894..127W}
\item yt \citep{2011ApJS..192....9T}
\item GlueViz \citep{2015ASPC..495..101B, robitaille_thomas_2017_1237692}
\item Houdini\footnote{\url{http://www.ytini.com/}} \citep[e.g.][]{2017PASP..129e8008N}
\item TOPCAT \citep{2005ASPC..347...29T}

\end{itemize}

\pagebreak

\section{Scenarios used for the inclusive collaboration breakouts} \label{sec:DeiScenarios}

%\chbox{This section gives the breakout scenarios for the \gls{DEI} sessions.}

The following four scenarios were used in the breakout sessions focused on inclusive collaboration. Participants were deliberately split into diverse groups.
Each group did a round table of introductions, then read two of the scenarios and discussed them.

The goal was stated as ``People come to these discussions with their own experiences and backgrounds. There may be some who have experiences very similar to the scenarios. Today's goal is to focus on the scenarios and to identify productive strategies based on the prompts that we all could potentially use to foster positive collaborations and avoid repeating prior mistakes.''

\subsection{Scenario 1 – Institutional Pressures} \label{sec:tutePressures}
Zahra is part of a large research team at an \gls{R1} (research intensive) institution developing a Research Inclusion plan for their research proposal. The team approaches Carl, a teaching institution astronomer that Zahra knows from a past \gls{AAS} meeting to join the team. Zahra thought Carl would be a good fit for the team because his dissertation research was on a similar topic as the topic they will study should their proposal get accepted.
Zahra emails Carl asking if he is interested in joining their team. Carl realizes this could be a good opportunity, as he is expected to publish (albeit minimally relative to an \gls{R1} institution) to qualify for tenure. However, Carl is apprehensive about joining the team because he doesn’t have much experience in large collaborations. He is also worried about the different institutional pressures they face as they work for different types of institutions.

{\bfseries Questions:}
\begin{enumerate}
\item What are some different pressures that researchers from research-intensive and teaching institutions may face?
\item What are some steps that Zahra’s collaborative team can take to make Carl’s participation on the team valuable for him?
\item What conversations could Zahra and Carl have during these early stages to better understand different institutional contexts, collaboration expectations, and collaborator capacities?
\item What questions might Zahra or Carl ask each other to begin laying a foundation for a successful collaboration?
\item Have you ever participated in a collaboration that spans multiple institutions of different sizes or types? What worked well in those collaborations? What was challenging?  If tensions arose based on different institution types, how were these tensions settled or resolved (if at all)?

\end{enumerate}

\subsection{Scenario 2 – Allocation of Credit} \label{sec:allocCredit}

A collaboration centered at a few large research institutions approaches Sarah, an assistant professor of astronomy at a teaching institution, to ask if she is interested in collaborating on a  proposal. The team explains that Sarah would be a good fit because of her expertise, and her involvement in the collaboration would contribute to the institutional diversity of the project as well. Sarah’s involvement would become a central pillar of a required research inclusion component to the project, which would make the proposal more competitive.
Sarah thinks this could be a great opportunity because she needs to publish more articles before going up for tenure. She knows that research collaboration tends to be more highly cited, more visible, more innovative, and more likely to have a greater impact than sole authored research. Funders often view collaboration more favorably than sole-investigator research as well. This opportunity could really strengthen her tenure file.
However, Sarah is also worried that it will be more difficult to get credit for her component of the work on such a large project.  Sarah’s hesitation is exacerbated when in preliminary discussions the team plans authorship on their first planned publication; a postdoc at the research-intensive university with similar expertise as Sarah is slated to be listed above her in authorship despite both of them being equally important to the development of the manuscript.

{\bfseries Questions:}
\begin{enumerate}
\item How and when do you typically navigate authorship on research publications? What strategies do you have for managing tensions that may arise around authorship?
\item Why might the team in the example above assume the postdoc would be higher in authorship order than Sarah even if their contributions are roughly equal?
\item What are some ways the team could ensure the collaboration will be meaningful for Sarah? What may be meaningful for Sarah’s type of institution? What types of conversations could the team have to illuminate what is meaningful for each collaborator, including Sarah, given their institutional contexts?
\item What are some ways the team could ensure collaborators are properly credited for their contributions?
\item What are some useful team publication conversations to have during this “inviting collaborator” stage? What issues might they address?
\item Have you ever had a negative research experience that arose due to the allocation of credit on a collaborative paper? If so, what lessons did you learn from that experience?

\end{enumerate}

\subsection{Scenario 3 – Inclusive Team Environment} \label{sec:inclusiveTeam}

Wei, an astrophysicist, is leading the development of a research proposal that (if accepted) would span research and teaching institutions, and consist of team members from different career stages (e.g. associate professors, assistant professors, postdocs, graduate students). The proposal requires a Research Inclusion plan alongside the Science and \gls{Data Management} plans. In drafting the Research Inclusion plan, Wei is prompted to describe how the team can foster an inclusive environment in which all team members feel respected and valued for their unique contributions to the project.

{\bfseries Questions:}
\begin{enumerate}
\item What are some examples of team dynamics that may foster an environment that is isolating to, or excludes, some members of the team?
\item What are some strategies that Wei and other team members could employ to foster an inclusive environment in which researchers of different ranks feel valued on the project? What are effective ways to implement these strategies?
\item What are some strategies that Wei and other team members could employ to foster an inclusive environment in which researchers from different types of institutions feel valued on the project? What are effective ways to implement these strategies?
\item Have you ever been on a team (or part of a department) in which the environment was not inclusive of everyone?  How were people excluded or made to feel like they were not fully integrated into that space? What were the impacts of the exclusionary environment? In what ways were these negative dynamics addressed (if at all)?
\end{enumerate}

\subsection{Scenario 4 – Student Contributions to Open-Source Software} \label{sec:studentOS}

An astronomer working at a teaching-intensive institution, named Ahmed, is collaborating with a team of astronomers from large, research-intensive institutions to build some open-source \gls{software} for data reduction. As part of this work, Ahmed wants to include some undergraduates from his institution on the project, as the project would provide these undergraduates with research and coding experiences typically unavailable to them. Including undergraduates would also benefit Ahmed; the administrators at Ahmed's institution reward this type of student engagement in tenure decisions.

{\bfseries Questions:}
\begin{enumerate}
\item What advice would you give those organizing the \gls{software} development effort to provide a clear and welcoming path to involvement from new contributors such as Ahmed's students?  For example, this might involve providing simple explanatory references, reducing use of jargon in the documentation and clearly defining it when it appears, or other measures.
\item How can Ahmed and the entire team ensure that students who invest significant short-term effort into developing open source \gls{software} as part of this code base get credit for their contributions?
\item How does the community ascribe credit for widely used, open-source software? What author-order considerations should be taken into account for software development credit? How does software development credit differ from other authorship credit scenarios? What about open-source software development?
\item When does writing a piece of open-source software convey authorship? Are there processes we can put in place to help guide this decision? How does software infrastructure (e.g. running software at scale) fit into this picture?

\end{enumerate}

\addcontentsline{toc}{section}{Glossary}
\printglossaries

\end{document}